\documentclass[11pt]{article}
\usepackage{amsmath,amsfonts,amssymb,graphicx,epsfig,pdflscape,multirow,theorem}
\usepackage{arydshln,cite}
\numberwithin{equation}{section}
\graphicspath{{./eps/}}
\usepackage{cite}
\usepackage[usenames]{color}
\usepackage{pstricks}
\usepackage[backref=false]{hyperref}
\hypersetup{
colorlinks=true,
citecolor=red,
linkcolor=darkblue
}
\definecolor{darkblue}{rgb}{0,0,.8}
\definecolor{red}{rgb}{1,0,0}
 
\long\def\ignore#1{}
\definecolor{purple}{rgb}{1,0,1}
\definecolor{coloroflink}{rgb}{0.7,0,1}
\definecolor{darkpurple}{rgb}{1,.2,1}
\definecolor{pink}{rgb}{1,.7,.7}
\def\blue#1{{\color{blue}#1}}

\theorembodyfont{\itshape} 
\theoremheaderfont{\scshape}
\theoremstyle{plain}  
\newtheorem{Lemme}{Lemma}[section]
\newtheorem{Theoreme}[Lemme]{Theorem}
\newtheorem{Proposition}[Lemme]{Proposition}
\newtheorem{Lemma}[Lemme]{Lemma}
\newtheorem{Corollaire}[Lemme]{Corollary}

\numberwithin{equation}{section}

\newcommand{\nc}{\newcommand}
\nc\disp{\displaystyle}
\nc{\fh}{\hat{f}}
\nc{\muh}{\hat{\mu}}
\nc{\nuh}{\hat{\nu}}
\nc{\spos}[2]{\makebox(0,0)[#1]{$\sm{#2}$}}
\nc{\sm}[1]{{\scriptstyle #1}}
\nc{\qbar}{\overline{q}}
\nc{\bib}{\bibitem}
\nc{\al}{\alpha}
\nc{\g}{\gamma}
\nc{\G}{\Gamma}
\nc{\D}{\Delta}
\nc{\eps}{\epsilon}
\nc{\la}{\lambda}
\nc{\La}{\Lambda}
\nc{\var}{\varphi}
\nc{\pa}{\partial}
\nc{\nn}{\nonumber \\ }
\nc{\hf}{\frac{1}{2}}
\nc{\dz}{\frac{dz}{2\pi i}}
\nc{\bin}[2]{\left(\!\!\!\begin{array}{c} {#1}\\ {#2} \end{array}\!\!\!\right)}
\nc{\be}{\begin{equation}}
\nc{\ee}{\end{equation}}
\nc{\bea}{\begin{eqnarray}}
\nc{\eea}{\end{eqnarray}}
\nc{\bra}[1]{\langle {#1}|}
\nc{\ket}[1]{|{#1}\rangle}
\nc{\ketw}[1]{({#1})^{\phantom{a}}_{{\cal W}}}
\nc{\chit}{\raisebox{0.25ex}{$\chi$}}
\nc{\chih}{\raisebox{0.25ex}{$\hat\chi$}}
\nc{\ir}{\mathrm{i}}
\nc{\Dbh}{\mbox{\boldmath $\hat D$}}
\nc{\Dbb}{\mbox{\boldmath $\bar D$}}
\nc{\Ibb}{\mbox{\boldmath $\bar I$}}
\nc{\Dbm}{\mbox{\boldmath $\mathcal D$}}
\nc{\db}{\mbox{\boldmath $d$}}
\nc{\Ab}{\mbox{\boldmath $A$}}
\nc{\Bb}{\mbox{\boldmath $B$}}
\nc{\Cb}{\mbox{\boldmath $C$}}
\nc{\Db}{\mbox{\boldmath $D$}}
\nc{\Dbt}{\mbox{\boldmath $\tilde{D}$}}
\nc{\Fb}{\mbox{\boldmath $F$}}
\nc{\Fbt}{\mbox{\boldmath $\tilde{F}$}}
\nc{\fb}{\mbox{\boldmath $f$}}
\nc{\fbt}{\mbox{\boldmath $\tilde{f}$}}
\nc{\Gb}{\mbox{\boldmath $G$}}
\nc{\Jb}{\mbox{\boldmath $J$}}
\nc{\Kb}{\mbox{\boldmath $K$}}
\nc{\Mb}{\mbox{\boldmath $M$}}
\nc{\Pb}{\mbox{\boldmath $P$}}
\nc{\Qb}{\mbox{\boldmath $Q$}}
\nc{\Tb}{\mbox{\boldmath $T$}}
\nc{\Tbb}{\mbox{\boldmath $\bar T$}}
\nc{\Tbm}{\mbox{\boldmath $\mathcal T$}}
\nc{\Tbt}{\mbox{\boldmath $\tilde{T}$}}
\nc{\tb}{\mbox{\boldmath $t$}}
\nc{\Ub}{\mbox{\boldmath $U$}}
\nc{\Vb}{\mbox{\boldmath $V$}}
\nc{\Wb}{\mbox{\boldmath $W$}}
\nc{\Xb}{\mbox{\boldmath $X$}}
\nc{\yb}{\mbox{\boldmath $y$}}
\nc{\Zb}{\mbox{\boldmath $Z$}}
\nc{\Hb}{\mbox{\boldmath $H$}}
\nc{\calH}{{\cal H}}
\nc{\calR}{{\cal R}}
\nc{\calL}{{\cal L}}
\nc{\calV}{{\cal V}}
\nc{\Hc}{{\cal H}}
\nc{\Rc}{{\cal R}}
\nc{\Lc}{{\cal L}}
\nc{\Vc}{{\cal V}}
\nc{\Ib}{\mbox{\boldmath $I$}}
\nc{\qb}{\bar{q}}
\nc{\Ac}{\mathcal{A}}
\nc{\Bc}{\mathcal{B}}
\nc{\Cc}{\mathcal{C}}
\nc{\Dc}{\mathcal{D}}
\nc{\Ec}{\mathcal{E}}
\nc{\Gc}{\mathcal{G}}
\nc{\Ic}{\mathcal{I}}
\nc{\Jc}{\mathcal{J}}
\nc{\Oc}{\mathcal{O}}
\nc{\Pc}{\mathcal{P}}
\nc{\Sc}{\mathcal{S}}
\nc{\Tc}{\mathcal{T}}
\nc{\Wc}{\mathcal{W}}
\nc{\Xc}{\mathcal{X}}
\nc{\Yc}{\mathcal{Y}}
\nc{\Zc}{\mathcal{Z}}
\nc{\fus}{\mbox{}\,\hat\otimes\,\mbox{}}
\nc{\Pch}{\hat{\Pc}}
\nc{\Rch}{\hat{\Rc}}
\nc{\Dh}{\hat{\Delta}}
\nc{\rh}{\hat{r}}
\nc{\sh}{\hat{s}}
\nc{\kb}{\bar{k}}
\nc{\taub}{\bar{\tau}}
\nc{\Jcb}{\Jc_{\mathrm{b}}}
\nc{\rtt}{\mathtt{r}}
\nc{\stt}{\mathtt{s}}
\nc{\cosR}{\cos\frac{\pi p'rr'}{p}}
\nc{\cosS}{\cos\frac{\pi pss'}{p'}}
\nc{\sinR}{\sin\frac{\pi p'rr'}{p}}
\nc{\sinS}{\sin\frac{\pi pss'}{p'}}
\nc{\hs}{\hspace{-0.08cm}}
\def\vvdots{\mathinner{\mkern1mu\raise1pt\vbox{\kern7pt\hbox{.}}\mkern2mu
  \raise4pt\hbox{.}\mkern2mu\raise7pt\hbox{.}\mkern1mu}}
\def \st#1{\raisebox{-6pt}{\rule{0pt}{18pt}}\makebox[16pt]{\small ${#1}$}}
\nc{\gauss}[2]{\left[\!\!\begin{array}{c} {#1}\\ {#2} \end{array}\!\!\right]}
\nc{\sbin}[2]{\left\{\!\!\!\begin{array}{c} {#1}\\ {#2} 
\end{array}\!\!\!\right\}}
\nc{\sbinlr}[2]{\Big\langle\!\!\begin{array}{c} {#1}\\ {#2} 
\end{array}\!\!\Big\rangle}
\nc{\bino}[2]{\left(\!\!\begin{array}{c} {#1}\\ {#2} \end{array}\!\!\right)}
\def\half {\mbox{$\textstyle \frac{1}{2}$}}
\def\vec#1{\mbox {\boldmath $#1$}}
\def\svec#1{\mbox {\scriptsize\boldmath $#1$}}
\definecolor{lightblue}{rgb}{.61,.61,1}
\definecolor{midblue}{rgb}{.7,.7,1}
\definecolor{lightlightblue}{rgb}{.9,.9,1}
\definecolor{lightestblue}{rgb}{.96,.96,1}
\definecolor{lightpurple}{rgb}{1,.65,1}
\def\leftarc#1{\psarc[linecolor=blue,linewidth=1.5pt]#1{.5}{90}{270}}
\def\rightarc#1{\psarc[linecolor=blue,linewidth=1.5pt]#1{.5}{-90}{90}}
\def\loopa{
\psframe[linewidth=.25pt](0,0)(1,1)
\psarc[linewidth=1.5pt,linecolor=blue](1,0){.5}{90}{180}
\psarc[linewidth=1.5pt,linecolor=blue](0,1){.5}{-90}{0}
}
\def\loopb{
\psframe[linewidth=.25pt](0,0)(1,1)
\psarc[linewidth=1.5pt,linecolor=blue](0,0){.5}{0}{90}
\psarc[linewidth=1.5pt,linecolor=blue](1,1){.5}{180}{270}
}
\def\floopa{
\psarc[linewidth=4pt,linecolor=blue]{-}(1,0){0.5}{90}{180}
\psarc[linewidth=2pt,linecolor=lightlightblue]{-}(1,0){0.5}{90}{180}
\psarc[linewidth=4pt,linecolor=blue]{-}(0,1){0.5}{-90}{0}
\psarc[linewidth=2pt,linecolor=lightlightblue]{-}(0,1){0.5}{-90}{0}
\psframe[linewidth=.3pt](0,0)(1,1)
}
\def\floopb{
\psarc[linewidth=4pt,linecolor=blue]{-}(0,0){0.5}{0}{90}
\psarc[linewidth=2pt,linecolor=lightlightblue]{-}(0,0){0.5}{0}{90}
\psarc[linewidth=4pt,linecolor=blue]{-}(1,1){0.5}{180}{-90}
\psarc[linewidth=2pt,linecolor=lightlightblue]{-}(1,1){0.5}{180}{-90}
\psframe[linewidth=.3pt](0,0)(1,1)
}
\def\emptysquare{\hspace{-.11\unitlength}
\begin{pspicture}(1,1)
\pspolygon[linewidth=.25pt](0,0)(1,0)(1,1)(0,1)(0,0)
\end{pspicture}}
\def\facegrid#1#2{
\psframe[fillstyle=solid,fillcolor=lightlightblue,linewidth=0pt]#1#2
\psgrid[gridlabels=0pt,subgriddiv=1]#1#2}
\def\facegridblue#1#2{
\psframe[fillstyle=solid,fillcolor=lightblue,linewidth=0pt]#1#2
\psgrid[gridlabels=0pt,subgriddiv=1]#1#2}
\def\biggerfacegrid#1#2{
\psframe[fillstyle=solid,fillcolor=lightlightblue,linewidth=0pt]#1#2
\psgrid[gridlabels=0pt,subgriddiv=1]#1#2
}
\def\conn#1#2{
\psframe[linewidth=1pt,fillstyle=solid,fillcolor=lightlightblue]#1#2}
\nc{\ch}{{\rm ch}}
\nc{\R}{{\cal R}}
\nc{\dkk}{\delta_{j,\{k,k'\}}^{(2)}}
\nc{\drr}{\delta_{j,\{r,r'\}}^{(2)}}
\nc{\ddkk}{\delta_{j,\{k,k'\}}^{(4)}}
\nc{\dddkk}{\delta_{j,\{k,k'\}}^{(8)}}
\nc{\dnn}{\delta_{j,\{n,n'\}}^{(2)}}
\nc{\ddnn}{\delta_{j,\{n,n'\}}^{(4)}}
\nc{\dddnn}{\delta_{j,\{n,n'\}}^{(8)}}
\nc{\ph}{\hat{p}}

\def\psq#1{\raisebox{-1.5\unitlength}{
\begin{pspicture}[shift=0](0,0)(4,4)
\psframe[linewidth=0pt,fillstyle=solid,fillcolor=lightlightblue](0,0)(4,4)
\pspolygon[linewidth=.25pt](0,0)(4,0)(4,4)(0,4)
\rput(2,2){\small $#1$}
\psarc(0,0){.35}{0}{90}
\end{pspicture}}}
\def\psqa#1{\raisebox{-1.5\unitlength}{
\begin{pspicture}[shift=0](0,0)(4,4)
\psframe[linewidth=0pt,fillstyle=solid,fillcolor=lightlightblue](0,0)(4,4)
\pspolygon[linewidth=.25pt](0,0)(4,0)(4,4)(0,4)
\psarc[linewidth=1pt](4,0){2}{90}{180}
\psarc[linewidth=1pt](0,4){2}{-90}{0}
\rput(2,2){\small $#1$}
\end{pspicture}}}
\def\psqb#1{\raisebox{-1.5\unitlength}{
\begin{pspicture}[shift=0](0,0)(4,4)
\psframe[linewidth=0pt,fillstyle=solid,fillcolor=lightlightblue](0,0)(4,4)
\pspolygon[linewidth=.25pt](0,0)(4,0)(4,4)(0,4)
\psarc[linewidth=1pt](0,0){2}{0}{90}
\psarc[linewidth=1pt](4,4){2}{180}{270}
\rput(2,2){\small $#1$}
\end{pspicture}}}
\def\psqbr#1{\raisebox{-1.5\unitlength}{
\begin{pspicture}[shift=0](0,0)(4,4)
\pspolygon[linewidth=.25pt](0,0)(4,0)(4,4)(0,4)
\rput(2,2){\small $#1$}
\psarc(4,0){.35}{90}{180}
\end{pspicture}}}

\definecolor{pink}{rgb}{1,.65,.65}

\makeatletter
\def\Ddots{\mathinner{\mkern1mu\raise\p@
\vbox{\kern7\p@\hbox{.}}\mkern2mu
\raise4\p@\hbox{.}\mkern2mu\raise7\p@\hbox{.}\mkern1mu}}
\makeatother

\thicklines
\def\punit#1{\hspace{#1\unitlength}}
\def\pvunit#1{\vspace{#1\unitlength}}

\def\diagface#1#2#3#4#5#6#7{\rule[-3.8\unitlength]{0in}{7.6\unitlength}
\begin{picture}(9,4)(-#6,-#7)
\put(1.5,0.5){\vector(1,-1){3}}
\put(4.5,3.5){\vector(1,-1){3}}
\put(4.5,3.5){\vector(-1,-1){3}}
\put(7.5,0.5){\vector(-1,-1){3}}
\put(1.2,0.5){\makebox(0,0)[r]{\smaller \mbox{$#1$}}}
\put(4.5,-2.8){\makebox(0,0)[t]{\smaller \mbox{$#2$}}}
\put(7.8,0.5){\makebox(0,0)[l]{\smaller \mbox{$#3$}}}
\put(4.5,3.8){\makebox(0,0)[b]{\smaller \mbox{$#4$}}}
\put(4.5,0.5){\makebox(0,0){\smaller \mbox{$#5$}}}
\end{picture}}

\def\sqface#1#2#3#4#5#6#7{\rule[-2.8\unitlength]{0in}{5.6\unitlength}
\begin{picture}(4,4)(-#6,-#7)
\put(0,-2){\vector(1,0){4}}
\put(4,2){\vector(0,-1){4}}
\put(0,2){\vector(1,0){4}}
\put(0,2){\vector(0,-1){4}}
\put(0,-3.2){\makebox(0,0)[b]{\smaller \mbox{$#1$}}}
\put(4,-3.2){\makebox(0,0)[b]{\smaller \mbox{$#2$}}}
\put(4,2.5){\makebox(0,0)[b]{\smaller \mbox{$#3$}}}
\put(0,2.5){\makebox(0,0)[b]{\smaller \mbox{$#4$}}}
\put(2,0){\makebox(0,0){\smaller \mbox{$#5$}}}
\end{picture}}

\def\sqedges#1#2#3#4{
\begin{picture}(0,0)(0,-.3)
\put(2,-3.2){\makebox(0,0)[b]{\smaller \mbox{$#1$}}}
\put(4.4,0){\makebox(0,0)[l]{\smaller \mbox{$#2$}}}
\put(2,2.5){\makebox(0,0)[b]{\smaller \mbox{$#3$}}}
\put(-.4,0){\makebox(0,0)[r]{\smaller \mbox{$#4$}}}
\end{picture}}

\def\cross#1#2#3#4#5{\rule[-2.8\unitlength]{0in}{5.6\unitlength}
\begin{picture}(4,4)(0,-.3)
\put(2,-2){\line(0,1){4}}
\put(0,0){\line(1,0){4}}
\put(0,-3.2){\makebox(0,0)[b]{\smaller \mbox{$#1$}}}
\put(4,-3.2){\makebox(0,0)[b]{\smaller \mbox{$#2$}}}
\put(4,2.5){\makebox(0,0)[b]{\smaller \mbox{$#3$}}}
\put(0,2.5){\makebox(0,0)[b]{\smaller \mbox{$#4$}}}
\put(2,0){\makebox(0,0){\smaller \mbox{$#5$}}}
\end{picture}}

\def\diagcross#1#2#3#4{\rule[-2.8\unitlength]{0in}{5.6\unitlength}
\begin{picture}(4,4)(0,0)
\put(0,-2){\line(1,1){4}}
\put(4,-2){\line(-1,1){4}}
\put(0,-3.2){\makebox(0,0)[b]{\smaller \mbox{$#1$}}}
\put(4,-3.2){\makebox(0,0)[b]{\smaller \mbox{$#2$}}}
\put(4,2.5){\makebox(0,0)[b]{\smaller \mbox{$#3$}}}
\put(0,2.5){\makebox(0,0)[b]{\smaller \mbox{$#4$}}}
\end{picture}}

\def\diamond#1#2#3#4#5{\rule[-6\unitlength]{0in}{12\unitlength}
\begin{picture}(2,5)(0,-.5)
\put(2,4){\vector(-1,-2){2}}
\put(0,0){\vector(1,-2){2}}
\put(2,4){\vector(1,-2){2}}
\put(4,0){\vector(-1,-2){2}}
\put(2,-5.2){\makebox(0,0)[b]{\smaller \mbox{$#1$}}}
\put(4.4,0){\makebox(0,0)[l]{\smaller \mbox{$#2$}}}
\put(2,4.5){\makebox(0,0)[b]{\smaller \mbox{$#3$}}}
\put(-.2,0){\makebox(0,0)[r]{\smaller \mbox{$#4$}}}
\put(2,0){\makebox(0,0){\smaller \mbox{$#5$}}}
\end{picture}}

\def\leftfaces#1#2#3#4#5#6#7{\rule[-6\unitlength]{0in}{12\unitlength}
\begin{picture}(4,5)(0,-.5)
\put(2,4){\vector(-1,-2){2}}
\put(0,0){\vector(1,-2){2}}
\put(2,-4){\vector(1,0){4}}
\put(2,4){\vector(1,0){4}}
\put(0,0){\vector(1,0){4}}
\put(4,0){\vector(1,-2){2}}
\put(6,4){\vector(-1,-2){2}}
\put(2,-5.2){\makebox(0,0)[b]{\smaller \mbox{$#1$}}}
\put(6,-5.2){\makebox(0,0)[b]{\smaller \mbox{$#2$}}}
\put(6,4.5){\makebox(0,0)[b]{\smaller \mbox{$#3$}}}
\put(2,4.5){\makebox(0,0)[b]{\smaller \mbox{$#4$}}}
\put(-.2,0){\makebox(0,0)[r]{\smaller \mbox{$#5$}}}
\put(3,-2){\makebox(0,0){\smaller \mbox{$#6$}}}
\put(3,2){\makebox(0,0){\smaller \mbox{$#7$}}}
\end{picture}}

\def\rightfaces#1#2#3#4#5#6#7{\rule[-6\unitlength]{0in}{12\unitlength}
\begin{picture}(4,5)(0,-.5)
\put(0,4){\vector(1,-2){2}}
\put(2,0){\vector(-1,-2){2}}
\put(0,-4){\vector(1,0){4}}
\put(0,4){\vector(1,0){4}}
\put(4,4){\vector(1,-2){2}}
\put(6,0){\vector(-1,-2){2}}
\put(2,0){\vector(1,0){4}}
\put(0,-5.2){\makebox(0,0)[b]{\smaller \mbox{$#1$}}}
\put(4,-5.2){\makebox(0,0)[b]{\smaller \mbox{$#2$}}}
\put(6.4,0){\makebox(0,0)[l]{\smaller \mbox{$#3$}}}
\put(4,4.5){\makebox(0,0)[b]{\smaller \mbox{$#4$}}}
\put(0,4.5){\makebox(0,0)[b]{\smaller \mbox{$#5$}}}
\put(3,-2){\makebox(0,0){\smaller \mbox{$#6$}}}
\put(3,2){\makebox(0,0){\smaller \mbox{$#7$}}}
\end{picture}}

\def\braid#1#2#3#4{\rule[-4\unitlength]{0in}{8\unitlength}
\begin{picture}(0,0)(-#1,-#2)
\put(0,0){\line(1,1){4}}
\put(0,4){\line(1,-1){1.5}}
\put(4,0){\line(-1,1){1.5}}
\put(0,-1){\makebox(0,0)[t]{\smaller \mbox{$#3$}}}
\put(4,-1){\makebox(0,0)[t]{\smaller \mbox{$#4$}}}
\end{picture}}

\def\invbraid#1#2#3#4{\rule[-4\unitlength]{0in}{8\unitlength}
\begin{picture}(0,0)(-#1,-#2)
\put(0,4){\line(1,-1){4}}
\put(0,0){\line(1,1){1.5}}
\put(4,4){\line(-1,-1){1.5}}
\put(0,-1){\makebox(0,0)[t]{\smaller \mbox{$#3$}}}
\put(4,-1){\makebox(0,0)[t]{\smaller \mbox{$#4$}}}
\end{picture}}

\def\Brauerbraid#1#2#3#4{\rule[-4\unitlength]{0in}{8\unitlength}
\begin{picture}(0,0)(-#1,-#2)
\put(0,0){\line(1,1){4}}
\put(0,4){\line(1,-1){4}}
\put(0,-1){\makebox(0,0)[t]{\smaller \mbox{$#3$}}}
\put(4,-1){\makebox(0,0)[t]{\smaller \mbox{$#4$}}}
\end{picture}}

\def\monoid#1#2#3#4{\rule[-4\unitlength]{0in}{8\unitlength}
\begin{picture}(0,0)(-#1,-#2)
\put(2,0){\oval(4,3.5)[t]}
\put(2,4){\oval(4,3.5)[b]}
\put(0,-1){\makebox(0,0)[t]{\smaller \mbox{$#3$}}}
\put(4,-1){\makebox(0,0)[t]{\smaller \mbox{$#4$}}}
\end{picture}}

\def\vert#1#2#3{\rule[-4\unitlength]{0in}{8\unitlength}
\begin{picture}(0,0)(-#1,-#2)
\put(0,0){\line(0,1){4}}
\put(0,-1){\makebox(0,0)[t]{\smaller \mbox{$#3$}}}
\end{picture}}

\def\vertl#1#2#3#4{
\begin{picture}(0,0)(-#1,-#2)
\put(0,0){\line(0,1){#4}}
\put(0,-1){\makebox(0,0)[t]{\smaller \mbox{$#3$}}}
\end{picture}}

\def\diagfaceup#1#2#3#4#5#6#7{\rule[-3.8\unitlength]{0in}{7.6\unitlength}
\begin{picture}(9,4)(-#6,-#7)
\put(1.5,0.5){\vector(1,1){3}}
\put(4.5,3.5){\vector(1,-1){3}}
\put(4.5,-2.5){\vector(1,1){3}}
\put(1.5,0.5){\vector(1,-1){3}}
\put(1.2,0.5){\makebox(0,0)[r]{\smaller \mbox{$#1$}}}
\put(4.5,-2.8){\makebox(0,0)[t]{\smaller \mbox{$#2$}}}
\put(7.8,0.5){\makebox(0,0)[l]{\smaller \mbox{$#3$}}}
\put(4.5,3.8){\makebox(0,0)[b]{\smaller \mbox{$#4$}}}
\put(4.5,0.5){\makebox(0,0){\smaller \mbox{$#5$}}}
\end{picture}}

\def\smaller{\small}

\def\wt#1#2#3#4{W\!\!\mbox{\small $\left(\matrix{#4&#3\cr #1&#2\cr}\right)$}}
\def\weight#1#2#3#4#5{#1\!\!\mbox{\small $\left(\matrix{#5&#4\cr #2&#3\cr}\right)$}}
\def\weightb#1#2#3#4#5#6#7#8#9{#1\!\!\mbox{\smaller $\left(\matrix{#8&#7&#6\cr#9&&#5\cr
#2&#3&#4\cr}\right)$}} 
\def\weightv#1#2#3#4#5{#1\!\!\mbox{\smaller $\left(#5\ \matrix{#4\cr\cr#2\cr}\ #3\right)$}}
\def\wtu#1#2#3#4#5{W\!\!\mbox{\small $\left\matrix{#4&#3\cr #1&#2\cr}\Bigm|\mbox{$#5$}\right)$}}
\def\weightu#1#2#3#4#5#6{\rule[-10pt]{0in}{20pt}
#1\!\mbox{\small $\left(\matrix{#5&#4\cr
#2&#3\cr}\Bigm|\mbox{$#6$}\right)$}} \def\weightbu#1#2#3#4#5#6#7#8#9{W\!\!\mbox{\smaller
$\left(\matrix{#7&#6&#5\cr#8&&#4\cr #1&#2&#3\cr}\biggm|\mbox{$#9$}\right)$}} 
\def\weightvu#1#2#3#4#5#6{#1\!\!\mbox{\smaller $\left(\matrix{&#4&\cr#5&&#3\cr
&#2&\cr}\biggm|\mbox{$#6$}\right)$}}

\begin{document}

\topmargin -15mm
\oddsidemargin 05mm

%%%%%%%%%%%%%%%%%%%%%
%
% Title page
%
%%%%%%%%%%%%%%%%%%%%%

\title{\mbox{}\vspace{-.5in}\bf 
Fusion hierarchies, $\boldsymbol T$-systems and $\boldsymbol Y$-systems\\[37pt]
of logarithmic minimal models
\vspace{-.2in}
}

%\date{}
\maketitle 

\begin{center}
{\vspace{-5mm}\Large Alexi Morin-Duchesne$^\ast$,\, Paul A. Pearce$^\dagger$,\, J{\o}rgen Rasmussen$^\ast$}
\\[.5cm]
{\em {}$^\ast$School of Mathematics and Physics, University of Queensland}\\
{\em St Lucia, Brisbane, Queensland 4072, Australia}
\\[.2cm]
{\em {}$^\dagger$Department of Mathematics and Statistics, University of Melbourne}\\
{\em Parkville, Victoria 3010, Australia}
\\[.4cm] 
{\tt alexi.morin.duchesne\,@\,gmail.com}
\quad
{\tt p.pearce\,@\,ms.unimelb.edu.au}
\quad
{\tt j.rasmussen\,@\,uq.edu.au}
\end{center}

%%%%%%%%%%%%%%%%%%%%%
%
% Abstract
%
%%%%%%%%%%%%%%%%%%%%%
 
\begin{abstract}
A Temperley-Lieb (TL) loop model is a Yang-Baxter integrable lattice model with nonlocal degrees of freedom. 
On a strip of width $N \in \mathbb N$, the evolution operator is the double-row transfer tangle $\Db(u)$,
an element of the TL algebra $T\!L_N(\beta)$ with loop fugacity $\beta = 2\cos \lambda$, $\lambda \in \mathbb R$. 
Similarly on a cylinder, the single-row transfer tangle $\Tb(u)$ is an element of the so-called enlarged periodic TL algebra. 
The logarithmic minimal models $\mathcal{LM}(p,p')$ comprise a subfamily of the TL loop models for which the 
crossing parameter $\lambda=(p'-p)\pi/p'$ is a rational multiple of $\pi$ parameterised by coprime integers $1\leq p<p'$. 
For these special values, additional symmetries allow for particular degeneracies in the spectra that account for the logarithmic nature of 
these theories. For critical dense polymers $\mathcal{LM}(1,2)$, with $\beta=0$, $\Db(u)$ and $\Tb(u)$ 
satisfy inversion identities that have led to the exact determination of the eigenvalues in any representation and for 
arbitrary finite system size $N$. The generalisation for $p'>2$ takes the form of functional relations for $\Db(u)$ and $\Tb(u)$ of 
polynomial degree $p'$. These derive from fusion hierarchies of commuting transfer tangles 
$\Db^{m,n}(u)$ and $\Tb^{m,n}(u)$ where $\Db(u) = \Db^{1,1}(u)$ and $\Tb(u) = \Tb^{1,1}(u)$. 
The fused transfer tangles are constructed from $(m,n)$-fused face operators involving Wenzl-Jones projectors 
$P_k$ on $k=m$ or $k=n$ nodes. Some projectors $P_k$ are singular for $k \ge p'$, but we argue that $\Db^{m,n}(u)$ 
and $\Tb^{m,n}(u)$ are nonsingular for every $m,n \in \mathbb N$ in certain cabled link state representations.
For generic $\lambda$, we derive the fusion hierarchies and the associated $T$- and $Y$-systems. 
For the logarithmic theories, the closure of the fusion hierarchies at $n=p'$
translates into functional relations of polynomial degree $p'$ for $\Db^{m,1}(u)$ and $\Tb^{m,1}(u)$.
We also derive the closure of the $Y$-systems for the logarithmic theories. 
The $T$- and $Y$-systems are the key to exact integrability and we observe that the underlying structure of these functional 
equations relate to Dynkin diagrams of affine Lie algebras.
\end{abstract}

%%%%%%%%%%%%%%%%%%%%%
%
% TOC
%
%%%%%%%%%%%%%%%%%%%%%

\newpage

\tableofcontents
\clearpage

%%%%%%%%%%%%%%%%%%%%
\section{Introduction}
\label{Sec:Introduction}
%%%%%%%%%%%%%%%%%%%%

The logarithmic minimal models ${\cal LM}(p,p')$~\cite{PRZ0607}, where $p,p'$ are coprime integers, are a family of Yang-Baxter 
integrable models built on the square lattice using the planar Temperley-Lieb (TL) algebra~\cite{TL,Jones}. 
These models are loop models that admit nonlocal degrees of freedom in the form of polymers or connectivities.
The first members of this series include critical dense polymers ${\cal LM}(1,2)$~\cite{PR2007,PRV0910,PRV1210,MDPRtorus} 
and critical (bond) percolation ${\cal LM}(2,3)$~\cite{BroadHamm57}. 
The motivation for introducing these models was (i) to initiate a lattice approach to the study of logarithmic Conformal Field Theory 
(CFT) from the continuum scaling limit of a Yang-Baxter integrable model, 
and (ii) to demonstrate that the Jordan blocks 
associated with reducible yet indecomposable representations of these theories are accessible through finite-size lattice calculations. 
The {\it logarithmic} minimal models are the simplest family of logarithmic CFTs 
(without extended symmetries) and are expected to play the same role for logarithmic 
CFTs that the {\it rational}\/ minimal models~\cite{BPZ} play for rational CFTs.
Recent reviews on logarithmic CFT can be found in~\cite{SpecialIssue}.

Although logarithmic minimal models are not rational and not unitary, they have many similarities with the rational minimal models 
${\cal M}(m,m')$ where $m,m'$ are coprime integers. 
The rational minimal models are realised as the continuum scaling limit of the Yang-Baxter integrable Restricted-Solid-on-Solid (RSOS) 
lattice models~\cite{ABF,FB}. For the {\it unitary} minimal models ${\cal M}(m,m+1)$~\cite{ABF}, 
a fusion hierarchy of functional equations satisfied by the periodic single-row transfer matrices of the level $n\in\mathbb{N}$ 
fused RSOS models ${\cal M}(m,m+1)_{n\times n}$ was obtained by Bazhanov and Reshetikhin~\cite{BR1989}.
Working within a lattice approach, it was shown, by 
Kl\"umper and Pearce~\cite{KlumperPearcePRL1991,KlumperPearce1991,KlumperPearce1992} and in subsequent work, 
that $T$- and $Y$-systems are the key to the analytic solution of these theories. The $T$-system is solved for the 
{\it non-universal}\/ properties of the statistical system such as bulk free energies~\cite{BaxInv82}, boundary and seam free 
energies~\cite{OPB95} and correlation lengths~\cite{OPB97}. The $Y$-system is solved, in the continuum scaling limit, 
for the {\it universal}\/ conformal quantities such as central charges, conformal dimensions and characters. 

To put $T$- and $Y$-systems into historical perspective, it was in the context of the $su(2)$ unitary minimal 
models~\cite{KlumperPearcePRL1991,KlumperPearce1991,KlumperPearce1992} that the first $T$-system appeared allowing the 
associated $Y$-system and nonlinear integral equations 
to be systematically derived and solved for all excitations. 
For integrable quantum systems, the $T$-systems take the form of discrete classical bilinear Hirota equations~\cite{KLWZ1997}. 
The first $Y$-systems appeared earlier in the work of Zamolodchikov~\cite{Zam1991a,Zam1991b} in relation to 
thermodynamic Bethe ansatz equations for massive scattering theories. $T$- and $Y$-systems for theories associated with Lie algebras 
other than $su(2)$ were obtained in~\cite{KunibaNS9309,KunibaNS9310,KunibaN1992,RTV1993}. 
$T$- and $Y$-systems appear in many contexts. Recently, they have been used to study integrable aspects of the AdS/CFT 
correspondence. A general review article on $T$- and $Y$-systems appears in~\cite{KunibaNS1010}. 
Historically, $T$- and $Y$-systems were first obtained for periodic single-row transfer matrices $\vec T^{m,n}(u)$. 
Later it was realised, at least for rational CFTs, that $Y$-systems are {\it universal}~\cite{universal} in the sense that precisely the same 
$Y$-system of equations holds for all topologies and all boundary conditions. This {\it miracle}, the fact that you are solving the same 
equations, explains why the same central charge and the same set of conformal weights appear in each topology with various 
boundary conditions. For unitary minimal models on the strip in the presence of boundary conditions, the $T$- and $Y$-systems were 
obtained~\cite{BehrendPearce} by working with double-row transfer matrices $\vec D^{m,n}(u)$. In the case of the tricritical Ising model 
on the strip, the functional equations have been solved~\cite{tricritIsing} for all conformal boundary conditions and all excitations, 
underscoring the power of these methods.

The fused minimal models ${\cal M}(m,m')_{n\times n}$ have a long history of being described, as CFTs, by models ${\cal M}(M,M';n)$ 
admitting diagonal GKO coset descriptions~\cite{GKO85,GKO86,ACT91}, 
where $M=M(m,m')$ and $M'=M'(m,m')$ are integers satisfying $2\le M<M'$. 
It has recently been argued~\cite{PR1305} that general logarithmic minimal models at higher fusion level can 
similarly be described as diagonal $su(2)$ GKO cosets
\be
 {\cal LM}(P,P';n)\simeq \mbox{COSET}\Big(\frac{nP}{P'\!-\!P}-2,n\Big),\quad \mathrm{gcd}\Big(P,\frac{P'\!-\!P}{n}\Big)=1,\quad 
    1\leq P<P',\quad n,P,P'\in\mathbb{N}
\label{Clog}
\ee
where 
$n$ is the integer fusion level. The diagonal GKO coset~\cite{GKO85,GKO86,ACT91} takes the form
\be
 \mbox{COSET}(k,n):\quad \frac{(A_1^{(1)})_k\oplus(A_1^{(1)})_n}{(A_1^{(1)})_{k+n}},
 \qquad k=\frac{\ph}{\ph'}-2,\qquad \mathrm{gcd}(\ph,\ph')=1,\qquad n,\ph,\ph'\in\mathbb{N}
\label{cosetAAA}
\ee
where $k$ is a fractional fusion level and the subscripts on the 
affine $su(2)$ current algebra $A_1^{(1)}$ denote the respective levels $k$, $n$ and $k+n$. 
The central charges of the logarithmic minimal models ${\cal LM}(P,P';n)$ are 
\be 
 c^{P,P';n}=\frac{3n}{n+2} \Big[1-\frac{2(n+2)(P'-P)^2}{n^2 P P'}\Big].
\label{centralcharges}
\ee

In~\cite{PRZ0607}, the non-fused lattice loop model ${\cal LM}(p,p')_{1\times 1}$ and the corresponding CFT ${\cal LM}(P,P';1)$
were studied and both referred to as the logarithmic minimal model ${\cal LM}(p,p')$. 
To emphasise the strong relation between the lattice model and the corresponding CFT, 
here we write 
\be
 {\cal LM}(p,p')_{1\times 1}\equiv {\cal LM}(P,P';1), \qquad P = p, \ \; P'=p'.
\ee
For $n>1$, it has been conjectured in~\cite{PRT2013} that the theories ${\cal LM}(P,P';n)$ are realised on the lattice by 
$n\times n$ fusions built from the elementary logarithmic minimal models ${\cal LM}(p,p')$, that is
\be
 {\cal LM}(p,p')_{n\times n}\equiv{\cal LM}(P,P';n)
\label{fusionIdentification}
\ee
for some $P=P(p,p')$, $P'=P'(p,p')$. In particular, for $n=2$, the theories are logarithmic superconformal field theories~\cite{PRT2013}
\be
 {\cal LM}(p,p')_{2\times 2}\equiv{\cal LM}(P,P';2),\qquad P=|2p-p'|,\ \; P'=p'
\label{super}
\ee
also denoted by ${\cal LSM}(p,p')$.
The first members of the superconformal series include superconformal dense polymers ${\cal LSM}(2,3)$ and superconformal 
percolation ${\cal LSM}(3,4)$ which are lattice models that are expected to be of independent interest in statistical mechanics. 
At present, it is not  known what relation between $P,P'$ and $p,p'$ is required to make the identification 
(\ref{fusionIdentification}) valid in general. 
This is also the case for the fused minimal models. 

According to~\cite{PR1305}, the coset construction of the general logarithmic minimal models ${\cal LM}(P,P';n)$ is completely 
analogous to the coset construction of the rational minimal models ${\cal M}(M,M';n)$, but based on more complicated representation 
theories. Furthermore, the chiral logarithmic minimal models can be described~\cite{PR1305} as the 
logarithmic limit~\cite{Ras0405,Ras0406} of the rational minimal models 
\be 
 {\cal LM}(P,P';n)=\lim_{M,M'\to\infty\atop M/M'\to P/P'}\!{\cal M}(M,M';n)
\label{logLimit}
\ee
where the limit is taken through a sequence of $M,M'$ values satisfying 
\be
\mathrm{gcd}\Big(M,\frac{M'-M}{n}\Big)=1,\quad
   2\leq M<M',\quad M,M'\in\mathbb{N}.
\label{pplimits}
\ee
It follows that the central charges (\ref{centralcharges}) of the logarithmic minimal models are given as limits of the 
central charges of the rational minimal models
\be
 c^{P,P';n}=\lim_{M,M'\to\infty\atop M/M'\to P/P'} c^{M,M';n}.
\ee
The coset construction (\ref{Clog}), along with the logarithmic limit (\ref{logLimit}), makes precise the sense in which the logarithmic 
minimal models ${\cal LM}(P,P';n)$ are extensions of the rational minimal models ${\cal M}(M,M';n)$.

The goal of this paper is to extend the fusion hierarchy and the $T$- and $Y$-systems to the commuting double-row and 
single-row transfer matrices of the fused logarithmic minimal models ${\cal LM}(p,p')_{n\times n}$. 
For the case of the double-row transfer matrices on the strip, we only consider the simplest vacuum boundary conditions conjugate to 
the identity operator. These transfer matrices are not actually matrices but rather diagrammatic objects 
(tangles~\cite{Conway,KauffmanLins1994}) defined in the corresponding planar TL algebra~\cite{TL,Jones}. 
Matrix representations are obtained by acting with these tangles on suitable vector spaces of link states~\cite{PRZ0607,PRT2013}. 
Our results mirror the results obtained in~\cite{KlumperPearce1992} for the corresponding rational RSOS  models~\cite{ABF,FB}, 
but here they are established directly in the planar algebra. Crucially, this ensures that the results are valid for {\em all\/} matrix 
representations of the transfer tangles. Since the logarithmic minimal models are $su(2)$ theories, we work throughout with  
$T$- and $Y$-systems of the same form as in~\cite{KlumperPearce1992}. 
Remarkably, we find that these functional equations hold for arbitrary coprime integers $p,p'$ and that the underlying structures are 
related to the Dynkin diagrams of the affine Lie algebras $A_{p'-1}^{(1)}$. Indeed, the determinantal structure of the polynomial 
functional equations of degree $p'$ is the same~\cite{functional} as for the Cyclic Solid-on-Solid (CSOS) 
models~\cite{PScyclicPRL,KYcyclic,PScyclic}. In contrast, the structure of the $T$- and $Y$-systems of rational minimal models are 
related to the Dynkin diagrams of classical Lie algebras.  

The layout of the paper is as follows. Section~\ref{sec:notation} contains notation and functions used repeatedly throughout the paper. 
In Section~\ref{Sec:TL}, we present a brief review of the ordinary (non-fused) TL loop models. 
As computations are later carried out in the planar TL algebras, we first discuss this framework, its relation with linear 
TL algebras and the construction of link state modules for these algebras. We then write down some fundamental
planar identities and define the transfer tangles on the strip and cylinder. 
The definition and key properties of Wenzl-Jones projectors are also reviewed.

In Section~\ref{sec:FusedTL}, we introduce $(m,n)$-fused face operators and study their decomposition in terms of 
generalised monoids. (The fusion index $m$ should not be confused with the minimal model label $m$ used above.)
The explicit decomposition coefficients generalise results of~\cite{FR2002,ZinnJ2007} for $(n,n)$-fusion and their computation is 
deferred to Appendix~\ref{App:FusedFaces}. Planar identities similar to the ones in the non-fused case are obtained for the fused faces. 
The generalisation of the boundary Yang-Baxter equation to the fused setting is nontrivial and a proof is presented in 
Appendix~\ref{app:YBEs}, following ideas developed in~\cite{BehrendPearce}.

In Section~\ref{Sec:Tangles}, we introduce fused transfer tangles. The projectors and fused faces are well defined for $\lambda$ 
generic, but in general not for $\frac{\lambda}{\pi}$ rational. However, we show that the expression for the transfer tangles can 
alternatively be given by replacing the $P_n$ projectors by effective projectors $Q_n$ which are well defined for all 
$\beta \in \mathbb C^\ast$. From this, we argue that singularities of the $P_n$ projectors can be eliminated from the transfer tangles.
A key property of the effective projectors is established in Appendix~\ref{app:EffectiveProjectors}.

In Section~\ref{Sec:CabledLinkStates}, we argue that singularities of the $P_m$ projectors can also be ignored by studying 
representations of the fused TL algebra built from cabled link states explored in Appendix~\ref{app:LatticePaths}. 
These representations are defined for generic 
$\lambda$ and we show that, in certain quotient representations, the limit to rational $\frac{\lambda}{\pi}$ is well defined.
The corresponding transfer matrix representations are therefore nonsingular and well defined.

In Section~\ref{Sec:FTY}, we derive the fusion hierarchies satisfied by the transfer tangles on the strip and cylinder, with
lengthy diagrammatic proofs deferred to Appendix~\ref{app:FusHier}.
The corresponding $T$- and $Y$-systems are subsequently worked out algebraically.

In Section~\ref{Sec:LMM}, we focus on fractional $\lambda=\frac{(p'-p)\pi}{p'}$. 
For $m=n$, this corresponds to the higher-level logarithmic minimal model ${\cal LM}(p,p')_{n\times n}$. For general $n$,
the fusion hierarchies close and we prove this diagrammatically in Appendix~\ref{app:Closure}. The closure translates into functional 
relations of polynomial degree $p'$ for the transfer tangles. In the cylinder case, the closure and functional relations are expressed
in terms of the so-called braid transfer tangle. The closure of the $Y$-systems is also worked out explicitly.

In Section~\ref{sec:ratcomp}, we compare our results for the fusion hierarchies, $T$-systems and $Y$-systems of the logarithmic 
minimal models with those obtained previously for rational models. We find that the fusion hierarchies and $T$-systems coincide with 
those of the critical $A$-$D$-$E$ models, as described in~\cite{universal}, 
but observe that the closure mechanisms of the fusion hierarchy and $Y$-systems differ significantly.

Finally, Section~\ref{Sec:Conclusion} contains some concluding remarks.

%%%%%%%%%%%%%%%%%%%%%%%%%
\subsection{Notation}  \label{sec:notation}
%%%%%%%%%%%%%%%%%%%%%%%%%

For ease of reference, definitions and conventions used repeatedly in the paper are listed below.
\\[.2cm]
Weight function and shifted spectral parameter:
\be
 s_k(u):=\frac{\sin (u+k\lambda)}{\sin\lambda} = s_0(u_k),\qquad u_k:=u+k\lambda.
\label{sk}
\ee
Coefficient functions arising from half-arc propagation:
\be
 q^m(u):= \prod_{i=0}^{m-1} s_{-i}(u)s_{2+i}(-u).
\label{eq:qkm}
\ee
Fused transfer tangles with shifted arguments:
\be
 \Db^{m,n}_k:=\Db^{m,n}(u+k \lambda), \qquad  \Tb^{m,n}_k:=\Tb^{m,n}(u+k \lambda).
\label{Dkmn}
\ee  
Renormalised double-row transfer tangles:
\be
 \Dbh^{m,n}_k:=s_{2k+n-1}(2u\!-\!\mu)\Big(\prod_{j= k}^{n+k-2} \!\!s_{2j+1}(2u\!-\!\mu)\Big)\Db^{m,n}_k.
\label{eq:Dhat}
\ee
Collapsed transfer and braid transfer tangles:
\be
 \Db^{m,0}(u)\equiv\Fb^{m,0}\equiv\Ib^m,\qquad \Tb^{m,0}(u)\equiv\Fbt^{m,0}\equiv\Ib^m, 
\label{DITI}
\ee
\be
 \Dbh^{m,-1}(u)\equiv\Fb^{m,-1}\equiv0,\qquad\Tb^{m,-1}(u)\equiv\Fbt^{m,-1}\equiv0. 
\ee
Fusion hierarchy coefficient functions:
\be
 f^m_k:= s_{2k+1}(2u\!-\!\mu) \Big(\prod_{i=0}^{m-1}s_{k-i}(u)s_{k+m-i}(u\!-\!\mu)\Big)^{\!N},\qquad
 h^m_k:= \Big(\prod_{j=0}^{m-1}(-\ir)s_{k-j}(u)\Big)^{\!N}.
\label{fkm}
\ee
$T$-system coefficient functions: 
\be
 \nu_k^{(n)}:= \prod_{j=k}^{n+k-1}f^m_{j-1}f^m_{j+1},\qquad
 \tilde{\nu}_k^{(n)}:= \prod_{j=k}^{n+k-1}h^m_{j-1}h^m_{j+1}.
\label{nu}
\ee
Fusion closure coefficient functions:
\be
 a:=\prod_{j=0}^{p'-1}f^m_j,\qquad 
 \tilde{a}:=e^{\ir\theta}\prod_{j=0}^{p'-1} h^m_{j},\qquad
 \theta:= \tfrac12 Nm(p'-p)\pi.
\label{aa}
\ee

%%%%%%%%%%%%%%%%%%%%%%%
\section{Temperley-Lieb loop models}
\label{Sec:TL}
%%%%%%%%%%%%%%%%%%%%%%%

%%%%%%%%%%%%%%%%%%%%%%%
\paragraph{Elementary face operators}
%%%%%%%%%%%%%%%%%%%%%%%
The dense loop model with crossing parameter $\lambda$ can be described by the planar TL 
algebra~\cite{TL,Jones,PRZ0607} generated by the elementary face operators
\be
\psset{unit=.9cm}
\begin{pspicture}[shift=-.42](1,1)
\facegrid{(0,0)}{(1,1)}
\psarc[linewidth=0.025]{-}(0,0){0.16}{0}{90}
\rput(.5,.5){$u$}
\end{pspicture}
\ :=\ 
s_1(-u)\ \begin{pspicture}[shift=-.45](1,1)
\facegrid{(0,0)}{(1,1)}
\rput[bl](0,0){\loopa}
\end{pspicture}
\;+\,s_0(u)\
\begin{pspicture}[shift=-.42](1,1)
\facegrid{(0,0)}{(1,1)}
\rput[bl](0,0){\loopb}
\end{pspicture}
\label{1x1}
\ee
This decomposition is in terms of the two possible internal connectivities with the coefficients indicating the 
corresponding (local Boltzmann) weights. 
The small quarter arc in the lower-left corner is a marker indicating the orientation of the diagrams in the decomposition. 
The weights are given in terms of the spectral parameter $u$
where the chosen normalisation in (\ref{sk}) requires a real crossing parameter to be of the form
\be 
\lambda\in\pi\big(\mathbb{R}\!\setminus\!\mathbb{Z}\big).
\label{laus}
\ee
The corresponding loop fugacity
\be
 \beta=2\cos\lambda= s_2(0)
\label{beta}
\ee 
is the (nonlocal Boltzmann) weight assigned to a (closed) contractible loop.
We note that
\be
 s_k(0)=U_{k-1}(\tfrac{\beta}{2}),\qquad k\in\mathbb{N}_0
\ee
where $U_k(x)$ is the $k$-th Chebyshev polynomial of the second kind.

The elements of the planar TL algebra are called tangles~\cite{Conway,KauffmanLins1994} and are diagrammatic objects 
formed by adding or linking together a number of elementary face operators. Noting that 
\be
\psset{unit=.9cm}
\begin{pspicture}[shift=-.45](1,1)
\facegrid{(0,0)}{(1,1)}
\rput[bl](0,0){\loopa}
\end{pspicture}
\ =\ \begin{pspicture}[shift=-.42](1,1)
\facegrid{(0,0)}{(1,1)}
\psarc[linewidth=0.025]{-}(0,0){0.16}{0}{90}
\rput(.5,.5){$0$}
\end{pspicture}
\ ,\qquad\quad
\begin{pspicture}[shift=-.42](1,1)
\facegrid{(0,0)}{(1,1)}
\rput[bl](0,0){\loopb}
\end{pspicture}
\ =\ \begin{pspicture}[shift=-.42](1,1)
\facegrid{(0,0)}{(1,1)}
\psarc[linewidth=0.025]{-}(0,0){0.16}{0}{90}
\rput(.5,.5){$\lambda$}
\end{pspicture}\ , 
\label{0lambda}
\ee
we stress that individual connectivity diagrams are themselves tangles. 
Tangles are linear combinations of planar connectivity diagrams with an even number of free nodes connected by 
non-intersecting loop segments. Here a node refers to the midpoint 
of an edge of an elementary face operator, and it is said to be free if it is not linked to any other node via the outside of the 
corresponding box. 
In the decomposition of a tangle in terms of connectivity diagrams, closed loops may be formed but every one of them is replaced by a 
multiplicative factor of $\beta$, the loop fugacity. Two tangles are equal if their decompositions in terms of connectivity diagrams are 
identical. A tangle with $2n$ free nodes is called an $n$-tangle and is a (possibly trivial) linear combination of connectivity
diagrams containing exactly $2n$ free nodes.

Multiplication of tangles is performed diagrammatically by linking or gluing together the tangles using non-intersecting loop segments.
Here is an example of a $4$-tangle, with its free nodes indicated by black dots, obtained as the product of two $3$-tangles:
\be
\psset{unit=.82cm}
\begin{pspicture}[shift=-1.7](-1.0,-1.4)(3.5,2.6)
\facegrid{(0,0)}{(1,1)}
\facegrid{(0,-1)}{(1,0)}
\facegrid{(1,-1)}{(2,0)}
\rput(-1,1){\facegrid{(0,0)}{(1,1)}}
\psarc[linewidth=1.5pt,linecolor=blue]{-}(0,1){0.5}{0}{90}
\psarc[linewidth=1.5pt,linecolor=blue]{-}(0,1){-0.5}{0}{90}
\psarc[linewidth=1.5pt,linecolor=blue]{-}(2,-1){0.5}{180}{90}
\psarc[linewidth=0.025]{-}(0,0){0.16}{0}{90}
\psarc[linewidth=0.025]{-}(1,0){-0.16}{0}{90}
\psarc[linewidth=0.025]{-}(-1,2){0.16}{-90}{0}
\psarc[linewidth=0.025]{-}(1,-1){0.16}{0}{90}
\psline[linewidth=1.5pt,linecolor=blue]{-}(1,0.5)(1.3,0.8)
\psline[linewidth=1.5pt,linecolor=blue]{-}(1.5,0)(1.8,0.3)
\psdots(0.5,-1)(-0.5,2)(0,-0.5)(-1,1.5)
\psdots[linecolor=blue](1,0.5)(1.5,0)
\rput(0.5,0.5){\small$u$}
\rput(0.5,-0.5){\small$w$}
\rput(-0.5,1.5){\small${2\lambda}$}
\rput(1.5,-0.5){\small$v$}
\rput(0.3,0.3){
\facegrid{(1,0)}{(3,1)}
\rput(1.5,0.5){\small$0$}
\rput(1.5,0.5){\small$0$}
\rput(2.5,0.5){\small${u\!+\!\lambda}$}
\psarc[linewidth=0.025]{-}(1,1){0.16}{-90}{0}
\rput{45}(2,1){\facegrid{(0,0)}{(1,1)}}
\rput(2,1.707){\small${v\!-\!\lambda}$}
\psarc[linewidth=0.025]{-}(2,1){0.16}{45}{135}
\psarc[linewidth=0.025]{-}(3,0){0.16}{90}{180}
\psarc[linewidth=1.5pt,linecolor=blue]{-}(2,1){0.5}{0}{45}
\psarc[linewidth=1.5pt,linecolor=blue]{-}(2,1){0.5}{135}{180}
\psdots(1.64645,2.06066)(2.354,2.06066)(2.5,0)(3,0.5)
\psdots[linecolor=blue](1,0.5)(1.5,0)
}
\end{pspicture}
\ee
where the extra free nodes of the two $3$-tangles are indicated in blue. 
In the decomposition of this $4$-tangle, the particular connectivity diagram
\be
\psset{unit=.82cm}
\begin{pspicture}[shift=-1.7](-1.0,-1.4)(3.5,2.6)
\facegrid{(0,0)}{(1,1)}
\facegrid{(0,-1)}{(1,0)}
\facegrid{(1,-1)}{(2,0)}
\rput(-1,1){\facegrid{(0,0)}{(1,1)}}
\psarc[linewidth=1.5pt,linecolor=blue]{-}(0,1){0.5}{0}{90}
\psarc[linewidth=1.5pt,linecolor=blue]{-}(0,1){-0.5}{0}{90}
\psarc[linewidth=1.5pt,linecolor=blue]{-}(2,-1){0.5}{180}{90}
\psline[linewidth=1.5pt,linecolor=blue]{-}(1,0.5)(1.3,0.8)
\psline[linewidth=1.5pt,linecolor=blue]{-}(1.5,0)(1.8,0.3)
\psdots(0.5,-1)(-0.5,2)(0,-0.5)(-1,1.5)
\psdots[linecolor=blue](1,0.5)(1.5,0)
\rput(0,0){\loopb}
\rput(0,-1){\loopb}
\rput(-1,1){\loopa}
\rput(1,-1){\loopa}
\rput(0.3,0.3){
\facegrid{(1,0)}{(3,1)}
\rput(1,0){\loopb}
\rput(2,0){\loopa}
\rput{45}(2,1){\facegrid{(0,0)}{(1,1)}}
\rput(2,1.707){\psarc[linewidth=1.5pt,linecolor=blue]{-}(0,0.707){0.5}{-135}{-45}
\psarc[linewidth=1.5pt,linecolor=blue]{-}(0,-0.707){0.5}{45}{135}}
\psarc[linewidth=1.5pt,linecolor=blue]{-}(2,1){0.5}{0}{45}
\psarc[linewidth=1.5pt,linecolor=blue]{-}(2,1){0.5}{135}{180}
\psdots(1.64645,2.06066)(2.354,2.06066)(2.5,0)(3,0.5)
\psdots[linecolor=blue](1,0.5)(1.5,0)
}
\end{pspicture}
\ee
appears with decomposition coefficient
\be
 s_0(2\lambda)s_0(u)s_0(w)s_1(-v)s_0(u+\lambda)s_0(v-\lambda)
  =-\beta s_0(u)s_1(u)[s_1(-v)]^2s_0(w).
\label{s0s0}
\ee
The removal of the three closed loops introduces an additional factor of $\beta^3$.

If the $2n$ free nodes of an $n$-tangle can be linked without intersections to $2n$ distinct points on a circle encircling the tangle, 
a direction of transfer can be defined by selecting $n$ of the free nodes, linked to $n$ consecutive points on the circle,
to represent an instate direction. In this case, the remaining $n$ (likewise consecutive) nodes represent the corresponding outstate 
direction. As tangles can also be defined on more complicated geometries such as annuli and cylinders, the notion of a direction of 
transfer has to be adjusted in some cases.

%%%%%%%%%%%%%%%%%%%%%%%%%%
\paragraph{Linear Temperley-Lieb algebras}
%%%%%%%%%%%%%%%%%%%%%%%%%%
With the direction of transfer fixed, a planar $N$-tangle can be expressed in terms of words of the ordinary (linear) TL algebra
\be
 TL_N(\beta)=\big\langle I,\,e_j ;\,j=1,\ldots,N-1\big\rangle,\qquad
 I=\!
\begin{pspicture}[shift=-0.55](-0.1,-0.65)(2.0,0.45)
\rput(1.4,0.0){\small$...$}
\psline[linecolor=blue,linewidth=1.5pt]{-}(0.2,0.35)(0.2,-0.35)\rput(0.2,-0.55){$_1$}
\psline[linecolor=blue,linewidth=1.5pt]{-}(0.6,0.35)(0.6,-0.35)\rput(0.6,-0.55){$_2$}
\psline[linecolor=blue,linewidth=1.5pt]{-}(1.0,0.35)(1.0,-0.35)\rput(1.0,-0.55){$_3$}
\psline[linecolor=blue,linewidth=1.5pt]{-}(1.8,0.35)(1.8,-0.35)\rput(1.8,-0.55){$_N$}
\end{pspicture} 
 ,\qquad
 e_j=\!
 \begin{pspicture}[shift=-0.55](-0.1,-0.65)(3.2,0.45)
\rput(0.6,0.0){\small$...$}
\rput(2.6,0.0){\small$...$}
\psline[linecolor=blue,linewidth=1.5pt]{-}(0.2,0.35)(0.2,-0.35)\rput(0.2,-0.55){$_1$}
\psline[linecolor=blue,linewidth=1.5pt]{-}(1.0,0.35)(1.0,-0.35)
\psline[linecolor=blue,linewidth=1.5pt]{-}(2.2,0.35)(2.2,-0.35)
\psline[linecolor=blue,linewidth=1.5pt]{-}(3.0,0.35)(3.0,-0.35)\rput(3.0,-0.55){$_{N}$}
\psarc[linecolor=blue,linewidth=1.5pt]{-}(1.6,0.35){0.2}{180}{0}\rput(1.35,-0.55){$_j$}
\psarc[linecolor=blue,linewidth=1.5pt]{-}(1.6,-0.35){0.2}{0}{180}\rput(1.85,-0.55){$_{j+1}$}
\end{pspicture} 
\ee
whose defining relations are
\be
 IA=AI=A,\qquad e_j^2=\beta e_j, \qquad e_j e_{j\pm1} e_j = e_j, \qquad e_i e_j = e_j e_i \qquad (|i-j|>1),
\label{eq:TLdef}
\ee
where $A$ is any element of $TL_N(\beta)$.
Multiplication in the diagrammatic realisation is by vertical concatenation placing $e_j$ atop $e_i$ to form the product $e_ie_j$,
for example. The diagrammatic realisation is faithful in the sense that the ensuing diagrammatic algebra is
isomorphic to the TL algebra. 

In the cylinder case, the relevant algebra is the enlarged periodic TL algebra, 
\be
 \mathcal EPTL_N(\alpha, \beta) = \big\langle I, \,\Omega, \,\Omega^{-1},\,e_j;\,j=1,\ldots,N\big\rangle,\qquad 
\ee 
where the three new generators are realised as
\be
\begin{pspicture}[shift=-0.55](-1.05,-0.55)(2.6,0.35)
\rput(-0.6,-0.04){$e_N=\,$}
\rput(1.4,0.0){\small$...$}
\rput(0.2,-0.55){$_1$}\rput(0.6,-0.55){$_2$}\rput(1.0,-0.55){$_3$}\rput(2.2,-0.55){$_N$}
\psarc[linecolor=blue,linewidth=1.5pt]{-}(0.0,0.35){0.2}{-90}{0}
\psarc[linecolor=blue,linewidth=1.5pt]{-}(0.0,-0.35){0.2}{0}{90}
\psline[linecolor=blue,linewidth=1.5pt]{-}(0.6,0.35)(0.6,-0.35)
\psline[linecolor=blue,linewidth=1.5pt]{-}(1.0,0.35)(1.0,-0.35)
\psline[linecolor=blue,linewidth=1.5pt]{-}(1.8,0.35)(1.8,-0.35)
\psarc[linecolor=blue,linewidth=1.5pt]{-}(2.4,-0.35){0.2}{90}{180}
\psarc[linecolor=blue,linewidth=1.5pt]{-}(2.4,0.35){0.2}{180}{-90}
\end{pspicture} 
\qquad\quad
\begin{pspicture}[shift=-0.55](-0.7,-0.55)(2.0,0.35)
\rput(0.2,-0.55){$_1$}\rput(0.6,-0.55){$_2$}\rput(1.0,-0.55){$_3$}\rput(1.4,-0.55){\small$...$}\rput(1.8,-0.55){$_N$}
\multiput(0,0)(0.4,0){6}{\psbezier[linecolor=blue,linewidth=1.5pt]{-}(-0.2,-0.35)(-0.2,-0.0)(0.2,0.0)(0.2,0.35)}
\psframe[fillstyle=solid,linecolor=white,linewidth=0pt](-0.3,-0.4)(0,0.4)
\psframe[fillstyle=solid,linecolor=white,linewidth=0pt](2.0,-0.4)(2.4,0.4)
\rput(-0.55,0.045){$\Omega=$}
\end{pspicture} 
\qquad\quad
\begin{pspicture}[shift=-0.55](-1.3,-0.55)(2.0,0.35)
\rput(0.2,-0.55){$_1$}\rput(0.6,-0.55){$_2$}\rput(1.0,-0.55){$_3$}\rput(1.4,-0.55){\small$...$}\rput(1.8,-0.55){$_N$}
\multiput(0,0)(0.4,0){6}{\psbezier[linecolor=blue,linewidth=1.5pt]{-}(-0.2,0.35)(-0.2,-0.0)(0.2,0.0)(0.2,-0.35)}
\psframe[fillstyle=solid,linecolor=white,linewidth=0pt](-0.3,-0.4)(0,0.4)
\psframe[fillstyle=solid,linecolor=white,linewidth=0pt](2.0,-0.4)(2.4,0.4)
\rput(-0.75,0.07){$\Omega^{-1}=$}
\end{pspicture} 
\ee
with periodic boundary conditions identifying the matching left and right nodes, thereby forming connectivity diagrams on the surface 
of a cylinder. The vertical line along which the identification is taking place is referred to as the {\em virtual boundary}.
The defining relations of $\mathcal EPTL_N(\alpha, \beta)$ are given by \eqref{eq:TLdef} supplemented by
\begin{alignat}{3}
\Omega e_i \Omega^{-1} &\,=\, e_{i-1},&&\nonumber\\
\Omega \Omega^{-1} &\,=\, \Omega^{-1} \Omega = I, \nonumber\\
(\Omega^{\pm 1} e_N)^{N-1} &\,=\, \Omega^{\pm N} (\Omega ^{\pm 1} e_N), 
\label{eq:EPTLdef}\\
\Omega^{\pm N} e_N \Omega^{\mp N} &\,=\, e_N,\nonumber\\
E \Omega^{\pm 1} E &\,=\, \alpha E, && 
E:=\, e_2e_4\ldots e_{N-2}e_N\qquad (N\ \mathrm{even}), \nonumber 
\end{alignat}
where the subscripts are defined modulo $N$. The parameter $\alpha$ is seen to be the (nonlocal Boltzmann) 
weight assigned to non-contractible loops formed around the cylinder. Note that such loops can only appear for $N$ even.
As for the linear TL algebra above, 
the diagrammatic realisation is faithful in the sense that the ensuing diagrammatic algebra is isomorphic to the 
enlarged periodic TL algebra. The enlarged periodic TL algebra is the quotient of the affine TL 
algebra~\cite{MartinSaleur1993,GrahamLehrer1998,Green1998,ErdmannGreen1999} by the last relation in (\ref{eq:EPTLdef}).

Even though vertical can be chosen as the natural direction of transfer in many of our considerations, 
see (\ref{Du}) and (\ref{Tu}) for example, it is often more convenient
to work in the planar or cylindrical setting without reference to any particular direction of transfer.

%%%%%%%%%%%%
\paragraph{Link states}
%%%%%%%%%%%%

A natural way to produce matrix representations of TL algebras is to let the tangles act on a suitable vector space of link states. 

A planar link state on $N$ nodes is a planar diagram that encodes connectivities between $N$ nodes equally spaced on a horizontal 
line. It consists of $d\in\{0,\ldots,N\}$ vertical loop segments (called defects) attached to individual nodes and $\frac{N-d}{2}$  
half-arcs, where $N-d=0$ mod $2$, connecting nodes pairwise by loop segments that live above the horizontal line
and cannot over-arch defects. 

We distinguish between planar link states and cylindrical link states.
An example of a planar link state on $N=10$ nodes with $d=2$ defects is given by
\be
\begin{pspicture}[shift=-0.35](-0.0,-0.1)(4.0,0.8)
\psline[linewidth=0.5pt]{-}(-0.2,0)(3.8,0)
\psarc[linewidth=1.5pt,linecolor=coloroflink]{-}(0.2,0){0.2}{0}{180}
\psarc[linewidth=1.5pt,linecolor=coloroflink]{-}(1.8,0){0.2}{0}{180}
\psarc[linewidth=1.5pt,linecolor=coloroflink]{-}(2.6,0){0.2}{0}{180}
\psbezier[linewidth=1.5pt,linecolor=coloroflink]{-}(1.2,0)(1.2,1)(3.2,1)(3.2,0)
\psline[linewidth=1.5pt,linecolor=coloroflink]{-}(0.8,0)(0.8,0.6)
\psline[linewidth=1.5pt,linecolor=coloroflink]{-}(3.6,0)(3.6,0.6)
\end{pspicture} 
\label{eq:stripLS}
\vspace{-0.0cm}
\ee 
while an example of a cylindrical link state on $N=10$ nodes with $d=4$ defects is given by
\be
\begin{pspicture}[shift=-0.35](-0.0,-0.1)(4.0,0.8)
\psline[linewidth=0.5pt]{-}(-0.2,0)(3.8,0)
\psarc[linewidth=1.5pt,linecolor=coloroflink]{-}(1.4,0){0.2}{0}{180}
\psarc[linewidth=1.5pt,linecolor=coloroflink]{-}(-0.2,0){0.2}{0}{90}
\psarc[linewidth=1.5pt,linecolor=coloroflink]{-}(-0.2,0){0.6}{0}{90}
\psarc[linewidth=1.5pt,linecolor=coloroflink]{-}(3.8,0){0.2}{90}{180}
\psarc[linewidth=1.5pt,linecolor=coloroflink]{-}(3.8,0){0.6}{90}{180}
\psline[linewidth=1.5pt,linecolor=coloroflink]{-}(0.8,0)(0.8,0.6)
\psline[linewidth=1.5pt,linecolor=coloroflink]{-}(2.0,0)(2.0,0.6)
\psline[linewidth=1.5pt,linecolor=coloroflink]{-}(2.4,0)(2.4,0.6)
\psline[linewidth=1.5pt,linecolor=coloroflink]{-}(2.8,0)(2.8,0.6)
\end{pspicture}
\label{eq:cylinderLS}
\vspace{-0.2cm}
\ee
Letting $V_N^d$ and $\tilde{V}_N^d$ respectively denote the linear spans of planar and cylindrical link states on 
$N$ nodes with $d$ defects, we have
\be
 \dim V_N^d=\left(\!\!\!\begin{array}{c} N \\ \frac{N-d}{2}\end{array}\!\!\!\right)
  -\left(\!\!\!\begin{array}{c} N \\ \frac{N-d-2}{2}\end{array}\!\!\!\right),\qquad 
  \dim \tilde{V}_N^d=\left(\!\!\!\begin{array}{c} N \\ \frac{N-d}{2}\end{array}\!\!\!\right).
\label{eq:LinkCounting}
\ee
We say that the planar and cylindrical link states form {\it canonical bases} of $V_N^d$ and $\tilde{V}_N^d$, respectively. 
To obtain concrete matrix representations, one still has to identify the subset of link states the tangles are meant to act on and 
specify the action itself. A particularly simple prescription yields the so-called standard modules reviewed in the following.

%%%%%%%%%%%%%%%%%%%%
\paragraph{Standard modules}
%%%%%%%%%%%%%%%%%%%%%

Both on the strip and cylinder, the simplest nontrivial modules are the {\em standard modules}.
For $N$ fixed, they yield representations of $TL_N(\beta)$ and $\mathcal EPTL_N(\alpha, \beta)$ on $\textrm{End}(V_N^d)$ and 
$\textrm{End}(\tilde V_N^d)$, respectively, with $d = 0, \dots, N$ subject to $d = N \,\textrm{mod}\, 2$.

For the strip case, let $c$ be a connectivity (diagrammatic realisation of a word) in 
$TL_N(\beta)$ and $w$ be a link state in the canonical basis of $V_N^d$. 
Viewing $w$ as the instate (see the discussion following 
(\ref{s0s0})), the action of $c$ on $w$ is computed from the composition diagram obtained by 
placing $w$ atop $c$ and linking the nodes of $w$ to the $N$ free nodes on the upper edge of $c$. 
If the result is not in $V_N^d$, it is readily set to zero. Otherwise,
the result $cw$ is a multiple scalar of a link state in the basis of $V_N^{d}$, called the outstate, and the scalar is given by 
$\beta^\#$ where $\#$ is the number of (closed) loops in the composition diagram.
By construction, this action preserves the number of defects. It is linearly extended to any tangle in $TL_N(\beta)$ and any element of 
$V_N^d$, and it is well known that it generates a representation of $TL_N(\beta)$, see~\cite{MartinBook,RSA1204}, for example. 
The matrix representative of $c \in TL_N(\beta)$ on the standard module with $d$ defects is denoted by $\rho_N^d(c)$.
We illustrate the action of connectivities on link states in the definition of standard modules on the strip for $N=10$ with the
two examples
\be
\begin{pspicture}[shift=-0.75](-0.0,-1.3)(4.0,0.8)
\psline[linewidth=0.5pt]{-}(-0.2,0)(3.8,0)
\psline[linewidth=0.5pt]{-}(-0.2,-1.2)(3.8,-1.2)
\psarc[linewidth=1.5pt,linecolor=coloroflink]{-}(0.2,0){0.2}{0}{180}
\psarc[linewidth=1.5pt,linecolor=coloroflink]{-}(1.8,0){0.2}{0}{180}
\psarc[linewidth=1.5pt,linecolor=coloroflink]{-}(2.6,0){0.2}{0}{180}
\psbezier[linewidth=1.5pt,linecolor=coloroflink]{-}(1.2,0)(1.2,1)(3.2,1)(3.2,0)
\psline[linewidth=1.5pt,linecolor=coloroflink]{-}(0.8,0)(0.8,0.6)
\psline[linewidth=1.5pt,linecolor=coloroflink]{-}(3.6,0)(3.6,0.6)
\psarc[linewidth=1.5pt,linecolor=blue]{-}(0.2,0){-0.2}{0}{180}
\psarc[linewidth=1.5pt,linecolor=blue]{-}(2.2,0){-0.2}{0}{180}
\psbezier[linewidth=1.5pt,linecolor=blue]{-}(1.6,0)(1.6,-0.7)(2.8,-0.7)(2.8,0)
\psbezier[linewidth=1.5pt,linecolor=blue]{-}(0.8,0)(0.8,-0.6)(0,-0.6)(0,-1.2)
\psbezier[linewidth=1.5pt,linecolor=blue]{-}(1.2,0)(1.2,-0.6)(0.4,-0.6)(0.4,-1.2)
\psarc[linewidth=1.5pt,linecolor=blue]{-}(1.4,-1.2){0.2}{0}{180}
\psarc[linewidth=1.5pt,linecolor=blue]{-}(2.6,-1.2){0.2}{0}{180}
\psline[linewidth=1.5pt,linecolor=blue]{-}(3.6,0)(3.6,-1.2)
\psline[linewidth=1.5pt,linecolor=blue]{-}(3.2,0)(3.2,-1.2)
\psbezier[linewidth=1.5pt,linecolor=blue]{-}(0.8,-1.2)(0.8,-0.5)(2.0,-0.5)(2.0,-1.2)
\end{pspicture} = \beta^2
\begin{pspicture}[shift=-0.25](-0.3,-0.1)(4.0,0.8)
\psline[linewidth=0.5pt]{-}(-0.2,0)(3.8,0)
\psarc[linewidth=1.5pt,linecolor=coloroflink]{-}(1.4,0){0.2}{0}{180}
\psarc[linewidth=1.5pt,linecolor=coloroflink]{-}(2.6,0){0.2}{0}{180}
\psbezier[linewidth=1.5pt,linecolor=coloroflink]{-}(0.8,0)(0.8,0.7)(2,0.7)(2,0)
\psbezier[linewidth=1.5pt,linecolor=coloroflink]{-}(0.4,0)(0.4,1.2)(3.2,1.2)(3.2,0)
\psline[linewidth=1.5pt,linecolor=coloroflink]{-}(0.0,0)(0.0,0.6)
\psline[linewidth=1.5pt,linecolor=coloroflink]{-}(3.6,0)(3.6,0.6)
\end{pspicture} 
\hspace{1.5cm}
\begin{pspicture}[shift=-0.75](-0.0,-1.3)(4.0,0.8)
\psline[linewidth=0.5pt]{-}(-0.2,0)(3.8,0)
\psline[linewidth=0.5pt]{-}(-0.2,-1.2)(3.8,-1.2)
\psarc[linewidth=1.5pt,linecolor=coloroflink]{-}(1.4,0){0.2}{0}{180}
\psarc[linewidth=1.5pt,linecolor=coloroflink]{-}(0.2,0){0.2}{0}{180}
\psarc[linewidth=1.5pt,linecolor=coloroflink]{-}(3.4,0){0.2}{0}{180}
\psline[linewidth=1.5pt,linecolor=coloroflink]{-}(0.8,0)(0.8,0.6)
\psline[linewidth=1.5pt,linecolor=coloroflink]{-}(2.0,0)(2.0,0.6)
\psline[linewidth=1.5pt,linecolor=coloroflink]{-}(2.4,0)(2.4,0.6)
\psline[linewidth=1.5pt,linecolor=coloroflink]{-}(2.8,0)(2.8,0.6)
\psline[linewidth=1.5pt,linecolor=coloroflink]{-}(0.8,0)(0.8,0.6)
\psarc[linewidth=1.5pt,linecolor=blue]{-}(0.2,0){-0.2}{0}{180}
\psarc[linewidth=1.5pt,linecolor=blue]{-}(2.2,0){-0.2}{0}{180}
\psbezier[linewidth=1.5pt,linecolor=blue]{-}(1.6,0)(1.6,-0.7)(2.8,-0.7)(2.8,0)
\psbezier[linewidth=1.5pt,linecolor=blue]{-}(0.8,0)(0.8,-0.6)(0,-0.6)(0,-1.2)
\psbezier[linewidth=1.5pt,linecolor=blue]{-}(1.2,0)(1.2,-0.6)(0.4,-0.6)(0.4,-1.2)
\psarc[linewidth=1.5pt,linecolor=blue]{-}(1.4,-1.2){0.2}{0}{180}
\psarc[linewidth=1.5pt,linecolor=blue]{-}(2.6,-1.2){0.2}{0}{180}
\psline[linewidth=1.5pt,linecolor=blue]{-}(3.6,0)(3.6,-1.2)
\psline[linewidth=1.5pt,linecolor=blue]{-}(3.2,0)(3.2,-1.2)
\psbezier[linewidth=1.5pt,linecolor=blue]{-}(0.8,-1.2)(0.8,-0.5)(2.0,-0.5)(2.0,-1.2)
\end{pspicture} = 0. 
\label{eq:standardstrip}
\vspace{-0.0cm}
\ee 

For standard modules on the cylinder, the action of a connectivity $c\in\mathcal EPTL_N(\alpha, \beta)$ on an element 
$w$ of the canonical basis of $\tilde V_N^d$ is likewise computed by the connection procedure described above for the strip case. 
The result $cw$ is a basis link state in $\tilde V_N^{d}$ multiplied by 
$\alpha^{\#(\alpha)} \beta^{\#(\beta)}$, where $\#(\alpha)$ and $\#(\beta)$ are the respective
numbers of non-contractible and contractible loops in the ensuing composition diagram, 
whereas it is zero if defects have been annihilated in the diagram. 
As an illustration, we have
\be
\begin{pspicture}[shift=-0.85](-0.0,-1.6)(4.0,1.0)
\psline[linewidth=0.5pt]{-}(-0.2,0)(3.8,0)
\psline[linewidth=0.5pt]{-}(-0.2,-1.5)(3.8,-1.5)
\psarc[linewidth=1.5pt,linecolor=blue]{-}(0.6,0){-0.2}{0}{180}
\psbezier[linewidth=1.5pt,linecolor=coloroflink]{-}(1.6,0)(1.6,0.7)(2.8,0.7)(2.8,0)
\psarc[linewidth=1.5pt,linecolor=coloroflink]{-}(2.2,0){0.2}{0}{180}
\multiput(0,0)(4,0){2}{\psbezier[linewidth=1.5pt,linecolor=coloroflink]{-}(-0.8,0)(-0.8,1.1)(1.2,1.1)(1.2,0)
\psbezier[linewidth=1.5pt,linecolor=coloroflink]{-}(-0.4,0)(-0.4,0.7)(0.8,0.7)(0.8,0)
\psbezier[linewidth=1.5pt,linecolor=blue]{-}(-1.6,0)(-1.6,-1.0)(2.0,-1.0)(2.0,0)}
\psarc[linewidth=1.5pt,linecolor=blue]{-}(-0.2,0){-0.2}{0}{180}
\psarc[linewidth=1.5pt,linecolor=blue]{-}(3.8,0){-0.2}{0}{180}
\psarc[linewidth=1.5pt,linecolor=coloroflink]{-}(0.2,0){0.2}{0}{180}
\psarc[linewidth=1.5pt,linecolor=blue]{-}(1.4,0){-0.2}{0}{180}
\psarc[linewidth=1.5pt,linecolor=blue]{-}(3.0,0){-0.2}{0}{180}
\rput(0,-1.5){
\psarc[linewidth=1.5pt,linecolor=blue]{-}(0.2,0){0.2}{0}{180}
\psarc[linewidth=1.5pt,linecolor=blue]{-}(1.8,0){0.2}{0}{180}
\psarc[linewidth=1.5pt,linecolor=blue]{-}(3.0,0){0.2}{0}{180}
\psbezier[linewidth=1.5pt,linecolor=blue]{-}(1.2,0)(1.2,0.7)(2.4,0.7)(2.4,0)
\multiput(0,0)(4,0){2}{\psbezier[linewidth=1.5pt,linecolor=blue]{-}(-0.4,0)(-0.4,0.7)(0.8,0.7)(0.8,0)}
}
\psframe[fillstyle=solid,linecolor=white,linewidth=0pt](-2.1,-1.5)(-0.2,1.0)
\psframe[fillstyle=solid,linecolor=white,linewidth=0pt](3.8,-1.5)(6.1,1.0)
\end{pspicture} 
\begin{pspicture}[shift=-0.25](-1.6,-0.1)(4.0,0.8)
\psline[linewidth=0.5pt]{-}(-0.2,0)(3.8,0)
\psarc[linewidth=1.5pt,linecolor=coloroflink]{-}(0.2,0){0.2}{0}{180}
\psarc[linewidth=1.5pt,linecolor=coloroflink]{-}(1.8,0){0.2}{0}{180}
\psarc[linewidth=1.5pt,linecolor=coloroflink]{-}(3.0,0){0.2}{0}{180}
\psbezier[linewidth=1.5pt,linecolor=coloroflink]{-}(1.2,0)(1.2,0.7)(2.4,0.7)(2.4,0)
\multiput(0,0)(4,0){2}{\psbezier[linewidth=1.5pt,linecolor=coloroflink]{-}(-0.4,0)(-0.4,0.7)(0.8,0.7)(0.8,0)}
\psframe[fillstyle=solid,linecolor=white,linewidth=0pt](-0.45,-0.1)(-0.2,0.5)
\psframe[fillstyle=solid,linecolor=white,linewidth=0pt](3.8,-0.1)(4.9,0.6)
\rput(-0.95,0.27){$=\alpha^2\beta$}
\end{pspicture} 
\label{eq:standardcylinder}
\ee 
This action defines a representation of $\mathcal EPTL_N(\alpha, \beta)$.
The matrix representative of $c \in \mathcal EPTL_N(\alpha, \beta)$ on the standard module with $d$ defects is denoted by 
$\tilde \rho_N^d(c)$.

%%%%%%%%%%%%%%%%%%%%%%%%%%
\paragraph{Fundamental tangle relations}
%%%%%%%%%%%%%%%%%%%%%%%%%
The face operators enjoy the crossing relations
\be
\psset{unit=.9cm}
\begin{pspicture}[shift=-.42](1,1)
\facegrid{(0,0)}{(1,1)}
\psarc[linewidth=0.025]{-}(0,0){0.16}{0}{90}
\rput(.5,.5){$u$}
\end{pspicture}
\ \ =\ \ 
\begin{pspicture}[shift=-.42](1,1)
\facegrid{(0,0)}{(1,1)}
\psarc[linewidth=0.025]{-}(1,0){0.16}{90}{180}
\rput(.5,.5){$\lambda\!-\!u$}
\end{pspicture}
\ \ =\ \ 
\begin{pspicture}[shift=-.42](1,1)
\facegrid{(0,0)}{(1,1)}
\psarc[linewidth=0.025]{-}(1,1){0.16}{180}{270}
\rput(.5,.5){$u$}
\end{pspicture}
\ \ =\ \ 
\begin{pspicture}[shift=-.42](1,1)
\facegrid{(0,0)}{(1,1)}
\psarc[linewidth=0.025]{-}(0,1){0.16}{-90}{0}
\rput(.5,.5){$\lambda\!-\!u$}
\end{pspicture}
\ \; , 
\label{crossing11}
\ee
commute in the sense that
\be
\psset{unit=0.6364cm}
\begin{pspicture}[shift=-0.95](-0.1,-1.1)(4.3,1.1)
\pspolygon[fillstyle=solid,fillcolor=lightlightblue](0,0)(1,1)(3,-1)(4,0)(3,1)(1,-1)
\psarc[linewidth=1.5pt,linecolor=blue](2,0){.707}{45}{135}
\psarc[linewidth=1.5pt,linecolor=blue](2,0){.707}{-135}{-45}
\psarc[linewidth=0.025]{-}(0,0){0.21}{-45}{45}
\psarc[linewidth=0.025]{-}(2,0){0.21}{-45}{45}
\rput(1,0){$u$}
\rput(3,0){$v$}
\end{pspicture}
=  \, \, \,
\begin{pspicture}[shift=-0.95](-0.1,-1.1)(4.3,1.1)
\pspolygon[fillstyle=solid,fillcolor=lightlightblue](0,0)(1,1)(3,-1)(4,0)(3,1)(1,-1)
\psarc[linewidth=1.5pt,linecolor=blue](2,0){.707}{45}{135}
\psarc[linewidth=1.5pt,linecolor=blue](2,0){.707}{-135}{-45}
\psarc[linewidth=0.025]{-}(0,0){0.21}{-45}{45}
\psarc[linewidth=0.025]{-}(2,0){0.21}{-45}{45}
\rput(1,0){$v$}
\rput(3,0){$u$}
\end{pspicture}
\label{eq:comm}
\ee

\noindent
and satisfy the local inversion relation
\be
\psset{unit=0.6364cm}
\begin{pspicture}[shift=-0.95](-0.1,-1.1)(4.3,1.1)
\pspolygon[fillstyle=solid,fillcolor=lightlightblue](0,0)(1,1)(3,-1)(4,0)(3,1)(1,-1)
\psarc[linewidth=1.5pt,linecolor=blue](2,0){.707}{45}{135}
\psarc[linewidth=1.5pt,linecolor=blue](2,0){.707}{-135}{-45}
\psarc[linewidth=0.025]{-}(0,0){0.21}{-45}{45}
\psarc[linewidth=0.025]{-}(2,0){0.21}{-45}{45}
\rput(1,0){$u$}
\rput(3,0){$-u$}
\end{pspicture}
= \ s_1(u)s_1(-u) \, \, \,
\begin{pspicture}[shift=-0.95](-0.1,-1.1)(2.3,1.1)
\pspolygon[fillstyle=solid,fillcolor=lightlightblue](0,0)(1,1)(2,0)(1,-1)(0,0)
\psarc[linewidth=1.5pt,linecolor=blue](1,-1){.707}{45}{135}
\psarc[linewidth=1.5pt,linecolor=blue](1,1){.707}{-135}{-45}
\end{pspicture}\ \ \,
\label{eq:inve}
\ee

\noindent 
the Yang-Baxter equation (YBE)
\be
\psset{unit=.82cm}
\begin{pspicture}[shift=-0.9](0,0)(3,2)
\facegrid{(2,0)}{(3,2)}
\pspolygon[fillstyle=solid,fillcolor=lightlightblue](0,1)(1,2)(2,1)(1,0)(0,1)
\psarc[linewidth=0.025]{-}(2,0){0.16}{0}{90}
\psarc[linewidth=0.025]{-}(2,1){0.16}{0}{90}
\psarc[linewidth=0.025]{-}(0,1){0.21}{-45}{45}
\rput(2.5,.5){$u$}
\rput(2.5,1.5){$v$}
\rput(1,1){$u-v$}
\psline[linecolor=blue,linewidth=1.5pt]{-}(2,0.5)(1.5,0.5)
\psline[linecolor=blue,linewidth=1.5pt]{-}(2,1.5)(1.5,1.5)
\end{pspicture}
\ \ =\ \ 
\begin{pspicture}[shift=-0.9](0,0)(3,2)
\facegrid{(0,0)}{(1,2)}
\pspolygon[fillstyle=solid,fillcolor=lightlightblue](1,1)(2,2)(3,1)(2,0)(1,1)
\psarc[linewidth=0.025]{-}(0,0){0.16}{0}{90}
\psarc[linewidth=0.025]{-}(0,1){0.16}{0}{90}
\psarc[linewidth=0.025]{-}(1,1){0.21}{-45}{45}
\rput(.5,.5){$v$}
\rput(.5,1.5){$u$}
\rput(2,1){$u-v$}
\psline[linecolor=blue,linewidth=1.5pt]{-}(1,0.5)(1.5,0.5)
\psline[linecolor=blue,linewidth=1.5pt]{-}(1,1.5)(1.5,1.5)
\end{pspicture}\ \
\label{eq:YBE}
\ee

\noindent 
and the boundary Yang-Baxter equations (BYBEs)
\bea
\psset{unit=0.6364cm}
\begin{pspicture}[shift=-1.9](-3,0)(0,4)
\psarc[linecolor=blue,linewidth=1.5pt]{-}(-2.8,1){-0.5}{-90}{90}
\psarc[linecolor=blue,linewidth=1.5pt]{-}(-2.8,3){-0.5}{-90}{90}
\psline[linecolor=blue,linewidth=1.5pt]{-}(0,0.5)(-0.6,0.5)
\psline[linecolor=blue,linewidth=1.5pt]{-}(0,1.5)(-0.6,1.5)
\psline[linecolor=blue,linewidth=1.5pt]{-}(0,2.5)(-1.6,2.5)
\psline[linecolor=blue,linewidth=1.5pt]{-}(0,3.5)(-2.8,3.5)
\psline[linecolor=blue,linewidth=1.5pt]{-}(-1.4,0.5)(-2.8,0.5)
\psline[linecolor=blue,linewidth=1.5pt]{-}(-2.4,1.5)(-2.8,1.5)
\psline[linecolor=blue,linewidth=1.5pt]{-}(-2.4,2.5)(-2.8,2.5)
\pspolygon[fillstyle=solid,fillcolor=lightlightblue](0,1)(-1,2)(-2,1)(-1,0)(0,1)
\pspolygon[fillstyle=solid,fillcolor=lightlightblue](-1,2)(-2,1)(-3,2)(-2,3)(-1,2)
\psline{-}(0,0)(0,4)
\psarc[linewidth=0.025]{-}(-2,1){0.21}{-45}{45}\rput(-1,1){$u$}
\psarc[linewidth=0.025]{-}(-3,2){0.21}{-45}{45}\rput(-2,2){$v$}
\end{pspicture}
\ \ \ =\ \ \ \
\begin{pspicture}[shift=-1.9](-3,0)(0,4)
\psarc[linecolor=blue,linewidth=1.5pt]{-}(-2.8,1){-0.5}{-90}{90}
\psarc[linecolor=blue,linewidth=1.5pt]{-}(-2.8,3){-0.5}{-90}{90}
\psline[linecolor=blue,linewidth=1.5pt]{-}(0,2.5)(-0.6,2.5)
\psline[linecolor=blue,linewidth=1.5pt]{-}(0,3.5)(-0.6,3.5)
\psline[linecolor=blue,linewidth=1.5pt]{-}(0,1.5)(-1.6,1.5)
\psline[linecolor=blue,linewidth=1.5pt]{-}(0,0.5)(-2.8,0.5)
\psline[linecolor=blue,linewidth=1.5pt]{-}(-1.4,3.5)(-2.8,3.5)
\psline[linecolor=blue,linewidth=1.5pt]{-}(-2.4,2.5)(-2.8,2.5)
\psline[linecolor=blue,linewidth=1.5pt]{-}(-2.4,1.5)(-2.8,1.5)
\pspolygon[fillstyle=solid,fillcolor=lightlightblue](0,3)(-1,4)(-2,3)(-1,2)(0,3)
\pspolygon[fillstyle=solid,fillcolor=lightlightblue](-1,2)(-2,1)(-3,2)(-2,3)(-1,2)
\psline{-}(0,0)(0,4)
\psarc[linewidth=0.025]{-}(-2,3){0.21}{-45}{45}\rput(-1,3){$u$}
\psarc[linewidth=0.025]{-}(-3,2){0.21}{-45}{45}\rput(-2,2){$v$}
\end{pspicture} \
\qquad \qquad \qquad 
\psset{unit=0.6364cm}
\begin{pspicture}[shift=-1.9](0,0)(3,4)
\pspolygon[fillstyle=solid,fillcolor=lightlightblue](0,1)(1,2)(2,1)(1,0)(0,1)
\pspolygon[fillstyle=solid,fillcolor=lightlightblue](1,2)(2,1)(3,2)(2,3)(1,2)
\psline{-}(0,0)(0,4)
\psarc[linecolor=blue,linewidth=1.5pt]{-}(2.8,1){0.5}{-90}{90}
\psarc[linecolor=blue,linewidth=1.5pt]{-}(2.8,3){0.5}{-90}{90}
\psline[linecolor=blue,linewidth=1.5pt]{-}(0,0.5)(0.5,0.5)
\psline[linecolor=blue,linewidth=1.5pt]{-}(0,1.5)(0.5,1.5)
\psline[linecolor=blue,linewidth=1.5pt]{-}(0,2.5)(1.5,2.5)
\psline[linecolor=blue,linewidth=1.5pt]{-}(0,3.5)(2.8,3.5)
\psline[linecolor=blue,linewidth=1.5pt]{-}(1.5,0.5)(2.8,0.5)
\psline[linecolor=blue,linewidth=1.5pt]{-}(2.5,1.5)(2.8,1.5)
\psline[linecolor=blue,linewidth=1.5pt]{-}(2.5,2.5)(2.8,2.5)
\psarc[linewidth=0.025]{-}(0,1){0.21}{-45}{45}\rput(1,1){$u$}
\psarc[linewidth=0.025]{-}(1,2){0.21}{-45}{45}\rput(2,2){$v$}
\end{pspicture}
\ \ \ \ =\ \ \
\begin{pspicture}[shift=-1.9](0,0)(3,4)
\pspolygon[fillstyle=solid,fillcolor=lightlightblue](0,3)(1,4)(2,3)(1,2)(0,3)
\pspolygon[fillstyle=solid,fillcolor=lightlightblue](1,2)(2,1)(3,2)(2,3)(1,2)
\psline{-}(0,0)(0,4)
\psarc[linecolor=blue,linewidth=1.5pt]{-}(2.8,1){0.5}{-90}{90}
\psarc[linecolor=blue,linewidth=1.5pt]{-}(2.8,3){0.5}{-90}{90}
\psline[linecolor=blue,linewidth=1.5pt]{-}(0,2.5)(0.5,2.5)
\psline[linecolor=blue,linewidth=1.5pt]{-}(0,3.5)(0.5,3.5)
\psline[linecolor=blue,linewidth=1.5pt]{-}(0,1.5)(1.5,1.5)
\psline[linecolor=blue,linewidth=1.5pt]{-}(0,0.5)(2.8,0.5)
\psline[linecolor=blue,linewidth=1.5pt]{-}(1.5,3.5)(2.8,3.5)
\psline[linecolor=blue,linewidth=1.5pt]{-}(2.5,2.5)(2.8,2.5)
\psline[linecolor=blue,linewidth=1.5pt]{-}(2.5,1.5)(2.8,1.5)
\psarc[linewidth=0.025]{-}(0,3){0.21}{-45}{45}\rput(1,3){$u$}
\psarc[linewidth=0.025]{-}(1,2){0.21}{-45}{45}\rput(2,2){$v$}
\end{pspicture} \ \ \ 
\label{eq:bybes}
\eea 

\noindent 
Because only vacuum boundary conditions are considered, the two relations in \eqref{eq:bybes} are simplified versions of the 
usual BYBE, see for example~\cite{PRZ0607}. 
In fact, the BYBEs (\ref{eq:bybes}) are equivalent to the commutation relation \eqref{eq:comm}.
The face operators also have the push-through properties
\be
\psset{unit=.9cm}
\begin{pspicture}[shift=-0.9](-0.5,0.0)(1.3,2)
\facegrid{(0,0)}{(1,2)}
\psarc[linewidth=0.025]{-}(0,0){0.16}{0}{90}
\psarc[linewidth=0.025]{-}(0,1){0.16}{0}{90}
\psarc[linewidth=1.5pt,linecolor=blue]{-}(0,1){0.5}{90}{-90}
\rput(0.5,.5){$u$}
\rput(0.5,1.5){\small$u\!-\!\lambda$}
\end{pspicture} 
= s_0(u)s_2(- u) \ \ 
\begin{pspicture}[shift=-0.9](-0.5,0.0)(1.3,2)
\facegrid{(0,0)}{(1,2)}
\psarc[linewidth=1.5pt,linecolor=blue]{-}(0,1){0.5}{90}{-90}
\psarc[linewidth=1.5pt,linecolor=blue]{-}(0,0){0.5}{0}{90}
\psarc[linewidth=1.5pt,linecolor=blue]{-}(1,1){0.5}{90}{-90}
\psarc[linewidth=1.5pt,linecolor=blue]{-}(0,2){0.5}{-90}{0}
\end{pspicture} \ \   \qquad\quad
\begin{pspicture}[shift=-0.9](-0.3,0.0)(1.5,2)
\facegrid{(0,0)}{(1,2)}
\psarc[linewidth=0.025]{-}(0,0){0.16}{0}{90}
\psarc[linewidth=0.025]{-}(0,1){0.16}{0}{90}
\psarc[linewidth=1.5pt,linecolor=blue]{-}(1,1){0.5}{-90}{90}
\rput(0.5,.5){$u$}
\rput(0.5,1.5){\small$u\!+\!\lambda$}
\end{pspicture} \ \ = s_1(u)s_1(- u) 
\begin{pspicture}[shift=-0.9](-0.3,0.0)(1.5,2)
\facegrid{(0,0)}{(1,2)}
\psarc[linewidth=1.5pt,linecolor=blue]{-}(1,1){0.5}{-90}{90}
\psarc[linewidth=1.5pt,linecolor=blue]{-}(1,0){0.5}{90}{180}
\psarc[linewidth=1.5pt,linecolor=blue]{-}(0,1){0.5}{-90}{90}
\psarc[linewidth=1.5pt,linecolor=blue]{-}(1,2){0.5}{180}{270}
\end{pspicture}\ \ 
\label{eq:pushthru}
\ee 
which one can regard as rewritings of the local inversion relation \eqref{eq:inve}.

%%%%%%%%%
\paragraph{Transfer tangles}
%%%%%%%%
On a strip of width $N$ and with trivial boundary conditions, the {\em double-row transfer tangle} $\Db(u)$ is defined by
\vspace{-0.3cm}

\be 
\Db (u):= \ 
\psset{unit=0.9}
\begin{pspicture}[shift=-1.7](-0.5,-0.8)(5.5,2.0)
\facegrid{(0,0)}{(5,2)}
\psarc[linewidth=0.025]{-}(0,0){0.16}{0}{90}
\psarc[linewidth=0.025]{-}(0,1){0.16}{0}{90}
\psarc[linewidth=0.025]{-}(1,0){0.16}{0}{90}
\psarc[linewidth=0.025]{-}(1,1){0.16}{0}{90}
\psarc[linewidth=0.025]{-}(4,0){0.16}{0}{90}
\psarc[linewidth=0.025]{-}(4,1){0.16}{0}{90}
\rput(2.5,0.5){$\ldots$}
\rput(2.5,1.5){$\ldots$}
\rput(3.5,0.5){$\ldots$}
\rput(3.5,1.5){$\ldots$}
\psarc[linewidth=1.5pt,linecolor=blue]{-}(0,1){0.5}{90}{-90}
\psarc[linewidth=1.5pt,linecolor=blue]{-}(5,1){0.5}{-90}{90}
\rput(0.5,.5){$u$}
\rput(0.5,1.5){\small$\lambda\!-\!u$}
\rput(1.5,.5){$u$}
\rput(1.5,1.5){\small$\lambda\!-\!u$}
\rput(4.5,.5){$u$}
\rput(4.5,1.5){\small$\lambda\!-\!u$}
\rput(2.5,-0.5){$\underbrace{\qquad \qquad \qquad \qquad \qquad  \qquad}_N$}
\end{pspicture}
\label{Du}
\ee   
Similarly on a cylinder, the {\em single-row transfer tangle} $\Tb(u)$ is defined as
\be 
\Tb (u):= \ 
\psset{unit=0.9}
\begin{pspicture}[shift=-1.1](-0.2,-0.7)(5.2,1.0)
\facegrid{(0,0)}{(5,1)}
\psarc[linewidth=0.025]{-}(0,0){0.16}{0}{90}
\psarc[linewidth=0.025]{-}(1,0){0.16}{0}{90}
\psarc[linewidth=0.025]{-}(4,0){0.16}{0}{90}
\psline[linewidth=1.5pt,linecolor=blue]{-}(0,0.5)(-0.2,0.5)
\psline[linewidth=1.5pt,linecolor=blue]{-}(5,0.5)(5.2,0.5)
\rput(2.5,0.5){$\ldots$}
\rput(3.5,0.5){$\ldots$}
\rput(0.5,.5){$u$}
\rput(1.5,.5){$u$}
\rput(4.5,.5){$u$}
\rput(2.5,-0.5){$\underbrace{\qquad \qquad \qquad \qquad  \qquad \qquad}_N$}
\end{pspicture} \ \ \ 
\label{Tu}
\ee  
where the periodic boundary conditions identify the left and right edges of $\Tb(u)$ thereby forming a band of length $N$
around the cylinder. Strictly speaking, $\Tb(u)$ is only {\em locally planar} as it is defined on the surface of a cylinder.
Transfer matrices on the strip and cylinder are obtained by fixing a representation of $TL_N(\beta)$ and 
$\mathcal EPTL_N(\alpha,\beta)$. For example, in the standard representations, the transfer matrices are 
$\rho_N^d(\Db(u))$ and $\tilde \rho_N^d(\Tb(u))$.

From \eqref{eq:inve}, \eqref{eq:YBE} and \eqref{eq:bybes}, these tangles can be shown to  
form distinct commuting families of tangles \cite{BaxBook,BehrendPearce}, in the sense that 
\be 
 [\Db(u),\Db(v)]=0,\qquad [\Tb(u),\Tb(v)]=0,\qquad u,v\in\mathbb{C}
\ee
where multiplication is by concatenation of diagrams placing $\Db(v)$ atop $\Db(u)$ in the product $\Db(u)\Db(v)$, for example.
Even though $\Db(u)$ and $\Tb(u)$ are planar tangles and hence not matrices, they are occasionally referred to as transfer 
matrices in the literature.

%%%%%%%%%
\paragraph{Crossing symmetry}
%%%%%%%%

Following arguments in~\cite{BehrendPearce}, it can be shown that $\Db(u)$ is crossing symmetric in the sense that
\be
 \Db(u) = \Db(\lambda - u).
\ee
The manifestation of crossing symmetry of the cylinder transfer tangle $\Tb(u)$ requires the extension of 
$\mathcal E PTL_N(\alpha, \beta)$ by a reflection generator $R$. Due to the periodicity of the cylinder, we can choose
$R=R_j$ for any $j=1,\ldots,N$ where $R_j$ is defined by the relations
\be
R_j \,e_j  R_j= e_j, \qquad R_j \,\Omega\, R_j = \Omega^{-1},\qquad R_j^2 = I.
\ee
Using relations such as $e_5=\Omega^{-3}e_2\,\Omega^3$, it follows that $R_2\,e_5R_2=e_{-1}\equiv e_{N-1}$ and more generally
\be
 R_i\,e_jR_i=e_{2i-j},\qquad i,j=1,\ldots,N
\ee
where it is recalled that $e_k\equiv e_\ell$ if $k\equiv \ell$ mod $N$.
The crossing symmetry then reads
\be
R \, \Tb(u) R = \Tb(\lambda -u).
\label{RTR}
\ee

%%%%%%%%%
\paragraph{Braid transfer tangles and braid limits}
%%%%%%%%

The {\em elementary braid tangles} are $2$-tangles defined by
\be
\psset{unit=.9cm}
\begin{pspicture}[shift=-.4](1,1)
\facegrid{(0,0)}{(1,1)}
\psline[linewidth=1.5pt,linecolor=blue]{-}(0,0.5)(1,0.5)
\psline[linewidth=1.5pt,linecolor=blue]{-}(0.5,0)(0.5,0.35)
\psline[linewidth=1.5pt,linecolor=blue]{-}(0.5,0.65)(0.5,1)
\end{pspicture}
\ := \ e^{-\ir\tfrac{\pi-\lambda}2}\;\begin{pspicture}[shift=-.4](1,1)
\facegrid{(0,0)}{(1,1)}
\rput[bl](0,0){\loopa}
\end{pspicture}\;+\,e^{\ir\tfrac{\pi-\lambda}2}\;
\begin{pspicture}[shift=-.4](1,1)
\facegrid{(0,0)}{(1,1)}
\rput[bl](0,0){\loopb}
\end{pspicture}
\qquad \qquad 
\psset{unit=.9cm}
\begin{pspicture}[shift=-.4](1,1)
\facegrid{(0,0)}{(1,1)}
\psline[linewidth=1.5pt,linecolor=blue]{-}(0,0.5)(0.35,0.5)
\psline[linewidth=1.5pt,linecolor=blue]{-}(0.65,0.5)(1,0.5)
\psline[linewidth=1.5pt,linecolor=blue]{-}(0.5,0)(0.5,1)
\end{pspicture}
\ =\
e^{\ir\tfrac{\pi-\lambda}2}\;
\begin{pspicture}[shift=-.4](1,1)
\facegrid{(0,0)}{(1,1)}
\rput[bl](0,0){\loopa}
\end{pspicture}\;+\,e^{-\ir\tfrac{\pi-\lambda}2}\;
\begin{pspicture}[shift=-.4](1,1)
\facegrid{(0,0)}{(1,1)}
\rput[bl](0,0){\loopb}
\end{pspicture}
\label{eq:braidvert}
\ee
These can be obtained in the {\em braid limit} of the face operators as
\be    
\psset{unit=.9cm}
\begin{pspicture}[shift=-.42](1,1)
\facegrid{(0,0)}{(1,1)}
\psline[linewidth=1.5pt,linecolor=blue]{-}(0,0.5)(1,0.5)
\psline[linewidth=1.5pt,linecolor=blue]{-}(0.5,0)(0.5,0.35)
\psline[linewidth=1.5pt,linecolor=blue]{-}(0.5,0.65)(0.5,1)
\end{pspicture}
\ = \lim_{u \rightarrow \ir \infty} \bigg(\frac{e^{\ir\tfrac{\pi-\lambda}2}}{s_k(u)} \ \ 
\begin{pspicture}[shift=-.42](1,1)
\facegrid{(0,0)}{(1,1)}
\psarc[linewidth=0.025]{-}(0,0){0.16}{0}{90}
\rput(.5,.5){$u_k$}
\end{pspicture}
\ \bigg), \qquad \quad
\begin{pspicture}[shift=-.42](1,1)
\facegrid{(0,0)}{(1,1)}
\psline[linewidth=1.5pt,linecolor=blue]{-}(0,0.5)(0.35,0.5)
\psline[linewidth=1.5pt,linecolor=blue]{-}(0.65,0.5)(1,0.5)
\psline[linewidth=1.5pt,linecolor=blue]{-}(0.5,0)(0.5,1)
\end{pspicture}
\ = \lim_{u \rightarrow \ir \infty}\bigg( \frac{e^{\ir\tfrac{\pi-\lambda}2}}{s_k(u)} \ \ 
\begin{pspicture}[shift=-.42](1,1)
\facegrid{(0,0)}{(1,1)}
\psarc[linewidth=0.025]{-}(1,0){0.16}{90}{180}
\rput(.5,.5){$u_k$}
\end{pspicture}
\ \bigg), \qquad k\in\mathbb{Z},
\ee 
and satisfy planar relations obtained as braid limits of usual planar identities, such as
\be
\psset{unit=0.6364cm}
\begin{pspicture}[shift=-0.95](-0.1,-1.1)(4.3,1.1)
\pspolygon[fillstyle=solid,fillcolor=lightlightblue](0,0)(1,1)(3,-1)(4,0)(3,1)(1,-1)
\psarc[linewidth=1.5pt,linecolor=blue](2,0){.707}{45}{135}
\psarc[linewidth=1.5pt,linecolor=blue](2,0){.707}{-135}{-45}
\psline[linewidth=1.5pt,linecolor=blue]{-}(0.5,0.5)(1.5,-0.5)
\psline[linewidth=1.5pt,linecolor=blue]{-}(0.5,-0.5)(0.85,-0.15)
\psline[linewidth=1.5pt,linecolor=blue]{-}(1.15,0.15)(1.5,0.5)
\psline[linewidth=1.5pt,linecolor=blue]{-}(2.5,0.5)(2.85,0.15)
\psline[linewidth=1.5pt,linecolor=blue]{-}(3.15,-0.15)(3.5,-0.5)
\psline[linewidth=1.5pt,linecolor=blue]{-}(2.5,-0.5)(3.5,0.5)
\end{pspicture}
= \ 
\begin{pspicture}[shift=-0.95](-0.1,-1.1)(2.3,1.1)
\pspolygon[fillstyle=solid,fillcolor=lightlightblue](0,0)(1,1)(2,0)(1,-1)(0,0)
\psarc[linewidth=1.5pt,linecolor=blue](1,-1){.707}{45}{135}
\psarc[linewidth=1.5pt,linecolor=blue](1,1){.707}{-135}{-45}
\end{pspicture} \qquad \qquad
\psset{unit=.82cm}
\begin{pspicture}[shift=-0.9](0,0)(3,2)
\facegrid{(2,0)}{(3,2)}
\pspolygon[fillstyle=solid,fillcolor=lightlightblue](0,1)(1,2)(2,1)(1,0)(0,1)
\psline[linecolor=blue,linewidth=1.5pt]{-}(2,0.5)(1.5,0.5)
\psline[linecolor=blue,linewidth=1.5pt]{-}(2,1.5)(1.5,1.5)
\rput(2,0){\psline[linewidth=1.5pt,linecolor=blue]{-}(0,0.5)(1,0.5)
\psline[linewidth=1.5pt,linecolor=blue]{-}(0.5,0)(0.5,0.35)
\psline[linewidth=1.5pt,linecolor=blue]{-}(0.5,0.65)(0.5,1)}
\rput(2,1){\psline[linewidth=1.5pt,linecolor=blue]{-}(0,0.5)(0.35,0.5)
\psline[linewidth=1.5pt,linecolor=blue]{-}(0.65,0.5)(1,0.5)
\psline[linewidth=1.5pt,linecolor=blue]{-}(0.5,0)(0.5,1)}
\rput(0,1){\psline[linewidth=1.5pt,linecolor=blue]{-}(0.5,0.5)(1.5,-0.5)
\psline[linewidth=1.5pt,linecolor=blue]{-}(0.5,-0.5)(0.85,-0.15)
\psline[linewidth=1.5pt,linecolor=blue]{-}(1.15,0.15)(1.5,0.5)}
\end{pspicture}
\ \ =\ \ 
\begin{pspicture}[shift=-0.9](0,0)(3,2)
\facegrid{(0,0)}{(1,2)}
\pspolygon[fillstyle=solid,fillcolor=lightlightblue](1,1)(2,2)(3,1)(2,0)(1,1)
\psline[linecolor=blue,linewidth=1.5pt]{-}(1,0.5)(1.5,0.5)
\psline[linecolor=blue,linewidth=1.5pt]{-}(1,1.5)(1.5,1.5)
\rput(0,1){\psline[linewidth=1.5pt,linecolor=blue]{-}(0,0.5)(1,0.5)
\psline[linewidth=1.5pt,linecolor=blue]{-}(0.5,0)(0.5,0.35)
\psline[linewidth=1.5pt,linecolor=blue]{-}(0.5,0.65)(0.5,1)}
\rput(0,0){\psline[linewidth=1.5pt,linecolor=blue]{-}(0,0.5)(0.35,0.5)
\psline[linewidth=1.5pt,linecolor=blue]{-}(0.65,0.5)(1,0.5)
\psline[linewidth=1.5pt,linecolor=blue]{-}(0.5,0)(0.5,1)}
\rput(1,1){\psline[linewidth=1.5pt,linecolor=blue]{-}(0.5,0.5)(1.5,-0.5)
\psline[linewidth=1.5pt,linecolor=blue]{-}(0.5,-0.5)(0.85,-0.15)
\psline[linewidth=1.5pt,linecolor=blue]{-}(1.15,0.15)(1.5,0.5)}

\end{pspicture}\ \
\label{eq:braidplanaridentities}
\ee
We define the {\em braid transfer tangles} in terms of these braid $2$-tangles as
\be
\Fb:= \ 
\psset{unit=.9cm}
\begin{pspicture}[shift=-0.9](-0.5,-0.0)(5.5,2.0)
\facegrid{(0,0)}{(5,2)}
\rput(2.5,0.5){$\ldots$}
\rput(2.5,1.5){$\ldots$}
\rput(3.5,0.5){$\ldots$}
\rput(3.5,1.5){$\ldots$}
\psarc[linewidth=1.5pt,linecolor=blue]{-}(0,1){0.5}{90}{-90}
\psarc[linewidth=1.5pt,linecolor=blue]{-}(5,1){0.5}{-90}{90}
\rput(0,0){\psline[linewidth=1.5pt,linecolor=blue]{-}(0.0,0.5)(1,0.5)
\psline[linewidth=1.5pt,linecolor=blue]{-}(0.5,0)(0.5,0.35)
\psline[linewidth=1.5pt,linecolor=blue]{-}(0.5,0.65)(0.5,1)}
\rput(1,0){\psline[linewidth=1.5pt,linecolor=blue]{-}(0.0,0.5)(1,0.5)
\psline[linewidth=1.5pt,linecolor=blue]{-}(0.5,0)(0.5,0.35)
\psline[linewidth=1.5pt,linecolor=blue]{-}(0.5,0.65)(0.5,1)}
\rput(4,0){\psline[linewidth=1.5pt,linecolor=blue]{-}(0.0,0.5)(1,0.5)
\psline[linewidth=1.5pt,linecolor=blue]{-}(0.5,0)(0.5,0.35)
\psline[linewidth=1.5pt,linecolor=blue]{-}(0.5,0.65)(0.5,1)}
\rput(0,1){\psline[linewidth=1.5pt,linecolor=blue]{-}(0.5,0.0)(0.5,1)
\psline[linewidth=1.5pt,linecolor=blue]{-}(0,0.5)(0.35,0.5)
\psline[linewidth=1.5pt,linecolor=blue]{-}(0.65,0.5)(1,0.5)}
\rput(1,1){\psline[linewidth=1.5pt,linecolor=blue]{-}(0.5,0.0)(0.5,1)
\psline[linewidth=1.5pt,linecolor=blue]{-}(0,0.5)(0.35,0.5)
\psline[linewidth=1.5pt,linecolor=blue]{-}(0.65,0.5)(1,0.5)}
\rput(4,1){\psline[linewidth=1.5pt,linecolor=blue]{-}(0.5,0.0)(0.5,1)
\psline[linewidth=1.5pt,linecolor=blue]{-}(0,0.5)(0.35,0.5)
\psline[linewidth=1.5pt,linecolor=blue]{-}(0.65,0.5)(1,0.5)}
\end{pspicture}
\qquad\qquad
 \Fbt:=\ \ 
\psset{unit=.9cm}
\begin{pspicture}[shift=-.42](-0.2,0)(5.2,1)
\facegrid{(-0,0)}{(5,1)}
\psline[linewidth=1.5pt,linecolor=blue]{-}(-0.2,0.5)(2,0.5)
\psline[linewidth=1.5pt,linecolor=blue]{-}(0.5,0)(0.5,0.35)
\psline[linewidth=1.5pt,linecolor=blue]{-}(0.5,0.65)(0.5,1)
\psline[linewidth=1.5pt,linecolor=blue]{-}(1.5,0)(1.5,0.35)
\psline[linewidth=1.5pt,linecolor=blue]{-}(1.5,0.65)(1.5,1)
\rput(2.53,.51){$\dots$}
\rput(3.53,.51){$\dots$}
\psline[linewidth=1.5pt,linecolor=blue]{-}(4,0.5)(5.2,0.5)
\psline[linewidth=1.5pt,linecolor=blue]{-}(4.5,0)(4.5,0.35)
\psline[linewidth=1.5pt,linecolor=blue]{-}(4.5,0.65)(4.5,1)
\end{pspicture}
\ee
The braid transfer tangles are central elements of $TL_N(\beta)$ and $\mathcal EPTL_N(\alpha, \beta)$, 
respectively~\cite{AMDYSA2011,AMDYSA2013}, and
can be obtained as braid limits of the corresponding transfer tangles as
\be   
 \Fb=\lim_{u\to\ir\infty}\bigg(\Big(\frac{e^{\ir\tfrac{\pi-\lambda}2}}{s_0(u)}\Big)^{\!2N}\!\Db(u)\bigg),\qquad
 \Fbt=\lim_{u\to\ir\infty}\bigg(\Big(\frac{e^{\ir\tfrac{\pi-\lambda}2}}{s_0(u)}\Big)^{\!N} \Tb(u)\bigg).
\ee

On standard modules, the braid transfer tangles $\Fb$ and $\Fbt$ act as scalar multiples of the identity, that is
\be
 \rho_N^d(\Fb)=\mathbb f^{d}_1\, I,\qquad \tilde\rho_N^d(\Fbt)=\tilde {\mathbb f}^{d}_1\, I,
\label{Ff}
\ee
where 
\be 
 \mathbb{f}^{d}_1=2\,(-1)^d\cos\big((d+1)\lambda\big), \qquad \tilde{\mathbb{f}}^{d}_1=2\cos\big(\tfrac d2(\pi-\lambda)\big)
\ee 
are special cases of the expressions 
in \eqref{eq:fvalues}, while $I$ is the identity matrix of dimension $\dim V_N^d$ or $\dim \tilde V_N^d$, respectively.

%%%%%%%%%
\paragraph{Characterisation of $\boldsymbol \lambda$ values}
%%%%%%%%
A primary goal of this paper is to find functional relations satisfied by the transfer tangles $\Db(u)$ and $\Tb(u)$ in the case the crossing 
parameter is a non-integer {\em rational multiple} of $\pi$. In this case, we write
\be
 \lambda=\lambda_{p,p'}:= \frac{(p'-p)\pi}{p'},\qquad \mathrm{gcd}(p,p')=1,\qquad p\in\mathbb{N},\qquad p' \in \mathbb{N}_{\geq2},
\label{rat}
\ee 
and we refer to these $\lambda$ values as {\em fractional}. Accordingly, the values 
$\lambda\in\pi\big(\mathbb{R}\!\setminus\!\mathbb{Q}\big)$ are referred to as {\em generic}. 
The {\em logarithmic minimal models} ${\cal LM}(p,p')$~\cite{PRZ0607} are defined for fractional $\lambda=\lambda_{p,p'}$
for which $1\leq p<p'$ and are the focus of Section~\ref{Sec:LMM}, in particular.

For $\lambda=\lambda_{p,1}\in\pi\mathbb{Z}$, the weights in \eqref{laus} diverge.
After these divergencies are removed by a renormalisation of the face operator, the relative weight of the 
two connectivity diagrams in (\ref{1x1}) is simply $\pm1$ since $s_0(u)$ and $s_1(-u)$ are equal up to a sign for 
$\lambda\in\pi\mathbb{Z}$. In this case, one can show that $\Db(u)$ is merely a scalar multiple of the {\em identity tangle}
\be
 \Ib :=
\begin{pspicture}[shift=-0.25](0.0,-0.35)(2.5,0.35)
\rput(1.8,0.0){\small$...$}
\psline[linecolor=blue,linewidth=1.5pt]{-}(0.2,0.35)(0.2,-0.35)
\psline[linecolor=blue,linewidth=1.5pt]{-}(0.6,0.35)(0.6,-0.35)
\psline[linecolor=blue,linewidth=1.5pt]{-}(1.0,0.35)(1.0,-0.35)
\psline[linecolor=blue,linewidth=1.5pt]{-}(1.4,0.35)(1.4,-0.35)
\psline[linecolor=blue,linewidth=1.5pt]{-}(2.2,0.35)(2.2,-0.35)
\end{pspicture}
\label{Ib}
\ee 
while the periodic transfer tangle $\Tb(u)$ is a scalar multiple of the single-row braid transfer tangle $\Fbt$.
We view these as trivial functional relations of degree $p'=1$.
As already indicated in (\ref{laus}), crossing parameters of the form $\lambda\in\pi\mathbb{Z}$ are therefore excluded in the remainder
of this paper.

%%%%%%%%%%%%%%%%%%%%
\paragraph{Wenzl-Jones projectors}
%%%%%%%%%%%%%%%%%%%%%

Let $\lambda$ be generic. The Wenzl-Jones (WJ) projector $P_n$\cite{Jones1983,Wenzl1988,KauffmanLins1994} 
is an $n$-tangle 
\be
 P_n\,=\,
 \begin{pspicture}[shift=-0.05](-0.03,0.00)(1.07,0.3)
 \pspolygon[fillstyle=solid,fillcolor=pink](0,0)(1,0)(1,0.3)(0,0.3)(0,0)\rput(0.5,0.15){$_n$}
 \end{pspicture},\qquad 
 n\in\mathbb{N}
\ee 
defined recursively by 
\be
\begin{pspicture}[shift=-0.05](-0.03,-0.15)(1.2,0.15)
\pspolygon[fillstyle=solid,fillcolor=pink](0,-0.15)(1.2,-0.15)(1.2,0.15)(0,0.15)(0,-0.15)
\rput(0.6,0){$_{n}$}
\end{pspicture}
\ = \ 
\begin{pspicture}[shift=-0.05](-0.03,-0.15)(1.2,0.15)
\pspolygon[fillstyle=solid,fillcolor=pink](0,-0.15)(1.0,-0.15)(1.0,0.15)(0,0.15)(0,0.15)
\rput(0.5,0){$_{n-1}$}
\psline[linecolor=blue,linewidth=1.5pt]{-}(1.15,-0.17)(1.15,0.17)
\end{pspicture} 
\ -\, \frac{s_{n-1}(0)}{s_{n}(0)} \ \,
\begin{pspicture}[shift=-0.45](0,-0.55)(1.24,0.55)
\pspolygon[fillstyle=solid,fillcolor=pink](0,-0.25)(1.0,-0.25)(1.0,-0.55)(0,-0.55)(0,-0.25)
\pspolygon[fillstyle=solid,fillcolor=pink](0,0.25)(1.0,0.25)(1.0,0.55)(0,0.55)(0,0.25)
\rput(0.5,0.4){$_{n-1}$}
\rput(0.5,-0.4){$_{n-1}$}
\psline[linecolor=blue,linewidth=1.5pt]{-}(0.1,-0.25)(0.1,0.25)
\psline[linecolor=blue,linewidth=1.5pt]{-}(0.25,-0.25)(0.25,0.25)
\rput(0.5,0){$_{\dots}$}
\psline[linecolor=blue,linewidth=1.5pt]{-}(0.75,-0.25)(0.75,0.25)
\psarc[linecolor=blue,linewidth=1.5pt]{-}(1.05,-0.25){0.15}{0}{180}
\psarc[linecolor=blue,linewidth=1.5pt]{-}(1.05,0.25){0.15}{180}{0}
\psline[linecolor=blue,linewidth=1.5pt]{-}(1.2,0.25)(1.2,0.55)
\psline[linecolor=blue,linewidth=1.5pt]{-}(1.2,-0.25)(1.2,-0.55)
\end{pspicture}\ \, , \qquad\quad
\begin{pspicture}(0,-0.12)(0.8,0.7)
\pspolygon[fillstyle=solid,fillcolor=pink](0,-0.15)(0.8,-0.15)(0.8,0.15)(0,0.15)(0,0.15)
\rput(0.4,0){$_{1}$}
\end{pspicture} 
\ =\,
\begin{pspicture}(0,-0.12)(0.2,0.7)
\psline[linecolor=blue,linewidth=1.5pt]{-}(0.1,-0.17)(0.1,0.17)
\end{pspicture} \, .
\label{eq:Prec}
\ee
The first nontrivial examples are
\be
\begin{pspicture}[shift=-0.05](0,-0.15)(0.8,0.15)
\pspolygon[fillstyle=solid,fillcolor=pink](0,-0.15)(0.8,-0.15)(0.8,0.15)(0,0.15)(0,0.15)
\rput(0.4,0){$_{2}$}
\end{pspicture} 
\ =\,
\begin{pspicture}[shift=-0.25](-0.0,-0.35)(0.8,0.35)
\psline[linecolor=blue,linewidth=1.5pt]{-}(0.2,0.35)(0.2,-0.35)
\psline[linecolor=blue,linewidth=1.5pt]{-}(0.6,0.35)(0.6,-0.35)
\end{pspicture} - \frac{1}{s_2(0)}
\begin{pspicture}[shift=-0.25](-0.,-0.35)(0.8,0.35)
\psarc[linecolor=blue,linewidth=1.5pt]{-}(0.4,0.35){0.2}{180}{0}
\psarc[linecolor=blue,linewidth=1.5pt]{-}(0.4,-0.35){0.2}{0}{180}
\end{pspicture}
\ee
and
\be
\begin{pspicture}[shift=-0.05](0,-0.15)(0.8,0.15)
\pspolygon[fillstyle=solid,fillcolor=pink](0,-0.15)(0.8,-0.15)(0.8,0.15)(0,0.15)(0,0.15)
\rput(0.4,0){$_{3}$}
\end{pspicture} 
\ =\,
\begin{pspicture}[shift=-0.25](-0.0,-0.35)(1.2,0.35)
\psline[linecolor=blue,linewidth=1.5pt]{-}(0.2,0.35)(0.2,-0.35)
\psline[linecolor=blue,linewidth=1.5pt]{-}(0.6,0.35)(0.6,-0.35)
\psline[linecolor=blue,linewidth=1.5pt]{-}(1.0,0.35)(1.0,-0.35)
\end{pspicture} - \frac{s_2(0)}{s_3(0)} \ \Big(
\begin{pspicture}[shift=-0.25](-0.,-0.35)(1.2,0.35)
\psarc[linecolor=blue,linewidth=1.5pt]{-}(0.4,0.35){0.2}{180}{0}
\psarc[linecolor=blue,linewidth=1.5pt]{-}(0.4,-0.35){0.2}{0}{180}
\psline[linecolor=blue,linewidth=1.5pt]{-}(1.0,0.35)(1.0,-0.35)
\end{pspicture} + 
\begin{pspicture}[shift=-0.25](-0.,-0.35)(1.2,0.35)
\psarc[linecolor=blue,linewidth=1.5pt]{-}(0.8,0.35){0.2}{180}{0}
\psarc[linecolor=blue,linewidth=1.5pt]{-}(0.8,-0.35){0.2}{0}{180}
\psline[linecolor=blue,linewidth=1.5pt]{-}(0.2,0.35)(0.2,-0.35)
\end{pspicture} \Big)
+ \frac{1}{s_3(0)} \ \Big(
\begin{pspicture}[shift=-0.25](-0.,-0.35)(1.2,0.35)
\psarc[linecolor=blue,linewidth=1.5pt]{-}(0.4,0.35){0.2}{180}{0}
\psarc[linecolor=blue,linewidth=1.5pt]{-}(0.8,-0.35){0.2}{0}{180}
\psbezier[linecolor=blue,linewidth=1.5pt]{-}(0.2,-0.35)(0.2,-0)(1.0,0)(1.0,0.35)
\end{pspicture} + 
\begin{pspicture}[shift=-0.25](-0.,-0.35)(1.2,0.35)
\psarc[linecolor=blue,linewidth=1.5pt]{-}(0.4,-0.35){-0.2}{180}{0}
\psarc[linecolor=blue,linewidth=1.5pt]{-}(0.8,0.35){-0.2}{0}{180}
\psbezier[linecolor=blue,linewidth=1.5pt]{-}(0.2,0.35)(0.2,-0)(1.0,0)(1.0,-0.35)
\end{pspicture}  \Big).
\ee

The WJ projectors have many remarkable properties. In particular, as an element of $TL_n(\beta)$, $P_n$ is indeed a projector,
that is
\be
\begin{pspicture}[shift=-0.2](-0.03,0.0)(1.07,0.6)
 \pspolygon[fillstyle=solid,fillcolor=pink](0,0)(1,0)(1,0.3)(0,0.3)(0,0)\rput(0.5,0.15){$_n$}
 \pspolygon[fillstyle=solid,fillcolor=pink](0,0.3)(1,0.3)(1,0.6)(0,0.6)(0,0.3)\rput(0.5,0.45){$_n$}
\end{pspicture}
\,=\,
\begin{pspicture}[shift=-0.05](-0.03,0.00)(1.07,0.3)
 \pspolygon[fillstyle=solid,fillcolor=pink](0,0)(1,0)(1,0.3)(0,0.3)(0,0)\rput(0.5,0.15){$_n$}
\end{pspicture}
.\label{eq:(i)}
\ee 
Second, $P_n$ is a half-arc annihilator in the sense that if two adjacent nodes from the top edge (or bottom edge) 
are connected by a half-arc, the resulting diagram vanishes as
\be
\begin{pspicture}[shift=-0.15](0,-0.3)(1.24,0.3)
\pspolygon[fillstyle=solid,fillcolor=pink](0,-0.15)(1.2,-0.15)(1.2,0.15)(0,0.15)(0,-0.15)
\rput(0.6,0){$_{n}$}
\psarc[linecolor=blue,linewidth=1.5pt]{-}(0.5,0.15){0.15}{0}{180}
\end{pspicture}
\  = \  
\begin{pspicture}[shift=-0.15](0,-0.3)(1.24,0.3)
\pspolygon[fillstyle=solid,fillcolor=pink](0,-0.15)(1.2,-0.15)(1.2,0.15)(0,0.15)(0,-0.15)
\rput(0.6,0){$_{n}$}
\psarc[linecolor=blue,linewidth=1.5pt]{-}(0.5,-0.15){0.15}{180}{0}
\end{pspicture} \ = 0.
\label{eq:(ii)}
\ee
These two properties are used in the definition of fused loop models in Section~\ref{sec:FusedTL} and repeatedly in the 
planar computations of Appendix~\ref{app:FusHier} and~\ref{app:Closure}. 
Moreover, $P_n$ is the unique $n$-tangle with properties \eqref{eq:(i)} and \eqref{eq:(ii)}. 
Two other important properties of WJ projectors are 
\be
\begin{pspicture}[shift=-.2](-0.03,0.0)(1.27,0.6)
 \pspolygon[fillstyle=solid,fillcolor=pink](0,0)(1.2,0)(1.2,0.3)(0,0.3)(0,0)\rput(0.6,0.15){$_n$}
 \pspolygon[fillstyle=solid,fillcolor=pink](0.1,0.3)(0.9,0.3)(0.9,0.6)(0.1,0.6)(0.1,0.3)\rput(0.5,0.45){$_m$}
\end{pspicture}
\;=\,
\begin{pspicture}[shift=-.2](-0.03,0.0)(1.27,0.6)
 \pspolygon[fillstyle=solid,fillcolor=pink](0,0.3)(1.2,0.3)(1.2,0.6)(0,0.6)(0,0.3)\rput(0.6,0.45){$_n$}
 \pspolygon[fillstyle=solid,fillcolor=pink](0.2,0)(1,0)(1,0.3)(0.2,0.3)(0.2,0)\rput(0.6,0.15){$_m$}
\end{pspicture}
\;=\;
\begin{pspicture}[shift=-0.05](-0.03,0.00)(1.27,0.3)
 \pspolygon[fillstyle=solid,fillcolor=pink](0,0)(1.2,0)(1.2,0.3)(0,0.3)(0,0)\rput(0.6,0.15){$_n$}
\end{pspicture}
, \qquad m\le n
\label{eq:(iv)}
\ee 
and 
\be
\begin{pspicture}[shift=-0.2](0,-0.3)(1.40,0.3)
\pspolygon[fillstyle=solid,fillcolor=pink](0,-0.15)(1.2,-0.15)(1.2,0.15)(0,0.15)(0,-0.15)
\rput(0.6,0){$_{n}$}
\psarc[linecolor=blue,linewidth=1.5pt]{-}(1.25,0.15){0.15}{0}{180}
\psline[linecolor=blue,linewidth=1.5pt]{-}(1.40,0.15)(1.40,-0.15)
\psarc[linecolor=blue,linewidth=1.5pt]{-}(1.25,-0.15){-0.15}{0}{180}
\end{pspicture}
\  = \; \frac{s_{n+1}(0)}{s_{n}(0)}
\begin{pspicture}(-0.2,-0.12)(1.15,0.15)
\pspolygon[fillstyle=solid,fillcolor=pink](0,-0.15)(1.0,-0.15)(1.0,0.15)(0,0.15)(0,-0.15)
\rput(0.5,0){$_{n-1}$}
\end{pspicture}\!. 
\label{eq:(v)}
\ee
Finally, $P_{n}$ can be realised in terms of face operators as 
\begin{equation}
P_{n}= \ 
\psset{unit=0.35}
\begin{pspicture}[shift=-4.8](0,-5)(6,5)
\pspolygon[fillstyle=solid,fillcolor=pink](0,0)(5,5)(6,4)(5,3)(6,2)(5,1)(6,0)(5,-1)(6,-2)(5,-3)(6,-4)(5,-5)(0,0)
\psline{-}(1,-1)(6,4)\psline{-}(1,1)(6,-4)
\psline{-}(2,-2)(6,2)\psline{-}(2,2)(6,-2)
\psline{-}(3,-3)(6,0)\psline{-}(3,3)(6,-0)
\psline{-}(4,-4)(6,-2)\psline{-}(4,4)(6,2)
\rput(5,4){$_1$}
\rput(5,2){$_1$}
\rput(5,0){$_1$}
\rput(5,-2){$_1$}
\rput(5,-4){$_1$}
\rput(4,3){$_2$}
\rput(4,1){$_2$}
\rput(4,-1){$_2$}
\rput(4,-3){$_2$}
\rput(3,2){$_{\dots}$}
\rput(3,0){$_{\dots}$}
\rput(3,-2){$_{\dots}$}
\rput(2,1){$_{n\!-\!2}$}
\rput(2,-1){$_{n\!-\!2}$}
\rput(1,0){$_{n\!-\!1}$}
\psarc[linecolor=blue,linewidth=1.5pt]{-}(5,-1){0.707}{-45}{45}
\psarc[linecolor=blue,linewidth=1.5pt]{-}(5,-3){0.707}{-45}{45}
\psarc[linecolor=blue,linewidth=1.5pt]{-}(5,1){0.707}{-45}{45}
\psarc[linecolor=blue,linewidth=1.5pt]{-}(5,3){0.707}{-45}{45}
\end{pspicture}
 \qquad \qquad
\begin{pspicture}[shift=-0.7](0,-1)(2,1)
\pspolygon[fillstyle=solid,fillcolor=pink](0,0)(1,1)(2,0)(1,-1)(0,0)
\rput(1,0){$_k$}  
\end{pspicture}\ =\, \frac1{s_{k+1}(0)} \
\psset{unit=1.5}
\begin{pspicture}[shift=-0.8](0,-1)(2,1)
\pspolygon[fillstyle=solid,fillcolor=lightlightblue](0,0)(1,1)(2,0)(1,-1)(0,0)
\psarc[linewidth=0.025]{-}(1,-1){0.21}{45}{135}
\rput(1,0){\small$-k\lambda$}  
\end{pspicture} \ = \ 
\begin{pspicture}[shift=-0.8](0,-1)(2,1)
\pspolygon[fillstyle=solid,fillcolor=lightlightblue](0,0)(1,1)(2,0)(1,-1)(0,0)
\psarc[linewidth=1.5pt,linecolor=blue]{-}(0,0){0.707}{-45}{45}
\psarc[linewidth=1.5pt,linecolor=blue]{-}(2,0){0.707}{135}{-135}
\end{pspicture} 
\ - \frac{s_k(0)}{s_{k+1}(0)} \ \
\begin{pspicture}[shift=-0.8](0,-1)(2,1)
\pspolygon[fillstyle=solid,fillcolor=lightlightblue](0,0)(1,1)(2,0)(1,-1)(0,0)
\psarc[linewidth=1.5pt,linecolor=blue]{-}(1,1){0.707}{-135}{-45}
\psarc[linewidth=1.5pt,linecolor=blue]{-}(1,-1){0.707}{45}{135}
\end{pspicture}
\label{eq:(vi)}
\end{equation}

For $\lambda=\lambda_{p,p'}$, the recursive definition (\ref{eq:Prec}) of 
$P_n$ is well defined for $n \in \{1, \dots, p'-1\}$, but breaks down for $n=p'$ since $s_{p'}(0)=0$.
In certain contexts, one can circumvent the non-existence of WJ projectors and still perform
the corresponding projection operations. For example, the general construction of boundary
conditions in logarithmic minimal models~\cite{PRZ0607} involves projection operations which can be handled by 
so-called generalised TL projectors~\cite{PRV1210}. 
In the remainder of this section, we only consider $\lambda$ for which the participating projectors exist.
Later, we will need to handle situations where some projectors do not exist. 
Thus, in Sections~\ref{Sec:Tangles} and~\ref{Sec:CabledLinkStates}, we will discuss how to implement the relevant properties of certain 
projectors for general $\lambda$ and, following~\cite{PRV1210}, why we may ignore certain other projectors in our analyses.

%%%%%%%%%%%%%%%%%%%%%%%%%%
\section{Fused Temperley-Lieb loop models}
\label{sec:FusedTL}
%%%%%%%%%%%%%%%%%%%%%%%%%%

In this section, we construct fused face operators and write down the generalised planar identities. The parameter $\lambda$ is 
taken to be generic, so that the projectors $P_m$ and $P_n$ exist for all $m$ and $n$.

%%%%%%%%%%%%%%%%%%%
\subsection{Fused face operators}
%%%%%%%%%%%%%%%%%%%

\paragraph{Definition of fusion}
Fusion of $m\times n$ blocks of loop faces is implemented diagrammatically by (i) forming a rectangular array of $m$ by $n$ 
elementary face operators with spectral parameters as in (\ref{eq:fusedface}), and (ii) applying $P_m$ 
and $P_n$ projectors along the horizontal and vertical edges, respectively. 
The ensuing {\em $(m,n)$-fused face operator} is thus the $(m+n)$-tangle given by
\bea
\psset{unit=.9cm} \qquad \qquad
\begin{pspicture}[shift=-2.5](0,-0.3)(5.3,5.3)
\pspolygon[fillstyle=solid,fillcolor=lightlightblue](-2.5,2)(-1.5,2)(-1.5,3)(-2.5,3)(-2.5,2)
\psarc[linewidth=0.025]{-}(-2.5,2){0.16}{0}{90}
\rput(-2,2.75){\tiny{$_{(m,n)}$}}
\rput(-2,2.5){$u$}
\rput(-1,2.5){$:=$}
\pspolygon[fillstyle=solid,fillcolor=pink](0.1,0)(4.9,0)(4.9,-0.3)(0.1,-0.3)(0.1,0)
\pspolygon[fillstyle=solid,fillcolor=pink](0,0.1)(0,4.9)(-0.3,4.9)(-0.3,0.1)(0,0.1)
\pspolygon[fillstyle=solid,fillcolor=pink](5,0.1)(5,4.9)(5.3,4.9)(5.3,0.1)(5,0.1)
\pspolygon[fillstyle=solid,fillcolor=pink](0.1,5)(4.9,5)(4.9,5.3)(0.1,5.3)(0.1,5)
\rput(2.5,-0.15){$_m$}
\rput(2.5,5.15){$_m$}
\rput(-0.15,2.5){$_n$}
\rput(5.15,2.5){$_n$}
\facegrid{(0,0)}{(5,5)}
\psarc[linewidth=0.025]{-}(0,0){0.16}{0}{90}
\psarc[linewidth=0.025]{-}(3,0){0.16}{0}{90}
\psarc[linewidth=0.025]{-}(4,0){0.16}{0}{90}
\psarc[linewidth=0.025]{-}(0,1){0.16}{0}{90}
\psarc[linewidth=0.025]{-}(3,1){0.16}{0}{90}
\psarc[linewidth=0.025]{-}(4,1){0.16}{0}{90}
\psarc[linewidth=0.025]{-}(0,4){0.16}{0}{90}
\psarc[linewidth=0.025]{-}(3,4){0.16}{0}{90}
\psarc[linewidth=0.025]{-}(4,4){0.16}{0}{90}
\rput(4.5,.5){\small $u_0$}
\rput(4.5,1.5){\small $u_1$}
\rput(4.5,2.6){\small $\vdots$}
\rput(4.5,3.6){\small $\vdots$}
\rput(4.5,4.5){\small $u_{n\!-\!1}$}
\rput(3.5,.5){\small $u_{-\!1}$}
\rput(3.5,1.5){\small $u_0$}
\rput(3.5,2.6){\small $\vdots$}
\rput(3.5,3.6){\small $\vdots$}
\rput(3.5,4.5){\small $u_{n\!-\!2}$}
\rput(.5,.5){\small $u_{1\!-\!m}$}
\rput(.5,1.5){\small $u_{2\!-\!m}$}
\rput(.5,2.6){\small $\vdots$}
\rput(.5,3.6){\small $\vdots$}
\rput(.5,4.5){\small $u_{n\!-\!m}$}
\multiput(0,0)(1,0){2}{\rput(1.5,.5){\small $\ldots$}}
\multiput(0,1)(1,0){2}{\rput(1.5,.5){\small $\ldots$}}
\multiput(0,4)(1,0){2}{\rput(1.5,.5){\small $\ldots$}}
\end{pspicture}
\label{eq:fusedface} 
\eea
A fused face with the small orientation-indicating 
quarter circle drawn in the lower-right, upper-right or upper-left corner corresponds to a 
counterclockwise rotation by $90^\circ, 180^\circ$ or $270^\circ$, respectively, of the diagram in \eqref{eq:fusedface}. 

We mention that fused face operators can be scaled by a factor $\eta^{m,n}$ that removes overall poles and zeroes, 
see for example~\cite{universal}. However, the absence of such normalisation factors makes the planar computations in 
Appendix~\ref{app:FusHier} and~\ref{app:Closure} slightly less cumbersome. A discussion of our results with the factors 
included is presented in Section~\ref{sec:ratcomp}.   

If an internal arc beginning and ending on the {\em same} edge appears in a configuration in the decomposition of 
the $mn$ elementary face operators in (\ref{eq:fusedface}), then there must be a small half-arc connecting neighbouring nodes 
somewhere along that edge and the configuration has weight zero due to property \eqref{eq:(ii)} of the WJ projectors. 
The fusion procedure thus projects out all faces with an internal arc attached to a single projector along the edges. 

A simplified albeit asymmetric 
realisation of the $(m,n)$-fused face is obtained by removing the $P_n$ projector acting on the right and the $P_m$ 
projector acting on the top of the fused block. Indeed, if these projectors are expanded in terms of connectivity diagrams, every one of 
these diagrams but $I$ (which is always present in the decomposition of a WJ projector)
contains one or more half-arcs that propagate through the fused face (because of the push-through 
properties (\ref{eq:pushthru})) to eventually be annihilated by the projectors
at the bottom or to the left, by property \eqref{eq:(ii)}. The only remaining connectivity diagram 
is $I$ and, from property \eqref{eq:(i)} or \eqref{eq:(vi)} of the projector, it readily follows that its weight is $1$. This implies that the 
original and simplified realisations are in fact equal as tangles,
\be
\psset{unit=.9cm}
\begin{pspicture}[shift=-2.5](-0.3,-0.3)(5.3,5.3)
\pspolygon[fillstyle=solid,fillcolor=pink](0.1,0)(4.9,0)(4.9,-0.3)(0.1,-0.3)(0.1,0)
\pspolygon[fillstyle=solid,fillcolor=pink](0,0.1)(0,4.9)(-0.3,4.9)(-0.3,0.1)(0,0.1)
\pspolygon[fillstyle=solid,fillcolor=pink](5,0.1)(5,4.9)(5.3,4.9)(5.3,0.1)(5,0.1)
\pspolygon[fillstyle=solid,fillcolor=pink](0.1,5)(4.9,5)(4.9,5.3)(0.1,5.3)(0.1,5)
\rput(2.5,-0.15){$_m$}
\rput(2.5,5.15){$_m$}
\rput(-0.15,2.5){$_n$}
\rput(5.15,2.5){$_n$}
\facegrid{(0,0)}{(5,5)}
\psarc[linewidth=0.025]{-}(0,0){0.16}{0}{90}
\psarc[linewidth=0.025]{-}(3,0){0.16}{0}{90}
\psarc[linewidth=0.025]{-}(4,0){0.16}{0}{90}
\psarc[linewidth=0.025]{-}(0,1){0.16}{0}{90}
\psarc[linewidth=0.025]{-}(3,1){0.16}{0}{90}
\psarc[linewidth=0.025]{-}(4,1){0.16}{0}{90}
\psarc[linewidth=0.025]{-}(0,4){0.16}{0}{90}
\psarc[linewidth=0.025]{-}(3,4){0.16}{0}{90}
\psarc[linewidth=0.025]{-}(4,4){0.16}{0}{90}
\rput(4.5,.5){\small $u_0$}
\rput(4.5,1.5){\small $u_1$}
\rput(4.5,2.6){\small $\vdots$}
\rput(4.5,3.6){\small $\vdots$}
\rput(4.5,4.5){\small $u_{n\!-\!1}$}
\rput(3.5,.5){\small $u_{-\!1}$}
\rput(3.5,1.5){\small $u_0$}
\rput(3.5,2.6){\small $\vdots$}
\rput(3.5,3.6){\small $\vdots$}
\rput(3.5,4.5){\small $u_{n\!-\!2}$}
\rput(.5,.5){\small $u_{1\!-\!m}$}
\rput(.5,1.5){\small $u_{2\!-\!m}$}
\rput(.5,2.6){\small $\vdots$}
\rput(.5,3.6){\small $\vdots$}
\rput(.5,4.5){\small $u_{n\!-\!m}$}
\multiput(0,0)(1,0){2}{\rput(1.5,.5){\small $\ldots$}}
\multiput(0,1)(1,0){2}{\rput(1.5,.5){\small $\ldots$}}
\multiput(0,4)(1,0){2}{\rput(1.5,.5){\small $\ldots$}}
\end{pspicture}
\ \ = \ \
\begin{pspicture}[shift=-2.5](-0.3,-0.3)(5.0,5.3)
\pspolygon[fillstyle=solid,fillcolor=pink](0.1,0)(4.9,0)(4.9,-0.3)(0.1,-0.3)(0.1,0)
\pspolygon[fillstyle=solid,fillcolor=pink](0,0.1)(0,4.9)(-0.3,4.9)(-0.3,0.1)(0,0.1)
\rput(2.5,-0.15){$_m$}
\rput(-0.15,2.5){$_n$}
\facegrid{(0,0)}{(5,5)}
\psarc[linewidth=0.025]{-}(0,0){0.16}{0}{90}
\psarc[linewidth=0.025]{-}(3,0){0.16}{0}{90}
\psarc[linewidth=0.025]{-}(4,0){0.16}{0}{90}
\psarc[linewidth=0.025]{-}(0,1){0.16}{0}{90}
\psarc[linewidth=0.025]{-}(3,1){0.16}{0}{90}
\psarc[linewidth=0.025]{-}(4,1){0.16}{0}{90}
\psarc[linewidth=0.025]{-}(0,4){0.16}{0}{90}
\psarc[linewidth=0.025]{-}(3,4){0.16}{0}{90}
\psarc[linewidth=0.025]{-}(4,4){0.16}{0}{90}
\rput(4.5,.5){\small $u_0$}
\rput(4.5,1.5){\small $u_1$}
\rput(4.5,2.6){\small $\vdots$}
\rput(4.5,3.6){\small $\vdots$}
\rput(4.5,4.5){\small $u_{n\!-\!1}$}
\rput(3.5,.5){\small $u_{-\!1}$}
\rput(3.5,1.5){\small $u_0$}
\rput(3.5,2.6){\small $\vdots$}
\rput(3.5,3.6){\small $\vdots$}
\rput(3.5,4.5){\small $u_{n\!-\!2}$}
\rput(.5,.5){\small $u_{1\!-\!m}$}
\rput(.5,1.5){\small $u_{2\!-\!m}$}
\rput(.5,2.6){\small $\vdots$}
\rput(.5,3.6){\small $\vdots$}
\rput(.5,4.5){\small $u_{n\!-\!m}$}
\multiput(0,0)(1,0){2}{\rput(1.5,.5){\small $\ldots$}}
\multiput(0,1)(1,0){2}{\rput(1.5,.5){\small $\ldots$}}
\multiput(0,4)(1,0){2}{\rput(1.5,.5){\small $\ldots$}}
\end{pspicture}
\ee

\paragraph{Monoid decomposition}
An $(m,n)$-fused face can be written as a linear combination of
$\textrm{min}(m,n)+1$ {\em generalised monoids} labeled by $a \in \{0, 1, \dots, \min(m,n)\}$ and given by
\be  
X^{m,n}_a = \ 
\psset{unit=0.6}
\begin{pspicture}[shift=-3.2](-0.2,-0.2)(7.7,7.7)
\pspolygon[fillstyle=solid,fillcolor=lightlightblue](0,3)(3,0)(7.5,4.5)(4.5,7.5)(0,3)
\pspolygon[fillstyle=solid,fillcolor=pink](0.2,2.8)(2.8,0.2)(2.4,-0.2)(-0.2,2.4)(0.2,2.8)
\pspolygon[fillstyle=solid,fillcolor=pink](5.1,7.7)(7.7,5.1)(7.3,4.7)(4.7,7.3)(5.1,7.7)
\rput(1.3,1.3){\small$_n$}
\rput(6.2,6.2){\small$_n$}
\pspolygon[fillstyle=solid,fillcolor=pink](3.2,0.2)(7.3,4.3)(7.7,3.9)(3.6,-0.2)(3.2,0.2)
\pspolygon[fillstyle=solid,fillcolor=pink](0.2,3.2)(4.3,7.3)(3.9,7.7)(-0.2,3.6)(0.2,3.2)
\rput(5.45,2.05){\small$_m$}
\rput(2.05,5.45){\small$_m$}
\psarc[linewidth=1.0pt,linecolor=blue](3,0){.4}{45}{135}
\psarc[linewidth=1.0pt,linecolor=blue](3,0){.7}{45}{135}
\psarc[linewidth=1.0pt,linecolor=blue](3,0){1.3}{45}{135}
\psarc[linewidth=1.0pt,linecolor=blue](3,0){1.6}{45}{135}
\psarc[linewidth=1.0pt,linecolor=blue](4.5,7.5){-.4}{45}{135}
\psarc[linewidth=1.0pt,linecolor=blue](4.5,7.5){-.7}{45}{135}
\psarc[linewidth=1.0pt,linecolor=blue](4.5,7.5){-1.3}{45}{135}
\psarc[linewidth=1.0pt,linecolor=blue](4.5,7.5){-1.6}{45}{135}
\psarc[linewidth=1.0pt,linecolor=blue](0,3){.4}{-45}{45}
\psarc[linewidth=1.0pt,linecolor=blue](0,3){.7}{-45}{45}
\psarc[linewidth=1.0pt,linecolor=blue](0,3){2.0}{-45}{45}
\psarc[linewidth=1.0pt,linecolor=blue](0,3){2.3}{-45}{45}
\psarc[linewidth=1.0pt,linecolor=blue](7.5,4.5){-.4}{-45}{45}
\psarc[linewidth=1.0pt,linecolor=blue](7.5,4.5){-.7}{-45}{45}
\psarc[linewidth=1.0pt,linecolor=blue](7.5,4.5){-2.0}{-45}{45}
\psarc[linewidth=1.0pt,linecolor=blue](7.5,4.5){-2.3}{-45}{45}
\psbezier[linewidth=1.0pt,linecolor=blue](4.3435,1.3435)(3.3435,2.3435)(2.83848,3.83848)(1.83848,4.83848)
\psbezier[linewidth=1.0pt,linecolor=blue](4.55563,1.55563)(3.55563,2.55563)(3.05061,4.05061)(2.05061,5.05061)
\psbezier[linewidth=1.0pt,linecolor=blue](5.66152,2.66152)(4.66152,3.66152)(4.1565,5.1565)(3.1565,6.1565)
\psbezier[linewidth=1.0pt,linecolor=blue](5.44939,2.44939)(4.44939,3.44939)(3.94437, 4.94437)(2.94437, 5.94437)
\rput(1.35,3){\tiny$_{(n-a)}$}
\rput(6.15,4.5){\tiny$_{(n-a)}$}
\rput(3,1.0){\tiny$_{(a)}$}
\rput(4.5,6.5){\tiny$_{(a)}$}
\rput(3.75,3.75){\tiny$_{(m-n)}$}
\rput(3.51,0.71){\tiny$.$}\rput(3.61,0.81){\tiny$.$}\rput(3.71,0.91){\tiny$.$}
\rput(2.49,0.71){\tiny$.$}\rput(2.39,0.81){\tiny$.$}\rput(2.29,0.91){\tiny$.$}
\rput(3.99,6.79){\tiny$.$}\rput(3.89,6.69){\tiny$.$}\rput(3.79,6.59){\tiny$.$}
\rput(5.01,6.79){\tiny$.$}\rput(5.11,6.69){\tiny$.$}\rput(5.21,6.59){\tiny$.$}
\rput(0.954594,3.754594){\tiny$.$}\rput(1.054594,3.854594){\tiny$.$}\rput(1.154594,3.954594){\tiny$.$}
\rput(0.954594,2.24541){\tiny$.$}\rput(1.054594,2.14541){\tiny$.$}\rput(1.154594,2.04541){\tiny$.$}
\rput(6.54541, 3.74541){\tiny$.$}\rput(6.44541, 3.64541){\tiny$.$}\rput(6.34541, 3.54541){\tiny$.$}
\rput(6.54541, 5.25459){\tiny$.$}\rput(6.44541, 5.35459){\tiny$.$}\rput(6.34541, 5.45459){\tiny$.$}
\rput(5.00858,2.20858){\tiny$.$}\rput(4.90858,2.10858){\tiny$.$}\rput(4.80858,2.00858){\tiny$.$}
\rput(2.49749,5.29749){\tiny$.$}\rput(2.59749,5.39749){\tiny$.$}\rput(2.69749,5.49749){\tiny$.$}
\end{pspicture} 
\qquad \qquad 
(m\ge n)
\label{Xmnk}
\ee 
where the diagram $X^{m,n}_a$ for $m<n$ is obtained by performing a left-right reflection of the diagram in (\ref{Xmnk})
and interchanging $m$ and $n$ in the reflected diagram.
The details of this decomposition is the content of the following proposition whose proof is presented in Appendix~\ref{App:FusedFaces}. 
\begin{Proposition}
The decomposition of an $(m,n)$-fused face in terms of generalised monoids is given by
\be 
\psset{unit=0.6364cm}
\begin{pspicture}[shift=-0.85](0,-1)(2,1)
\pspolygon[fillstyle=solid,fillcolor=lightlightblue](0,0)(1,1)(2,0)(1,-1)(0,0)
\psarc[linewidth=0.025]{-}(1,-1){0.21}{45}{135}
\rput(1,0){$u$}\rput(1,0.35){\tiny{$_{(m,n)}$}}
\end{pspicture} 
\ = \sum_{a=0}^{r} \alpha_a^{m,n} \, X^{m,n}_a,\qquad\ r:=\min(m,n)
\label{Xmn}
\ee
where 
\be   
 \frac{\alpha_a^{m,n}}{\alpha_0^{m,n}} 
 = (-1)^{(m+n)a} \Big(\prod_{j=1}^a \frac{s_{r-j+1}(0)}{s_{j}(0)} \Big)
  \Big(\prod_{i=0}^{a-1}\frac{s_{n-r+i}(u)}{s_{m-i}(-u)}\Big),  \qquad \
 \alpha_0^{m,n} = \prod_{i=0}^{m-1}\prod_{j=0}^{n-1}s_{i-j+1}(-u).
\label{alphaamn}
\ee
\label{prop:GenMonoids}
\end{Proposition}
For example, in the non-fused case where $m=n=1$, we simply have
\be
 X^{1,1}_0 = \
\psset{unit=0.6364cm}
\begin{pspicture}[shift=-0.85](0,-1)(2,1)
\pspolygon[fillstyle=solid,fillcolor=lightlightblue](0,0)(1,1)(2,0)(1,-1)(0,0)
\psarc[linewidth=1.5pt,linecolor=blue]{-}(0,0){0.707}{-45}{45}
\psarc[linewidth=1.5pt,linecolor=blue]{-}(2,0){0.707}{135}{-135}
\end{pspicture} 
\qquad\quad X^{1,1}_1 = \
\begin{pspicture}[shift=-0.85](0,-1)(2,1)
\pspolygon[fillstyle=solid,fillcolor=lightlightblue](0,0)(1,1)(2,0)(1,-1)(0,0)
\psarc[linewidth=1.5pt,linecolor=blue]{-}(1,1){0.707}{-135}{-45}
\psarc[linewidth=1.5pt,linecolor=blue]{-}(1,-1){0.707}{45}{135}
\end{pspicture}
\ \ \qquad \alpha_0^{1,1} = s_1(-u), \qquad \alpha_1^{1,1} = s_0(u),
\ee
corresponding to the decomposition \eqref{1x1} of the elementary face operators,
while in the case of $(2,2)$-fused faces, the decomposition reads 
\be
\psset{unit=0.6364cm}
\begin{pspicture}[shift=-0.9](0,-1)(2,1)
\pspolygon[fillstyle=solid,fillcolor=lightlightblue](0,0)(1,1)(2,0)(1,-1)(0,0)
\psarc[linewidth=0.025]{-}(1,-1){0.21}{45}{135}
\rput(1,0){$u$}\rput(1,0.35){\tiny{$_{(2,2)}$}}
\end{pspicture} 
\ =  s_{-1}(u)s_{0}(u)\Big(s_{-2}(u)s_{-1}(u)X_0^{2,2}-\beta s_{-1}(u)s_0(u)X_1^{2,2}+s_{0}(u)s_{1}(u)X_2^{2,2}\Big)
\ee
and involves the three generalised monoids
\be
 X^{2,2}_0=
 \begin{pspicture}[shift=-0.9](0,-1)(2,1)
\pspolygon[fillstyle=solid,fillcolor=lightlightblue, linewidth=1.2pt](0,0)(1,1)(2,0)(1,-1)(0,0)
\psarc[linewidth=1.5pt,linecolor=blue](0,0){0.471}{-45}{45}
\psbezier[linewidth=1.5pt,linecolor=blue](0.66,-0.66)(0.86,-0.46)(0.86,0.46)(0.66,0.66)
\psarc[linewidth=1.5pt,linecolor=blue](2,0){-0.471}{-45}{45}
\psbezier[linewidth=1.5pt,linecolor=blue](1.34,-0.66)(1.14,-0.46)(1.14,0.46)(1.34,0.66)
\pspolygon[fillstyle=solid,fillcolor=pink](0.2,0.2)(-0.0,0.4)(0.6,1.0)(0.8,0.8)(0.2,0.2)\rput(0.4,0.6){\small$_2$}
\pspolygon[fillstyle=solid,fillcolor=pink](0.2,-0.2)(-0.0,-0.4)(0.6,-1.0)(0.8,-0.8)(0.2,-0.2)\rput(0.4,-0.6){\small$_2$}
\pspolygon[fillstyle=solid,fillcolor=pink](1.8,0.2)(2,0.4)(1.4,1.0)(1.2,0.8)(1.8,0.2)\rput(1.6,0.6){\small$_2$}
\pspolygon[fillstyle=solid,fillcolor=pink](1.8,-0.2)(2,-0.4)(1.4,-1.0)(1.2,-0.8)(1.8,-0.2)\rput(1.6,-0.6){\small$_2$}
\end{pspicture}
 \qquad \qquad
  X^{2,2}_1=
 \begin{pspicture}[shift=-0.9](0,-1)(2,1)
\pspolygon[fillstyle=solid,fillcolor=lightlightblue, linewidth=1.2pt](0,0)(1,1)(2,0)(1,-1)(0,0)
\psarc[linewidth=1.5pt,linecolor=blue](0,0){0.471}{-45}{45}
\psarc[linewidth=1.5pt,linecolor=blue](2,0){-0.471}{-45}{45}
\psarc[linewidth=1.5pt,linecolor=blue](1,1){0.471}{-135}{-45}
\psarc[linewidth=1.5pt,linecolor=blue](1,-1){-0.471}{-135}{-45}
\pspolygon[fillstyle=solid,fillcolor=pink](0.2,0.2)(-0.0,0.4)(0.6,1.0)(0.8,0.8)(0.2,0.2)\rput(0.4,0.6){\small$_2$}
\pspolygon[fillstyle=solid,fillcolor=pink](0.2,-0.2)(-0.0,-0.4)(0.6,-1.0)(0.8,-0.8)(0.2,-0.2)\rput(0.4,-0.6){\small$_2$}
\pspolygon[fillstyle=solid,fillcolor=pink](1.8,0.2)(2,0.4)(1.4,1.0)(1.2,0.8)(1.8,0.2)\rput(1.6,0.6){\small$_2$}
\pspolygon[fillstyle=solid,fillcolor=pink](1.8,-0.2)(2,-0.4)(1.4,-1.0)(1.2,-0.8)(1.8,-0.2)\rput(1.6,-0.6){\small$_2$}
\end{pspicture}
 \qquad \qquad
   X^{2,2}_2=
 \begin{pspicture}[shift=-0.9](0,-1)(2,1)
\pspolygon[fillstyle=solid,fillcolor=lightlightblue, linewidth=1.2pt](0,0)(1,1)(2,0)(1,-1)(0,0)
\psarc[linewidth=1.5pt,linecolor=blue](1,1){0.471}{-135}{-45}
\psarc[linewidth=1.5pt,linecolor=blue](1,-1){-0.471}{-135}{-45}
\psbezier[linewidth=1.5pt,linecolor=blue](0.34,0.34)(0.54,0.14)(1.46,0.14)(1.66,0.34)
\psbezier[linewidth=1.5pt,linecolor=blue](0.34,-0.34)(0.54,-0.14)(1.46,-0.14)(1.66,-0.34)
\pspolygon[fillstyle=solid,fillcolor=pink](0.2,0.2)(-0.0,0.4)(0.6,1.0)(0.8,0.8)(0.2,0.2)\rput(0.4,0.6){\small$_2$}
\pspolygon[fillstyle=solid,fillcolor=pink](0.2,-0.2)(-0.0,-0.4)(0.6,-1.0)(0.8,-0.8)(0.2,-0.2)\rput(0.4,-0.6){\small$_2$}
\pspolygon[fillstyle=solid,fillcolor=pink](1.8,0.2)(2,0.4)(1.4,1.0)(1.2,0.8)(1.8,0.2)\rput(1.6,0.6){\small$_2$}
\pspolygon[fillstyle=solid,fillcolor=pink](1.8,-0.2)(2,-0.4)(1.4,-1.0)(1.2,-0.8)(1.8,-0.2)\rput(1.6,-0.6){\small$_2$}
\end{pspicture}
\label{X2x2}
\ee
Along with the corresponding coefficients $\alpha_a^{2,2}\!$, 
diagrams similar to (\ref{X2x2}) were introduced in~\cite{FR2002} and later generalised to the $n\times n$ case in~\cite{ZinnJ2007}. 
The $(2,2)$-fused loop model and its conformal properties are investigated in~\cite{PRT2013}.

\paragraph{Fused planar identities}
The $(m+n)$-tangle
\be
I^{m,n} = \ 
\begin{pspicture}[shift=-0.55](0.1,-0.65)(3.3,0.65)
\pspolygon[fillstyle=solid,fillcolor=pink](0.1,0.15)(0.1,-0.15)(1.9,-0.15)(1.9,0.15)(0.1,0.15)
\pspolygon[fillstyle=solid,fillcolor=pink](2.1,0.15)(2.1,-0.15)(3.4,-0.15)(3.4,0.15)(2.1,0.15)
\rput(1,0){$_m$}
\rput(2.75,0){$_n$}
\rput(1.4,0.4){\small$...$}\rput(1.4,-0.4){\small$...$}
\rput(2.95,0.4){\small$...$}\rput(2.95,-0.4){\small$...$}
\psline[linecolor=blue,linewidth=1.5pt]{-}(0.2,0.65)(0.2,0.15)
\psline[linecolor=blue,linewidth=1.5pt]{-}(0.6,0.65)(0.6,0.15)
\psline[linecolor=blue,linewidth=1.5pt]{-}(1.0,0.65)(1.0,0.15)
\psline[linecolor=blue,linewidth=1.5pt]{-}(1.8,0.65)(1.8,0.15)
\psline[linecolor=blue,linewidth=1.5pt]{-}(2.2,0.65)(2.2,0.15)
\psline[linecolor=blue,linewidth=1.5pt]{-}(2.6,0.65)(2.6,0.15)
\psline[linecolor=blue,linewidth=1.5pt]{-}(3.3,0.65)(3.3,0.15)
\psline[linecolor=blue,linewidth=1.5pt]{-}(0.2,-0.65)(0.2,-0.15)
\psline[linecolor=blue,linewidth=1.5pt]{-}(0.6,-0.65)(0.6,-0.15)
\psline[linecolor=blue,linewidth=1.5pt]{-}(1.0,-0.65)(1.0,-0.15)
\psline[linecolor=blue,linewidth=1.5pt]{-}(1.8,-0.65)(1.8,-0.15)
\psline[linecolor=blue,linewidth=1.5pt]{-}(2.2,-0.65)(2.2,-0.15)
\psline[linecolor=blue,linewidth=1.5pt]{-}(2.6,-0.65)(2.6,-0.15)
\psline[linecolor=blue,linewidth=1.5pt]{-}(3.3,-0.65)(3.3,-0.15)
\end{pspicture} 
\label{Ijmn}
\ee

\noindent 
acts from below as the identity on the generalised monoid $X_a^{m,n}$\!, while $I^{n,m}$ acts as the identity from above, that is
\be
 I^{m,n} X_a^{m,n}= X_a^{m,n} I^{n,m} = X_a^{m,n}, \qquad a=0,1,\ldots,r.
\ee
In the case $m=n$, we have $X^{n,n}_0=I^{n,n}$.

The fused face operators satisfy generalised versions of the various identities for the elementary $1\times1$ boxes discussed
in Section~\ref{Sec:TL}. The generalised YBEs and crossing relations are thus discussed in Sections~\ref{Sec:YBE} 
and~\ref{Sec:Crossing}, while here we note the local inversion relation
\be
\psset{unit=0.6364cm}
\begin{pspicture}[shift=-0.85](0,-1)(4,1)
\psarc[linewidth=4pt,linecolor=blue](2,0){.707}{45}{135}\psarc[linewidth=2pt,linecolor=white](2,0){.707}{45}{135}
\psarc[linewidth=4pt,linecolor=blue](2,0){.707}{-135}{-45}\psarc[linewidth=2pt,linecolor=white](2,0){.707}{-135}{-45}
\pspolygon[fillstyle=solid,fillcolor=lightlightblue](0,0)(1,1)(3,-1)(4,0)(3,1)(1,-1)
\psarc[linewidth=0.025]{-}(0,0){0.21}{-45}{45}
\psarc[linewidth=0.025]{-}(2,0){0.21}{-45}{45}
\rput(1,0){$u$}\rput(1,0.35){\tiny{$_{(m,n)}$}}
\rput(3,0){$-u$}\rput(3,0.35){\tiny{$_{(n,m)}$}}
\end{pspicture}   
\ =\,g^{m,n}(u) \, \ 
\psset{unit=0.6364cm}
\begin{pspicture}[shift=-1.2](-0.6,-1.4)(0.6,1.2)
\pspolygon[fillstyle=solid,fillcolor=pink](-0.2,0.2)(-0.2,1.2)(0.2,1.2)(0.2,0.2)(-0.2,0.2) \rput(0,0.7){\tiny{$_n$}}
\pspolygon[fillstyle=solid,fillcolor=pink](-0.2,0)(-0.2,-1.4)(0.2,-1.4)(0.2,0)(-0.2,0) \rput(0,-0.7){\tiny{$_m$}}
\psline[linecolor=blue]{-}(-0.2,0.3)(-0.6,0.3)
\psline[linecolor=blue]{-}(-0.2,0.90)(-0.6,0.90)
\rput(-0.4,0.5){\tiny .}\rput(-0.4,0.6){\tiny .}\rput(-0.4,0.7){\tiny .}
\psline[linecolor=blue]{-}(-0.2,1.1)(-0.6,1.1)
\psline[linecolor=blue]{-}(0.2,0.3)(0.6,0.3)
\psline[linecolor=blue]{-}(0.2,0.9)(0.6,0.9)
\rput(0.4,0.5){\tiny .}\rput(0.4,0.6){\tiny .}\rput(0.4,0.7){\tiny .}
\psline[linecolor=blue]{-}(0.2,1.1)(0.6,1.1)
\psline[linecolor=blue]{-}(0.2,-0.1)(0.6,-0.1)
\psline[linecolor=blue]{-}(0.2,-0.3)(0.6,-0.3)
\psline[linecolor=blue]{-}(0.2,-0.5)(0.6,-0.5)
\rput(0.4,-0.8){\tiny .}\rput(0.4,-0.9){\tiny .}\rput(0.4,-1){\tiny .}
\psline[linecolor=blue]{-}(0.2,-1.3)(0.6,-1.3)
\psline[linecolor=blue]{-}(-0.2,-0.1)(-0.6,-0.1)
\psline[linecolor=blue]{-}(-0.2,-0.3)(-0.6,-0.3)
\psline[linecolor=blue]{-}(-0.2,-0.5)(-0.6,-0.5)
\rput(-0.4,-0.8){\tiny .}\rput(-0.4,-0.9){\tiny .}\rput(-0.4,-1){\tiny .}
\psline[linecolor=blue]{-}(-0.2,-1.3)(-0.6,-1.3)
\end{pspicture}\ ,\qquad\quad 
g^{m,n}(u):=\prod_{i=0}^{m-1}\prod_{j=0}^{n-1} s_{j-i+1}(u)s_{i-j+1}(-u),
\label{eq:fusedinv}
\ee
\vspace{-0.1cm} 

\noindent
where the diagram to the right represents the identity tangle $I^{m,n}$ in (\ref{Ijmn}) rotated by $90^\circ$ in the
counterclockwise direction.
Pairs of half-arcs as in the diagram to the left in (\ref{eq:fusedinv}) will be used to indicate that {\em multiple half-arcs} 
connect the fused faces. In the concrete example in (\ref{eq:fusedinv}), the upper and lower pairs represent $m$ and $n$ nested 
half-arcs, respectively. Another example of multiple half-arcs 
is given by comparing equations \eqref{eq:transfermatrix} and \eqref{eq:transfermatrix2} below.

The extension of the push-through properties (\ref{eq:pushthru}) to the $(m,1)$-fused faces is given by
\be
\begin{pspicture}[shift=-0.9](-0.5,-0.0)(1,2.0)
\facegrid{(0,0)}{(1,2)}
\psarc[linewidth=0.025]{-}(0,0){0.16}{0}{90}
\psarc[linewidth=0.025]{-}(0,1){0.16}{0}{90}
\psarc[linewidth=1.5pt,linecolor=blue]{-}(0,1){0.5}{90}{-90}
\rput(0.5,.5){$u$}\rput(0.5,0.75){\tiny{$_{(m,1)}$}}
\rput(0.5,1.5){\small$u\!-\!\lambda$}\rput(0.5,1.75){\tiny{$_{(m,1)}$}}
\end{pspicture} \ =  q^m(u)  \ 
\begin{pspicture}[shift=-1.2](-0.6,-0.3)(1,2.3)
\facegrid{(0,0)}{(1,2)}
\pspolygon[fillstyle=solid,fillcolor=pink](0.2,0)(0.8,0)(0.8,-0.3)(0.2,-0.3)(0.2,0)\rput(0.5,-0.15){\small{$_m$}}
\pspolygon[fillstyle=solid,fillcolor=pink](0.2,2)(0.8,2)(0.8,2.3)(0.2,2.3)(0.2,2)\rput(0.5,2.16){\small{$_m$}}
\psarc[linewidth=1pt,linecolor=blue]{-}(0,1){0.56}{90}{-90}
\psarc[linewidth=1pt,linecolor=blue]{-}(0,1){0.44}{90}{-90}
\psarc[linewidth=1pt,linecolor=blue]{-}(0,0){0.44}{0}{90}
\psarc[linewidth=1pt,linecolor=blue]{-}(0,0){0.56}{0}{90}
\psarc[linewidth=1.5pt,linecolor=blue]{-}(1,1){0.5}{90}{-90}
\psarc[linewidth=1pt,linecolor=blue]{-}(0,2){0.44}{-90}{0}
\psarc[linewidth=1pt,linecolor=blue]{-}(0,2){0.56}{-90}{0}
\end{pspicture}
\,= q^m(u)\  
\begin{pspicture}[shift=-0.05](0.1,0.83)(1.5,1.13)
\pspolygon[fillstyle=solid,fillcolor=pink](0.2,1.13)(0.8,1.13)(0.8,0.83)(0.2,0.83)(0.2,1.13)\rput(0.5,0.98){\small{$_m$}}
\psarc[linewidth=1.5pt,linecolor=blue]{-}(1.5,1){0.5}{90}{-90}
\end{pspicture}
\ \ ,\qquad
\begin{pspicture}[shift=-0.9](0.0,0.0)(1.5,2)
\facegrid{(0,0)}{(1,2)}
\psarc[linewidth=0.025]{-}(0,0){0.16}{0}{90}
\psarc[linewidth=0.025]{-}(0,1){0.16}{0}{90}
\psarc[linewidth=1.5pt,linecolor=blue]{-}(1,1){0.5}{-90}{90}
\rput(0.5,.5){$u$}\rput(0.5,0.75){\tiny{$_{(m,1)}$}}
\rput(0.5,1.5){\small$u\!+\!\lambda$}\rput(0.5,1.75){\tiny{$_{(m,1)}$}}
\end{pspicture} \ = q^m(u_1) \ \,
\begin{pspicture}[shift=-1.2](0,-0.3)(1.5,2.3)
\facegrid{(0,0)}{(1,2)}
\pspolygon[fillstyle=solid,fillcolor=pink](0.2,0)(0.8,0)(0.8,-0.3)(0.2,-0.3)(0.2,0)\rput(0.5,-0.15){\small{$_m$}}
\pspolygon[fillstyle=solid,fillcolor=pink](0.2,2)(0.8,2)(0.8,2.3)(0.2,2.3)(0.2,2)\rput(0.5,2.16){\small{$_m$}}
\psarc[linewidth=1pt,linecolor=blue]{-}(1,1){0.44}{-90}{90}
\psarc[linewidth=1pt,linecolor=blue]{-}(1,1){0.56}{-90}{90}
\psarc[linewidth=1pt,linecolor=blue]{-}(1,0){0.44}{90}{180}
\psarc[linewidth=1pt,linecolor=blue]{-}(1,0){0.56}{90}{180}
\psarc[linewidth=1.5pt,linecolor=blue]{-}(0,1){0.5}{-90}{90}
\psarc[linewidth=1pt,linecolor=blue]{-}(1,2){0.44}{180}{270}
\psarc[linewidth=1pt,linecolor=blue]{-}(1,2){0.56}{180}{270}
\end{pspicture}\ 
=q^m(u_1) \ 
\begin{pspicture}[shift=-0.4](-0.5,0.5)(0.8,1.5)
\pspolygon[fillstyle=solid,fillcolor=pink](0.2,1.13)(0.8,1.13)(0.8,0.83)(0.2,0.83)(0.2,1.13)\rput(0.5,0.98){\small{$_m$}}
\psarc[linewidth=1.5pt,linecolor=blue]{-}(-0.5,1){0.5}{-90}{90}
\end{pspicture}\
\label{eq:fusedpushthru}
\ee 
where $q^m(u)$ is defined in (\ref{eq:qkm}).

%%%%%%%%%%%%%%%%%%%%%%%%
\subsection{Yang-Baxter equations}
\label{Sec:YBE}
%%%%%%%%%%%%%%%%%%%%%%%%%

The YBE for fused face operators is
\be 
\psset{unit=0.8}
\begin{pspicture}[shift=-.9](0,0)(3,2)
\psline[linecolor=blue,linewidth=4pt]{-}(2,0.5)(1.4,0.5)\psline[linecolor=white,linewidth=2pt]{-}(2,0.5)(1.4,0.5)
\psline[linecolor=blue,linewidth=4pt]{-}(2,1.5)(1.4,1.5)\psline[linecolor=white,linewidth=2pt]{-}(2,1.5)(1.4,1.5)
\facegrid{(2,0)}{(3,2)}
\pspolygon[fillstyle=solid,fillcolor=lightlightblue](0,1)(1,2)(2,1)(1,0)(0,1)
\rput(1,1.3){\tiny{$_{(n,\ell)}$}}
\psarc[linewidth=0.025]{-}(2,0){0.16}{0}{90}
\psarc[linewidth=0.025]{-}(2,1){0.16}{0}{90}
\psarc[linewidth=0.025]{-}(0,1){0.21}{-45}{45}
\rput(2.5,.45){$u$}\rput(2.5,0.75){\tiny{$_{(m,\ell)}$}}
\rput(2.5,1.45){$v$}\rput(2.5,1.75){\tiny{$_{(m,n)}$}}
\rput(1,1){$u-v$}
\end{pspicture}
\ \ =\ \ 
\begin{pspicture}[shift=-.9](0,0)(3,2)
\psline[linecolor=blue,linewidth=4pt]{-}(1,0.5)(1.6,0.5)\psline[linecolor=white,linewidth=2pt]{-}(1,0.5)(1.6,0.5)
\psline[linecolor=blue,linewidth=4pt]{-}(1,1.5)(1.6,1.5)\psline[linecolor=white,linewidth=2pt]{-}(1,1.5)(1.6,1.5)
\facegrid{(0,0)}{(1,2)}
\pspolygon[fillstyle=solid,fillcolor=lightlightblue](1,1)(2,2)(3,1)(2,0)(1,1)
\psarc[linewidth=0.025]{-}(0,0){0.16}{0}{90}
\psarc[linewidth=0.025]{-}(0,1){0.16}{0}{90}
\psarc[linewidth=0.025]{-}(1,1){0.21}{-45}{45}
\rput(.5,.45){$v$}\rput(0.5,0.75){\tiny{$_{(m,n)}$}}
\rput(.5,1.45){$u$}\rput(0.5,1.75){\tiny{$_{(m,\ell)}$}}
\rput(2,1){$u-v$}\rput(2,1.3){\tiny{$_{(n,\ell)}$}}
\end{pspicture} \  
\label{eq:fusedYBE}
\ee
For general $\ell,m,n\in\mathbb{N}$, this identity follows readily from the YBE \eqref{eq:YBE}
for the elementary $1\times 1$ face operators and properties of the projectors. 
The fused face operators also satisfy~\cite{BehrendPearce} the left and right BYBEs
\be 
\psset{unit=0.6364cm}
\begin{pspicture}[shift=-1.9](-3.2,0)(0,4)
\psarc[linecolor=blue,linewidth=4pt]{-}(-2.8,1){-0.5}{-90}{90}\psarc[linecolor=white,linewidth=2pt]{-}(-2.8,1){-0.5}{-90}{90}
\psarc[linecolor=blue,linewidth=4pt]{-}(-2.8,3){-0.5}{-90}{90}\psarc[linecolor=white,linewidth=2pt]{-}(-2.8,3){-0.5}{-90}{90}
\psline[linecolor=blue,linewidth=4pt]{-}(0,0.5)(-0.6,0.5)\psline[linecolor=white,linewidth=2pt]{-}(0,0.5)(-0.6,0.5)
\psline[linecolor=blue,linewidth=4pt]{-}(0,1.5)(-0.6,1.5)\psline[linecolor=white,linewidth=2pt]{-}(0,1.5)(-0.6,1.5)
\psline[linecolor=blue,linewidth=4pt]{-}(0,2.5)(-1.6,2.5)\psline[linecolor=white,linewidth=2pt]{-}(0,2.5)(-1.6,2.5)
\psline[linecolor=blue,linewidth=4pt]{-}(0,3.5)(-2.8,3.5)\psline[linecolor=white,linewidth=2pt]{-}(0,3.5)(-2.8,3.5)
\psline[linecolor=blue,linewidth=4pt]{-}(-1.4,0.5)(-2.8,0.5)\psline[linecolor=white,linewidth=2pt]{-}(-1.4,0.5)(-2.8,0.5)
\psline[linecolor=blue,linewidth=4pt]{-}(-2.4,1.5)(-2.8,1.5)\psline[linecolor=white,linewidth=2pt]{-}(-2.4,1.5)(-2.8,1.5)
\psline[linecolor=blue,linewidth=4pt]{-}(-2.4,2.5)(-2.8,2.5)\psline[linecolor=white,linewidth=2pt]{-}(-2.4,2.5)(-2.8,2.5)
\pspolygon[fillstyle=solid,fillcolor=lightlightblue](0,1)(-1,2)(-2,1)(-1,0)(0,1)
\pspolygon[fillstyle=solid,fillcolor=lightlightblue](-1,2)(-2,1)(-3,2)(-2,3)(-1,2)
\psline{-}(0,0)(0,4)
\psarc[linewidth=0.025]{-}(-2,1){0.21}{-45}{45}\rput(-1,1){$u$}\rput(-1,1.35){\tiny{$_{(m,n)}$}}
\psarc[linewidth=0.025]{-}(-3,2){0.21}{-45}{45}\rput(-2,2){$v$}\rput(-2,2.35){\tiny{$_{(m,n)}$}}
\end{pspicture}
\ \  =\ \
\begin{pspicture}[shift=-1.9](-3.2,0)(0,4)
\psarc[linecolor=blue,linewidth=4pt]{-}(-2.8,1){-0.5}{-90}{90}\psarc[linecolor=white,linewidth=2pt]{-}(-2.8,1){-0.5}{-90}{90}
\psarc[linecolor=blue,linewidth=4pt]{-}(-2.8,3){-0.5}{-90}{90}\psarc[linecolor=white,linewidth=2pt]{-}(-2.8,3){-0.5}{-90}{90}
\psline[linecolor=blue,linewidth=4pt]{-}(0,2.5)(-0.6,2.5)\psline[linecolor=white,linewidth=2pt]{-}(0,2.5)(-0.6,2.5)
\psline[linecolor=blue,linewidth=4pt]{-}(0,3.5)(-0.6,3.5)\psline[linecolor=white,linewidth=2pt]{-}(0,3.5)(-0.6,3.5)
\psline[linecolor=blue,linewidth=4pt]{-}(0,1.5)(-1.6,1.5)\psline[linecolor=white,linewidth=2pt]{-}(0,1.5)(-1.6,1.5)
\psline[linecolor=blue,linewidth=4pt]{-}(0,0.5)(-2.8,0.5)\psline[linecolor=white,linewidth=2pt]{-}(0,0.5)(-2.8,0.5)
\psline[linecolor=blue,linewidth=4pt]{-}(-1.4,3.5)(-2.8,3.5)\psline[linecolor=white,linewidth=2pt]{-}(-1.4,3.5)(-2.8,3.5)
\psline[linecolor=blue,linewidth=4pt]{-}(-2.4,2.5)(-2.8,2.5)\psline[linecolor=white,linewidth=2pt]{-}(-2.4,2.5)(-2.8,2.5)
\psline[linecolor=blue,linewidth=4pt]{-}(-2.4,1.5)(-2.8,1.5)\psline[linecolor=white,linewidth=2pt]{-}(-2.4,1.5)(-2.8,1.5)
\pspolygon[fillstyle=solid,fillcolor=lightlightblue](0,3)(-1,4)(-2,3)(-1,2)(0,3)
\pspolygon[fillstyle=solid,fillcolor=lightlightblue](-1,2)(-2,1)(-3,2)(-2,3)(-1,2)
\psline{-}(0,0)(0,4)
\psarc[linewidth=0.025]{-}(-2,3){0.21}{-45}{45}\rput(-1,3){\scriptsize$u_{n\!-\!m}$}\rput(-1,3.35){\tiny{$_{(n,m)}$}}
\psarc[linewidth=0.025]{-}(-3,2){0.21}{-45}{45}\rput(-2,2){\scriptsize$v_{n\!-\!m}$}\rput(-2,2.35){\tiny{$_{(n,m)}$}}
\end{pspicture} \
\qquad \qquad \qquad 
\begin{pspicture}[shift=-1.9](0,0)(3.2,4)
\psarc[linecolor=blue,linewidth=4pt]{-}(2.8,1){0.5}{-90}{90}\psarc[linecolor=white,linewidth=2pt]{-}(2.8,1){0.5}{-90}{90}
\psarc[linecolor=blue,linewidth=4pt]{-}(2.8,3){0.5}{-90}{90}\psarc[linecolor=white,linewidth=2pt]{-}(2.8,3){0.5}{-90}{90}
\psline[linecolor=blue,linewidth=4pt]{-}(0,0.5)(0.6,0.5)\psline[linecolor=white,linewidth=2pt]{-}(0,0.5)(0.6,0.5)
\psline[linecolor=blue,linewidth=4pt]{-}(0,1.5)(0.6,1.5)\psline[linecolor=white,linewidth=2pt]{-}(0,1.5)(0.6,1.5)
\psline[linecolor=blue,linewidth=4pt]{-}(0,2.5)(1.6,2.5)\psline[linecolor=white,linewidth=2pt]{-}(0,2.5)(1.6,2.5)
\psline[linecolor=blue,linewidth=4pt]{-}(0,3.5)(2.8,3.5)\psline[linecolor=white,linewidth=2pt]{-}(0,3.5)(2.8,3.5)
\psline[linecolor=blue,linewidth=4pt]{-}(1.4,0.5)(2.8,0.5)\psline[linecolor=white,linewidth=2pt]{-}(1.4,0.5)(2.8,0.5)
\psline[linecolor=blue,linewidth=4pt]{-}(2.4,1.5)(2.8,1.5)\psline[linecolor=white,linewidth=2pt]{-}(2.4,1.5)(2.8,1.5)
\psline[linecolor=blue,linewidth=4pt]{-}(2.4,2.5)(2.8,2.5)\psline[linecolor=white,linewidth=2pt]{-}(2.4,2.5)(2.8,2.5)
\pspolygon[fillstyle=solid,fillcolor=lightlightblue](0,1)(1,2)(2,1)(1,0)(0,1)
\pspolygon[fillstyle=solid,fillcolor=lightlightblue](1,2)(2,1)(3,2)(2,3)(1,2)
\psline{-}(0,0)(0,4)
\psarc[linewidth=0.025]{-}(0,1){0.21}{-45}{45}\rput(1,1){$u$}\rput(1,1.35){\tiny{$_{(m,n)}$}}
\psarc[linewidth=0.025]{-}(1,2){0.21}{-45}{45}\rput(2,2){$v$}\rput(2,2.35){\tiny{$_{(m,n)}$}}
\end{pspicture}
\ \ =\ \ 
\begin{pspicture}[shift=-1.9](0,0)(3.2,4)
\psarc[linecolor=blue,linewidth=4pt]{-}(2.8,1){0.5}{-90}{90}\psarc[linecolor=white,linewidth=2pt]{-}(2.8,1){0.5}{-90}{90}
\psarc[linecolor=blue,linewidth=4pt]{-}(2.8,3){0.5}{-90}{90}\psarc[linecolor=white,linewidth=2pt]{-}(2.8,3){0.5}{-90}{90}
\psline[linecolor=blue,linewidth=4pt]{-}(0,2.5)(0.6,2.5)\psline[linecolor=white,linewidth=2pt]{-}(0,2.5)(0.6,2.5)
\psline[linecolor=blue,linewidth=4pt]{-}(0,3.5)(0.6,3.5)\psline[linecolor=white,linewidth=2pt]{-}(0,3.5)(0.6,3.5)
\psline[linecolor=blue,linewidth=4pt]{-}(0,1.5)(1.6,1.5)\psline[linecolor=white,linewidth=2pt]{-}(0,1.5)(1.6,1.5)
\psline[linecolor=blue,linewidth=4pt]{-}(0,0.5)(2.8,0.5)\psline[linecolor=white,linewidth=2pt]{-}(0,0.5)(2.8,0.5)
\psline[linecolor=blue,linewidth=4pt]{-}(1.4,3.5)(2.8,3.5)\psline[linecolor=white,linewidth=2pt]{-}(1.4,3.5)(2.8,3.5)
\psline[linecolor=blue,linewidth=4pt]{-}(2.4,2.5)(2.8,2.5)\psline[linecolor=white,linewidth=2pt]{-}(2.4,2.5)(2.8,2.5)
\psline[linecolor=blue,linewidth=4pt]{-}(2.4,1.5)(2.8,1.5)\psline[linecolor=white,linewidth=2pt]{-}(2.4,1.5)(2.8,1.5)
\pspolygon[fillstyle=solid,fillcolor=lightlightblue](0,3)(1,4)(2,3)(1,2)(0,3)
\pspolygon[fillstyle=solid,fillcolor=lightlightblue](1,2)(2,1)(3,2)(2,3)(1,2)
\psline{-}(0,0)(0,4)
\psarc[linewidth=0.025]{-}(0,3){0.21}{-45}{45}\rput(1,3){\scriptsize$u_{n\!-\!m}$}\rput(1,3.35){\tiny{$_{(n,m)}$}}
\psarc[linewidth=0.025]{-}(1,2){0.21}{-45}{45}\rput(2,2){\scriptsize$v_{n\!-\!m}$}\rput(2,2.35){\tiny{$_{(n,m)}$}}
\end{pspicture} \ \ \ 
\label{eq:fusedBYBE}
\ee 

\noindent  As in the non-fused case \eqref{eq:bybes}, the simplicity of the BYBE (\ref{eq:fusedBYBE}) is due to the fact that we are 
only considering vacuum boundary conditions. 
Proofs of \eqref{eq:fusedBYBE}, adapted from~\cite{BehrendPearce} to the 
present loop model context, are given in Appendix~\ref{app:YBEs}.

%%%%%%%%%%%%%%%%%%%%%%
\subsection{Crossing symmetry}
\label{Sec:Crossing}
%%%%%%%%%%%%%%%%%%%%%%

Fused faces of the form $(1,n)$ or $(m,1)$ decompose in terms of two generalised monoids only. This is merely a special
case of the more general situation where
a vertical array of elementary faces sandwiched between two $P_n$ projectors decomposes as
\be 
\psset{unit=0.9}
\begin{pspicture}[shift=-1.4](-0.3,0)(1.3,3)
\facegrid{(0,0)}{(1,3)}
\rput(0.55,0.55){$v^{(1)}$}
\rput(0.5,1.6){$\vdots$}
\rput(0.55,2.55){$v^{(n)}$}
\pspolygon[fillstyle=solid,fillcolor=pink](0,0.1)(0,2.9)(-0.3,2.9)(-0.3,0.1)(0,0.1)\rput(-0.15,1.5){\scriptsize{$_n$}}
\pspolygon[fillstyle=solid,fillcolor=pink](1,0.1)(1,2.9)(1.3,2.9)(1.3,0.1)(1,0.1)\rput(1.15,1.5){\scriptsize{$_n$}}
\psarc[linewidth=0.025]{-}(0,0){0.16}{0}{90}
\psarc[linewidth=0.025]{-}(0,2){0.16}{0}{90}
\end{pspicture}
\ \ = \; \prod_{j=1}^n s_0(v^{(j)}) \ \
\begin{pspicture}[shift=-1.4](-0.3,0)(1.3,3)
\facegrid{(0,0)}{(1,3)}
\pspolygon[fillstyle=solid,fillcolor=pink](0,0.1)(0,2.9)(-0.3,2.9)(-0.3,0.1)(0,0.1)\rput(-0.15,1.5){\scriptsize{$_n$}}
\pspolygon[fillstyle=solid,fillcolor=pink](1,0.1)(1,2.9)(1.3,2.9)(1.3,0.1)(1,0.1)\rput(1.15,1.5){\scriptsize{$_n$}}
\psarc[linecolor=blue,linewidth=1.5pt]{-}(0,0){0.5}{0}{90}
\psarc[linecolor=blue,linewidth=1.5pt]{-}(0,1){0.5}{0}{90}
\psarc[linecolor=blue,linewidth=1.5pt]{-}(0,2){0.5}{0}{90}
\psarc[linecolor=blue,linewidth=1.5pt]{-}(1,1){0.5}{180}{-90}
\psarc[linecolor=blue,linewidth=1.5pt]{-}(1,2){0.5}{180}{-90}
\psarc[linecolor=blue,linewidth=1.5pt]{-}(1,3){0.5}{180}{-90}
\end{pspicture} 
\ \ + \; \prod_{j=1}^n s_1(-v^{(j)}) \ \
\begin{pspicture}[shift=-1.4](-0.3,0)(1.3,3)
\facegrid{(0,0)}{(1,3)}
\pspolygon[fillstyle=solid,fillcolor=pink](0,0.1)(0,2.9)(-0.3,2.9)(-0.3,0.1)(0,0.1)\rput(-0.15,1.5){\scriptsize{$_n$}}
\pspolygon[fillstyle=solid,fillcolor=pink](1,0.1)(1,2.9)(1.3,2.9)(1.3,0.1)(1,0.1)\rput(1.15,1.5){\scriptsize{$_n$}}
\psarc[linecolor=blue,linewidth=1.5pt]{-}(1,0){0.5}{90}{180}
\psarc[linecolor=blue,linewidth=1.5pt]{-}(1,1){0.5}{90}{180}
\psarc[linecolor=blue,linewidth=1.5pt]{-}(1,2){0.5}{90}{180}
\psarc[linecolor=blue,linewidth=1.5pt]{-}(0,1){0.5}{-90}{0}
\psarc[linecolor=blue,linewidth=1.5pt]{-}(0,2){0.5}{-90}{0}
\psarc[linecolor=blue,linewidth=1.5pt]{-}(0,3){0.5}{-90}{0}
\end{pspicture}
\label{eq:2termexpansion}
\ee
It readily follows that fusion of single columns or rows yields fully symmetric functions of the spectral parameters,
\be 
\psset{unit=0.9}
\begin{pspicture}[shift=-1.4](-0.3,0)(1.3,3)
\facegrid{(0,0)}{(1,3)}
\rput(0.55,0.55){$v^{(1)}$}
\rput(0.5,1.6){$\vdots$}
\rput(0.55,2.55){$v^{(n)}$}
\pspolygon[fillstyle=solid,fillcolor=pink](0,0.1)(0,2.9)(-0.3,2.9)(-0.3,0.1)(0,0.1)\rput(-0.15,1.5){\scriptsize{$_n$}}
\pspolygon[fillstyle=solid,fillcolor=pink](1,0.1)(1,2.9)(1.3,2.9)(1.3,0.1)(1,0.1)\rput(1.15,1.5){\scriptsize{$_n$}}
\psarc[linewidth=0.025]{-}(0,0){0.16}{0}{90}
\psarc[linewidth=0.025]{-}(0,2){0.16}{0}{90}
\end{pspicture}
\ \  = \ \ 
\begin{pspicture}[shift=-1.4](-0.3,0)(1.3,3)
\facegrid{(0,0)}{(1,3)}
\rput(0.55,0.55){$w^{(1)}$}
\rput(0.5,1.6){$\vdots$}
\rput(0.55,2.55){$w^{(n)}$}
\pspolygon[fillstyle=solid,fillcolor=pink](0,0.1)(0,2.9)(-0.3,2.9)(-0.3,0.1)(0,0.1)\rput(-0.15,1.5){\scriptsize{$_n$}}
\pspolygon[fillstyle=solid,fillcolor=pink](1,0.1)(1,2.9)(1.3,2.9)(1.3,0.1)(1,0.1)\rput(1.15,1.5){\scriptsize{$_n$}}
\psarc[linewidth=0.025]{-}(0,0){0.16}{0}{90}
\psarc[linewidth=0.025]{-}(0,2){0.16}{0}{90}
\end{pspicture}\ \ ,
\quad\qquad
(w^{(1)},\ldots,w^{(n)})=\mathcal{P}_n(v^{(1)},\ldots,v^{(n)})
\label{vw}
\ee
where the face weights are related by a permutation operator $\mathcal{P}_n$ on $n$ items.
Similar relations obviously hold for horizontal arrays of elementary faces. 
As an immediate consequence, we have the crossing relation
\be \psset{unit=0.9}
\begin{pspicture}[shift=-0.4](0,0)(1,1)
\facegrid{(0,0)}{(1,1)}
\psarc[linewidth=0.025]{-}(0,0){0.16}{0}{90}
\rput(0.5,.45){$u$}\rput(0.5,0.75){\tiny{$_{(1,n)}$}}
\end{pspicture}
 \ \ = \ \ 
\begin{pspicture}[shift=-0.4](0,0)(1,1)
\facegrid{(0,0)}{(1,1)}
\psarc[linewidth=0.025]{-}(1,1){0.16}{180}{-90}
\rput(0.5,.45){$u$}\rput(0.5,0.75){\tiny{$_{(1,n)}$}}
\end{pspicture}
\ee
Generally, the fused faces enjoy the crossing relations
\be
\psset{unit=0.9}
\begin{pspicture}[shift=-0.4](0,0)(1,1)
\facegrid{(0,0)}{(1,1)}
\psarc[linewidth=0.025]{-}(0,0){0.16}{0}{90}
\rput(0.5,.45){$u$}\rput(0.5,0.75){\tiny{$_{(m,n)}$}}
\end{pspicture}
\ \ =\ \ 
\begin{pspicture}[shift=-0.4](0,0)(1,1)
\facegrid{(0,0)}{(1,1)}
\psarc[linewidth=0.025]{-}(1,0){0.16}{90}{180}
\rput(0.5,.45){\small$\lambda\!-\!u$}\rput(0.5,0.75){\tiny{$_{(n,m)}$}}
\end{pspicture}
\ \ =\ \ 
\begin{pspicture}[shift=-0.4](0,0)(1,1)
\facegrid{(0,0)}{(1,1)}
\psarc[linewidth=0.025]{-}(1,1){0.16}{180}{-90}
\rput(0.5,.45){$u$}\rput(0.5,0.75){\tiny{$_{(m,n)}$}}
\end{pspicture}
\ \ =\ \ 
\begin{pspicture}[shift=-0.4](0,0)(1,1)
\facegrid{(0,0)}{(1,1)}
\psarc[linewidth=0.025]{-}(0,1){0.16}{270}{0}
\rput(0.5,.45){\small$\lambda\!-\!u$}\rput(0.5,0.75){\tiny{$_{(n,m)}$}}
\end{pspicture}
\label{eq:fusedcrossing}
\ee
To prove the crossing relations for $m>1$, extra projectors are inserted inside the fused faces. This is made possible because of the 
push-through properties \eqref{eq:pushthru} and property \eqref{eq:(ii)} of the WJ projectors. 
The proof then uses the crossing property already established for $(1,n)$, from which the similar results for $(m,1)$ follow. For example, 
for $(m,n) = (2,3)$,

\begin{align}
\psset{unit=0.9}
\begin{pspicture}[shift=-0.4](0,0)(1,1)
\facegrid{(0,0)}{(1,1)}
\psarc[linewidth=0.025]{-}(0,0){0.16}{0}{90}
\rput(0.5,.45){$u$}\rput(0.5,0.75){\tiny{$_{(2,3)}$}}
\end{pspicture}
\ \ &= \ \
\psset{unit=0.9}
\begin{pspicture}[shift=-1.7](-0.3,-0.3)(2.3,3.3)
\facegrid{(0,0)}{(2,3)}
\pspolygon[fillstyle=solid,fillcolor=pink](0,0.1)(0,2.9)(-0.3,2.9)(-0.3,0.1)(0,0.1)\rput(-0.15,1.5){\small{$_3$}}
\pspolygon[fillstyle=solid,fillcolor=pink](2,0.1)(2,2.9)(2.3,2.9)(2.3,0.1)(2,0.1)\rput(2.15,1.5){\small{$_3$}}
\pspolygon[fillstyle=solid,fillcolor=pink](0.1,3)(0.1,3.3)(1.9,3.3)(1.9,3)(0.1,3)\rput(1,3.15){\small{$_2$}}
\pspolygon[fillstyle=solid,fillcolor=pink](0.1,0)(0.1,-0.3)(1.9,-0.3)(1.9,0)(0.1,0)\rput(1,-0.15){\small{$_2$}}
\psarc[linewidth=0.025]{-}(0,0){0.16}{0}{90}
\psarc[linewidth=0.025]{-}(0,1){0.16}{0}{90}
\psarc[linewidth=0.025]{-}(0,2){0.16}{0}{90}
\psarc[linewidth=0.025]{-}(1,0){0.16}{0}{90}
\psarc[linewidth=0.025]{-}(1,1){0.16}{0}{90}
\psarc[linewidth=0.025]{-}(1,2){0.16}{0}{90}
\rput(0.5,1.5){$u_0$}
\rput(0.5,2.5){$u_1$}
\rput(0.5,0.5){$u_{-\!1}$}
\rput(1.5,0.5){$u_0$}
\rput(1.5,1.5){$u_1$}
\rput(1.5,2.5){$u_2$}
\end{pspicture} \ \ = \ \
\begin{pspicture}[shift=-1.7](-0.3,-0.3)(2.6,3.3)
\facegrid{(0,0)}{(1,3)}
\pspolygon[fillstyle=solid,fillcolor=lightlightblue](1.3,0)(2.3,0)(2.3,3)(1.3,3)(1.3,0)\psline{-}(1.3,1)(2.3,1)\psline{-}(1.3,2)(2.3,2)
\pspolygon[fillstyle=solid,fillcolor=pink](0,0.1)(0,2.9)(-0.3,2.9)(-0.3,0.1)(0,0.1)\rput(-0.15,1.5){\small{$_3$}}
\pspolygon[fillstyle=solid,fillcolor=pink](2.3,0.1)(2.3,2.9)(2.6,2.9)(2.6,0.1)(2.3,0.1)\rput(2.45,1.5){\small{$_3$}}
\pspolygon[fillstyle=solid,fillcolor=pink](1.0,0.1)(1.0,2.9)(1.3,2.9)(1.3,0.1)(1.0,0.1)\rput(1.15,1.5){\small{$_3$}}
\pspolygon[fillstyle=solid,fillcolor=pink](0.1,3)(0.1,3.3)(2.2,3.3)(2.2,3)(0.1,3)\rput(1.15,3.15){\small{$_2$}}
\pspolygon[fillstyle=solid,fillcolor=pink](0.1,0)(0.1,-0.3)(2.2,-0.3)(2.2,0)(0.1,0)\rput(1.15,-0.15){\small{$_2$}}
\psarc[linewidth=0.025]{-}(0,0){0.16}{0}{90}
\psarc[linewidth=0.025]{-}(0,1){0.16}{0}{90}
\psarc[linewidth=0.025]{-}(0,2){0.16}{0}{90}
\psarc[linewidth=0.025]{-}(1.3,0){0.16}{0}{90}
\psarc[linewidth=0.025]{-}(1.3,1){0.16}{0}{90}
\psarc[linewidth=0.025]{-}(1.3,2){0.16}{0}{90}
\rput(0.5,1.5){$u_0$}
\rput(0.5,2.5){$u_1$}
\rput(0.5,0.5){$u_{-\!1}$}
\rput(1.8,0.5){$u_0$}
\rput(1.8,1.5){$u_1$}
\rput(1.8,2.5){$u_2$}
\end{pspicture} \ \ = \ \
\begin{pspicture}[shift=-1.7](-0.3,-0.3)(2.6,3.3)
\facegrid{(0,0)}{(1,3)}
\pspolygon[fillstyle=solid,fillcolor=lightlightblue](1.3,0)(2.3,0)(2.3,3)(1.3,3)(1.3,0)\psline{-}(1.3,1)(2.3,1)\psline{-}(1.3,2)(2.3,2)
\pspolygon[fillstyle=solid,fillcolor=pink](0,0.1)(0,2.9)(-0.3,2.9)(-0.3,0.1)(0,0.1)\rput(-0.15,1.5){\small{$_3$}}
\pspolygon[fillstyle=solid,fillcolor=pink](2.3,0.1)(2.3,2.9)(2.6,2.9)(2.6,0.1)(2.3,0.1)\rput(2.45,1.5){\small{$_3$}}
\pspolygon[fillstyle=solid,fillcolor=pink](1.0,0.1)(1.0,2.9)(1.3,2.9)(1.3,0.1)(1.0,0.1)\rput(1.15,1.5){\small{$_3$}}
\pspolygon[fillstyle=solid,fillcolor=pink](0.1,3)(0.1,3.3)(2.2,3.3)(2.2,3)(0.1,3)\rput(1.15,3.15){\small{$_2$}}
\pspolygon[fillstyle=solid,fillcolor=pink](0.1,0)(0.1,-0.3)(2.2,-0.3)(2.2,0)(0.1,0)\rput(1.15,-0.15){\small{$_2$}}
\psarc[linewidth=0.025]{-}(0,0){0.16}{0}{90}
\psarc[linewidth=0.025]{-}(0,1){0.16}{0}{90}
\psarc[linewidth=0.025]{-}(0,2){0.16}{0}{90}
\psarc[linewidth=0.025]{-}(1.3,0){0.16}{0}{90}
\psarc[linewidth=0.025]{-}(1.3,1){0.16}{0}{90}
\psarc[linewidth=0.025]{-}(1.3,2){0.16}{0}{90}
\rput(0.5,0.5){$u_1$}
\rput(0.5,1.5){$u_0$}
\rput(0.5,2.5){$u_{-\!1}$}
\rput(1.8,2.5){$u_0$}
\rput(1.8,1.5){$u_1$}
\rput(1.8,0.5){$u_2$}
\end{pspicture} \nonumber \\ \\
& = \ \ \psset{unit=0.9}
\begin{pspicture}[shift=-1.55](-0.3,-0.3)(2.3,3.0)
\facegrid{(0,0)}{(2,3)}
\pspolygon[fillstyle=solid,fillcolor=pink](0,0.1)(0,2.9)(-0.3,2.9)(-0.3,0.1)(0,0.1)\rput(-0.15,1.5){\small{$_3$}}
\pspolygon[fillstyle=solid,fillcolor=pink](2,0.1)(2,2.9)(2.3,2.9)(2.3,0.1)(2,0.1)\rput(2.15,1.5){\small{$_3$}}
\pspolygon[fillstyle=solid,fillcolor=pink](0.1,3)(0.1,3.3)(1.9,3.3)(1.9,3)(0.1,3)\rput(1,3.15){\small{$_2$}}
\pspolygon[fillstyle=solid,fillcolor=pink](0.1,0)(0.1,-0.3)(1.9,-0.3)(1.9,0)(0.1,0)\rput(1,-0.15){\small{$_2$}}
\psarc[linewidth=0.025]{-}(0,0){0.16}{0}{90}
\psarc[linewidth=0.025]{-}(0,1){0.16}{0}{90}
\psarc[linewidth=0.025]{-}(0,2){0.16}{0}{90}
\psarc[linewidth=0.025]{-}(1,0){0.16}{0}{90}
\psarc[linewidth=0.025]{-}(1,1){0.16}{0}{90}
\psarc[linewidth=0.025]{-}(1,2){0.16}{0}{90}
\rput(0.5,2.5){$u_{-\!1}$}
\rput(0.5,1.5){$u_0$}
\rput(0.5,0.5){$u_1$}
\rput(1.5,2.5){$u_0$}
\rput(1.5,1.5){$u_1$}
\rput(1.5,0.5){$u_2$}
\end{pspicture} 
\, \ \  = \ \ \, 
\begin{pspicture}[shift=-1.55](-0.3,-0.3)(2.3,3.0)
\facegrid{(0,0)}{(2,3)}
\pspolygon[fillstyle=solid,fillcolor=pink](0,0.1)(0,2.9)(-0.3,2.9)(-0.3,0.1)(0,0.1)\rput(-0.15,1.5){\small{$_3$}}
\pspolygon[fillstyle=solid,fillcolor=pink](2,0.1)(2,2.9)(2.3,2.9)(2.3,0.1)(2,0.1)\rput(2.15,1.5){\small{$_3$}}
\pspolygon[fillstyle=solid,fillcolor=pink](0.1,3)(0.1,3.3)(1.9,3.3)(1.9,3)(0.1,3)\rput(1,3.15){\small{$_2$}}
\pspolygon[fillstyle=solid,fillcolor=pink](0.1,0)(0.1,-0.3)(1.9,-0.3)(1.9,0)(0.1,0)\rput(1,-0.15){\small{$_2$}}
\psarc[linewidth=0.025]{-}(1,0){0.16}{90}{180}
\psarc[linewidth=0.025]{-}(1,1){0.16}{90}{180}
\psarc[linewidth=0.025]{-}(1,2){0.16}{90}{180}
\psarc[linewidth=0.025]{-}(2,0){0.16}{90}{180}
\psarc[linewidth=0.025]{-}(2,1){0.16}{90}{180}
\psarc[linewidth=0.025]{-}(2,2){0.16}{90}{180}
\rput(0.5,2.5){\tiny$\lambda\!-\!u_{-\!1}$}
\rput(0.5,1.5){\scriptsize$\lambda\!-\!u_0$}
\rput(0.5,0.5){\scriptsize$\lambda\!-\!u_1$}
\rput(1.5,2.5){\scriptsize$\lambda\!-\!u_0$}
\rput(1.5,1.5){\scriptsize$\lambda\!-\!u_1$}
\rput(1.5,0.5){\scriptsize$\lambda\!-\!u_2$}
\end{pspicture} 
\ \ = \ \ 
\begin{pspicture}[shift=-0.4](0,0)(1,1)
\facegrid{(0,0)}{(1,1)}
\psarc[linewidth=0.025]{-}(1,0){0.16}{90}{180}
\rput(0.5,.45){\small$\lambda\!-\!u$}\rput(0.5,0.75){\tiny{$_{(3,2)}$}}
\end{pspicture}
\ \, \nonumber
\end{align}

\noindent
This is readily extended from $(2,3)$ to general $(m,n)$, and
the remaining equalities in \eqref{eq:fusedcrossing} are proved using similar arguments.

%%%%%%%%%%%%%%%%%
\section{Transfer tangles}
\label{Sec:Tangles}
%%%%%%%%%%%%%%%%%

For $\lambda$ generic, a set of $m$ cabled strands is obtained by inserting a $P_m$ projector acting on these $m$ strands,
\be
 \underbrace{
\begin{pspicture}(0.1,-0.8)(1.9,0.7)
\rput(1.4,0){\small$...$}
\psline[linecolor=blue,linewidth=1.5pt]{-}(0.2,-0.65)(0.2,0.65)
\psline[linecolor=blue,linewidth=1.5pt]{-}(0.6,-0.65)(0.6,0.65)
\psline[linecolor=blue,linewidth=1.5pt]{-}(1.0,-0.65)(1.0,0.65)
\psline[linecolor=blue,linewidth=1.5pt]{-}(1.8,-0.65)(1.8,0.65)
\end{pspicture} 
}_{{\scriptsize \begin{array}{c} m\; \mathrm{free}\\ \mathrm{strands}\end{array}}}
\qquad\qquad
 \underbrace{
\begin{pspicture}(0.1,-0.8)(1.9,0.7)
\pspolygon[fillstyle=solid,fillcolor=pink](0.1,0.15)(0.1,-0.15)(1.9,-0.15)(1.9,0.15)(0.1,0.15)
\rput(1,0){$_m$}
\rput(1.4,0.4){\small$...$}\rput(1.4,-0.4){\small$...$}
\psline[linecolor=blue,linewidth=1.5pt]{-}(0.2,0.65)(0.2,0.15)
\psline[linecolor=blue,linewidth=1.5pt]{-}(0.6,0.65)(0.6,0.15)
\psline[linecolor=blue,linewidth=1.5pt]{-}(1.0,0.65)(1.0,0.15)
\psline[linecolor=blue,linewidth=1.5pt]{-}(1.8,0.65)(1.8,0.15)
\psline[linecolor=blue,linewidth=1.5pt]{-}(0.2,-0.65)(0.2,-0.15)
\psline[linecolor=blue,linewidth=1.5pt]{-}(0.6,-0.65)(0.6,-0.15)
\psline[linecolor=blue,linewidth=1.5pt]{-}(1.0,-0.65)(1.0,-0.15)
\psline[linecolor=blue,linewidth=1.5pt]{-}(1.8,-0.65)(1.8,-0.15)
\end{pspicture} 
}_{{\scriptsize \begin{array}{c} m\; \mathrm{cabled}\\ \mathrm{strands}\end{array}}}
\label{strands}
\ee
As discussed at the end of Section~\ref{Sec:TL}, some projectors fail to exist for fractional $\lambda$ and in this case, 
an alternative scheme must be used to project out
unwanted connections. This is described in Section~\ref{Sec:CabledLinkStates}.
For now, we keep $\lambda$ generic and note that on
an array of $N$ sets of $m$-cabled strands as in (\ref{strands}), the identity tangle is given by
\be
\Ib^{m}\,=\! 
\begin{pspicture}[shift=-0.6](-0.2,-0.7)(7.2,0.7)
\pspolygon[fillstyle=solid,fillcolor=pink](0.1,0.15)(0.1,-0.15)(1.9,-0.15)(1.9,0.15)(0.1,0.15)
\pspolygon[fillstyle=solid,fillcolor=pink](2.1,0.15)(2.1,-0.15)(3.9,-0.15)(3.9,0.15)(2.1,0.15)
\pspolygon[fillstyle=solid,fillcolor=pink](5.1,0.15)(5.1,-0.15)(6.9,-0.15)(6.9,0.15)(5.1,0.15)
\rput(1,0){$_m$}
\rput(3,0){$_m$}
\rput(6.0,0){$_m$}
\rput(1.4,0.4){\small$...$}\rput(1.4,-0.4){\small$...$}
\rput(3.4,0.4){\small$...$}\rput(3.4,-0.4){\small$...$}
\rput(6.4,0.4){\small$...$}\rput(6.4,-0.4){\small$...$}
\rput(4.5,0.23){\small$...$}\rput(4.5,-0.23){\small$...$}
\psline[linecolor=blue,linewidth=1.5pt]{-}(0.2,0.65)(0.2,0.15)
\psline[linecolor=blue,linewidth=1.5pt]{-}(0.6,0.65)(0.6,0.15)
\psline[linecolor=blue,linewidth=1.5pt]{-}(1.0,0.65)(1.0,0.15)
\psline[linecolor=blue,linewidth=1.5pt]{-}(1.8,0.65)(1.8,0.15)
\psline[linecolor=blue,linewidth=1.5pt]{-}(2.2,0.65)(2.2,0.15)
\psline[linecolor=blue,linewidth=1.5pt]{-}(2.6,0.65)(2.6,0.15)
\psline[linecolor=blue,linewidth=1.5pt]{-}(3.0,0.65)(3.0,0.15)
\psline[linecolor=blue,linewidth=1.5pt]{-}(3.8,0.65)(3.8,0.15)
\psline[linecolor=blue,linewidth=1.5pt]{-}(5.2,0.65)(5.2,0.15)
\psline[linecolor=blue,linewidth=1.5pt]{-}(5.6,0.65)(5.6,0.15)
\psline[linecolor=blue,linewidth=1.5pt]{-}(6.0,0.65)(6.0,0.15)
\psline[linecolor=blue,linewidth=1.5pt]{-}(6.8,0.65)(6.8,0.15)
\psline[linecolor=blue,linewidth=1.5pt]{-}(0.2,-0.65)(0.2,-0.15)
\psline[linecolor=blue,linewidth=1.5pt]{-}(0.6,-0.65)(0.6,-0.15)
\psline[linecolor=blue,linewidth=1.5pt]{-}(1.0,-0.65)(1.0,-0.15)
\psline[linecolor=blue,linewidth=1.5pt]{-}(1.8,-0.65)(1.8,-0.15)
\psline[linecolor=blue,linewidth=1.5pt]{-}(2.2,-0.65)(2.2,-0.15)
\psline[linecolor=blue,linewidth=1.5pt]{-}(2.6,-0.65)(2.6,-0.15)
\psline[linecolor=blue,linewidth=1.5pt]{-}(3.0,-0.65)(3.0,-0.15)
\psline[linecolor=blue,linewidth=1.5pt]{-}(3.8,-0.65)(3.8,-0.15)
\psline[linecolor=blue,linewidth=1.5pt]{-}(5.2,-0.65)(5.2,-0.15)
\psline[linecolor=blue,linewidth=1.5pt]{-}(5.6,-0.65)(5.6,-0.15)
\psline[linecolor=blue,linewidth=1.5pt]{-}(6.0,-0.65)(6.0,-0.15)
\psline[linecolor=blue,linewidth=1.5pt]{-}(6.8,-0.65)(6.8,-0.15)
\end{pspicture}
\label{eq:I}
\ee 
where the number of $P_m$ projectors is $N$. 
For $m=1$, this reduces to the identity tangle in \eqref{Ib} as $\Ib^1=\Ib$. 
For $m>1$, on the other hand, $\Ib^m$ does not act as the identity on free strands.
Instead, it is the identity of the subalgebras $FTL_{N,m}(\beta)\subset TL_{Nm}(\beta)$ and
$F\mathcal E PTL_{N,m}(\alpha,\beta)\subset \mathcal E PTL_{Nm}(\alpha,\beta)$
defined in Section~\ref{Sec:CabledLinkStates}.

%%%%%%%%%%%%%%%%%
\subsection[Fused transfer tangles for generic $\lambda$]{Fused transfer tangles for generic $\boldsymbol \lambda$}
\label{Sec:nonrational}
%%%%%%%%%%%%%%%%%

For $\lambda$ generic, the {\em multi-row transfer tangles} are defined by
\be 
\Db^{m,n}(u):= 
\begin{pspicture}[shift=-1.6](-0.7,-0.7)(5.5,2)
\psarc[linewidth=4pt,linecolor=blue]{-}(0,1){0.5}{90}{-90}\psarc[linewidth=2pt,linecolor=white]{-}(0,1){0.5}{90}{-90}
\psarc[linewidth=4pt,linecolor=blue]{-}(5,1){0.5}{-90}{90}\psarc[linewidth=2pt,linecolor=white]{-}(5,1){0.5}{-90}{90}
\facegrid{(0,0)}{(5,2)}
\psarc[linewidth=0.025]{-}(0,0){0.16}{0}{90}
\psarc[linewidth=0.025]{-}(0,1){0.16}{0}{90}
\psarc[linewidth=0.025]{-}(1,0){0.16}{0}{90}
\psarc[linewidth=0.025]{-}(1,1){0.16}{0}{90}
\psarc[linewidth=0.025]{-}(4,0){0.16}{0}{90}
\psarc[linewidth=0.025]{-}(4,1){0.16}{0}{90}
\rput(0.5,0.75){\tiny{$_{(m,n)}$}}\rput(0.5,1.75){\tiny{$_{(m,n)}$}}
\rput(1.5,0.75){\tiny{$_{(m,n)}$}}\rput(1.5,1.75){\tiny{$_{(m,n)}$}}
\rput(4.5,0.75){\tiny{$_{(m,n)}$}}\rput(4.5,1.75){\tiny{$_{(m,n)}$}}
\rput(2.5,0.5){$\ldots$}
\rput(2.5,1.5){$\ldots$}
\rput(3.5,0.5){$\ldots$}
\rput(3.5,1.5){$\ldots$}
\rput(0.5,.5){$u$}
\rput(0.52,1.45){\scriptsize$\mu\!\!-\!\!u_{n\!-\!1}$}
\rput(1.5,.5){$u$}
\rput(1.52,1.45){\scriptsize$\mu\!\!-\!\!u_{n\!-\!1}$}
\rput(4.5,.5){$u$}
\rput(4.52,1.45){\scriptsize$\mu\!\!-\!\!u_{n\!-\!1}$}
\rput(2.5,-0.5){$\underbrace{\qquad \qquad \qquad \qquad \qquad \quad \qquad}_N$}
\end{pspicture} \ \ \ 
\label{eq:transfermatrix}
\ee  
and
\be 
\Tb^{m,n} (u):= \
\begin{pspicture}[shift=-0.5](-0.2,-0.1)(5.2,1.2)
\psline[linewidth=4pt,linecolor=blue]{-}(0,0.5)(-0.2,0.5)\psline[linewidth=2pt,linecolor=white]{-}(0,0.5)(-0.2,0.5)
\psline[linewidth=4pt,linecolor=blue]{-}(5,0.5)(5.2,0.5)\psline[linewidth=2pt,linecolor=white]{-}(5,0.5)(5.2,0.5)
\facegrid{(0,0)}{(5,1)}
\psarc[linewidth=0.025]{-}(0,0){0.16}{0}{90}
\psarc[linewidth=0.025]{-}(1,0){0.16}{0}{90}
\psarc[linewidth=0.025]{-}(4,0){0.16}{0}{90}
\rput(2.5,0.5){$\ldots$}
\rput(3.5,0.5){$\ldots$}
\rput(0.5,.5){$u$}
\rput(1.5,.5){$u$}
\rput(4.5,.5){$u$}
\rput(0.5,0.75){\tiny{$_{(m,n)}$}}
\rput(1.5,0.75){\tiny{$_{(m,n)}$}}
\rput(4.5,0.75){\tiny{$_{(m,n)}$}}
\rput(2.5,-0.5){$\underbrace{\qquad \qquad \qquad \qquad \qquad \quad \qquad}_N$}
\end{pspicture} \ \ \ 
\label{eq:Ttransfermatrix}
\ee  
\vspace{0.5cm}

\noindent 
On the strip, the construction of fused transfer tangles involves an additional {\em fused crossing parameter} $\mu \in \mathbb C$. 
One should note that (i) $\mu$ only enters in the weight of the face operators in the top multi-row of $\Db^{m,n}(u)$, 
and (ii) the transfer tangle still depends on $\lambda$ through the fugacity of closed loops $\beta = 2\cos\lambda$ and the 
definition \eqref{1x1} of the face operators.
This mimics the analogous construction for rational models \cite{BehrendPearce}. 
For $m=1$ and $n=1$, the logarithmic minimal models ${\cal LM}(p,p')$ correspond to $\mu = \lambda$, 
but in general, $\mu$ may depend on the fusion indices. 
For example, the infinite family of higher-level logarithmic minimal models~\cite{PR1305}
are believed to correspond to $m=n$ and $\mu=n\lambda$. In the case $m=n=2$, in particular, setting $\mu = 2\lambda$ 
indeed gives rise to logarithmic superconformal minimal models~\cite{PRT2013}.

It is occasionally convenient to indicate explicitly the dependence on the crossing parameter $\lambda$ by writing
\be
 \Db^{m,n}(u)=\Db^{m,n}(u,\lambda), \qquad  \Tb^{m,n}(u)=\Tb^{m,n}(u,\lambda).
\ee
It is also observed that 
$\Db^{m,n}(u)$ and $\Tb^{m,n}(u)$ have the crossing and periodicity properties
\be
\Db^{m,n}(\mu-u_{n-1}) = \Db^{m,n}(u), \qquad R\, \Tb^{m,n}(\lambda - u) R = \Tb^{m,n}(u)
\ee
and
\be
 \Db^{m,n}(u + \pi) = \Db^{m,n}(u), \qquad  \Tb^{m,n}(u + \pi) = (-1)^{Nmn}\,\Tb^{m,n}(u),
\ee
where $R$ is the reflection generator appearing in (\ref{RTR}).
For convenience, we recall the shorthand notations (\ref{Dkmn})
\be 
 \Db^{m,n}_k:=\Db^{m,n}(u+k \lambda), \qquad  \Tb^{m,n}_k:=\Tb^{m,n}(u+k \lambda).
\ee  

Crucially, the transfer tangles (\ref{eq:transfermatrix}) and (\ref{eq:Ttransfermatrix}) form two separate {\em commuting families},
\be 
 [\Db^{m,n}(u),\Db^{m,n'}\!(v)] = 0, \qquad  [\Tb^{m,n}(u),\Tb^{m,n'}\!(v)] = 0.
\label{eq:Dcomm}
\ee
Using diagrammatic arguments as in~\cite{BehrendPearce}, this commutativity property follows from 
equations \eqref{eq:fusedinv}, \eqref{eq:fusedYBE} and \eqref{eq:fusedBYBE}.

As a consequence of the crossing relations \eqref{crossing11} and the push-through properties \eqref{eq:pushthru}, 
the multi-row transfer tangles $\Db^{m,n}(u)$ and 
$\Tb^{m,n}(u)$ can be written in terms of $(m,1)$- and $(1,m)$-fused face operators and a {\em single} $P_n$ projector, 
\be 
\psset{unit=0.9}
\Db^{m,n}_0 = \ 
\begin{pspicture}[shift=-3.9](-1.7,0)(7,8)
\facegrid{(0,0)}{(5,8)}
\psarc[linewidth=0.025]{-}(0,0){0.16}{0}{90}
\psarc[linewidth=0.025]{-}(0,1){0.16}{0}{90}
\psarc[linewidth=0.025]{-}(0,3){0.16}{0}{90}
\psarc[linewidth=0.025]{-}(1,4){0.16}{90}{180}
\psarc[linewidth=0.025]{-}(1,5){0.16}{90}{180}
\psarc[linewidth=0.025]{-}(1,7){0.16}{90}{180}
\psarc[linewidth=0.025]{-}(1,0){0.16}{0}{90}
\psarc[linewidth=0.025]{-}(1,1){0.16}{0}{90}
\psarc[linewidth=0.025]{-}(1,3){0.16}{0}{90}
\psarc[linewidth=0.025]{-}(2,4){0.16}{90}{180}
\psarc[linewidth=0.025]{-}(2,5){0.16}{90}{180}
\psarc[linewidth=0.025]{-}(2,7){0.16}{90}{180}
\psarc[linewidth=0.025]{-}(4,0){0.16}{0}{90}
\psarc[linewidth=0.025]{-}(4,1){0.16}{0}{90}
\psarc[linewidth=0.025]{-}(4,3){0.16}{0}{90}
\psarc[linewidth=0.025]{-}(5,4){0.16}{90}{180}
\psarc[linewidth=0.025]{-}(5,5){0.16}{90}{180}
\psarc[linewidth=0.025]{-}(5,7){0.16}{90}{180}
\psarc[linewidth=1.5pt,linecolor=blue]{-}(-0.3,4){0.5}{90}{-90}
\psarc[linewidth=1.5pt,linecolor=blue]{-}(5,4){0.5}{-90}{90}
\psbezier[linewidth=1.5pt,linecolor=blue]{-}(-0.3,2.5)(-1.3,2.5)(-1.3,5.5)(-0.3,5.5)
\psbezier[linewidth=1.5pt,linecolor=blue]{-}(-0.3,1.5)(-1.7,1.5)(-1.7,6.5)(-0.3,6.5)
\psbezier[linewidth=1.5pt,linecolor=blue]{-}(-0.3,0.5)(-2.1,0.5)(-2.1,7.5)(-0.3,7.5)
\psbezier[linewidth=1.5pt,linecolor=blue]{-}(5,2.5)(6,2.5)(6,5.5)(5,5.5)
\psbezier[linewidth=1.5pt,linecolor=blue]{-}(5,1.5)(6.4,1.5)(6.4,6.5)(5,6.5)
\psbezier[linewidth=1.5pt,linecolor=blue]{-}(5,0.5)(6.8,0.5)(6.8,7.5)(5,7.5)
\psline[linewidth=1.5pt,linecolor=blue]{-}(-0.3,4.5)(0,4.5)
\psline[linewidth=1.5pt,linecolor=blue]{-}(-0.3,5.5)(0,5.5)
\psline[linewidth=1.5pt,linecolor=blue]{-}(-0.3,6.5)(0,6.5)
\psline[linewidth=1.5pt,linecolor=blue]{-}(-0.3,7.5)(0,7.5)
\pspolygon[fillstyle=solid,fillcolor=pink](0,0.1)(0,3.9)(-0.3,3.9)(-0.3,0.1)(0,0.1)
\rput(-0.15,2){$_n$}
\rput(0.5,.4){\small$u_0$}
\rput(1.5,.4){\small$u_0$}
\rput(4.5,.4){\small$u_0$}
\rput(0.5,1.4){\small$u_1$}
\rput(1.5,1.4){\small$u_1$}
\rput(4.5,1.4){\small$u_1$}
\rput(0.5,2.6){$\vdots$}
\rput(1.5,2.6){$\vdots$}
\rput(4.5,2.6){$\vdots$}
\rput(0.5,3.4){\small$u_{n\!-\!1}$}
\rput(1.5,3.4){\small$u_{n\!-\!1}$}
\rput(4.5,3.4){\small$u_{n\!-\!1}$}
\rput(0.5,4.4){\scriptsize$u_{1}\hs-\hs\mu$}
\rput(1.5,4.4){\scriptsize$u_{1}\hs-\hs\mu$}
\rput(4.5,4.4){\scriptsize$u_{1}\hs-\hs\mu$}
\rput(0.5,5.4){\scriptsize$u_{2}\hs-\hs\mu$}
\rput(1.5,5.4){\scriptsize$u_{2}\hs-\hs\mu$}
\rput(4.5,5.4){\scriptsize$u_{2}\hs-\hs\mu$}
\rput(0.5,6.6){$\vdots$}
\rput(1.5,6.6){$\vdots$}
\rput(4.5,6.6){$\vdots$}
\rput(0.5,7.4){\scriptsize$u_{n}\hs-\hs\mu$}
\rput(1.5,7.4){\scriptsize$u_{n}\hs-\hs\mu$}
\rput(4.5,7.4){\scriptsize$u_{n}\hs-\hs\mu$}
\rput(2.5,0.5){$\ldots$}
\rput(2.5,1.5){$\ldots$}
\rput(3.5,0.5){$\ldots$}
\rput(3.5,1.5){$\ldots$}%
\rput(2.5,3.5){$\ldots$}
\rput(2.5,4.5){$\ldots$}
\rput(3.5,3.5){$\ldots$}
\rput(3.5,4.5){$\ldots$}
\rput(2.5,5.5){$\ldots$}
\rput(2.5,7.5){$\ldots$}
\rput(3.5,5.5){$\ldots$}
\rput(3.5,7.5){$\ldots$}
\rput(0.5,0.75){\tiny{$_{(m,1)}$}}\rput(0.5,1.75){\tiny{$_{(m,1)}$}}\rput(0.5,3.75){\tiny{$_{(m,1)}$}}\rput(0.5,4.75){\tiny{$_{(1,m)}$}}
\rput(0.5,5.75){\tiny{$_{(1,m)}$}}\rput(0.5,7.75){\tiny{$_{(1,m)}$}}
\rput(1.5,0.75){\tiny{$_{(m,1)}$}}\rput(1.5,1.75){\tiny{$_{(m,1)}$}}\rput(1.5,3.75){\tiny{$_{(m,1)}$}}\rput(1.5,4.75){\tiny{$_{(1,m)}$}}
\rput(1.5,5.75){\tiny{$_{(1,m)}$}}\rput(1.5,7.75){\tiny{$_{(1,m)}$}}
\rput(4.5,0.75){\tiny{$_{(m,1)}$}}\rput(4.5,1.75){\tiny{$_{(m,1)}$}}\rput(4.5,3.75){\tiny{$_{(m,1)}$}}\rput(4.5,4.75){\tiny{$_{(1,m)}$}}
\rput(4.5,5.75){\tiny{$_{(1,m)}$}}\rput(4.5,7.75){\tiny{$_{(1,m)}$}}
\end{pspicture} 
\quad \Tb^{m,n}_0 = \
\begin{pspicture}[shift=-1.9](-0.5,0)(5.3,4)
\facegrid{(0,0)}{(5,4)}
\psarc[linewidth=0.025]{-}(0,0){0.16}{0}{90}
\psarc[linewidth=0.025]{-}(0,1){0.16}{0}{90}
\psarc[linewidth=0.025]{-}(0,3){0.16}{0}{90}
\psarc[linewidth=0.025]{-}(1,0){0.16}{0}{90}
\psarc[linewidth=0.025]{-}(1,1){0.16}{0}{90}
\psarc[linewidth=0.025]{-}(1,3){0.16}{0}{90}
\psarc[linewidth=0.025]{-}(4,0){0.16}{0}{90}
\psarc[linewidth=0.025]{-}(4,1){0.16}{0}{90}
\psarc[linewidth=0.025]{-}(4,3){0.16}{0}{90}
\psline[linecolor=blue,linewidth=1.5pt]{-}(-0.3,0.5)(-0.5,0.5)
\psline[linecolor=blue,linewidth=1.5pt]{-}(-0.3,1.5)(-0.5,1.5)
\psline[linecolor=blue,linewidth=1.5pt]{-}(-0.3,2.5)(-0.5,2.5)
\psline[linecolor=blue,linewidth=1.5pt]{-}(-0.3,3.5)(-0.5,3.5)
\psline[linecolor=blue,linewidth=1.5pt]{-}(5.0,0.5)(5.2,0.5)
\psline[linecolor=blue,linewidth=1.5pt]{-}(5.0,1.5)(5.2,1.5)
\psline[linecolor=blue,linewidth=1.5pt]{-}(5.0,2.5)(5.2,2.5)
\psline[linecolor=blue,linewidth=1.5pt]{-}(5.0,3.5)(5.2,3.5)
\pspolygon[fillstyle=solid,fillcolor=pink](0,0.1)(0,3.9)(-0.3,3.9)(-0.3,0.1)(0,0.1)
\rput(-0.15,2){$_n$}
\rput(0.5,.4){\small$u_0$}
\rput(1.5,.4){\small$u_0$}
\rput(4.5,.4){\small$u_0$}
\rput(0.5,1.4){\small$u_1$}
\rput(1.5,1.4){\small$u_1$}
\rput(4.5,1.4){\small$u_1$}
\rput(0.5,2.6){$\vdots$}
\rput(1.5,2.6){$\vdots$}
\rput(4.5,2.6){$\vdots$}
\rput(0.5,3.4){\small$u_{n\!-\!1}$}
\rput(1.5,3.4){\small$u_{n\!-\!1}$}
\rput(4.5,3.4){\small$u_{n\!-\!1}$}
\rput(2.5,0.5){$\ldots$}
\rput(2.5,1.5){$\ldots$}
\rput(3.5,0.5){$\ldots$}
\rput(3.5,1.5){$\ldots$}%
\rput(2.5,3.5){$\ldots$}
\rput(3.5,3.5){$\ldots$}
\rput(0.5,0.75){\tiny{$_{(m,1)}$}}\rput(0.5,1.75){\tiny{$_{(m,1)}$}}\rput(0.5,3.75){\tiny{$_{(m,1)}$}}
\rput(1.5,0.75){\tiny{$_{(m,1)}$}}\rput(1.5,1.75){\tiny{$_{(m,1)}$}}\rput(1.5,3.75){\tiny{$_{(m,1)}$}}
\rput(4.5,0.75){\tiny{$_{(m,1)}$}}\rput(4.5,1.75){\tiny{$_{(m,1)}$}}\rput(4.5,3.75){\tiny{$_{(m,1)}$}}
\end{pspicture}  
\label{eq:transfermatrix2}
\ee  
It is noted that the connections indicated in blue are all single (not multiple) connections. Using the definition \eqref{eq:fusedface}
in \eqref{eq:transfermatrix}, push-through properties allow half-arc propagation towards the left both in the top 
$n$ and bottom $n$ layers of $\Db^{m,n}(u)$ 
(~\begin{pspicture}(0.1,.125)(0.425,0.5)
\psline[linewidth=0.7pt]{->}(0.5,0.35)(0,0.35)
\psline[linewidth=0.7pt]{->}(0.5,0.15)(0,0.15)
\end{pspicture}~), 
whereas in \eqref{eq:transfermatrix2}, half-arcs propagate towards the right in the top fused faces, but towards the left in the 
bottom faces 
(~\begin{pspicture}(0.1,.125)(0.425,0.5)
\psline[linewidth=0.7pt]{->}(0,0.35)(0.5,0.35)
\psline[linewidth=0.7pt]{->}(0.5,0.15)(0,0.15)
\end{pspicture}~).

%%%%%%%%%%%%%%%%%%%%%%%%%%%%%%%%
\subsection[Effective projectors and fused transfer tangles for fractional $\lambda$]{Effective projectors and fused transfer tangles for fractional $\boldsymbol \lambda$}
\label{sec:Dcrit}
%%%%%%%%%%%%%%%%%%%%%%%%%%%%%%%%

It is recalled from Section~\ref{Sec:TL} that the WJ projectors fail to exist for some fractional $\lambda$. 
However, there is an important distinction between the roles of the WJ projectors $P_n$ and $P_m$ located along the vertical and 
horizontal interfaces (and edges), respectively, of the multi-row transfer tangles. We 
can actually ignore the projectors $P_m$ along the horizontal interfaces and edges due to the push-through properties and 
because the transfer tangles are ultimately meant to act on so-called cabled link states whose characterisation will implement the 
projection offered by $P_m$, see Section~\ref{Sec:CabledLinkStates}. 
We will therefore not be concerned with questions about the existence of $P_m$ until Section~\ref{Sec:CabledLinkStates}.

Regarding $P_n$, the above constructions of the multi-row transfer tangles $\Db^{m,n}(u,\lambda)$ and $\Tb^{m,n}(u,\lambda)$ 
are ill-defined for some fractional $\lambda$.
However, as we will discuss below, there exist alternative expressions for $\Db^{m,n}(u,\lambda)$ and $\Tb^{m,n}(u,\lambda)$ 
which are manifestly equivalent to our original definition for generic $\lambda$, but are well defined for fractional $\lambda$ as 
well (modulo potential issues with $P_m$). 
We view these alternative expressions as suitable starting points for generalisations to fractional $\lambda$.
Alternatively, we could have taken any of these expressions and used it as our definition of the corresponding strip or cylinder transfer 
tangle. In doing so, however, many features and manipulations of the transfer tangles would become more involved and less 
transparent. This is why we prefer to work with the projector dependent constructions of Section~\ref{Sec:nonrational}.

To appreciate that we can avoid the projector $P_n$ in (\ref{eq:transfermatrix2}), we consider the transfer tangle 
$\Db^{m,n}(u)$ in the case 
$\mu = \lambda$, $m=1$ and $n=4$, where it readily follows from \eqref{eq:Prec} and the push-through properties that 
\bea   
& \hspace{-3cm}\Db^{1,4}_0=
\psset{unit=0.60}
\begin{pspicture}[shift=-3.9](-1.5,0)(4.35,8)
\facegrid{(0,0)}{(3,8)}
\psarc[linewidth=0.025]{-}(0,0){0.16}{0}{90}\psarc[linewidth=0.025]{-}(2,0){0.16}{0}{90}
\psarc[linewidth=0.025]{-}(0,1){0.16}{0}{90}\psarc[linewidth=0.025]{-}(2,1){0.16}{0}{90}
\psarc[linewidth=0.025]{-}(0,2){0.16}{0}{90}\psarc[linewidth=0.025]{-}(2,2){0.16}{0}{90}
\psarc[linewidth=0.025]{-}(0,3){0.16}{0}{90}\psarc[linewidth=0.025]{-}(2,3){0.16}{0}{90}
\psarc[linewidth=0.025]{-}(1,4){0.16}{90}{180}\psarc[linewidth=0.025]{-}(3,4){0.16}{90}{180}
\psarc[linewidth=0.025]{-}(1,5){0.16}{90}{180}\psarc[linewidth=0.025]{-}(3,5){0.16}{90}{180}
\psarc[linewidth=0.025]{-}(1,6){0.16}{90}{180}\psarc[linewidth=0.025]{-}(3,6){0.16}{90}{180}
\psarc[linewidth=0.025]{-}(1,7){0.16}{90}{180}\psarc[linewidth=0.025]{-}(3,7){0.16}{90}{180}
\psarc[linewidth=1.5pt,linecolor=blue]{-}(0,4){-0.5}{-90}{90}
\psarc[linewidth=1.5pt,linecolor=blue]{-}(3.0,4){0.5}{-90}{90}
\psbezier[linewidth=1.5pt,linecolor=blue]{-}(0,2.5)(-1.0,2.5)(-1.0,5.5)(0.0,5.5)
\psbezier[linewidth=1.5pt,linecolor=blue]{-}(0,1.5)(-1.4,1.5)(-1.4,6.5)(0.0,6.5)
\psbezier[linewidth=1.5pt,linecolor=blue]{-}(0,0.5)(-1.8,0.5)(-1.8,7.5)(0.0,7.5)
\psbezier[linewidth=1.5pt,linecolor=blue]{-}(3.0,2.5)(4.0,2.5)(4.0,5.5)(3.0,5.5)
\psbezier[linewidth=1.5pt,linecolor=blue]{-}(3.0,1.5)(4.4,1.5)(4.4,6.5)(3.0,6.5)
\psbezier[linewidth=1.5pt,linecolor=blue]{-}(3.0,0.5)(4.8,0.5)(4.8,7.5)(3.0,7.5)
\rput(0.5,0.5){\small$u_0$}
\rput(2.5,0.5){\small$u_0$}
\rput(0.5,1.5){\small$u_1$}
\rput(2.5,1.5){\small$u_1$}
\rput(0.5,2.5){\small$u_2$}
\rput(2.5,2.5){\small$u_2$}
\rput(0.5,3.5){\small$u_3$}
\rput(2.5,3.5){\small$u_3$}
\rput(0.5,4.5){\small$u_0$}
\rput(2.5,4.5){\small$u_0$}
\rput(0.5,5.5){\small$u_1$}
\rput(2.5,5.5){\small$u_1$}
\rput(0.5,6.5){\small$u_2$}
\rput(2.5,6.5){\small$u_2$}
\rput(0.5,7.5){\small$u_3$}
\rput(2.5,7.5){\small$u_3$}
\rput(1.5,0.5){$\ldots$}
\rput(1.5,1.5){$\ldots$}
\rput(1.5,2.5){$\ldots$}
\rput(1.5,3.5){$\ldots$}
\rput(1.5,4.5){$\ldots$}
\rput(1.5,5.5){$\ldots$}
\rput(1.5,6.5){$\ldots$}
\rput(1.5,7.5){$\ldots$}
\end{pspicture}
\ \ - \big[q^1(u_2)q^1(-u_0)\big]^{N} \ \ 
\begin{pspicture}[shift=-1.9](-0.7,2)(3.95,6)
\facegrid{(0,2)}{(3,6)}
\psarc[linewidth=0.025]{-}(0,2){0.16}{0}{90}\psarc[linewidth=0.025]{-}(2,2){0.16}{0}{90}
\psarc[linewidth=0.025]{-}(0,3){0.16}{0}{90}\psarc[linewidth=0.025]{-}(2,3){0.16}{0}{90}
\psarc[linewidth=0.025]{-}(1,4){0.16}{90}{180}\psarc[linewidth=0.025]{-}(3,4){0.16}{90}{180}
\psarc[linewidth=0.025]{-}(1,5){0.16}{90}{180}\psarc[linewidth=0.025]{-}(3,5){0.16}{90}{180}
\psarc[linewidth=1.5pt,linecolor=blue]{-}(0,4){-0.5}{-90}{90}
\psarc[linewidth=1.5pt,linecolor=blue]{-}(3.0,4){0.5}{-90}{90}
\psbezier[linewidth=1.5pt,linecolor=blue]{-}(0,2.5)(-1.0,2.5)(-1.0,5.5)(0.0,5.5)
\psbezier[linewidth=1.5pt,linecolor=blue]{-}(3.0,2.5)(4.0,2.5)(4.0,5.5)(3.0,5.5)
\rput(0.5,2.5){\small$u_0$}
\rput(2.5,2.5){\small$u_0$}
\rput(0.5,3.5){\small$u_3$}
\rput(2.5,3.5){\small$u_3$}
\rput(0.5,4.5){\small$u_0$}
\rput(2.5,4.5){\small$u_0$}
\rput(0.5,5.5){\small$u_3$}
\rput(2.5,5.5){\small$u_3$}
\rput(1.5,2.5){$\ldots$}
\rput(1.5,3.5){$\ldots$}
\rput(1.5,4.5){$\ldots$}
\rput(1.5,5.5){$\ldots$}
\end{pspicture} 
\nonumber \\ \label{eq:D140}
\\ 
&\psset{unit=0.60}
- \big[q^1(u_3)q^1(-u_{-1})\big]^N \ 
\begin{pspicture}[shift=-1.75](-0.7,2)(3.65,5.5)
\facegrid{(0,2)}{(3,6)}
\psarc[linewidth=0.025]{-}(0,2){0.16}{0}{90}\psarc[linewidth=0.025]{-}(2,2){0.16}{0}{90}
\psarc[linewidth=0.025]{-}(0,3){0.16}{0}{90}\psarc[linewidth=0.025]{-}(2,3){0.16}{0}{90}
\psarc[linewidth=0.025]{-}(1,4){0.16}{90}{180}\psarc[linewidth=0.025]{-}(3,4){0.16}{90}{180}
\psarc[linewidth=0.025]{-}(1,5){0.16}{90}{180}\psarc[linewidth=0.025]{-}(3,5){0.16}{90}{180}
\psarc[linewidth=1.5pt,linecolor=blue]{-}(0,4){-0.5}{-90}{90}
\psarc[linewidth=1.5pt,linecolor=blue]{-}(3.0,4){0.5}{-90}{90}
\psbezier[linewidth=1.5pt,linecolor=blue]{-}(0,2.5)(-1.0,2.5)(-1.0,5.5)(0.0,5.5)
\psbezier[linewidth=1.5pt,linecolor=blue]{-}(3.0,2.5)(4.0,2.5)(4.0,5.5)(3.0,5.5)
\rput(0.5,2.5){\small$u_0$}
\rput(2.5,2.5){\small$u_0$}
\rput(0.5,3.5){\small$u_1$}
\rput(2.5,3.5){\small$u_1$}
\rput(0.5,4.5){\small$u_2$}
\rput(2.5,4.5){\small$u_2$}
\rput(0.5,5.5){\small$u_3$}
\rput(2.5,5.5){\small$u_3$}
\rput(1.5,2.5){$\ldots$}
\rput(1.5,3.5){$\ldots$}
\rput(1.5,4.5){$\ldots$}
\rput(1.5,5.5){$\ldots$}
\end{pspicture}  
\ \ -\big[q^1(u_1)q^1(-u_1)\big]^N  \ \
\begin{pspicture}[shift=-1.75](-0.6,2)(3.65,5.5)
\facegrid{(0,2)}{(3,6)}
\psarc[linewidth=0.025]{-}(0,2){0.16}{0}{90}\psarc[linewidth=0.025]{-}(2,2){0.16}{0}{90}
\psarc[linewidth=0.025]{-}(0,3){0.16}{0}{90}\psarc[linewidth=0.025]{-}(2,3){0.16}{0}{90}
\psarc[linewidth=0.025]{-}(1,4){0.16}{90}{180}\psarc[linewidth=0.025]{-}(3,4){0.16}{90}{180}
\psarc[linewidth=0.025]{-}(1,5){0.16}{90}{180}\psarc[linewidth=0.025]{-}(3,5){0.16}{90}{180}
\psarc[linewidth=1.5pt,linecolor=blue]{-}(0,4){-0.5}{-90}{90}
\psarc[linewidth=1.5pt,linecolor=blue]{-}(3.0,4){0.5}{-90}{90}
\psbezier[linewidth=1.5pt,linecolor=blue]{-}(0,2.5)(-1.0,2.5)(-1.0,5.5)(0.0,5.5)
\psbezier[linewidth=1.5pt,linecolor=blue]{-}(3.0,2.5)(4.0,2.5)(4.0,5.5)(3.0,5.5)
\rput(0.5,2.5){\small$u_2$}
\rput(2.5,2.5){\small$u_2$}
\rput(0.5,3.5){\small$u_3$}
\rput(2.5,3.5){\small$u_3$}
\rput(0.5,4.5){\small$u_0$}
\rput(2.5,4.5){\small$u_0$}
\rput(0.5,5.5){\small$u_1$}
\rput(2.5,5.5){\small$u_1$}
\rput(1.5,2.5){$\ldots$}
\rput(1.5,3.5){$\ldots$}
\rput(1.5,4.5){$\ldots$}
\rput(1.5,5.5){$\ldots$}
\end{pspicture} \ \nonumber\\ \nonumber \\ 
&+ \big[q^1(u_1)q^1(u_3)q^1(-u_{-1})q^1(-u_1)\big]^N \Ib. 
\nonumber
\eea 
It is emphasized that this expression is {\em independent} of WJ projectors and thus well defined for all $N\in\mathbb{N}$ 
and $\beta\in\mathbb{C}$. It can therefore serve as a definition of $\Db^{1,4}_0$ for all 
$\lambda\in\pi\big(\mathbb{R}\!\setminus\!\mathbb{Z}\big)$, in particular for fractional $\lambda$.

%%%%%%%%%%%%%%%%%%%%%%%%%%%%%
\paragraph{Effective projectors}
%%%%%%%%%%%%%%%%%%%%%%%%%%%%%
For $\beta\neq0$, there is a neat way to encode the above result for $\Db_0^{1,4}$. In the definition \eqref{eq:transfermatrix2} 
of $\Db_0^{1,4}$, one can replace the $P_4$ projector, 
$
\begin{pspicture}[shift=-0.05](-0,-0.15)(0.8,0.15)
\pspolygon[fillstyle=solid,fillcolor=pink](0,-0.15)(0.8,-0.15)(0.8,0.15)(0,0.15)(0,0.15)
\rput(0.4,0){$_{4}$}
\end{pspicture} 
$\,,
by the 4-tangle 
\be 
\begin{pspicture}[shift=-0.05](-0.15,-0.15)(0.95,0.15)
\pspolygon[fillstyle=solid,fillcolor=orange](0,-0.15)(0.8,-0.15)(0.95,0)(0.8,0.15)(0,0.15)(0,0.15)(-0.15,0)
\rput(0.4,0){$_{4}$}
\end{pspicture} 
\, :=\,
\begin{pspicture}[shift=-0.25](-0.0,-0.35)(1.6,0.35)
\psline[linecolor=blue,linewidth=1.5pt]{-}(0.2,0.35)(0.2,-0.35)
\psline[linecolor=blue,linewidth=1.5pt]{-}(0.6,0.35)(0.6,-0.35)
\psline[linecolor=blue,linewidth=1.5pt]{-}(1.0,0.35)(1.0,-0.35)
\psline[linecolor=blue,linewidth=1.5pt]{-}(1.4,0.35)(1.4,-0.35)
\end{pspicture} - \frac1\beta  \Big(
\begin{pspicture}[shift=-0.25](-0.0,-0.35)(1.6,0.35)
\psarc[linecolor=blue,linewidth=1.5pt]{-}(0.4,0.35){0.2}{180}{0}
\psarc[linecolor=blue,linewidth=1.5pt]{-}(0.4,-0.35){0.2}{0}{180}
\psline[linecolor=blue,linewidth=1.5pt]{-}(1.0,0.35)(1.0,-0.35)
\psline[linecolor=blue,linewidth=1.5pt]{-}(1.4,0.35)(1.4,-0.35)
\end{pspicture} + 
\begin{pspicture}[shift=-0.25](-0.0,-0.35)(1.6,0.35)
\psarc[linecolor=blue,linewidth=1.5pt]{-}(0.8,0.35){0.2}{180}{0}
\psarc[linecolor=blue,linewidth=1.5pt]{-}(0.8,-0.35){0.2}{0}{180}
\psline[linecolor=blue,linewidth=1.5pt]{-}(0.2,0.35)(0.2,-0.35)
\psline[linecolor=blue,linewidth=1.5pt]{-}(1.4,0.35)(1.4,-0.35)
\end{pspicture} + 
\begin{pspicture}[shift=-0.25](-0.0,-0.35)(1.6,0.35)
\psarc[linecolor=blue,linewidth=1.5pt]{-}(1.2,0.35){0.2}{180}{0}
\psarc[linecolor=blue,linewidth=1.5pt]{-}(1.2,-0.35){0.2}{0}{180}
\psline[linecolor=blue,linewidth=1.5pt]{-}(0.2,0.35)(0.2,-0.35)
\psline[linecolor=blue,linewidth=1.5pt]{-}(0.6,0.35)(0.6,-0.35)
\end{pspicture} 
\Big)
+ \frac{1}{\beta^2} 
\begin{pspicture}[shift=-0.25](-0.0,-0.35)(1.6,0.35)
\psarc[linecolor=blue,linewidth=1.5pt]{-}(0.4,0.35){0.2}{180}{0}
\psarc[linecolor=blue,linewidth=1.5pt]{-}(0.4,-0.35){0.2}{0}{180}
\psarc[linecolor=blue,linewidth=1.5pt]{-}(1.2,0.35){0.2}{180}{0}
\psarc[linecolor=blue,linewidth=1.5pt]{-}(1.2,-0.35){0.2}{0}{180}
\end{pspicture} 
\ee
Indeed, it is readily verified that this prescription correctly reproduces the decomposition \eqref{eq:D140}.

For $\beta\neq0$, this construction extends to general $n\in\mathbb{N}$, replacing the WJ projector $P_n$ by the 
{\it effective projector} 
\be
 Q_n=\ 
\begin{pspicture}[shift=-0.05](-0.15,-0.15)(0.95,0.15)
\pspolygon[fillstyle=solid,fillcolor=orange](0,-0.15)(0.8,-0.15)(0.95,0)(0.8,0.15)(0,0.15)(0,0.15)(-0.15,0)
\rput(0.4,0){$_{n}$}
\end{pspicture} 
\ ,\qquad n\in\mathbb{N}
\ee 
defined recursively as
\be
\begin{pspicture}[shift=-0.05](-0.15,-0.15)(0.95,0.15)
\pspolygon[fillstyle=solid,fillcolor=orange](0,-0.15)(0.8,-0.15)(0.95,0)(0.8,0.15)(0,0.15)(0,0.15)(-0.15,0)
\rput(0.4,0){$_{n}$}
\end{pspicture} 
\ =\
\begin{pspicture}[shift=-0.05](-0.15,-0.15)(1.25,0.15)
\pspolygon[fillstyle=solid,fillcolor=orange](0,-0.15)(0.8,-0.15)(0.95,0)(0.8,0.15)(0,0.15)(0,0.15)(-0.15,0)
\rput(0.4,0){$_{n-1}$}
\psline[linecolor=blue,linewidth=1.5pt]{-}(1.15,-0.17)(1.15,0.17)
\end{pspicture} - \tfrac{1}\beta\
\begin{pspicture}[shift=-0.05](-0.15,-0.15)(1.25,0.15)
\pspolygon[fillstyle=solid,fillcolor=orange](0,-0.15)(0.8,-0.15)(0.95,0)(0.8,0.15)(0,0.15)(0,0.15)(-0.15,0)
\rput(0.4,0){$_{n-2}$}
\psarc[linecolor=blue,linewidth=1.5pt]{-}(1.25,-0.17){0.125}{0}{180}
\psarc[linecolor=blue,linewidth=1.5pt]{-}(1.25,0.17){-0.125}{0}{180}
\end{pspicture} \ \ ,
\qquad \quad
\begin{pspicture}[shift=-0.05](-0.15,-0.15)(0.95,0.15)
\pspolygon[fillstyle=solid,fillcolor=orange](0,-0.15)(0.8,-0.15)(0.95,0)(0.8,0.15)(0,0.15)(0,0.15)(-0.15,0)
\rput(0.4,0){$_{1}$}
\end{pspicture} \, :=\
\begin{pspicture}[shift=-0.07](0,-0.17)(0.2,0.17)
\psline[linecolor=blue,linewidth=1.5pt]{-}(0.1,-0.17)(0.1,0.17)
\end{pspicture} 
,\qquad \quad
\begin{pspicture}[shift=-0.05](-0.15,-0.15)(0.95,0.15)
\pspolygon[fillstyle=solid,fillcolor=orange](0,-0.15)(0.8,-0.15)(0.95,0)(0.8,0.15)(0,0.15)(0,0.15)(-0.15,0)
\rput(0.4,0){$_{2}$}
\end{pspicture} \, :=\
\begin{pspicture}[shift=-0.07](0,-0.17)(0.2,0.17)
\psline[linecolor=blue,linewidth=1.5pt]{-}(0.1,-0.17)(0.1,0.17)
\psline[linecolor=blue,linewidth=1.5pt]{-}(0.35,-0.17)(0.35,0.17)
\end{pspicture} \ \;- \tfrac1\beta\
\begin{pspicture}[shift=-0.07](0,-0.17)(0.2,0.17)
\psarc[linecolor=blue,linewidth=1.5pt]{-}(0.225,-0.17){0.125}{0}{180}
\psarc[linecolor=blue,linewidth=1.5pt]{-}(0.225,0.17){-0.125}{0}{180}
\end{pspicture}
\label{Qn}
\ee
Written in terms of elements of $TL_n(\beta)$, $Q_n$ is simply given by
\be
 Q_n=I+\sum_{k=1}^{\lfloor\frac{n}{2}\rfloor}\;
  \sum_{1\leq j_1\ll j_2\ll\ldots\ll j_k\leq n-1}\big(\!-\tfrac{1}{\beta}\big)^k e_{j_1}e_{j_2}\ldots e_{j_k}
  \label{eq:Qndec}
\ee
where $a\ll b$ here means that $a\leq b-2$. 

It is emphasised that, for $n>2$, $Q_n$ is not a projector as it violates the condition for 
property \eqref{eq:(i)} of the WJ projectors.
It likewise does not have the fundamental properties \eqref{eq:(ii)} and \eqref{eq:(iv)} of the WJ projectors.
However, the following proposition, which we prove in Appendix~\ref{app:EffectiveProjectors},
states that the expressions (\ref{eq:transfermatrix2}) for $\Db^{m,n}_0$\! and\, $\Tb^{m,n}_0$ are equal to 
the corresponding tangles with the projectors $P_n$ replaced by $Q_n$.
This result thus explains why we refer to the $n$-tangle $Q_n$ as an effective projector.
\begin{Proposition}
For $\lambda$ generic, the transfer tangles can be expressed in terms of effective projectors as

\be 
\psset{unit=0.9}
\qquad\qquad
\begin{pspicture}[shift=-3.9](-1.7,0)(7,8)
\rput(-2.8,4){$\Db^{m,n}_0 =$}
\facegrid{(0,0)}{(5,8)}
\psarc[linewidth=0.025]{-}(0,0){0.16}{0}{90}
\psarc[linewidth=0.025]{-}(0,1){0.16}{0}{90}
\psarc[linewidth=0.025]{-}(0,3){0.16}{0}{90}
\psarc[linewidth=0.025]{-}(1,4){0.16}{90}{180}
\psarc[linewidth=0.025]{-}(1,5){0.16}{90}{180}
\psarc[linewidth=0.025]{-}(1,7){0.16}{90}{180}
\psarc[linewidth=0.025]{-}(1,0){0.16}{0}{90}
\psarc[linewidth=0.025]{-}(1,1){0.16}{0}{90}
\psarc[linewidth=0.025]{-}(1,3){0.16}{0}{90}
\psarc[linewidth=0.025]{-}(2,4){0.16}{90}{180}
\psarc[linewidth=0.025]{-}(2,5){0.16}{90}{180}
\psarc[linewidth=0.025]{-}(2,7){0.16}{90}{180}
\psarc[linewidth=0.025]{-}(4,0){0.16}{0}{90}
\psarc[linewidth=0.025]{-}(4,1){0.16}{0}{90}
\psarc[linewidth=0.025]{-}(4,3){0.16}{0}{90}
\psarc[linewidth=0.025]{-}(5,4){0.16}{90}{180}
\psarc[linewidth=0.025]{-}(5,5){0.16}{90}{180}
\psarc[linewidth=0.025]{-}(5,7){0.16}{90}{180}
\psarc[linewidth=1.5pt,linecolor=blue]{-}(-0.3,4){0.5}{90}{-90}
\psarc[linewidth=1.5pt,linecolor=blue]{-}(5,4){0.5}{-90}{90}
\psbezier[linewidth=1.5pt,linecolor=blue]{-}(-0.3,2.5)(-1.3,2.5)(-1.3,5.5)(-0.3,5.5)
\psbezier[linewidth=1.5pt,linecolor=blue]{-}(-0.3,1.5)(-1.7,1.5)(-1.7,6.5)(-0.3,6.5)
\psbezier[linewidth=1.5pt,linecolor=blue]{-}(-0.3,0.5)(-2.1,0.5)(-2.1,7.5)(-0.3,7.5)
\psbezier[linewidth=1.5pt,linecolor=blue]{-}(5,2.5)(6,2.5)(6,5.5)(5,5.5)
\psbezier[linewidth=1.5pt,linecolor=blue]{-}(5,1.5)(6.4,1.5)(6.4,6.5)(5,6.5)
\psbezier[linewidth=1.5pt,linecolor=blue]{-}(5,0.5)(6.8,0.5)(6.8,7.5)(5,7.5)
\psline[linewidth=1.5pt,linecolor=blue]{-}(-0.3,4.5)(0,4.5)
\psline[linewidth=1.5pt,linecolor=blue]{-}(-0.3,5.5)(0,5.5)
\psline[linewidth=1.5pt,linecolor=blue]{-}(-0.3,6.5)(0,6.5)
\psline[linewidth=1.5pt,linecolor=blue]{-}(-0.3,7.5)(0,7.5)
\pspolygon[fillstyle=solid,fillcolor=orange](0,0.15)(0,3.85)(-0.15,4)(-0.3,3.85)(-0.3,0.15)(-0.15,0)(0,0.15)
\rput(-0.15,2){$_n$}
\rput(0.5,.4){\small$u_0$}
\rput(1.5,.4){\small$u_0$}
\rput(4.5,.4){\small$u_0$}
\rput(0.5,1.4){\small$u_1$}
\rput(1.5,1.4){\small$u_1$}
\rput(4.5,1.4){\small$u_1$}
\rput(0.5,2.6){$\vdots$}
\rput(1.5,2.6){$\vdots$}
\rput(4.5,2.6){$\vdots$}
\rput(0.5,3.4){\small$u_{n\!-\!1}$}
\rput(1.5,3.4){\small$u_{n\!-\!1}$}
\rput(4.5,3.4){\small$u_{n\!-\!1}$}
\rput(0.5,4.4){\scriptsize$u_{1}\hs-\hs\mu$}
\rput(1.5,4.4){\scriptsize$u_{1}\hs-\hs\mu$}
\rput(4.5,4.4){\scriptsize$u_{1}\hs-\hs\mu$}
\rput(0.5,5.4){\scriptsize$u_{2}\hs-\hs\mu$}
\rput(1.5,5.4){\scriptsize$u_{2}\hs-\hs\mu$}
\rput(4.5,5.4){\scriptsize$u_{2}\hs-\hs\mu$}
\rput(0.5,6.6){$\vdots$}
\rput(1.5,6.6){$\vdots$}
\rput(4.5,6.6){$\vdots$}
\rput(0.5,7.4){\scriptsize$u_{n}\hs-\hs\mu$}
\rput(1.5,7.4){\scriptsize$u_{n}\hs-\hs\mu$}
\rput(4.5,7.4){\scriptsize$u_{n}\hs-\hs\mu$}
\rput(2.5,0.5){$\ldots$}
\rput(2.5,1.5){$\ldots$}
\rput(3.5,0.5){$\ldots$}
\rput(3.5,1.5){$\ldots$}%
\rput(2.5,3.5){$\ldots$}
\rput(2.5,4.5){$\ldots$}
\rput(3.5,3.5){$\ldots$}
\rput(3.5,4.5){$\ldots$}
\rput(2.5,5.5){$\ldots$}
\rput(2.5,7.5){$\ldots$}
\rput(3.5,5.5){$\ldots$}
\rput(3.5,7.5){$\ldots$}
\rput(0.5,0.75){\tiny{$_{(m,1)}$}}\rput(0.5,1.75){\tiny{$_{(m,1)}$}}\rput(0.5,3.75){\tiny{$_{(m,1)}$}}\rput(0.5,4.75){\tiny{$_{(1,m)}$}}
\rput(0.5,5.75){\tiny{$_{(1,m)}$}}\rput(0.5,7.75){\tiny{$_{(1,m)}$}}
\rput(1.5,0.75){\tiny{$_{(m,1)}$}}\rput(1.5,1.75){\tiny{$_{(m,1)}$}}\rput(1.5,3.75){\tiny{$_{(m,1)}$}}\rput(1.5,4.75){\tiny{$_{(1,m)}$}}
\rput(1.5,5.75){\tiny{$_{(1,m)}$}}\rput(1.5,7.75){\tiny{$_{(1,m)}$}}
\rput(4.5,0.75){\tiny{$_{(m,1)}$}}\rput(4.5,1.75){\tiny{$_{(m,1)}$}}\rput(4.5,3.75){\tiny{$_{(m,1)}$}}\rput(4.5,4.75){\tiny{$_{(1,m)}$}}
\rput(4.5,5.75){\tiny{$_{(1,m)}$}}\rput(4.5,7.75){\tiny{$_{(1,m)}$}}
\end{pspicture} 
\qquad \qquad\
 \psset{unit=0.9}
\begin{pspicture}[shift=-1.9](-0.5,0)(5.3,4)
\rput(-1.6,2){$\Tb^{m,n}_0 =$}
\facegrid{(0,0)}{(5,4)}
\psarc[linewidth=0.025]{-}(0,0){0.16}{0}{90}
\psarc[linewidth=0.025]{-}(0,1){0.16}{0}{90}
\psarc[linewidth=0.025]{-}(0,3){0.16}{0}{90}
\psarc[linewidth=0.025]{-}(1,0){0.16}{0}{90}
\psarc[linewidth=0.025]{-}(1,1){0.16}{0}{90}
\psarc[linewidth=0.025]{-}(1,3){0.16}{0}{90}
\psarc[linewidth=0.025]{-}(4,0){0.16}{0}{90}
\psarc[linewidth=0.025]{-}(4,1){0.16}{0}{90}
\psarc[linewidth=0.025]{-}(4,3){0.16}{0}{90}
\psline[linecolor=blue,linewidth=1.5pt]{-}(-0.3,0.5)(-0.5,0.5)
\psline[linecolor=blue,linewidth=1.5pt]{-}(-0.3,1.5)(-0.5,1.5)
\psline[linecolor=blue,linewidth=1.5pt]{-}(-0.3,2.5)(-0.5,2.5)
\psline[linecolor=blue,linewidth=1.5pt]{-}(-0.3,3.5)(-0.5,3.5)
\psline[linecolor=blue,linewidth=1.5pt]{-}(5.0,0.5)(5.2,0.5)
\psline[linecolor=blue,linewidth=1.5pt]{-}(5.0,1.5)(5.2,1.5)
\psline[linecolor=blue,linewidth=1.5pt]{-}(5.0,2.5)(5.2,2.5)
\psline[linecolor=blue,linewidth=1.5pt]{-}(5.0,3.5)(5.2,3.5)
\pspolygon[fillstyle=solid,fillcolor=orange](0,0.15)(0,3.85)(-0.15,4)(-0.3,3.85)(-0.3,0.15)(-0.15,0)(0,0.15)
\rput(-0.15,2){$_n$}
\rput(0.5,.4){\small$u_0$}
\rput(1.5,.4){\small$u_0$}
\rput(4.5,.4){\small$u_0$}
\rput(0.5,1.4){\small$u_1$}
\rput(1.5,1.4){\small$u_1$}
\rput(4.5,1.4){\small$u_1$}
\rput(0.5,2.6){$\vdots$}
\rput(1.5,2.6){$\vdots$}
\rput(4.5,2.6){$\vdots$}
\rput(0.5,3.4){\small$u_{n\!-\!1}$}
\rput(1.5,3.4){\small$u_{n\!-\!1}$}
\rput(4.5,3.4){\small$u_{n\!-\!1}$}
\rput(2.5,0.5){$\ldots$}
\rput(2.5,1.5){$\ldots$}
\rput(3.5,0.5){$\ldots$}
\rput(3.5,1.5){$\ldots$}%
\rput(2.5,3.5){$\ldots$}
\rput(3.5,3.5){$\ldots$}
\rput(0.5,0.75){\tiny{$_{(m,1)}$}}\rput(0.5,1.75){\tiny{$_{(m,1)}$}}\rput(0.5,3.75){\tiny{$_{(m,1)}$}}
\rput(1.5,0.75){\tiny{$_{(m,1)}$}}\rput(1.5,1.75){\tiny{$_{(m,1)}$}}\rput(1.5,3.75){\tiny{$_{(m,1)}$}}
\rput(4.5,0.75){\tiny{$_{(m,1)}$}}\rput(4.5,1.75){\tiny{$_{(m,1)}$}}\rput(4.5,3.75){\tiny{$_{(m,1)}$}}
\end{pspicture}  
\label{eq:DTQ}
\ee  
\label{prop:DTQ}
\end{Proposition}

\noindent
Since the expressions (\ref{eq:DTQ}) are based on the effective projector $Q_n$, they are well defined for
all $\beta\in\mathbb{C}^\ast$ and can therefore serve 
as definitions of the transfer tangles for fractional $\lambda$ for which $2\cos\lambda\neq0$. 
The argument extends to $\beta = 0$ in the following way. For each tangle in the decomposition \eqref{eq:Qndec} applied to 
\eqref{eq:DTQ}, the factor $(\tfrac 1\beta)^k$ is exactly cancelled by a factor $\beta^{k}$ 
coming from closed loops that appear in the simplification of the tangle. As in \eqref{eq:D140}, 
the decomposition coefficients are thereby rendered nonsingular at $\beta = 0$.

As we will discuss in Sections~\ref{Sec:Fusion} and~\ref{Sec:FusionCylinder}, 
there exists yet another way, valid for all $\lambda\in\pi\big(\mathbb{R}\!\setminus\!\mathbb{Z}\big)$, 
to write $\Db^{m,n}(u)$ and $\Tb^{m,n}(u)$ without the use of $P_n$, namely the determinant forms in (\ref{Ddet}) and (\ref{Tdet}).

Having established that $\Db^{m,n}(u)$ in (\ref{eq:transfermatrix}) and $\Tb^{m,n}(u)$ in (\ref{eq:Ttransfermatrix}) for generic 
$\lambda$ can be rewritten in a form independent of the projector $P_n$, 
we may view the transfer tangles for fractional $\lambda = \lambda_{p,p'}$ and $m<p'$ as 
\be
 \Db^{m,n}(u,\lambda_{p,p'})=\lim_{\lambda\to\lambda_{p,p'}}\Db^{m,n}(u,\lambda),\qquad
 \Tb^{m,n}(u,\lambda_{p,p'})=\lim_{\lambda\to\lambda_{p,p'}}\Tb^{m,n}(u,\lambda), \quad \ 1\le m\le p'-1,\ \ \, n \in \mathbb N.
\label{eq:existinglimits1}
\ee
We note that the arguments of these limits are given by projector {\em dependent} transfer tangles. 
The existence of the limits implies that every coefficient in the decomposition of these tangles in elementary connectivity diagrams is 
nonsingular at $\lambda = \lambda_{p,p'}$. 
As already indicated, potential issues with the projectors along the horizontal edges disappear for the fused link states representations 
$\rho_{N,m}^d$ and $\tilde \rho_{N,m}^d$ introduced in Section~\ref{Sec:CabledLinkStates}, so the limits 
\be
\begin{array}{l}
\rho_{N,m}^d \big[\Db^{m,n}(u,\lambda_{p,p'})\big]
 =\displaystyle \lim_{\lambda\to\lambda_{p,p'}}\rho_{N,m}^d\big[\Db^{m,n}(u,\lambda)\big],\\[0.4cm]
  \tilde \rho_{N,m}^d\big[ \Tb^{m,n}(u,\lambda_{p,p'})\big] 
  =\displaystyle\lim_{\lambda\to\lambda_{p,p'}}\tilde \rho_{N,m}^d\big[\Tb^{m,n}(u,\lambda)\big],
\end{array} \qquad m,n \in \mathbb N
\label{eq:existinglimits}
\ee
for the corresponding {\em transfer matrices} are well defined for all $m,n\in\mathbb{N}$. This further implies that relations between 
transfer {\em tangles} obtained by diagrammatic manipulations for $\lambda$ generic are valid for the transfer {\em matrices} in the 
$\lambda \rightarrow \lambda_{p,p'}$ limit. As we will discuss in Section~\ref{Sec:LMM}, 
additional special relations hold for fractional $\lambda$.

%%%%%%%%%%%%%
\subsection{Braid transfer tangles}
%%%%%%%%%%%%%

{\em $(m,n)$-fused braid operators} are defined by
\be
\psset{unit=.9cm}
\begin{pspicture}[shift=-2.5](-2.5,-0.3)(5.3,5.3)
\pspolygon[fillstyle=solid,fillcolor=lightlightblue](-2.5,2)(-1.5,2)(-1.5,3)(-2.5,3)(-2.5,2)
\rput(-2.5,2.0){
\psline[linewidth=1.0pt,linecolor=blue]{-}(0,0.558)(1,0.558)
\psline[linewidth=1.0pt,linecolor=blue]{-}(0,0.442)(1,0.442)
\psline[linewidth=1.0pt,linecolor=blue]{-}(0.558,0)(0.558,0.342)
\psline[linewidth=1.0pt,linecolor=blue]{-}(0.558,0.658)(0.558,1)
\psline[linewidth=1.0pt,linecolor=blue]{-}(0.442,0.658)(0.442,1)
\psline[linewidth=1.0pt,linecolor=blue]{-}(0.442,0)(0.442,0.342)}
\rput(-1,2.5){$:=$}
\pspolygon[fillstyle=solid,fillcolor=pink](0.1,0)(4.9,0)(4.9,-0.3)(0.1,-0.3)(0.1,0)
\pspolygon[fillstyle=solid,fillcolor=pink](0,0.1)(0,4.9)(-0.3,4.9)(-0.3,0.1)(0,0.1)
\pspolygon[fillstyle=solid,fillcolor=pink](5,0.1)(5,4.9)(5.3,4.9)(5.3,0.1)(5,0.1)
\pspolygon[fillstyle=solid,fillcolor=pink](0.1,5)(4.9,5)(4.9,5.3)(0.1,5.3)(0.1,5)
\rput(2.5,-0.15){$_m$}
\rput(2.5,5.15){$_m$}
\rput(-0.15,2.5){$_n$}
\rput(5.15,2.5){$_n$}
\facegrid{(0,0)}{(5,5)}
\multiput(0,0)(1,0){2}
{
\multiput(0,0)(0,1){2}{
\multiput(0,0)(0,3){2}{
\multiput(0,0)(3,0){2}{
\psline[linewidth=1.5pt,linecolor=blue]{-}(0,0.5)(1,0.5)
\psline[linewidth=1.5pt,linecolor=blue]{-}(0.5,0)(0.5,0.35)
\psline[linewidth=1.5pt,linecolor=blue]{-}(0.5,0.65)(0.5,1)
}}}}
\multiput(2,0)(0,1){2}{\rput(.5,.5){\small $\ldots$}}
\multiput(2,3)(0,1){2}{\rput(.5,.5){\small $\ldots$}}
\multiput(0,2)(1,0){2}{\rput(.5,.6){\small $\vdots$}}
\multiput(3,2)(1,0){2}{\rput(.5,.6){\small $\vdots$}}
\end{pspicture} 
\qquad\qquad
\begin{pspicture}[shift=-2.5](-2.5,-0.3)(5.3,5.3)
\pspolygon[fillstyle=solid,fillcolor=lightlightblue](-2.5,2)(-1.5,2)(-1.5,3)(-2.5,3)(-2.5,2)
\rput(-2.5,2.0){
\psline[linewidth=1.0pt,linecolor=blue]{-}(0.558,0)(0.558,1)
\psline[linewidth=1.0pt,linecolor=blue]{-}(0.442,0)(0.442,1)
\psline[linewidth=1.0pt,linecolor=blue]{-}(0,0.558)(0.342,0.558)
\psline[linewidth=1.0pt,linecolor=blue]{-}(0.658,0.558)(1,0.558)
\psline[linewidth=1.0pt,linecolor=blue]{-}(0.658,0.442)(1,0.442)
\psline[linewidth=1.0pt,linecolor=blue]{-}(0,0.442)(0.342,0.442)}
\rput(-1,2.5){$:=$}
\pspolygon[fillstyle=solid,fillcolor=pink](0.1,0)(4.9,0)(4.9,-0.3)(0.1,-0.3)(0.1,0)
\pspolygon[fillstyle=solid,fillcolor=pink](0,0.1)(0,4.9)(-0.3,4.9)(-0.3,0.1)(0,0.1)
\pspolygon[fillstyle=solid,fillcolor=pink](5,0.1)(5,4.9)(5.3,4.9)(5.3,0.1)(5,0.1)
\pspolygon[fillstyle=solid,fillcolor=pink](0.1,5)(4.9,5)(4.9,5.3)(0.1,5.3)(0.1,5)
\rput(2.5,-0.15){$_m$}
\rput(2.5,5.15){$_m$}
\rput(-0.15,2.5){$_n$}
\rput(5.15,2.5){$_n$}
\facegrid{(0,0)}{(5,5)}
\multiput(0,0)(1,0){2}
{
\multiput(0,0)(0,1){2}{
\multiput(0,0)(0,3){2}{
\multiput(0,0)(3,0){2}{
\psline[linewidth=1.5pt,linecolor=blue]{-}(0.5,0)(0.5,1)
\psline[linewidth=1.5pt,linecolor=blue]{-}(0,0.5)(0.35,0.5)
\psline[linewidth=1.5pt,linecolor=blue]{-}(0.65,0.5)(1,0.5)
}}}}
\multiput(2,0)(0,1){2}{\rput(.5,.5){\small $\ldots$}}
\multiput(2,3)(0,1){2}{\rput(.5,.5){\small $\ldots$}}
\multiput(0,2)(1,0){2}{\rput(.5,.6){\small $\vdots$}}
\multiput(3,2)(1,0){2}{\rput(.5,.6){\small $\vdots$}}
\end{pspicture} 
\label{eq:fusedbraid} 
\ee
and appear in the braid limit of $(m,n)$-fused faces,
\be 
\psset{unit=.9cm}
\begin{pspicture}[shift=-0.4](0,0.0)(1,1)
\pspolygon[fillstyle=solid,fillcolor=lightlightblue](0,0)(1,0)(1,1)(0,1)(0,0)
\psline[linewidth=1.0pt,linecolor=blue]{-}(1,0.558)(0,0.558)
\psline[linewidth=1.0pt,linecolor=blue]{-}(1,0.442)(0,0.442)
\psline[linewidth=1.0pt,linecolor=blue]{-}(0.558,0)(0.558,0.342)
\psline[linewidth=1.0pt,linecolor=blue]{-}(0.558,0.658)(0.558,1)
\psline[linewidth=1.0pt,linecolor=blue]{-}(0.442,0.658)(0.442,1)
\psline[linewidth=1.0pt,linecolor=blue]{-}(0.442,0)(0.442,0.342)
\end{pspicture}
\ =\lim_{u\to\ir\infty}\Big(
\prod_{i=0}^{m-1}\prod_{j=0}^{n-1} \frac{e^{\ir\tfrac{\pi-\lambda}2}}{s_{j-i}(u_k)}
\
\begin{pspicture}[shift=-0.4](0,0.0)(1,1)
\facegrid{(0,0)}{(1,1)}
\psarc[linewidth=0.025]{-}(0,0){0.16}{0}{90}
\rput(0.5,.45){$u_k$}\rput(0.5,0.75){\tiny{$_{(m,n)}$}}
\end{pspicture} 
\ \Big)
,\qquad 
\begin{pspicture}[shift=-0.4](0,0.0)(1,1)
\pspolygon[fillstyle=solid,fillcolor=lightlightblue](0,0)(1,0)(1,1)(0,1)(0,0)
\psline[linewidth=1.0pt,linecolor=blue]{-}(0.558,0)(0.558,1)
\psline[linewidth=1.0pt,linecolor=blue]{-}(0.442,0)(0.442,1)
\psline[linewidth=1.0pt,linecolor=blue]{-}(0,0.558)(0.342,0.558)
\psline[linewidth=1.0pt,linecolor=blue]{-}(0.658,0.558)(1,0.558)
\psline[linewidth=1.0pt,linecolor=blue]{-}(0.658,0.442)(1,0.442)
\psline[linewidth=1.0pt,linecolor=blue]{-}(0,0.442)(0.342,0.442)
\end{pspicture}
\ =\lim_{u\to \ir\infty}\Big(
\prod_{i=0}^{m-1}\prod_{j=0}^{n-1} \frac{e^{\ir\tfrac{\pi-\lambda}2}}{s_{i-j}(u_k)}
\ 
\begin{pspicture}[shift=-0.4](0,0.0)(1,1)
\facegrid{(0,0)}{(1,1)}
\psarc[linewidth=0.025]{-}(1,0){0.16}{90}{180}
\rput(0.5,.45){$u_k$}\rput(0.5,0.75){\tiny{$_{(n,m)}$}}
\end{pspicture}
\ \Big).
\ee 
In terms of these, the {\em fused braid transfer tangles} are defined as
\be 
 \Fb^{m,n}:=  
\begin{pspicture}[shift=-1.1](-0.7,-0.2)(5.7,2)
\psarc[linewidth=4pt,linecolor=blue]{-}(0,1){0.5}{90}{-90}\psarc[linewidth=2pt,linecolor=white]{-}(0,1){0.5}{90}{-90}
\psarc[linewidth=4pt,linecolor=blue]{-}(5,1){0.5}{-90}{90}\psarc[linewidth=2pt,linecolor=white]{-}(5,1){0.5}{-90}{90}
\facegrid{(0,0)}{(5,2)}
\rput(2.5,0.5){$\ldots$}
\rput(2.5,1.5){$\ldots$}
\rput(3.5,0.5){$\ldots$}
\rput(3.5,1.5){$\ldots$}
\multiput(0,0)(1,0){2}{
\psline[linewidth=1.0pt,linecolor=blue]{-}(0,0.558)(1,0.558)
\psline[linewidth=1.0pt,linecolor=blue]{-}(0,0.442)(1,0.442)
\psline[linewidth=1.0pt,linecolor=blue]{-}(0.558,0)(0.558,0.342)
\psline[linewidth=1.0pt,linecolor=blue]{-}(0.558,0.658)(0.558,1)
\psline[linewidth=1.0pt,linecolor=blue]{-}(0.442,0.658)(0.442,1)
\psline[linewidth=1.0pt,linecolor=blue]{-}(0.442,0)(0.442,0.342)}
\rput(4,0){
\psline[linewidth=1.0pt,linecolor=blue]{-}(0,0.558)(1,0.558)
\psline[linewidth=1.0pt,linecolor=blue]{-}(0,0.442)(1,0.442)
\psline[linewidth=1.0pt,linecolor=blue]{-}(0.558,0)(0.558,0.342)
\psline[linewidth=1.0pt,linecolor=blue]{-}(0.558,0.658)(0.558,1)
\psline[linewidth=1.0pt,linecolor=blue]{-}(0.442,0.658)(0.442,1)
\psline[linewidth=1.0pt,linecolor=blue]{-}(0.442,0)(0.442,0.342)}
\rput(4,1.0){
\psline[linewidth=1.0pt,linecolor=blue]{-}(0.558,0)(0.558,1)
\psline[linewidth=1.0pt,linecolor=blue]{-}(0.442,0)(0.442,1)
\psline[linewidth=1.0pt,linecolor=blue]{-}(0,0.558)(0.342,0.558)
\psline[linewidth=1.0pt,linecolor=blue]{-}(0.658,0.558)(1,0.558)
\psline[linewidth=1.0pt,linecolor=blue]{-}(0.658,0.442)(1,0.442)
\psline[linewidth=1.0pt,linecolor=blue]{-}(0,0.442)(0.342,0.442)}
\multiput(0,1.0)(1,0){2}{
\psline[linewidth=1.0pt,linecolor=blue]{-}(0.558,0)(0.558,1)
\psline[linewidth=1.0pt,linecolor=blue]{-}(0.442,0)(0.442,1)
\psline[linewidth=1.0pt,linecolor=blue]{-}(0,0.558)(0.342,0.558)
\psline[linewidth=1.0pt,linecolor=blue]{-}(0.658,0.558)(1,0.558)
\psline[linewidth=1.0pt,linecolor=blue]{-}(0.658,0.442)(1,0.442)
\psline[linewidth=1.0pt,linecolor=blue]{-}(0,0.442)(0.342,0.442)}
\end{pspicture}
\qquad
\Fbt^{m,n}:= 
\begin{pspicture}[shift=-0.6](-0.3,-0.2)(5.3,1)
\facegrid{(0,0)}{(5,1)}
\psline[linewidth=1.0pt,linecolor=blue]{-}(0,0.558)(-0.2,0.558)
\psline[linewidth=1.0pt,linecolor=blue]{-}(0,0.442)(-0.2,0.442)
\psline[linewidth=1.0pt,linecolor=blue]{-}(5,0.558)(5.2,0.558)
\psline[linewidth=1.0pt,linecolor=blue]{-}(5,0.442)(5.2,0.442)
\multiput(0,0)(1,0){2}{
\psline[linewidth=1.0pt,linecolor=blue]{-}(0,0.558)(1,0.558)
\psline[linewidth=1.0pt,linecolor=blue]{-}(0,0.442)(1,0.442)
\psline[linewidth=1.0pt,linecolor=blue]{-}(0.558,0)(0.558,0.342)
\psline[linewidth=1.0pt,linecolor=blue]{-}(0.558,0.658)(0.558,1)
\psline[linewidth=1.0pt,linecolor=blue]{-}(0.442,0.658)(0.442,1)
\psline[linewidth=1.0pt,linecolor=blue]{-}(0.442,0)(0.442,0.342)}
\rput(4,0){
\psline[linewidth=1.0pt,linecolor=blue]{-}(0,0.558)(1,0.558)
\psline[linewidth=1.0pt,linecolor=blue]{-}(0,0.442)(1,0.442)
\psline[linewidth=1.0pt,linecolor=blue]{-}(0.558,0)(0.558,0.342)
\psline[linewidth=1.0pt,linecolor=blue]{-}(0.558,0.658)(0.558,1)
\psline[linewidth=1.0pt,linecolor=blue]{-}(0.442,0.658)(0.442,1)
\psline[linewidth=1.0pt,linecolor=blue]{-}(0.442,0)(0.442,0.342)}
\rput(2.5,0.5){$\ldots$}
\rput(3.5,0.5){$\ldots$}
\end{pspicture}% \ \ \ 
\label{eq:Ftangle}
\ee  
and appear in the braid limit as
\be
 \Fb^{m,n}=\lim_{u\to\ir\infty}\bigg(
  \Big(\prod_{i=0}^{m-1}\prod_{j=0}^{n-1} \frac{e^{\ir(\pi-\lambda)}}{s_{j-i}(u)s_{i-j}(u_{n}\!-\!\mu)}\Big)^{\!N}\Db^{m,n}_0\bigg), 
 \qquad
 \Fbt^{m,n}=\lim_{u\to\ir\infty}\bigg(
  \Big(\prod_{i=0}^{m-1}\prod_{j=0}^{n-1} \frac{e^{\ir\tfrac{\pi-\lambda}2}}{s_{j-i}(u)}\Big)^{\!N}\Tb^{m,n}_0\bigg).
\ee

%%%%%%%%%%%%%%%
\section{Matrix representations} 
\label{Sec:CabledLinkStates}
%%%%%%%%%%%%%%%

In this section, we identify states and modules 
on which the fused transfer tangles can act. A primary goal is to resolve the issue with 
the non-existence of some $P_m$ projectors for fractional $\lambda$.
Initially, we let $\lambda$ be generic and we commence by making some algebraic observations before discussing the 
construction of states and representations. To build these for $m>1$ and generic $\lambda$, 
we adopt the defect preserving diagrammatic action of tangles on link states used in the construction of standard modules 
for $m=1$ in Section~\ref{Sec:TL}. We conclude by turning our attention to fractional $\lambda$,
by showing that the $P_m$ projectors play no role in the matrix representations for the transfer tangles.

\paragraph{Transfer tangles without $P_m$ projectors}
The push-through properties allow us to write
\be
\Db^{m,n}(u) = \Ib^m \,\Dbb^{m,n}(u)\, \Ib^m, \qquad \Tb^{m,n}(u) = \Ib^m\,\Tbb^{m,n}(u)\,\Ib^m,
\label{eq:DTbar}
\ee
where $\Dbb^{m,n}(u)$ and $\Tbb^{m,n}(u)$ are identical to \eqref{eq:transfermatrix} and \eqref{eq:Ttransfermatrix}, but with 
all $P_m$ projectors removed. That is,
\be 
\Dbb^{m,n}(u):=
\begin{pspicture}[shift=-0.9](-0.7,-0)(5.3,2)
\psarc[linewidth=4pt,linecolor=blue]{-}(0,1){0.5}{90}{-90}\psarc[linewidth=2pt,linecolor=white]{-}(0,1){0.5}{90}{-90}
\psarc[linewidth=4pt,linecolor=blue]{-}(5,1){0.5}{-90}{90}\psarc[linewidth=2pt,linecolor=white]{-}(5,1){0.5}{-90}{90}
\facegrid{(0,0)}{(5,2)}
\psline[linewidth=0.025]{-}(0,0.16)(0.16,0.16)(0.16,0)
\psline[linewidth=0.025]{-}(1,0.16)(1.16,0.16)(1.16,0)
\psline[linewidth=0.025]{-}(4,0.16)(4.16,0.16)(4.16,0)
\psline[linewidth=0.025]{-}(0,1.16)(0.16,1.16)(0.16,1)
\psline[linewidth=0.025]{-}(1,1.16)(1.16,1.16)(1.16,1)
\psline[linewidth=0.025]{-}(4,1.16)(4.16,1.16)(4.16,1)
\rput(0.5,0.75){\tiny{$_{(m,n)}$}}\rput(0.5,1.75){\tiny{$_{(m,n)}$}}
\rput(1.5,0.75){\tiny{$_{(m,n)}$}}\rput(1.5,1.75){\tiny{$_{(m,n)}$}}
\rput(4.5,0.75){\tiny{$_{(m,n)}$}}\rput(4.5,1.75){\tiny{$_{(m,n)}$}}
\rput(2.5,0.5){$\ldots$}
\rput(2.5,1.5){$\ldots$}
\rput(3.5,0.5){$\ldots$}
\rput(3.5,1.5){$\ldots$}
\rput(0.5,.5){$u$}
\rput(0.52,1.45){\scriptsize$\mu\!\!-\!\!u_{n\!-\!1}$}
\rput(1.5,.5){$u$}
\rput(1.52,1.45){\scriptsize$\mu\!\!-\!\!u_{n\!-\!1}$}
\rput(4.5,.5){$u$}
\rput(4.52,1.45){\scriptsize$\mu\!\!-\!\!u_{n\!-\!1}$}
\end{pspicture}
\label{eq:Dbar}
\ee
\noindent and
\be 
\Tbb^{m,n} (u):= \hspace{0.4cm}
\begin{pspicture}[shift=-0.6](-0.0,-0.2)(5.0,1.2)
\psline[linewidth=4pt,linecolor=blue]{-}(0,0.5)(-0.2,0.5)\psline[linewidth=2pt,linecolor=white]{-}(0,0.5)(-0.2,0.5)
\psline[linewidth=4pt,linecolor=blue]{-}(5,0.5)(5.2,0.5)\psline[linewidth=2pt,linecolor=white]{-}(5,0.5)(5.2,0.5)
\facegrid{(0,0)}{(5,1)}
\psline[linewidth=0.025]{-}(0,0.16)(0.16,0.16)(0.16,0)
\psline[linewidth=0.025]{-}(1,0.16)(1.16,0.16)(1.16,0)
\psline[linewidth=0.025]{-}(4,0.16)(4.16,0.16)(4.16,0)
\rput(2.5,0.5){$\ldots$}
\rput(3.5,0.5){$\ldots$}
\rput(0.5,.5){$u$}
\rput(1.5,.5){$u$}
\rput(4.5,.5){$u$}
\rput(0.5,0.75){\tiny{$_{(m,n)}$}}
\rput(1.5,0.75){\tiny{$_{(m,n)}$}}
\rput(4.5,0.75){\tiny{$_{(m,n)}$}}
\end{pspicture}
\label{eq:Tbar}
\ee
where
\be
\psset{unit=0.9}
\begin{pspicture}[shift=-2.6](-2.5,-0.0)(5.3,5.0)
\pspolygon[fillstyle=solid,fillcolor=lightlightblue](-2.5,2)(-1.5,2)(-1.5,3)(-2.5,3)(-2.5,2)
\psline[linewidth=0.025]{-}(-2.5,2.16)(-2.34,2.16)(-2.34,2)
\rput(-2,2.75){\tiny{$_{(m,n)}$}}
\rput(-2,2.5){$u$}
\rput(-1,2.5){$:=$}
\pspolygon[fillstyle=solid,fillcolor=pink](0,0.1)(0,4.9)(-0.3,4.9)(-0.3,0.1)(0,0.1)
\pspolygon[fillstyle=solid,fillcolor=pink](5,0.1)(5,4.9)(5.3,4.9)(5.3,0.1)(5,0.1)
\rput(-0.15,2.5){$_n$}
\rput(5.15,2.5){$_n$}
\facegrid{(0,0)}{(5,5)}
\psarc[linewidth=0.025]{-}(0,0){0.16}{0}{90}
\psarc[linewidth=0.025]{-}(3,0){0.16}{0}{90}
\psarc[linewidth=0.025]{-}(4,0){0.16}{0}{90}
\psarc[linewidth=0.025]{-}(0,1){0.16}{0}{90}
\psarc[linewidth=0.025]{-}(3,1){0.16}{0}{90}
\psarc[linewidth=0.025]{-}(4,1){0.16}{0}{90}
\psarc[linewidth=0.025]{-}(0,4){0.16}{0}{90}
\psarc[linewidth=0.025]{-}(3,4){0.16}{0}{90}
\psarc[linewidth=0.025]{-}(4,4){0.16}{0}{90}
\rput(4.5,.5){\small $u_0$}
\rput(4.5,1.5){\small $u_1$}
\rput(4.5,2.6){\small $\vdots$}
\rput(4.5,3.6){\small $\vdots$}
\rput(4.5,4.5){\small $u_{n\!-\!1}$}
\rput(3.5,.5){\small $u_{-\!1}$}
\rput(3.5,1.5){\small $u_0$}
\rput(3.5,2.6){\small $\vdots$}
\rput(3.5,3.6){\small $\vdots$}
\rput(3.5,4.5){\small $u_{n\!-\!2}$}
\rput(.5,.5){\small $u_{1\!-\!m}$}
\rput(.5,1.5){\small $u_{2\!-\!m}$}
\rput(.5,2.6){\small $\vdots$}
\rput(.5,3.6){\small $\vdots$}
\rput(.5,4.5){\small $u_{n\!-\!m}$}
\multiput(0,0)(1,0){2}{\rput(1.5,.5){\small $\ldots$}}
\multiput(0,1)(1,0){2}{\rput(1.5,.5){\small $\ldots$}}
\multiput(0,4)(1,0){2}{\rput(1.5,.5){\small $\ldots$}}
\end{pspicture}
\label{eq:FusedBar}
\ee  
To distinguish these fused faces with projectors only on the left and right, we indicate the orientation with a small square instead of the 
usual small arc. In fact, the push-through properties allow us to simplify \eqref{eq:DTbar} and write
\be
 \Db^{m,n}(u) = \Ib^m\, \Dbb^{m,n}(u), \qquad \Tb^{m,n}(u) = \Ib^m\,\Tbb^{m,n}(u).
\label{eq:IDTbar}
\ee
We also note that 
\be
 \Ibb^{\,m} = \Ib
\label{eq:Ibb}
\ee
and that we can elevate the limit (\ref{eq:existinglimits}) from matrices to tangles as 
\be
 \Dbb^{m,n}(u,\lambda_{p,p'})=\lim_{\lambda\to\lambda_{p,p'}}\Dbb^{m,n}(u,\lambda),\qquad
 \Tbb^{m,n}(u,\lambda_{p,p'})=\lim_{\lambda\to\lambda_{p,p'}}\Tbb^{m,n}(u,\lambda), \qquad m,n\in\mathbb{N}.
 \label{eq:existinglimitsBar}
\ee

%%%%%%%%%%%%%%%%%%%%%%%%%%%
\paragraph{Fused Temperley-Lieb algebra}
%%%%%%%%%%%%%%%%%%%%%%%%%%%

Focusing on the situation on the strip, we introduce
\be 
 FTL_{N,m}(\beta):=\Big\langle x\in TL_{Nm}(\beta)\big|\, \exists\, \bar{x}\in TL_{Nm}(\beta)\;\mathrm{such\;that}\;
  x=\Ib^m\bar{x}\Ib^m=\Ib^m\bar{x}\Big\rangle
\label{STL}
\ee
which is a subalgebra of $TL_{Nm}(\beta)$ since 
\be 
 x_1,x_2\in FTL_{N,m}(\beta)\ \Rightarrow\ \left\{\!\!
 \begin{array}{ll} x_1 + x_2 = \Ib^m(\bar x_1 + \bar x_2) \Ib^m = \Ib^m(\bar x_1 + \bar x_2) \quad &\in FTL_{N,m}(\beta),\\[.3cm]
 x_1x_2
  =\begin{cases}
  \big(\Ib^m\bar{x}_1\Ib^m\big)\big(\Ib^m\bar{x}_2\Ib^m\big)=\Ib^m(\bar{x}_1\bar{x}_2)\Ib^m \\[.2cm]
  \big(\Ib^m\bar{x}_1\big)\big(\Ib^m\bar{x}_2\big)=\Ib^m(\bar{x}_1\bar{x}_2)\end{cases} &\in FTL_{N,m}(\beta).\end{array} \right.
\ee
Tacitly assuming that the direction of transfer is vertical, the transfer tangle $\Db^{m,n}(u)$ is an element of $FTL_{N,m}(\beta)$.
It is also noted that $\Ib^m$ is 
in the algebra $FTL_{N,m}(\beta)$ and that it plays the role of the identity. On the cylinder, we similarly introduce
\be
 F \mathcal EPTL_{N,m}(\alpha,\beta):=\Big\langle x\in \mathcal EPTL_{Nm}(\alpha,\beta)\big|\, \exists\, \bar{x}\in 
  \mathcal EPTL_{Nm}(\alpha,\beta)\;\mathrm{such\;that}\;
  x=\Ib^m\bar{x}\Ib^m=\Ib^m\bar{x}\Big\rangle
\label{ESTL}
\ee
which is a subalgebra of $\mathcal EPTL_{Nm}(\alpha,\beta)$ and includes $\Tb^{m,n}(u)$ as one of its elements.

%%%%%%%%%%%%%%%%%%%%%%%%%%
\paragraph{Cabled link states}
%%%%%%%%%%%%%%%%%%%%%%%%%%

On the strip, we define an {\em $m$-cabled link state} on $Nm$ nodes with $d$ defects as a (canonical basis)
link state in $V^d_{N\times m}$ that does not have a half-arc linking a pair of nodes located between positions 
$\ell m+1$ and $(\ell+1)m$ for any $\ell\in\{0,\ldots,N-1\}$.
This generalises the construction for $m=2$ in~\cite{PRT2013}.
The linear span of $m$-cabled link states is denoted by $V_{N,m}^d$ and its complement $W_{N,m}^d$ in $V_{N\times m}^d$
is generated by the link states having at least one half-arc linking a pair of nodes located between positions
$\ell m+1$ and $(\ell+1)m$ for some $\ell\in\{0,\ldots,N-1\}$. 
We illustrate this characterisation of link states with a link state in $V_{4\times3}^4$ and one in $V_{4\times3}^2$, namely
\be 
\begin{pspicture}[shift=-0.35](-0.0,-0.4)(4.7,0.6)
\psarc[linewidth=1.5pt,linecolor=blue]{-}(0.2,0){0.2}{0}{180}
\psarc[linewidth=1.5pt,linecolor=blue]{-}(1.8,0){0.2}{0}{180}
\psarc[linewidth=1.5pt,linecolor=blue]{-}(3.8,0){0.2}{0}{180}
\psbezier[linewidth=1.5pt,linecolor=blue]{-}(1.2,0)(1.2,0.8)(2.4,0.8)(2.4,0)
\psline[linewidth=1.5pt,linecolor=blue]{-}(0.8,0)(0.8,0.6)
\psline[linewidth=1.5pt,linecolor=blue]{-}(4.4,0)(4.4,0.6)
\psline[linewidth=1.5pt,linecolor=blue]{-}(3.2,0)(3.2,0.6)
\psline[linewidth=1.5pt,linecolor=blue]{-}(2.8,0)(2.8,0.6)
\multiput(-0.1,-0.05)(1.2,0){4}{$\underbrace{\hspace{1cm}}_3$}
\end{pspicture} 
\in W_{4,3}^4,\qquad\qquad\qquad
\begin{pspicture}[shift=-0.35](-0.0,-0.4)(4.8,0.6)
\psarc[linewidth=1.5pt,linecolor=blue]{-}(1.0,0){0.2}{0}{180}
\psarc[linewidth=1.5pt,linecolor=blue]{-}(2.2,0){0.2}{0}{180}
\psarc[linewidth=1.5pt,linecolor=blue]{-}(3.4,0){0.2}{0}{180}
\psbezier[linewidth=1.5pt,linecolor=blue]{-}(0.4,0)(0.4,0.8)(1.6,0.8)(1.6,0)
\psbezier[linewidth=1.5pt,linecolor=blue]{-}(2.8,0)(2.8,0.8)(4,0.8)(4,0)
\psline[linewidth=1.5pt,linecolor=blue]{-}(0,0)(0,0.6)
\psline[linewidth=1.5pt,linecolor=blue]{-}(4.4,0)(4.4,0.6)
\multiput(-0.1,-0.05)(1.2,0){4}{$\underbrace{\hspace{1cm}}_3$}
\end{pspicture} 
\in V_{4,3}^2.
\vspace{0.05cm}
\ee
As indicated, only the second of these is a $3$-cabled link state. It is also noted that the defect number $d$ is subject to the constraints
\be
 0 \le d \le Nm, \qquad  Nm - d = 0 \ \textrm{mod}\ 2.
\label{dNm}
\ee

Similarly on the cylinder, an $m$-cabled link state on $Nm$ nodes with $d$ defects is defined as a link state in the canonical basis of
$\tilde V^d_{N\times m}$ with no half-arc linking, via the front of the cylinder, two nodes located between 
positions $\ell m+1$ and $(\ell+1)m$ for any $\ell\in\{0,\ldots,N-1\}$.
The linear span of these $m$-cabled link states is denoted by $\tilde V_{N,m}^d$ and its complement $\tilde W_{N,m}^d$ in
$\tilde V_{N\times m}^d$ is generated by the link states having at least one half-arc linking, via the front of the cylinder, 
a pair of nodes located between positions $\ell m+1$ and $(\ell+1)m$ for some $\ell\in\{0,\ldots,N-1\}$. Examples are
\be 
\begin{pspicture}[shift=-0.35](-0.0,-0.4)(4.7,0.6)
\psarc[linewidth=1.5pt,linecolor=blue]{-}(0.2,0){0.2}{0}{180}
\psarc[linewidth=1.5pt,linecolor=blue]{-}(3.8,0){0.2}{0}{180}
\psarc[linewidth=1.5pt,linecolor=blue]{-}(2.2,0){0.2}{0}{180}
\psbezier[linewidth=1.5pt,linecolor=blue]{-}(-0.4,0)(-0.4,0.8)(0.9,0.8)(0.8,0)
\rput(4.8,0){\psbezier[linewidth=1.5pt,linecolor=blue]{-}(-0.4,0)(-0.4,0.8)(0.9,0.8)(0.8,0)}
\psbezier[linewidth=1.5pt,linecolor=blue]{-}(1.6,0)(1.6,0.8)(2.8,0.8)(2.8,0)
\psline[linewidth=1.5pt,linecolor=blue]{-}(1.2,0)(1.2,0.6)
\psline[linewidth=1.5pt,linecolor=blue]{-}(3.2,0)(3.2,0.6)
\multiput(-0.1,-0.05)(1.2,0){4}{$\underbrace{\hspace{1cm}}_3$}
\psframe[fillstyle=solid,linecolor=white,linewidth=0pt](-2.2,-0.1)(-0.2,1.5)
\psframe[fillstyle=solid,linecolor=white,linewidth=0pt](4.6,-0.1)(7.6,1.5)
\end{pspicture} 
\in \tilde W_{4,3}^2,\qquad\qquad\qquad
\begin{pspicture}[shift=-0.35](-0.0,-0.4)(4.8,1.3)
\psarc[linewidth=1.5pt,linecolor=blue]{-}(1.0,0){0.2}{0}{180}
\psarc[linewidth=1.5pt,linecolor=blue]{-}(-0.2,0){0.2}{0}{90}
\psarc[linewidth=1.5pt,linecolor=blue]{-}(4.6,0){0.2}{90}{180}
\psarc[linewidth=1.5pt,linecolor=blue]{-}(3.4,0){0.2}{0}{180}
\psbezier[linewidth=1.5pt,linecolor=blue]{-}(0.4,0)(0.4,0.8)(1.6,0.8)(1.6,0)
\psbezier[linewidth=1.5pt,linecolor=blue]{-}(2.8,0)(2.8,1.8)(7.2,1.8)(7.2,0)
\psbezier[linewidth=1.5pt,linecolor=blue]{-}(4.0,0)(4.0,1.3)(6.8,1.3)(6.8,0)
\psbezier[linewidth=1.5pt,linecolor=blue]{-}(-0.8,0)(-0.8,1.3)(2.0,1.3)(2.0,0)
\psbezier[linewidth=1.5pt,linecolor=blue]{-}(-2.0,0)(-2.0,1.8)(2.4,1.8)(2.4,0)
\multiput(-0.1,-0.05)(1.2,0){4}{$\underbrace{\hspace{1cm}}_3$}
\psframe[fillstyle=solid,linecolor=white,linewidth=0pt](-2.2,-0.1)(-0.2,1.5)
\psframe[fillstyle=solid,linecolor=white,linewidth=0pt](4.6,-0.1)(7.6,1.5)
\end{pspicture} 
\in \tilde V_{4,3}^0.
\vspace{0.05cm}
\ee 

The spaces $V^d_{N\times m}$ and $\tilde V^d_{N \times m}$ thus decompose as
\be
 V^d_{N\times m}=V_{N,m}^d\oplus W_{N,m}^d,\qquad
 \tilde V^d_{N\times m}=\tilde V_{N,m}^d\oplus \tilde W_{N,m}^d,
\label{VVW}
\ee
and we let 
$\mathcal{P}_{N,m}^d$ and $\tilde{\mathcal{P}}_{N,m}^d$ be the corresponding 
matrix projectors from $V^d_{N\times m}$ and $\tilde V^d_{N\times m}$ onto $V^d_{N,m}$ and $\tilde V^d_{N,m}$, respectively.
With respect to the decompositions (\ref{VVW}), these matrices are of the form
\be
 {\mathcal P}_{N,m}^d=\begin{pmatrix} I&0\\ 0&0\end{pmatrix},\qquad 
 \tilde {\mathcal P}_{N,m}^d=\begin{pmatrix} I&0\\ 0&0\end{pmatrix}.
 \label{eq:matproj}
\ee
It is stressed that the decompositions (\ref{VVW}) and the matrix projectors are well defined for all $m\in\mathbb{N}$, 
independently of $\lambda$. 

In Appendix~\ref{app:LatticePaths}, we prove that the numbers 
of $m$-cabled link states on the strip and cylinder, respectively, for $N$ and $d$ subject to (\ref{dNm}) are given by
\be
 \dim V_{N,m}^d=\left(\!\!\!\begin{array}{c} N\\[.1cm] \frac{Nm-d}{2}\end{array}\!\!\!\right)_{\!\!m}
   -\,\left(\!\!\!\begin{array}{c} N\\[.1cm] \frac{Nm-d-2}{2}\end{array}\!\!\!\right)_{\!\!m},\qquad  \quad
  \dim \tilde{V}_{N,m}^d=\left(\!\!\!\begin{array}{c} N\\[.1cm] \frac{Nm-d}{2}\end{array}\!\!\!\right)_{\!\!m},
\ee
see Propositions~\ref{prop:LinkCountingCylinder} and~\ref{prop:LinkCountingStrip}.
Here the $(m+1)$-nomial 
coefficients $\left(\!\begin{smallmatrix}N\\ k\end{smallmatrix}\!\right)_m$, with $k \in \{0, \dots, Nm \}$,
are defined as the expansion coefficients of the generating function
\be 
\Big[\sum_{i=0}^{m} z^i\Big]^N = \sum_{k=0}^{Nm} \begin{pmatrix}N\\ k\end{pmatrix}_{\!\!m}\!z^k,
  \qquad N,m\in\mathbb{N}_0.
  \label{eq:mnomialgen}
\ee
For $m=1,2$, these coefficients are recognised as the usual binomial and trinomial coefficients, respectively.
In Appendix~\ref{app:LatticePaths}, we also discuss how these sets of link states are related to certain classes of lattice paths.

\paragraph{Generalised standard modules}
By construction of the link states in $W_{N,m}^d$, 
the diagrammatic action of $x\in FTL_{N,m}(\beta)$ on $w\in W_{N,m}^d$ yields $xw=0$ due to
the presence of $\Ib^m$ in $x$. The space $W_{N,m}^d$ is therefore trivially stable under the action of $FTL_{N,m}(\beta)$.
As a similar argument applies in the cylinder case, we can view the quotient spaces
\be 
 V^d_{N,m}=V^d_{N\times m}\,\big/\,W^d_{N,m}, \qquad \qquad 
 \tilde{V}^d_{N,m}=\tilde{V}^d_{N\times m}\,\big/\,\tilde{W}^d_{N,m}
\label{eq:quotients}
\ee 
as vector spaces on which representations of $FTL_{N,m}(\beta)$ and its cylinder counterpart are built.
Decomposing the matrix representative $\rho_{N\times m}^d(x)$ of an element $x \in FTL_{N,m}(\beta)$ on $V_{N\times m}^d$ 
with respect to (\ref{VVW}) as
\be
 \rho_{N\times m}^d(x)=\begin{pmatrix} A_x& 0\\ C_x&0\end{pmatrix},
\ee
the corresponding representation $\rho_{N,m}^d$ of $FTL_{N,m}(\beta)$
on $V_{N,m}^d$ is obtained by using the matrix projector $\mathcal P^d_{N,m}$ introduced in \eqref{eq:matproj}, or equivalent by 
taking the upper-left matrix minor $M_{N,m}^d$ of the appropriate size,
\be
 \rho_{N,m}^d(x):=M_{N,m}^d\big[\rho_{N\times m}^d(x)\big]=A_x.
\ee
From
\be
 \rho_{N\times m}^d(x_1x_2)=\begin{pmatrix} A_{x_1}A_{x_2} & 0 \\[.1cm] C_{x_1}A_{x_2} & 0\end{pmatrix},
\ee
it readily follows that $\rho_{N,m}^d$ is a representation of $FTL_{N,m}(\beta)$, that is, 
\be
 \rho_{N,m}^d(x_1x_2)= A_{x_1}A_{x_2} =\rho_{N,m}^d(x_1)\rho_{N,m}^d(x_2).
\ee
In stark contrast, the matrices  $M_{N,m}^d\big[\rho_{N\times m}^d(y)\big]$, $y \in TL_{Nm}(\beta)$,
do not constitute representations of $TL_{N\times m}(\beta)$ for $m\ge2$. 

The construction of matrix representations $\tilde \rho_{N,m}^d$ of $F\mathcal EPTL_N(\alpha,\beta)$ is similar. 
Thus, for simplicity, the arguments are given for $\rho_{N,m}^d$ only.

%%%%%%%%%%%%%%%%%%%%%%%%%%%%%%
\paragraph{Representations for fractional $\lambda$} 
%%%%%%%%%%%%%%%%%%%%%%%%%%%%%%
Let $v$ be an element of $V_{N\times m}^d$ and decompose it as
\be
 v=v'+w,\qquad v'\in V_{N,m}^d,\qquad w\in W_{N,m}^d.
\ee
From the definition of $\Ib^m$, it satisfies
\be
 \Ib^m v=\Ib^m v'=v'+w',\qquad w'\in W_{N,m}^d
\label{Imv}
\ee
from which it follows that
\be
\rho_{N\times m}^d(\Ib^{m}) = \begin{pmatrix} I & 0 \\ S & 0\end{pmatrix}, \qquad \rho^d_{N,m}(\Ib^{m}) = I,
\ee
where $I$ is the identity matrix of size $\dim V_{N,m}^{d}$ and $S$ is a $\lambda$-dependent rectangular matrix that can have 
singularities for fractional $\lambda$. Then, for some $x \in FTL_{N,m}(\beta)$, the corresponding $\bar x$ satisfies
\be
\rho^d_{N\times m}(\bar x) =  
\begin{pmatrix}  A_{\bar x}  &  B_{\bar x} \\[.1cm] C_{\bar x} & D_{\bar x}\end{pmatrix}
\ee
and because $x = \Ib^{m} \bar x$,
\be
 A_x = \rho^d_{N,m}(x)  = M^d_{N,m}\left[\begin{pmatrix} I&0 \\ S& 0 \end{pmatrix}
  \begin{pmatrix}A_{\bar x}  &  B_{\bar x} \\[.1cm] C_{\bar x} & D_{\bar x} \end{pmatrix}\right] =  A_{\bar x}.
\label{eq:withoutWJ}
\ee
This shows that, for fractional $\lambda$, if $\bar x$ exists, $\rho_{N,m}^d(x)$ is free of singularities. 
These arguments carry over to the cylinder case.
We thus conclude that, for the action on cabled link states, it is equivalent to work with $\Dbb^{m,n}(u)$ 
and $\Tbb^{m,n}(u)$ instead of the transfer tangles $\Db^{m,n}(u)$ and $\Tb^{m,n}(u)$. 
The expression on the righthand side of \eqref{eq:withoutWJ}, in particular, does not involve any $P_m$ projectors.
Furthermore, recalling that the potential singularities arising from $P_n$ projectors were dealt with using the corresponding
effective projectors $Q_n$ in Section~\ref{sec:Dcrit}, 
we see that we can compute matrix representations of the transfer tangles {\em without employing any projectors at all}.
It follows that the proposed matrix representatives of $\Db^{m,n}(u)$ and $\Tb^{m,n}(u)$ 
are well behaved for all $m,n\in\mathbb{N}$ and $\lambda$ generic or fractional alike.

The modules built from the defect preserving diagrammatic action on $V_{N,m}^d$ or $\tilde V_{N,m}^d$ resemble
the standard modules reviewed in Section~\ref{Sec:TL}.
This class of modules is certainly not exhaustive, though, as more general modules and representations can be envisaged
and are known to exist for $m=1$.
For example, representations of $TL_N(\beta)$ where defects can be annihilated in pairs appear naturally in the study of transfer 
tangles with nontrivial boundary conditions~\cite{PRZ0607,PRR0803,Ras1012,PRV1210}.
The above construction of matrix representations without projectors can be extended to this larger class of modules.

As a final remark, we stress that we have defined the cabled link states {\em without} the use of projectors. Although it might 
seem natural a priori to include projectors in their definition by considering $\{\Ib^m v;\,v\in V_{N\times m}^d\}$ instead of $V_{N,m}^d$, 
and similarly on the cylinder, it is generally not appropriate in scenarios with $\lambda = \lambda_{p,p'}$ and  $m\ge p'$. 
However, if one makes specific assumptions about $\lambda_{p,p'}$ (or $\beta$) to ensure the existence of the projector $P_m$, 
one can of course choose to work with cabled link states defined using these projectors. This is indeed the situation in~\cite{PRT2013},
where the main focus is on the $(2,2)$-fused case and $\beta$ is assumed non-vanishing such that $P_2$ exists.

%%%%%%%%%%%%%%%%%%%%%%%%%
\paragraph{Braid transfer matrices}
%%%%%%%%%%%%%%%%%%%%%%%%%

Using arguments as in~\cite{AMDYSA2011} and \cite{AMDYSA2013}, one can show that the braid transfer tangles 
$\Fb^{m,n}$ and $\Fbt^{m,n}$\! are in the center of 
$FTL_{N,m}(\beta)$ and $F\mathcal{E}PTL_{N,m}(\alpha,\beta)$, respectively. 
They furthermore act as scalar multiples of the identity on $V_{N,m}^d$ and $\tilde V_{N,m}^d$, respectively, that is 
\be   
 \rho_{N,m}^d(\Fb^{m,n})=\mathbb f_n^d\, I,\qquad \tilde\rho_{N,m}^d(\Fbt^{m,n})=\tilde  {\mathbb f}^{d}_n\, I 
\label{eq:Ff}
\ee
where $I$ is the identity matrix of dimension dim$V_{N,m}^d$ or dim$\tilde V_{N,m}^d$, respectively, and the constants 
$\mathbb{f}^{d}_n$ and $\tilde{\mathbb{f}}^{d}_n$ are real, independent of $N$ and $m$ and given by
\be
\mathbb f^{d}_n = (-1)^{nd}\, \frac{\sin\big((n+1)(d+1)\lambda\big)}{\sin\big((d+1)\lambda\big)}, \qquad 
 \tilde {\mathbb f}^{d}_n = \frac{\sin\big((n+1)\tfrac d 2 (\pi - \lambda)\big)}{\sin\big(\tfrac d 2 (\pi - \lambda)\big)}.
\label{eq:fvalues}
\ee

%%%%%%%%%%%%%%%%%%%%%%%%%%%%%%%
\section{Fusion hierarchies and $\boldsymbol T$- and $\boldsymbol Y$-systems}
\label{Sec:FTY}
%%%%%%%%%%%%%%%%%%%%%%%%%%%%%%%%

In this section, we show that for fixed $m$ and $\lambda$, the commuting fused transfer tangles \eqref{eq:transfermatrix2} with different 
vertical fusion index $n$ are not all independent. 
Both on the strip and cylinder in Sections~\ref{Sec:Fusion} and~\ref{Sec:FusionCylinder}, we present a set of relations, the 
{\em fusion hierarchy}, relating transfer tangles with different vertical stacks of face operators. 
In Sections~\ref{Sec:Tsystem} and~\ref{Sec:Ysystem}, we then show how the fusion hierarchy in each case
translates into $T$- and $Y$-systems for the transfer tangles. 

The proofs of Propositions~\ref{prop:FusHierD} and~\ref{prop:TmHier}
are provided in Appendices~\ref{app:FusHierStrip} and~\ref{app:FusHierCylinder}, respectively. There, the 
diagrammatic manipulations in the planar algebra are made for $\lambda$ generic.
However, it is recalled that the barred transfer tangles $\Dbb^{m,n}(u)$ and $\Tbb^{m,n}(u)$ are well defined for all 
$\lambda\in\pi\big(\mathbb{R}\!\setminus\!\mathbb{Z}\big)$ (see Sections~\ref{sec:Dcrit} and~\ref{Sec:CabledLinkStates}).
The results presented in the five Propositions~\ref{prop:FusHierD}--\ref{prop:Y} for the transfer tangles
$\Db^{m,n}(u)$ and $\Tb^{m,n}(u)$ are therefore valid for the barred transfer tangles for $\lambda$ generic and fractional alike.
The results are likewise valid for the transfer tangles themselves for all fractional $\lambda$ for which $P_m$ exists.
In all cases, the results are valid for the corresponding transfer matrices 
in the representations discussed in Section \ref{Sec:CabledLinkStates}.

%%%%%%%%%%%%%%%%%%%%%%%%%
\subsection{Fusion hierarchy on the strip}
\label{Sec:Fusion}
%%%%%%%%%%%%%%%%%%%%%%%%%

\begin{Proposition} On the strip, the fusion hierarchy for $m,n\in\mathbb{N}$ is given by
\bea
\Db^{m,n}_0 \Db^{m,1}_n &\!\!\!=\!\!\!& \frac{s_{n}(2u-\mu)s_{2n-1}(2u-\mu)}{s_{n-1}(2u-\mu) s_{2n}(2u-\mu)}\,  \Db^{m,n+1}_0   \nn
  &&\hspace{2cm}+\,\frac{s_{n-2}(2u-\mu) s_{2n+1}(2u-\mu)}{s_{n-1}(2u-\mu) s_{2n}(2u-\mu)}
   \big[q^m(u_n)q^m(\mu-u_{n-1})\big]^{\!N}\!\Db^{m,n-1}_{0} 
\label{eq:DmHier}
\eea
where $q^m(u)$ is defined in \eqref{eq:qkm}. 
\label{prop:FusHierD}
\end{Proposition}
In terms of the renormalised transfer tangles $\Dbh^{m,n}_k$ defined in (\ref{eq:Dhat}),
equation \eqref{eq:DmHier} can be expressed as
\be
 \Dbh^{m,n}_0 {\Dbh}^{m,1}_n =  \Dbh^{m,n+1}_0 + f^m_n f^m_{n-2}\, \Dbh^{m,n-1}_0 
\label{eq:simplerH}
\ee
where $f_k^m$ is defined in (\ref{fkm}) and it is noted that
\be
 \Dbh_k^{m,n}=\Dbh_0^{m,n}(u+k\lambda),\qquad f_k^m=f_0^m(u+k\lambda).
\ee
Equivalently, we have the determinant expression
\be
\Dbh^{m,n+1}_0 = \left| \begin{matrix}
\Dbh^{m,1}_0 & f^m_{-1} & 0 & 0 & 0 \\
f^m_1 & \Dbh^{m,1}_1 & f^m_{0} & 0 & 0 \\
0 & f^m_2 & \ddots & \ddots & 0 \\
0 & 0 & \ddots & \Dbh^{m,1}_{n-1} & f^m_{n-2} \\
0 & 0 & 0 & f^m_n & \Dbh^{m,1}_{n}
\end{matrix}\right|
\label{Ddet}
\ee
where non-tangle entries are understood to be multiplied by $\Ib^m$. 
Likewise, the braid transfer tangles satisfy 
\be
\Fb^{m,n+1} = \Fb^{m,n}\Fb^{m,1} - \Fb^{m,n-1}
= U_{n+1}\big(\tfrac12 \Fb^{m,1}\big) = 
\left|\begin{matrix}
\Fb^{m,1} & 1  & 0 & 0 & 0 \\
1 & \Fb^{m,1} & 1 & 0 & 0 \\
0 & 1 & \ddots & \ddots & 0 \\
0 & 0 & \ddots & \Fb^{m,1} & 1 \\
0 & 0 & 0 & 1 & \Fb^{m,1}
\end{matrix}\right|
\label{eq:Fdet}
\ee
where $U_{n}(x)$ are Chebyshev polynomials of the second kind, and the matrix on the righthand side has dimension $n+1$. 
This result for the braid transfer tangles can be shown either by the diagrammatic arguments underlying 
Proposition~\ref{prop:FusHierD} in Appendix~\ref{app:FusHierStrip}, or by taking the 
appropriate braid limit of \eqref{eq:DmHier} and \eqref{Ddet}.

%%%%%%%%%%%%%%%%%%%%%%%%%%%%%%
\subsection{Fusion hierarchy on the cylinder}
\label{Sec:FusionCylinder}
%%%%%%%%%%%%%%%%%%%%%%%%%%%%%%

\begin{Proposition}
On the cylinder, the fusion hierarchy for $m,n\in\mathbb{N}$ is given by
\be
\Tb_0^{m,n} \Tb^{m,1}_n = \Tb_0^{m,n+1} + h^m_nh^m_{n-2}\,\Tb_0^{m,n-1},
\label{eq:TmHier}
\ee
where $h^m_k$ is defined in (\ref{fkm}).
\label{prop:TmHier}
\end{Proposition}
The general fusion hierarchy can be rewritten in the determinant form
\be
\Tb^{m,n+1}_0 = \left|\begin{matrix}
\Tb^{m,1}_0 & h^m_{-1}  & 0 & 0 & 0 \\
h^m_1 & \Tb^{m,1}_1 & h^m_0 & 0 & 0 \\
0 & h^m_2 & \ddots & \ddots & 0 \\
0 & 0 & \ddots & \Tb^{m,1}_{n-1} & h^m_{n-2} \\
0 & 0 & 0 & h^m_n & \Tb^{m,1}_{n}
\end{matrix}\right| .
\label{Tdet}
\ee
The braid transfer tangles on the cylinder form a hierarchy identical to the one on the strip (\ref{eq:Fdet}), namely
\be
\Fbt^{m,n+1} = \Fbt^{m,n}\Fbt^{m,1} - \Fbt^{m,n-1}
= U_{n+1}(\tfrac12 \Fbt^{m,1}) = 
\left|\begin{matrix} 
\Fbt^{m,1} & 1  & 0 & 0 & 0 \\
1 & \Fbt^{m,1} & 1 & 0 & 0 \\
0 & 1 & \ddots & \ddots & 0 \\
0 & 0 & \ddots & \Fbt^{m,1} & 1 \\
0 & 0 & 0 & 1 & \Fbt^{m,1}
\end{matrix}\right|.
\label{eq:Ftdet}
\ee

%%%%%%%%%%%%%%%%%%%%%%%%%%%%%
\subsection[$T$-systems]{$\boldsymbol T$-systems}
\label{Sec:Tsystem}
%%%%%%%%%%%%%%%%%%%%%%%%%%%%%

From the fusion hierarchy \eqref{eq:simplerH}, we derive the $T$-system on the strip.
This is the content of the following proposition.
\begin{Proposition} 
For $m,n\in\mathbb{N}$, the $T$-system on the strip is given by
\be
 \Dbh^{m,n}_0 \Dbh^{m,n}_1 = \Dbh^{m,n+1}_0 \Dbh^{m,n-1}_1 + \nu_0^{(n)} \Ib^m,
\label{eq:TsysD}
\ee
where $\nu_0^{(n)}$\! is defined in (\ref{nu}).
\label{prop:TsysD}
\end{Proposition}
\noindent{\scshape Proof:} 
For $n=1$, equation \eqref{eq:TsysD} is trivially true as it coincides with \eqref{eq:simplerH}. 
Following arguments given in~\cite{KlumperPearce1992}, we prove \eqref{eq:TsysD} recursively.
First, we use the commutativity of the transfer tangles and equation \eqref{eq:simplerH} to obtain
\begin{align}
 \Dbh^{m,n}_0 \big(\Dbh^{m,n-1}_1\Dbh^{m,1}_n\big) &= \big(\Dbh^{m,n}_0 \Dbh^{m,1}_n\big)\Dbh^{m,n-1}_1, \nonumber \\
 \Dbh^{m,n}_0 \big(\Dbh^{m,n}_1+ f^m_n f^m_{n-2}\,\Dbh^{m,n-2}_1\big) 
 &=\big(\Dbh^{m,n+1}_0+ f^m_n f^m_{n-2}\,\Dbh^{m,n-1}_0\big)\Dbh^{m,n-1}_1.
\end{align}
Rearranging terms and assuming equation \eqref{eq:TsysD} for $n-1$ subsequently gives
\begin{align}
 \Dbh^{m,n}_0 \Dbh^{m,n}_1 &= \Dbh^{m,n+1}_0 \Dbh^{m,n-1}_1 + f^m_n f^m_{n-2} \big(\Dbh^{m,n-1}_0\Dbh^{m,n-1}_1 
  - \Dbh^{m,n}_0\Dbh^{m,n-2}_1\big) \nonumber\\
& = \Dbh^{m,n+1}_0 \Dbh^{m,n-1}_1 + f^m_n f^m_{n-2}\, \nu_0^{(n-1)} \Ib^m
  =  \Dbh^{m,n+1}_0 \Dbh^{m,n-1}_1 +\nu_0^{(n)} \Ib^m \end{align}
as required. 
\hfill $\square$
\medskip

\noindent Taking the appropriate braid limit of the $T$-system in Proposition~\ref{prop:TsysD} yields the following $T$-system for
the braid transfer tangles on the strip,
\be
 \Fb^{m,n}\Fb^{m,n}=\Fb^{m,n+1}\Fb^{m,n-1}+\Ib^m.
\label{TsysF}
\ee

From the fusion hierarchy in Proposition~\ref{prop:TmHier}, we derive the $T$-system on the cylinder.
This is the content of the following proposition.
\begin{Proposition} 
For $m,n\in\mathbb{N}$, the $T$-system on the cylinder is given by
\be
\Tb^{m,n}_0 \Tb^{m,n}_1 = \Tb^{m,n+1}_0 \Tb^{m,n-1}_1 + \tilde{\nu}_0^{(n)} \Ib^m,
\label{eq:TsysT}
\ee
where $\tilde{\nu}_0^{(n)}$\! is defined in (\ref{nu}).
\label{prop:TsysT}
\end{Proposition}
\noindent{\scshape Proof:}
The proof is identical to the proof of Proposition~\ref{prop:TsysD}.
\hfill $\square$
\medskip

\noindent The $T$-system for the braid transfer tangles on the cylinder is obtained by taking the appropriate braid limit of the 
$T$-system in Proposition~\ref{prop:TsysT} and is given by
\be
 \Fbt^{m,n}\Fbt^{m,n}=\Fbt^{m,n+1}\Fbt^{m,n-1}+\Ib^m.
\ee
This is seen to be identical to the $T$-system (\ref{TsysF}) for the braid transfer tangles on the strip.

%%%%%%%%%%%%%%%%%%%%%%%%%%%%%%
\subsection[$Y$-systems]{$\boldsymbol Y$-systems} 
\label{Sec:Ysystem}
%%%%%%%%%%%%%%%%%%%%%%%%%%%%%%

For $m\in\mathbb{N}$ and $n\in\mathbb{N}_0$, we define 
\be
 \db^{m,n}_k:= \frac{\Dbh^{m,n-1}_{k+1}\Dbh^{m,n+1}_k}{\nu_k^{(n)}},\qquad
 \tb^{m,n}_k:= \frac{\Tb^{m,n-1}_{k+1}\Tb^{m,n+1}_k}{\tilde{\nu}_k^{(n)}},\qquad k\in\mathbb{Z}.
\label{dt}
\ee
Since $\Dbh_{k+1}^{m,-1}\equiv0$ and $\Tb_{k+1}^{m,-1}\equiv0$, it follows that $\db_k^{m,0}\equiv0$ and $\tb_k^{m,0}\equiv0$.
As demonstrated in the following proposition, the infinite $Y$-system generated by (\ref{dt})
is {\em universal} in the sense that it is the same on the strip as on the cylinder.
\begin{Proposition}
For $m,n\in\mathbb{N}$, the $Y$-system for $\yb\in\{\db,\,\tb\}$ is universally given by 
\be
 \yb^{m,n}_0 \yb^{m,n}_1=\big(\Ib^m+ \yb^{m,n-1}_1\big)\big(\Ib^m+ \yb^{m,n+1}_0\big).
\ee
\label{prop:Y}
\end{Proposition}
\noindent{\scshape Proof:} 
For $\yb=\db$, we have
\begin{align}
 \db^{m,n}_0 \db^{m,n}_1 
 &=\frac{\big(\Dbh^{m,n-1}_1\Dbh^{m,n-1}_2\big)\big(\Dbh^{m,n+1}_0\Dbh^{m,n+1}_1\big)}{\nu_0^{(n)}\nu_1^{(n)}} \nonumber\\
&=\frac{\big(\nu_1^{(n-1)} \Ib^m+ \Dbh^{m,n-2}_2\Dbh^{m,n}_1\big)
  \big(\nu_0^{(n+1)} \Ib^m+\Dbh^{m,n}_1\Dbh^{m,n+2}_0\big)}{\nu_0^{(n)}\nu_1^{(n)}} \nonumber\\
&=\Big(\Ib^m+ \frac{\Dbh^{m,n-2}_2\Dbh^{m,n}_1}{\nu_1^{(n-1)}}\Big)
 \Big(\Ib^m+ \frac{\Dbh^{m,n}_1\Dbh^{m,n+2}_0}{\nu_0^{(n+1)}}\Big) \nonumber\\
&=\big(\Ib^m+ \db^{m,n-1}_1\big)\big(\Ib^m+ \db^{m,n+1}_0\big)
\end{align}
where the third equality follows from
\be 
 \nu_1^{(n-1)}\nu_0^{(n+1)} = \nu_0^{(n)}\nu_1^{(n)}.
\ee
The proof for $\yb=\tb$ is identical to the proof for $\yb=\db$.
\hfill $\square$
\medskip

\noindent It is noted that a shift in the spectral parameter $u$ by a multiple of the crossing parameter $\lambda$ readily yields
\be
 \yb^{m,n}_k \yb^{m,n}_{k+1}=\big(\Ib^m+ \yb^{m,n-1}_{k+1}\big)\big(\Ib^m+ \yb^{m,n+1}_k\big).
\ee
The corresponding $Y$-system for the braid transfer tangles is given by
\be
 \yb^{m,n}\yb^{m,n}=\big(\Ib^m+\yb^{m,n-1}\big)\big(\Ib^m+\yb^{m,n+1}\big),\qquad \yb\in\{\fb,\,\fbt\}
\label{yf}
\ee
where
\be
 \fb^{m,n}\!:=\Fb^{m,n-1}\Fb^{m,n+1},\qquad \fbt^{\,m,n}\!:=\Fbt^{m,n-1}\Fbt^{m,n+1}
\ee
with $\fb^{m,0}\equiv0$ and $\fbt^{\,m,0}\equiv0$ following from $\Fb^{m,-1}\equiv0$ and $\Fbt^{m,-1}\equiv0$.

%%%%%%%%
\section{Functional relations in logarithmic minimal models}
\label{Sec:LMM}
%%%%%%%%

In this section, we consider fractional $\lambda = \lambda_{p,p'}$ with $1\leq p<p'$. The corresponding TL loop model 
($m=1$) is referred to as the {\em logarithmic minimal model}
${\cal LM}(p,p')$ and, as argued in~\cite{PRZ0607} and subsequent works, gives rise to logarithmic conformal field theories in the 
continuum scaling limit. Here we establish that the transfer tangles $\Db(u)$ and $\Tb(u)$ satisfy certain functional relations of 
polynomial degree $p'$. More generally, we find that the transfer tangles $\Db^{m,1}(u)$ and $\Tb^{m,1}(u)$ satisfy such relations for all 
$m\in\mathbb{N}$ for which $P_m$ exists, 
and we emphasise that the derivation of these relations is performed in the planar algebra. 
If $P_m$ does not exist, the relations are still valid provided the transfer tangles are replaced by their barred counterparts, 
see Section~\ref{Sec:CabledLinkStates}. 
For ease of presentation, however, the various results are given in terms of the transfer tangles themselves. 
The functional relations we obtain generalise the inversion identities (of polynomial degree $p'=2$) found
in critical dense polymers ${\cal LM}(1,2)$~\cite{PR2007,PRV0910}.

We also discuss how the $Y$-systems on the strip and cylinder close on {\em finite} sets of generators.
The universality of the $Y$-system, as presented in Proposition~\ref{prop:Y}, is broken by the closure mechanism.
The finite systems on the two geometries can nevertheless be described in a uniform way as outlined in 
Corollary~\ref{cor:Yfolding} and Theorem~\ref{prop:Yfinite} below.

%%%%%%%%%%%%%%%%%%%%%%%%%%%%%%%%%%%%%%%
\subsection{Closure of the fusion hierarchy on the strip}
\label{Sec:ClosureStrip}
%%%%%%%%%%%%%%%%%%%%%%%%%%%%%%%%%%%%%%%

First, we establish that the fusion hierarchy on the strip closes for fractional $\lambda$.
\noindent 
\begin{Proposition} 
For $\lambda = \lambda_{p,p'}$ and $m\in\mathbb{N}$, the fusion hierarchy on the strip closes as
\be
 \Db^{m,p'}_0 =\big[q^m(u_0)q^m(\mu-u_{-1})\big]^N\Db^{m,p'-2}_1+ 2\, (-1)^{p'-p} 
  \Big(\prod_{j=0}^{p'-1}s_j(u)s_{j+1}(u-\mu) \Big)^{\!Nm}\Ib^m.
\label{eq:closure}
\ee
\label{prop:closureD}
\end{Proposition}
A proof of this proposition is given in Appendix~\ref{app:ClosureStrip}. 

In terms of the renormalised transfer tangles $\Dbh$ defined in (\ref{eq:Dhat}), the fusion closure in
equation \eqref{eq:closure} can be written as
\be
 \Dbh^{m,p'}_0 =f^m_0 f^m_{p'-2}\, \Dbh^{m,p'-2}_1+  2a\,\Ib^m
\label{eq:simplerclosure}
\ee
where $a$ is defined in (\ref{aa}) and
where it is noted that
\be
 \Dbh_{k+p'}^{m,n}=\Dbh_k^{m,n},\qquad f_{k+p'}^m=f_k^m,\qquad \nu_{k+p'}^{(n)}=\nu_k^{(n)},\qquad
   \nu_k^{(p')}=a^2,\qquad k\in\mathbb{Z}.
\ee
Combined with the fusion hierarchy derived in Section~\ref{Sec:Fusion}, equation \eqref{eq:simplerclosure} implies the 
{\em functional relations} given in the following theorem. They are expressed in terms of the double-row transfer tangles
for which 
$\Dbh^{m,1}_k=s_{2k}(2u\!-\!\mu)\Db^{m,1}(u+k\lambda)$.
\begin{Theoreme}
For $\lambda = \lambda_{p,p'}$ and $m\in\mathbb{N}$, the double-row transfer tangle on the strip satisfies
the functional relation
\be
p'>2:\ \ 
\left|\begin{matrix}
\Dbh^{m,1}_0 & f^m_{-1} & 0 & 0 & (-1)^{p'}f^m_0 \\[0.12cm]
f^m_1 & \Dbh^{m,1}_1 & f^m_{0} & 0 & 0 \\
0 & f^m_2 & \ddots & \ddots & 0 \\
0 & 0 & \ddots & \Dbh^{m,1}_{p'-2} & f^m_{p'-3} \\[0.12cm]
(-1)^{p'}f^m_{p'-2} & 0 & 0 & f^m_{p'-1} & \Dbh^{m,1}_{p'-1} 
\end{matrix}\right| = 0,\qquad
p'=2:\ \
\left|\begin{matrix}
\Dbh^{m,1}_0 & f^m_0 + f^m_1 \\[.2cm]
f^m_0 + f^m_1 & \Dbh^{m,1}_1
\end{matrix}\right| = 0,
\label{eq:funcdet}
\ee
where non-tangle entries are understood to be multiplied by $\Ib^m$. 
\label{theorem:DFR}
\end{Theoreme}
The functional relations in Theorem~\ref{theorem:DFR}, along with the similar ones on the cylinder (\ref{Tdetclosure}), are among 
the main results of this paper. We stress that they are valid in the planar algebra and therefore in any representation,
for example those described in Section~\ref{Sec:CabledLinkStates}.

For $p'=2$, the functional relation (\ref{eq:funcdet}) is an {\em inversion identity}. It was investigated in~\cite{PR2007} for 
$p=1$, $m=1$ and $\mu = \lambda$ corresponding to critical dense polymers ${\cal LM}(1,2)$. 
For $m=1$ and $\mu \in \mathbb C$, we find
\be
 \Db^{1,1}(u)\Db^{1,1}(u+\tfrac{\pi}{2}) 
 = \left(\displaystyle{\frac{[s_0(u)s_1(u\!-\!\mu)]^N \!- [s_1(u)s_2(u\!-\!\mu)]^N}{s_0(2u\!-\!\mu)}}\right)^{\!2} \!\Ib.
\ee
For $\lambda=\lambda_{1,2}$ and $m>1$, the tangle $\Ib^m$ does not exist in general. 
An inversion identity nevertheless exists for every $m$, but only for the corresponding barred transfer tangle \eqref{eq:Dbar}. 
Recalling from \eqref{eq:Ibb} that $\Ibb^{\,m} = \Ib$, this inversion relation reads
\be
 \Dbb^{m,1}(u)\Dbb^{m,1}(u+\tfrac{\pi}{2})
 =\left\{\!\!\begin{array}{ll}
  \gamma(u)\gamma(u+\tfrac{\pi}{2})\left(\displaystyle{\frac{[s_0(u)s_1(u\!-\!\mu)]^N \!- [s_1(u)s_2(u\!-\!\mu)]^N}{s_0(2u\!-\!\mu)}}\right)^{\!2}
     \!\Ib ,\ & m\;\mathrm{odd,}
  \\[.30cm]
  0, & m\;\mathrm{even,} 
 \end{array}\right.
\label{DbDb}
\ee
where
\be
 \gamma(u)=s_1(2u\!-\!\mu)\Big(s_0(u)s_1(u)s_1(u\!-\!\mu)s_2(u\!-\!\mu)\Big)^{N(m-1)/2}.
\ee

In the case of critical percolation ${\cal LM}(2,3)$, for which $m=1$ and $\mu=\lambda$, the functional relation for the double-row 
transfer tangle $\Db(u)=\Db^{1,1}(u)$\! on the strip can be written as
\be
 \mathfrak{D}(u)\mathfrak{D}(u+\tfrac{\pi}{3})\mathfrak{D}(u+\tfrac{2\pi}{3})=\mathfrak{D}(u)+\mathfrak{D}(u+\tfrac{\pi}{3})
  +\mathfrak{D}(u+\tfrac{2\pi}{3})-2\, \Ib,\qquad
  \mathfrak{D}(u):=\frac{\Db(u)}{[s_1(u)]^{2N}}\;.
\ee

The closure of the fusion hierarchy, as described in Proposition~\ref{prop:closureD}, can be viewed as a {\em folding} of 
$\Dbh^{m,n}_0$ for $n=p'$ onto a subset of $\{\Dbh^{m,j}_k;\;j,k= 0, \dots, p'-1\}$. 
A similar folding property holds for general $n\in\mathbb{N}$ as described by the following proposition.
\begin{Proposition}
Let $n = y p' + j$ with $y \in \mathbb N_0$ and $j\in \{0, \dots, p'-1\}$. Then
\be
 \Dbh_0^{m,n} = (y+1) a^y  \Dbh_0^{m,j} + y a^{y-1} \nu_{-1}^{(j+1)}\Dbh_{j+1}^{m,p'-j-2}.
\label{Dfold}
\ee
\label{prop:Dfolding}
\end{Proposition}
\noindent{\scshape Proof:}
This is trivially true for $n=1,\ldots,p'-1$ and it reduces to \eqref{eq:simplerclosure} for $n=p'$.
For $n>p'$, it follows recursively in $n$ from the relations
\be
 \Dbh_k^{m,n} \Dbh^{m,1}_{n+k} - f^m_{n+k}f^m_{n+k-2}\,\Dbh_k^{m,n-1} =
 \Dbh_k^{m,n+1}=\Dbh_k^{m,1}\Dbh_{k+1}^{m,n} - f^m_{k+1}f^m_{k-1}\, \Dbh_{k+2}^{m,n-1}.
\ee
The second of these relations is obtained by summing over entries of the first column and row in an expansion of the determinant 
expression in \eqref{Ddet}, and by adding a shift in the spectral parameter. 
\hfill $\square$
\medskip

\noindent For $p'=2$, it readily follows from Proposition~\ref{prop:Dfolding} that
\be
 \Dbh_0^{m,n}=\begin{cases}\! (y+1)a^y\Dbh_0^{m,1},\ &\  n=2y+1,\\[.15cm]
  \!\big(yf_0^m+(y+1)f_1^m\big)f_0^ma^{y-1}\Ib^m, &\ n=2y.  \end{cases}
\label{DDI}
\ee
As already indicated at the beginning of Section~\ref{Sec:LMM}, for $m>1$, we stress that \eqref{DDI} only makes sense in general 
with $\hat \Db$ replaced by the barred and renormalised transfer tangle $\hat {\bar \Db}$ and with $\Ib^m$ replaced by $\Ib$.
\begin{Corollaire}
Let $y\in\mathbb{N}_0$ and $j\in \{0, \dots, p'-1\}$. With the definition
\be
 \Dbm_\pm^{(y,j)}:= a\, \Dbh^{m,yp'+j}_0 \pm \nu_{-1}^{(j+1)} \Dbh^{m,yp'+p'-j-2}_{j+1},
\ee
the general folding can be expressed as
\be
 \Dbm_\pm^{(y,j)} = (y+1\pm y)\,a^y \, \Dbm_\pm^{(0,j)}.
\ee
\end{Corollaire}

By taking the appropriate braid limit of \eqref{eq:closure}, one finds a closure relation for the braid transfer tangles,
\be
 \Fb^{m,p'} = \Fb^{m,p'-2} + 2\, (-1)^{p'}\sigma\Ib^m, \qquad \sigma:=(-1)^{(Nm+1)p}.
\label{FFIsigma}
\ee 
From (\ref{eq:Fdet}), we have $\Fb^{m,n}\!=U_n\big(\frac{1}{2}\Fb^{m,1}\big)$ which together with (\ref{FFIsigma}) implies that
\be
 T_{p'}\big(\tfrac{1}{2}\Fb^{m,1}\big)=\epsilon\Ib^m,\qquad \epsilon:=(-1)^{p'}\sigma
\ee
where $T_k(x)$ is the $k$-th Chebyshev polynomial of the first kind and we have used the relation
\be
 U_k(x)-U_{k-2}(x)=2\,T_k(x),\qquad k\in\mathbb{N}.
\label{UUT}
\ee
In determinant form, the closure reads
\be 
\left|\begin{matrix}
\Fb^{m,1} & 1 & 0 & 0 & \sigma \\
1 & \Fb^{m,1} & 1 & 0 & 0 \\
0 & 1 & \ddots & \ddots & 0 \\
0 & 0 & \ddots & \Fb^{m,1} & 1 \\[0.12cm]
\sigma & 0 & 0 & 1 & \Fb^{m,1}
\end{matrix}\right| = 0,
\label{detF}
\ee
where the dimension of the matrix is $p'$.
Because we can rewrite the determinant relation in (\ref{detF}) as
\be
\prod_{k=0}^{p'-1} \Big(\Fb^{m,1} - 2 \cos\tfrac{(2k+\hat\epsilon)\pi}{p'}\Ib^m \Big) = 0, \qquad 
  \hat\epsilon:= \frac{1-\epsilon}{2},
\ee
the eigenvalues of $\Fb^{m,1}$ in any representation (where $\Ib^m$ acts as the identity) are of the form 
$2 \cos\frac{(2k+\hat\epsilon)\pi}{p'}$ for some integers $k$. 
It also follows that the Jordan cells associated to these eigenvalues have rank at most $2$, and that the rank is $1$ if the eigenvalue 
$2 \cos\frac{(2k+\hat\epsilon)\pi}{p'}$ is $\pm 2$.
For $y\in\mathbb{N}_0$ and $j\in \{0, \dots, p'-1\}$, the general folding property of the set of braid transfer tangles on the strip 
is given by 
\be
 \Fb^{m,yp'+j}=U_y(\epsilon)\Fb^{m,j}+U_{y-1}(\epsilon)\Fb^{m,p'-j-2}
  =\epsilon^y(y+1)\Fb^{m,j}+\epsilon^{y-1} y\Fb^{m,p'-j-2}.
\ee

%%%%%%%%%%%%%%%%%%%%%%%%%%%%%%%
\subsection{Closure of the fusion hierarchy on the cylinder}
\label{Sec:ClosureCylinder}
%%%%%%%%%%%%%%%%%%%%%%%%%%%%%%%%

The fusion hierarchy on the cylinder closes for fractional $\lambda$.
\begin{Proposition} 
For $\lambda = \lambda_{p,p'}$ and $m\in\mathbb{N}$, the fusion hierarchy on the cylinder closes as
\be
 \Tb_0^{m,p'} =h^m_0h^m_{p'-2}\, \Tb_1^{m,p'-2}+2\tilde{a}\,\Jb^m,
\label{eq:Tclosurem}
\ee
where $\tilde{a}$ is defined in (\ref{aa}), and
where $\Jb^m$\! is a $u$-independent tangle.
\label{prop:ClosureTm}
\end{Proposition}
The proof of Proposition~\ref{prop:ClosureTm} for $m=1$ is provided in Appendix~\ref{app:ClosureCylinder} where an
explicit expression for the tangle $\Jb^1$\! is given in (\ref{Jdef}). 
Arguments outlining the proof for general $m$ follow Proposition~\ref{prop:closureTmapp}.
It is furthermore observed that for $\lambda=\lambda_{p,p'}$ and $k\in\mathbb{Z}$,
\be
 \Tb_{k+p'}^{m,n}=e^{2\ir n\theta}\Tb_k^{m,n},\qquad h_{k+p'}^m=e^{2\ir\theta}h_k^m,\qquad 
  \tilde{\nu}_{k+p'}^{(n)}=\tilde{\nu}_k^{(n)},\qquad
   \tilde{\nu}_k^{(p')}=e^{2\ir\theta}\tilde{a}^2
\label{TTpp}
\ee
where $\theta$ is defined in (\ref{aa}) and
where it is noted that $e^{2\ir\theta}\in\{-1,\,+1\}$.

The $Nm$-tangle $\Jb^m$ appearing in Proposition~\ref{prop:ClosureTm} can be written in terms of braid transfer tangles. 
This is done by considering the closure properties of the braid transfer tangles
and results in the following proposition where $T_k(x)$ is the $k$-th Chebyshev polynomial of the first kind.
\begin{Proposition} 
For $\lambda = \lambda_{p,p'}$ and $m\in\mathbb{N}$, the 
tangle 
$\Jb^m$\! appearing in (\ref{eq:Tclosurem}) is given by
\be
 \Jb^m= \tfrac{1}{2}\big(\Fbt^{m,p'}\!-\Fbt^{m,p'-2}\big)=T_{p'}\big(\tfrac{1}{2}\Fbt^{\,m,1}\big).
\label{eq:J}
\ee
\label{prop:J}
\end{Proposition}
\noindent{\scshape Proof:}
Applying the {\it braid limit evaluator}
\be 
 \mathcal B(f(u)) = \lim_{u \rightarrow\ir\infty} \bigg(
  \Big(\prod_{i=0}^{m-1}\prod_{j=0}^{p'-1}\frac{e^{\ir\frac{\pi - \lambda}2}}{s_{j-i}(u)}\Big)^{\!N}  f(u)\bigg)
\ee
to both sides of \eqref{eq:Tclosurem}, and using
\be
 \mathcal B(\Tb^{m,p'}_0) = \Fbt^{m,p'}, \qquad 
\mathcal B\big(h^m_0h^m_{p'-2}\, \Tb_1^{m,p'-2}\big) = \Fbt^{m,p'-2},
\qquad
 \mathcal B(\tilde{a}) =1,
\ee
yields the first equality in (\ref{eq:J}). The second equality follows from $\Fbt^{m,n}\!=U_n\big(\frac{1}{2}\Fbt^{m,1}\big)$, as given in
(\ref{eq:Ftdet}), and the relation (\ref{UUT}).
\hfill $\square$
\medskip

\noindent As on the strip, the fusion closure in Proposition~\ref{prop:ClosureTm} gives rise to a functional relation which can be written 
neatly in determinant form. On the cylinder, this expression involves the braid transfer tangle.
\begin{Theoreme}
For $\lambda = \lambda_{p,p'}$ and $m\in\mathbb{N}$, the single-row transfer tangle on the cylinder
satisfies the functional relation
\be
\left|
\begin{matrix}
\Tb^{m,1}_0 & h^m_{-1}  & 0 & 0 & h^m_0 \\
h^m_1 & \Tb^{m,1}_1 & h^m_0 & 0 & 0 \\
0 & h^m_2 & \ddots & \ddots & 0 \\
0 & 0 & \ddots & \Tb^{m,1}_{p'-2} & h^m_{p'-3} \\
h^m_{p'-2} & 0 & 0 & h^m_{p'-1} & \Tb^{m,1}_{p'-1}
\end{matrix}\right| 
=\tilde{a}
\left|\begin{matrix} 
\Fbt^{m,1} & 1  & 0 & 0 & e^{-\ir\theta} \\
1 & \Fbt^{m,1} & 1 & 0 & 0 \\
0 & 1 & \ddots & \ddots & 0 \\
0 & 0 & \ddots & \Fbt^{m,1} & 1 \\
e^{\ir\theta} & 0 & 0 & 1& \Fbt^{m,1}
\end{matrix} \right| ,\qquad
p'>2, 
\label{Tdetclosure}
\ee
\be
 \left|\begin{matrix}
 \Tb^{m,1}_0 & h^m_0 + h^m_{-1} \\[.1cm]
 h^m_0 + h^m_1 & \Tb^{m,1}_1
 \end{matrix}\right| 
 =\tilde{a}
 \left|\begin{matrix}
 \Fbt^{m,1} & 1+e^{-\ir \theta}\\[.1cm]
 1+e^{\ir \theta} & \Fbt^{m,1}
 \end{matrix}\right|, \qquad 
 p'=2,
\ee
where non-tangle entries are understood to be multiplied by $\Ib^m$. 
\label{theorem:TFR}
\end{Theoreme}

The case $(p,p')=(1,2)$, for which $\theta = \tfrac12{Nm\pi}$, was investigated in~\cite{PRV0910} for $m=1$
corresponding to critical dense polymers ${\cal LM}(1,2)$. 
For critical percolation ${\cal LM}(2,3)$, the functional relation
for the single-row transfer tangle $\Tb=\Tb^{1,1}$\! on the cylinder can be written as
\be
 \mathfrak{T}(u)\mathfrak{T}(u+\tfrac{\pi}{3})\mathfrak{T}(u+\tfrac{2\pi}{3})
  =\mathfrak{T}(u)+\mathfrak{T}(u+\tfrac{\pi}{3})+\mathfrak{T}(u+\tfrac{2\pi}{3})
   +\ir^N\big(\Fbt^3-3\Fbt\big),\qquad \mathfrak{T}(u):=\frac{\Tb(u)}{[s_1(u)]^N}
\ee
where it is recalled that $\Fbt=\Fbt^{1,1}$\!.

As in the strip case, the closure of the fusion hierarchy on the cylinder can be viewed as a folding procedure.
For general $n\in\mathbb{N}$, this is the content of the following proposition.
\begin{Proposition} 
Let $n = y p' + j$ with $y \in \mathbb N_0$ and $j\in \{0, \dots, p'-1\}$. Then
\be
 \Tb_0^{m,n}=e^{\ir \phi(y,j)}\,\tilde{a}^{y-1}\!\left(\tilde a\, U_y(\Jb^m)\,\Tb_0^{m,j}
  +e^{2\ir \theta} \tilde{\nu}_{-1}^{(j+1)}U_{y-1}(\Jb^m)\,\Tb_{j+1}^{m,p'-j-2}\right),
\label{eq:Tfolding}
\ee
where $U_0(\Jb^m)=\Ib^m$\! and
\be
  \phi(y,j):=\big((y-1)p'+2j\big)y\theta.
\ee
\label{prop:Tfolding}
\end{Proposition}
\noindent{\scshape Proof:} 
This is trivially true for $n=1,\ldots,p'-1$ and is the content of Proposition~\ref{prop:ClosureTm} for $n=p'$.
For $n>p'$, it follows recursively in $n$ from the relations
\be
  \Tb_k^{m,n} \Tb^{m,1}_{n+k} - h^m_{n+k}h^m_{n+k-2}\,\Tb_k^{m,n-1} 
  =\Tb_k^{m,n+1} =\Tb_k^{m,1}\Tb_{k+1}^{m,n} - h^m_{k+1}h^m_{k-1}\, \Tb_{k+2}^{m,n-1}, 
\label{eq:otherHT}
\ee
the second of which is obtained by summing over entries of the first column and row in an expansion of the determinant 
expression in \eqref{Tdet}. 
\hfill $\square$
\medskip

\noindent For $p'=2$, it readily follows from Proposition~\ref{prop:Tfolding} that
\be
 \Tb_0^{m,n}=\begin{cases}  (-1)^{Nmy}\,\tilde{a}^{y}\, U_y(\Jb^m)\,\Tb_0^{m,1},\ &\  n=2y+1,\\[.2cm]
  \tilde{a}^{y-1}\, h_0 \, 
  \Big(e^{\ir\theta}h_1 U_y(\Jb^m)+h_0 U_{y-1}(\Jb^m)\Big), &\ n=2y,  \end{cases}
\label{TTI}
\ee
where the comment following \eqref{DDI} about barred transfer tangles on the strip for $m>1$ applies here too, 
for $\Tb^{m,n}$ and $\Jb^{m}$.
It is also recalled that results like (\ref{TTI}) are valid for the corresponding matrix representations for all $m$.
\begin{Corollaire} 
Let $n = y p' + j$ with $y \in \mathbb N_0$ and $j\in \{0, \dots, p'-1\}$. Then
\be
\Fbt^{m,p'-1} \Tb_0^{m,n} = e^{\ir \phi(y,j)}\,\tilde{a}^{y-1}\left(\tilde a \,\Fbt^{m,yp'+p'-1}   
   \Tb_0^{m,j} +  e^{2\ir \theta} \tilde{\nu}_{-1}^{(j+1)} \Fbt^{m,yp'-1}\, \Tb_{j+1}^{m,p'-j-2}\right).
\label{eq:Tfolding2}
\ee
\label{cor:Tfolding}
\end{Corollaire}
\noindent{\scshape Proof:} 
This follows from multiplying both sides of (\ref{eq:Tfolding}) by $\Fbt^{m,p'-1}$\! and subsequently using
$\Fbt^{m,n}\!=U_n\big(\frac{1}{2}\Fbt^{m,1}\big)$ and $\Jb^m\!=T_{p'}\big(\tfrac{1}{2}\Fbt^{\,m,1}\big)$ 
combined with the Chebyshev relation
\be 
 U_{\ell k+k-1}(x)=U_{k-1}(x)\,U_\ell\big(T_k(x)\big),
 \qquad k\in\mathbb{N},\quad \ell\in\mathbb{N}_0.
\ee
\hfill $\square$
\begin{Corollaire}
Let $y\in\mathbb{N}_0$ and $j\in \{0, \dots, p'-1\}$. With the definition
\be
 \Tbm_\pm^{(y,j)}:= \tilde a\, e^{-\ir \phi(y,j)} \Tb^{m,yp'+j}_0 
  \pm e^{-\ir \phi(y+1,1)} \tilde{\nu}_{-1}^{(j+1)}\, \Tb^{m,yp'+p'-j-2}_{j+1},
\ee
the general folding can be written as
\be
 \Fbt^{m,p'-1} \Tbm_\pm^{(y,j)} = \tilde a^y \,\Big( \Fbt^{m,yp'+p'-1} \pm \Fbt^{m,yp'-1}\Big)\, \Tbm_\pm^{(0,j)}.
\ee
\end{Corollaire}
For $y\in\mathbb{N}_0$ and $j\in \{0, \dots, p'-1\}$, the folding of the set of braid transfer tangles on the cylinder is given by
\be
 \Fbt^{m,yp'+j}=U_y(\Jb^m)\Fbt^{m,j}+U_{y-1}(\Jb^m)\Fbt^{m,p'-j-2}.
\ee

%%%%%%%%%%%%%%%%%
\subsection[Closure of the $Y$-systems]{Closure of the $\boldsymbol Y$-systems} 
\label{Sec:TYrat}
%%%%%%%%%%%%%%%%%

For fractional $\lambda=\lambda_{p,p'}$, the $Y$-system generators (\ref{dt}) have the translational symmetry properties
\be
 \db_{k+p'}^{m,n}=\db_{k}^{m,n},\qquad
 \tb_{k+p'}^{m,n}=\tb_{k}^{m,n},\qquad k\in\mathbb{Z}.
\ee
\begin{Proposition}
For $\lambda = \lambda_{p,p'}$, $m\in\mathbb{N}$, $y\in\mathbb{N}_0$ and $j\in\{0,\ldots,p'-1\}$, the set of 
$Y$-system generators on the strip folds as
\be
 \db_0^{m,yp'+j}-\db_{j+1}^{m,yp'+p'-j-2}=(2y+1)\big(\db_0^{m,j}-\db_{j+1}^{m,p'-j-2}\big)
\label{dYfold}
\ee
where $\db_k^{m,-1}\equiv-\Ib^m$.
\label{prop:dYfolding}
\end{Proposition}
\noindent{\scshape Proof:} 
For $p'=2$, the relation follows from $\db_0^{m,2}=\db_1^{m,2}$, itself a consequence of (\ref{DDI}).
For $p'>2$, the folding rule (\ref{Dfold}) in Proposition~\ref{prop:Dfolding} is used to evaluate
the lefthand side of (\ref{dYfold}), and this is done separately in the four cases $j=0$, $1\leq j\leq p'-3$, $j=p'-2$ and $j=p'-1$.  
The expression on the righthand side subsequently follows from relations such as
\be
 \tilde{\nu}_{0}^{(j)}\tilde{\nu}_{j}^{(p'-j)}=a^2=\tilde{\nu}_{-1}^{(j+2)}\tilde{\nu}_{j+1}^{(p'-j-2)}.
\ee
\hfill$\Box$
\begin{Proposition}
For $\lambda = \lambda_{p,p'}$, $m\in\mathbb{N}$, $y\in\mathbb{N}_0$ and $j\in\{0,\ldots,p'-1\}$, the set of 
$Y$-system generators on the cylinder folds as
\be
 \tb_0^{m,yp'+j}-\tb_{j+1}^{m,yp'+p'-j-2}=U_{2y}(\Jb^m)
  \big(\tb_0^{m,j}-\tb_{j+1}^{m,p'-j-2}\big)
\label{eq:tfold}
\ee
where $\tb_k^{m,-1}\equiv-\Ib^m$.
\label{prop:tYfolding}
\end{Proposition}
\noindent{\scshape Proof:} 
The proof is similar to the one of Proposition~\ref{prop:dYfolding}, but is now based on the folding properties 
(\ref{TTI}) and (\ref{eq:Tfolding}), supplemented by the Chebyshev relation
\be
 U_{2y}(x)=\big[U_{y}(x)\big]^2-\big[U_{y-1}(x)\big]^2,\qquad y\in\mathbb{N}_0.
\ee
\hfill$\Box$
\begin{Corollaire}
For $\lambda = \lambda_{p,p'}$, $m\in\mathbb{N}$, $y\in\mathbb{N}_0$ and $j\in\{0,\ldots,p'-1\}$, the two sets of 
$Y$-system generators fold as
\be
 \yb_0^{m,yp'+j}-\yb_{j+1}^{m,yp'+p'-j-2}=U_{2y}(\Kb^m)
  \big(\yb_0^{m,j}-\yb_{j+1}^{m,p'-j-2}\big),\qquad
   \Kb^m=\begin{cases} \Ib^m,\ & \yb=\db, \\[.15cm] 
    \Jb^m,\ & \yb=\tb. \end{cases}
\label{eq:yfold}
\ee
\label{cor:Yfolding}
\end{Corollaire}
\noindent{\scshape Proof:} 
This follows readily from $U_{2y}(1)=2y+1$.
\hfill$\Box$
\begin{Proposition}
For $\lambda = \lambda_{p,p'}$ and $m\in\mathbb{N}$, the $Y$-systems close as 
\be
 \yb_0^{m,\lfloor\frac{3p'}{2}\rfloor}=\yb_{\lfloor\frac{p'+2}{2}\rfloor}^{m,\lfloor\frac{3p'-3}{2}\rfloor}
  +U_{2}(\Kb^m)\Big(\yb_0^{m,\lfloor\frac{p'}{2}\rfloor}-\yb_{\lfloor\frac{p'+2}{2}\rfloor}^{m,\lfloor\frac{p'-3}{2}\rfloor}\Big)
\label{y3p2}
\ee
where $\Kb^m$ is as in (\ref{eq:yfold}).
\label{prop:Ytruncation}
\end{Proposition}
\noindent{\scshape Proof:} 
The folding property (\ref{eq:yfold}) implies that the $Y$-systems close at the minimum value of $\max\{yp'+j,\,yp'+p'-j-2\}$,
for which $yp'+j\neq yp'+p'-j-2$, evaluated over $y\in\mathbb{N}$ and $j=\{0,\ldots,p'-1\}$.
\hfill$\Box$
\medskip

\noindent
For $p'$ even, the folding relation
\be
 \yb_0^{m,yp'+\frac{p'-2}{2}}-\yb_{p'/2}^{m,yp'+\frac{p'-2}{2}}=U_{2y}(\Kb^m)\Big(\yb_0^{m,\frac{p'-2}{2}}
   -\yb_{p'/2}^{m,\frac{p'-2}{2}}\Big)
\ee
relates a pair of generators at the same fusion level $n=yp'+\frac{p'}{2}-1$ and is therefore not suited to close the system.
For $p'=2$, for example, we simply obtain $\yb_0^{m,2}=\yb_1^{m,2}$\!.
Recalling that $U_2(x)=4x^2-1$, the closure of the $Y$-system for small $p'$ is given by
\begin{align}
 p'=2:&\qquad\yb_0^{m,3}=4(\Kb^m)^2\yb_0^{m,1}+4(\Kb^m)^2-\Ib^m,
 \\[.15cm]
 p'=3:&\qquad\yb_0^{m,4}=\yb_2^{m,3}+\big(4(\Kb^m)^2-\Ib^m\big)\yb_0^{m,1},
 \\[.15cm]
 p'=4:&\qquad\yb_0^{m,6}=\yb_3^{m,4}+\big(4(\Kb^m)^2-\Ib^m\big)\yb_0^{m,2},
 \\[.15cm]
 p'=5:&\qquad\yb_0^{m,7}=\yb_3^{m,6}+\big(4(\Kb^m)^2-\Ib^m\big)\big(\yb_0^{m,2}-\yb_3^{m,1}\big).
\end{align}
As an immediate consequence of the propositions above, we obtain the following theorem.
\begin{Theoreme}
For $\lambda = \lambda_{p,p'}$ and $m\in\mathbb{N}$, the $Y$-systems on the strip and cylinder
are finite as they close on the sets
\be
 \big\{\db^{m,n};\,n=1,\ldots,\lfloor\tfrac{3p'-2}{2}\rfloor\big\},\qquad\big\{\tb^{m,n};\,n=1,\ldots,\lfloor\tfrac{3p'-2}{2}\rfloor\big\},
\ee
respectively, with relations given by
\be
 \yb^{m,n}_0 \yb^{m,n}_1=\big(\Ib^m+ \yb^{m,n-1}_1\big)\big(\Ib^m+ \yb^{m,n+1}_0\big),\qquad
  n=1,\ldots,\lfloor\tfrac{3p'-2}{2}\rfloor,
\ee
where $\yb_0^{m,\lfloor\frac{3p'}{2}\rfloor}$\! is given in (\ref{y3p2}) and where $\yb\in\{\db,\tb\}$.
\label{prop:Yfinite}
\end{Theoreme}

To obtain the corresponding finite $Y$-system for the braid transfer tangles, we first establish the folding property
\be
 \yb^{m,yp'+j}-\yb^{m,yp'+p'-j-2}=U_{2y}(\Kb^m)\big(\yb^{m,j}-\yb^{m,p'-j-2}\big),\qquad
  \Kb^m=\begin{cases} \Ib^m,\ & \yb=\fb, \\[.15cm] 
    \Jb^m,\ & \yb=\fbt, \end{cases}
    \label{eq:yfclosure}
\ee
using $U_{2y}(\epsilon)=U_{2y}(1)$. This folding relation implies a closure of the $Y$-system
(\ref{yf}) similar to the one in Proposition~\ref{prop:Ytruncation}. However, the simplicity of the braid transfer
tangles implies the stronger closure 
\be
 \yb^{m,p'}=2\,\yb^{m,p'-1}-\yb^{m,p'-2}+U_2(\Kb^m)-\Ib^m.
\label{ypp}
\ee
The finite $Y$-system for the braid transfer tangles is thus given by 
\be
 \yb^{m,n} \yb^{m,n}=\big(\Ib^m+ \yb^{m,n-1}\big)\big(\Ib^m+ \yb^{m,n+1}\big),\qquad  n=1,\ldots,p'-1, 
\ee
and the closure relation \eqref{ypp}.

%%%%%%%%%%%%%%%%%%%%%%%%%
\section{Comparison with rational models} 
\label{sec:ratcomp}
%%%%%%%%%%%%%%%%%%%%%%%%%

In this section, we compare our results for the logarithmic fusion hierarchies, $T$-systems and $Y$-systems to those obtained 
previously for rational models. To this end, we renormalise the $(m,n)$-fused face operator (\ref{eq:fusedface}) by defining
\be
\psset{unit=0.9}
\begin{pspicture}[shift=-0.4](0,0)(1,1)
\facegrid{(0,0)}{(1,1)}
\psarc[linewidth=0.025]{-}(0,0){0.16}{0}{90}
\rput(0.5,0.2){$\sim$}
\rput(0.5,.45){$u$}\rput(0.5,0.75){\tiny{$_{(m,n)}$}}
\end{pspicture}
\; :=\frac{1}{\eta^{m,n}(u)}\ \,
\begin{pspicture}[shift=-0.4](0,0)(1,1)
\facegrid{(0,0)}{(1,1)}
\psarc[linewidth=0.025]{-}(0,0){0.16}{0}{90}
\rput(0.5,.45){$u$}\rput(0.5,0.75){\tiny{$_{(m,n)}$}}
\end{pspicture}
\label{ren}
\ee
where
\be 
  \eta^{m,n}(u):=(-\ir)^{mn}\prod_{k=1}^{n-1}\prod_{j=1}^{m}s_{k-j}(u).
\label{eta}
\ee
Recalling the decomposition (\ref{Xmn}) in terms of generalised monoids, 
the renormalised decomposition coefficients are given by
\be
 \tilde\alpha_a^{m,n}(u):=\frac{\alpha_a^{m,n}(u)}{\eta^{m,n}(u)}
 =(-\ir)^{mn} (-1)^{(m+n)a}\Big(\prod_{j=1}^a\frac{s_{r-j+1}(0)}{s_j(0)}\Big)
  \Big(\prod_{j=0}^{a-1}s_{r-n-j}(-u)\Big)\Big(\prod_{j=1}^{m-a}s_{-j}(u)\Big).
\label{alphat}
\ee

In terms of the renormalised fused faces (\ref{ren}), we define renormalised fused transfer tangles by 
\begin{alignat}{2}
\Dbt^{m,n}(u):= 
\begin{pspicture}[shift=-1.6](-0.7,-0.7)(5.5,2)
\psarc[linewidth=4pt,linecolor=blue]{-}(0,1){0.5}{90}{-90}\psarc[linewidth=2pt,linecolor=white]{-}(0,1){0.5}{90}{-90}
\psarc[linewidth=4pt,linecolor=blue]{-}(5,1){0.5}{-90}{90}\psarc[linewidth=2pt,linecolor=white]{-}(5,1){0.5}{-90}{90}
\facegrid{(0,0)}{(5,2)}
\psarc[linewidth=0.025]{-}(0,0){0.16}{0}{90}
\psarc[linewidth=0.025]{-}(0,1){0.16}{0}{90}
\psarc[linewidth=0.025]{-}(1,0){0.16}{0}{90}
\psarc[linewidth=0.025]{-}(1,1){0.16}{0}{90}
\psarc[linewidth=0.025]{-}(4,0){0.16}{0}{90}
\psarc[linewidth=0.025]{-}(4,1){0.16}{0}{90}
\rput(0.5,0.75){\tiny{$_{(m,n)}$}}\rput(0.5,1.75){\tiny{$_{(m,n)}$}}
\rput(1.5,0.75){\tiny{$_{(m,n)}$}}\rput(1.5,1.75){\tiny{$_{(m,n)}$}}
\rput(4.5,0.75){\tiny{$_{(m,n)}$}}\rput(4.5,1.75){\tiny{$_{(m,n)}$}}
\rput(2.5,0.5){$\ldots$}
\rput(2.5,1.5){$\ldots$}
\rput(3.5,0.5){$\ldots$}
\rput(3.5,1.5){$\ldots$}
\rput(0.5,.45){$u$}
\rput(0.52,1.45){\scriptsize$\mu\!\!-\!\!u_{n\!-\!1}$}
\rput(1.5,.45){$u$}
\rput(1.52,1.45){\scriptsize$\mu\!\!-\!\!u_{n\!-\!1}$}
\rput(4.5,.45){$u$}
\rput(4.52,1.45){\scriptsize$\mu\!\!-\!\!u_{n\!-\!1}$}
\rput(2.5,-0.5){$\underbrace{\qquad \qquad \qquad \qquad \qquad \quad \qquad}_N$}
\rput(0.5,0.2){$\sim$}
\rput(1.5,0.2){$\sim$}
\rput(4.5,0.2){$\sim$}
\rput(0.5,1.2){$\sim$}
\rput(1.5,1.2){$\sim$}
\rput(4.5,1.2){$\sim$}
\end{pspicture} \ \,
&=\left(\frac{1}{\eta^{m,n}(u)\,\eta^{m,n}(\mu\!-\!u_{n-1})}\right)^{\!N}\!\!\Db^{m,n}(u) 
\nonumber 
\\[-0.8cm]
& = (-1)^{Nmn}\Big(\prod_{k=0}^{n-2}\frac{1}{\tilde f_k^m}\Big)\, \Db^{m,n}(u)
\end{alignat}  
and
\be 
\Tbt^{m,n} (u):= \
\begin{pspicture}[shift=-0.75](-0.2,-0.5)(5.2,1.2)
\psline[linewidth=4pt,linecolor=blue]{-}(0,0.5)(-0.2,0.5)\psline[linewidth=2pt,linecolor=white]{-}(0,0.5)(-0.2,0.5)
\psline[linewidth=4pt,linecolor=blue]{-}(5,0.5)(5.2,0.5)\psline[linewidth=2pt,linecolor=white]{-}(5,0.5)(5.2,0.5)
\facegrid{(0,0)}{(5,1)}
\psarc[linewidth=0.025]{-}(0,0){0.16}{0}{90}
\psarc[linewidth=0.025]{-}(1,0){0.16}{0}{90}
\psarc[linewidth=0.025]{-}(4,0){0.16}{0}{90}
\rput(2.5,0.5){$\ldots$}
\rput(3.5,0.5){$\ldots$}
\rput(0.5,.45){$u$}
\rput(1.5,.45){$u$}
\rput(4.5,.45){$u$}
\rput(0.5,0.75){\tiny{$_{(m,n)}$}}
\rput(1.5,0.75){\tiny{$_{(m,n)}$}}
\rput(4.5,0.75){\tiny{$_{(m,n)}$}}
\rput(2.5,-0.5){$\underbrace{\qquad \qquad \qquad \qquad \qquad \quad \qquad}_N$}
\rput(0.5,0.2){$\sim$}
\rput(1.5,0.2){$\sim$}
\rput(4.5,0.2){$\sim$}
\end{pspicture} \ \,   
=\left(\frac{1}{\eta^{m,n}(u)}\right)^{\!N}\!\Tb^{m,n}(u) = \ir^{Nm}\Big(\prod_{k=0}^{n-2}\frac{1}{h_k^m}\Big)\, \Tb^{m,n}(u).
\ee  
Here we have set 
\be
 \Dbt^{m,0}_0\equiv\tilde f^m_{-1}\Ib^m,\qquad\Dbt^{m,-1}_0\equiv0,\qquad \Tbt^{m,0}_0\equiv \tilde h_{-1}^m\Ib^m,\qquad
  \Tbt^{m,-1}_0\equiv0
\ee
and introduced
\be
 \tilde f_k^m = (-1)^{Nm}\Big(\prod_{j=0}^{m-1}s_{k-j}(u)s_{k+m-j}(u-\mu)\Big)^{\!N},
 \qquad
 \tilde h_k^m:=\ir^{Nm}h_k^m=\Big(\prod_{j=0}^{m-1}s_{k-j}(u)\Big)^{\!N}.
\ee 
These renormalised transfer tangles also form fusion hierarchies $T$-systems and $Y$-systems, readily obtained from the
similar structures satisfied by the original fused transfer tangles. Indeed, we find the fusion hierarchies
\be
 \Dbt^{m,n}_0\Dbt^{m,1}_n=\frac{s_{n-2}(2u\!-\!\mu)s_{2n+1}(2u\!-\!\mu)}{s_{n-1}(2u\!-\!\mu)s_{2n}(2u\!-\!\mu)}
  \,\tilde f_n^m\, \Dbt^{m,n-1}_0 + \frac{s_{n}(2u\!-\!\mu)s_{2n-1}(2u\!-\!\mu)}{s_{n-1}(2u\!-\!\mu)s_{2n}(2u\!-\!\mu)}
  \,\tilde f_{n-1}^m\, \Dbt^{m,n+1}_0
\label{fusratD}
\ee
and
\be 
\Tbt^{m,n}_0\Tbt^{m,1}_n =  \tilde h_n^m\, \Tbt^{m,n-1}_0 +  \tilde h_{n-1}^m\, \Tbt^{m,n+1}_0.
\label{fusratT}
\ee
The corresponding $T$-systems are given by 
\be
 \Dbt^{m,n}_0\Dbt^{m,n}_1 = \frac{s_{-1}(2u\!-\!\mu)s_{2n+1}(2u\!-\!\mu)}{s_{n-1}(2u\!-\!\mu)s_{n+1}(2u\!-\!\mu)}
  \, \tilde f^m_{-1}\tilde f^m_n \Ib^m +  \frac{[s_{n}(2u\!-\!\mu)]^2}{s_{n-1}(2u\!-\!\mu)s_{n+1}(2u\!-\!\mu)} 
   \Dbt^{m,n+1}_0\Dbt^{m,n-1}_1
\label{Drat}
\ee
and
\be
\Tbt^{m,n}_0 \Tbt^{m,n}_1 =  \tilde h_{-1}^m \tilde h_n^m \, \Ib^m +\Tbt^{m,n+1}_0 \Tbt^{m,n-1}_1,
\label{Trat}
\ee
while the $Y$-systems are the same as in Section~\ref{Sec:Ysystem}.
The closure relations for fractional $\lambda=\lambda_{p,p'}$ now read
\be
 \Dbt^{m,p'}_0 = \Dbt^{m,p'-2}_1 + 2(-1)^{p'-p} \tilde f^m_{-1} \Ib^m,\qquad
 \Tbt^{m,p'}_0 = \Tbt^{m,p'-2}_1 + 2e^{-i \theta} \tilde h^m_{-1} \Jb^m.
\label{Yrat}
\ee

The fusion hierarchies (\ref{fusratD})-(\ref{fusratT}) and $T$-systems (\ref{Drat})-(\ref{Trat}) coincide with those of the (rational) critical
$A$-$D$-$E$ models, as described in~\cite{universal}, for example. 
The closure relations (\ref{Yrat}), on the other hand, differ significantly
from the closure relations in the rational models. 
Indeed, in the rational models, the fusion hierarchies truncate at $n=p'-1$, in the sense that the fused transfer matrices with fusion index 
$n=p'-1$ are identically zero. Consequently, the corresponding $Y$-systems truncate at $n=p'-2$, so the way in 
which our $Y$-systems fold for fractional $\lambda=\lambda_{p,p'}$, as described in Section~\ref{Sec:TYrat}, is therefore also new.

%%%%%%%%%%%%%%
\section{Conclusion}
\label{Sec:Conclusion}
%%%%%%%%%%%%%%%

The (higher fusion level) logarithmic minimal models ${\cal LM}(P,P';n)$~\cite{PR1305,PRT2013} constitute a very large class 
of $su(2)$ theories describing the critical behaviour of statistical systems with nonlocal degrees of freedom such as generalised 
polymer systems and generalised percolation processes.  
In this paper, we have initiated the study of the integrable algebraic structures of the critical fused lattice models 
${\cal LM}(p,p')_{n\times n}$~\cite{PRT2013} realising the general logarithmic minimal models ${\cal LM}(P,P';n)$ 
described as coset CFTs in~\cite{PR1305}. Specifically, we have obtained the fusion hierarchies, $T$-systems and $Y$-systems for 
these models and demonstrated their finite closure both on a strip and on a cylinder. The $T$-systems and $Y$-systems are the key to 
the analytic calculation of non-universal statistical and universal conformal quantities, respectively.
The coset construction of the logarithmic minimal models brings these theories within standard schemes for constructing and classifying 
CFTs based on Lie algebras. The explicit manifestation of the $T$- and $Y$-systems further imposes standard integrable algebraic 
structures on the underlying Yang-Baxter integrable lattice models. 
The logarithmic minimal models are not rational and not unitary and 
the accompanying representation theory, with the profusion of reducible yet indecomposable structures, is clearly very intricate. 
Nevertheless, it is becoming increasingly clear, at least as far as the study of spectra is concerned, that these logarithmic theories are 
amenable to the standard approaches of mathematical physics as applied to rational CFTs.

The structure of the logarithmic $T$- and $Y$-systems we obtain closely mirrors the $su(2)$ $T$-systems and $Y$-systems of the 
(rational) critical RSOS $A$-$D$-$E$ models~\cite{KlumperPearce1992,BehrendPearce,universal}. 
Remarkably, we find that these functional equations hold for arbitrary coprime integers $p,p'$ and that the underlying structures are 
related to the Dynkin diagrams of the affine Lie algebras $A_{p'-1}^{(1)}$. Indeed, the determinantal structure of the polynomial 
functional equations of degree $p'$ is the same~\cite{functional} as for the CSOS 
models~\cite{PScyclicPRL,KYcyclic,PScyclic}. In contrast, the structure of the $T$- and $Y$-systems of rational unitary minimal models 
are related to Dynkin diagrams of classical Lie algebras. Perhaps it should not be too surprising to see affine Lie algebras enter in 
the logarithmic setting. After all, it has been argued~\cite{PR1010} that, with ${\cal W}$-extended symmetries on the strip or torus, 
certain aspects of the $n=1$ theories are in fact encoded by twisted affine coset graphs 
$A_{p,p'}^{(2)}=A_p^{(2)}\otimes A_{p'}^{(2)}/\mathbb{Z}_2$.

The next stage in this program is to solve the $Y$-system for the central charges and conformal weights of the logarithmic minimal 
models in terms of Rogers dilogarithms. Since this system of functional equations closes but does not truncate in the usual way, some 
new techniques will probably need to be developed. One hope is that, if ${\cal W}$-extended symmetry can be properly incorporated 
into the functional equations, then perhaps the associated $Y$-systems would exhibit the twisted affine coset structure and naturally 
truncate. In this regard, we stress that the only boundary condition on the strip considered here, namely, the vacuum boundary condition 
conjugate to the identity operator in the so-called Virasoro picture, in general breaks the ${\cal W}$-extended symmetry. 
It is clearly of interest to extend the derivation of the logarithmic $T$- and $Y$-systems on the strip to more general boundary conditions 
including ${\cal W}$-symmetric boundary conditions. 
Lastly, we note that, since the logarithmic minimal models are also exactly solvable off-criticality~\cite{PS1207}, it should also be 
possible to obtain both massless (critical) and massive (off-critical) thermodynamic Bethe ansatz 
equations for these theories with their concomitant implications for the associated relativistic two-particle scattering theories.

\subsection*{Acknowledgments}

AMD is supported by the National Sciences and Engineering Research Council of Canada
as a postdoctoral fellow. He is also grateful for support from the University of Queensland. 
JR is supported by the Australian Research Council under the Future Fellowship scheme, project number FT100100774. 
The authors thank Adam Ong, Yvan Saint-Aubin, Hubert Saleur and Elena Tartaglia for comments and discussions. 
AMD and JR also thank Jean-Bernard Zuber and the LPTHE at Universit\'e Pierre et Marie Curie,
where part of this work was done, for their kind hospitality.

%%%%%%%%%%%%
%
%      Appendices
%
%%%%%%%%%%%%

\bigskip

\bigskip

\noindent{\LARGE\bfseries Appendices}

\appendix

%%%%%%%%%%%%
\section{Decomposition of fused face operators}
\label{App:FusedFaces}
%%%%%%%%%%%%

Here we recall and prove Proposition~\ref{prop:GenMonoids}.
\\[.2cm]
\noindent{\scshape Proposition~\ref{prop:GenMonoids}}
{\em The decomposition of an $(m,n)$-fused face in terms of generalised monoids is given by
\be 
\psset{unit=0.6364cm}
\begin{pspicture}[shift=-0.85](0,-1)(2,1)
\pspolygon[fillstyle=solid,fillcolor=lightlightblue](0,0)(1,1)(2,0)(1,-1)(0,0)
\psarc[linewidth=0.025]{-}(1,-1){0.21}{45}{135}
\rput(1,0){$u$}\rput(1,0.35){\tiny{$_{(m,n)}$}}
\end{pspicture} 
\ = \sum_{a=0}^{r} \alpha_a^{m,n} \, X^{m,n}_a,\qquad\ r=\min(m,n)
\ee
where 
\be  
 \frac{\alpha_a^{m,n}}{\alpha_0^{m,n}} 
 = (-1)^{(m+n)a} \Big(\prod_{j=1}^a \frac{s_{r-j+1}(0)}{s_{j}(0)}\Big)
  \Big(\prod_{i=0}^{a-1}\frac{s_{n-r+i}(u)}{s_{m-i}(-u)}\Big),  \qquad \
 \alpha_0^{m,n} = \prod_{i=0}^{m-1}\prod_{j=0}^{n-1}s_{i-j+1}(-u).
\label{alphaamn-app}
\ee
}
\noindent  
\!\!{\scshape Proof:}
Because $X_0^{m,n}$ and $X_r^{m,n}$ can only be obtained from the $(m,n)$-fused face operator by respectively fixing every 
elementary face operator to\,
$\psset{unit=0.25cm}
\begin{pspicture}(0,-0.5)(2,0)
\pspolygon[fillstyle=solid,fillcolor=lightlightblue](0,0)(1,1)(2,0)(1,-1)(0,0)
\psarc[linewidth=1.5pt, linecolor=blue]{-}(0,0){0.7071}{-45}{45}
\psarc[linewidth=1.5pt, linecolor=blue]{-}(2,0){-0.7071}{-45}{45}
\end{pspicture}$
\,or\, 
$\psset{unit=0.25cm}
\begin{pspicture}(0,-0.5)(2,0)
\pspolygon[fillstyle=solid,fillcolor=lightlightblue](0,0)(1,1)(2,0)(1,-1)(0,0)
\psarc[linewidth=1.5pt, linecolor=blue]{-}(1,1){-0.7071}{45}{135}
\psarc[linewidth=1.5pt, linecolor=blue]{-}(1,-1){0.7071}{45}{135}
\end{pspicture}$,
their coefficients $\alpha_0^{m,n}$ and $\alpha_r^{m,n}$
are seen to be given by
\be 
\alpha_0^{m,n} = \prod_{i=0}^{m-1}\prod_{j=0}^{n-1}s_{i-j+1}(-u), \qquad
\alpha_r^{m,n} = \prod_{i=0}^{m-1}\prod_{j=0}^{n-1}s_{j-i}(u).
\label{alpha0r}
\ee
This is correctly reproduced by \eqref{alphaamn-app}, with the usual convention 
$\prod_{j=1}^{0} f(j) = 1$. 
The coefficients $\alpha^{m,n}_a$ for $a=1,\ldots,r-1$
are found recursively by using the following three relations. The first one of these relations,
\be
\alpha_a^{m,n}(u) = \alpha^{n,m}_{r-a}(\lambda - u),
\ee
stems from the crossing relation and allows one to relate a coefficient $\alpha^{m,n}_a$ with $m\le n$ to 
a coefficient $\alpha^{m',n'}_{a'}\!$ with $m'\ge n'$. The other two relations,
\begin{alignat}{3}
\alpha_a^{m,n} &= \alpha^{m,n-1}_{a} \ \hspace{-0.1cm}\prod_{j=n-m}^{n-m+a-1}\hspace{-0.3cm} s_{j}(u) \hspace{-0.1cm} 
\prod_{k=n-m+a}^{n-1}\hspace{-0.4cm} s_{1-k}(-u)  
&& \qquad (m<n),
\label{eq:alpharec1}\\
\alpha_a^{n,n} &= (\alpha^{n,n-1}_{a}+ \alpha^{n,n-1}_{a-1}) \, \prod_{j=0}^{a-1} s_{j}(u)  \,\prod_{k=a}^{n-1}s_{1-k}(-u) 
&& \qquad (0<a<n),
\label{eq:alpharec2}
\end{alignat}
are proven diagrammatically in the following. For $m < n$, we have
\be
\psset{unit=0.6364cm}
\begin{pspicture}(-0.1,-0.1)(2.1,0.7)
\pspolygon[fillstyle=solid,fillcolor=lightlightblue](0,0)(1,1)(2,0)(1,-1)(0,0)
\psarc[linewidth=0.025]{-}(1,-1){0.21}{45}{135}
\rput(1,0){$u$}\rput(1,0.35){\tiny{$_{(m,n)}$}}
\end{pspicture} 
\,=
\begin{pspicture}[shift=-0.85](0.5,1)(4.5,4)
\rput{90}(5,0){
\pspolygon[fillstyle=solid,fillcolor=lightlightblue](2,1)(1,2)(2,3)(3,2)(2,1)
\pspolygon[fillstyle=solid,fillcolor=lightlightblue](3,2)(2,3)(3,4)(4,3)(3,2)
\pspolygon[fillstyle=solid,fillcolor=pink](1.1,2.1)(0.85,2.35)(2.65,4.15)(2.9,3.9)(1.1,2.1)\rput(1.875,3.125){\rput{-90}(0,0){\tiny{$_n$}}}
\pspolygon[fillstyle=solid,fillcolor=pink](2.1,1.1)(2.35,0.85)(4.15,2.65)(3.9,2.9)(2.1,1.1)\rput(3.125,1.875){\rput{-90}(0,0){\tiny{$_n$}}}
\psarc[linewidth=0.025]{-}(1,2){0.21}{-45}{45}
\psarc[linewidth=0.025]{-}(2,3){0.21}{-45}{45}}
\rput(3,1.94){\small$u_0$}\rput(3,2.32){\tiny{$_{(\!m\!,n\!-\!1\!)}$}}
\rput(2,2.94){\small$u_{n\!-\!1}$}\rput(2,3.32){\tiny{$_{(m,1)}$}}
\end{pspicture}
=\, \sum_{a = 0}^{m} \alpha^{m,n-1}_a 
\begin{pspicture}[shift=-3](0.5,-0.2)(9.5,9.45)
\rput{90}(9,0)
{\pspolygon[fillstyle=solid,fillcolor=lightlightblue](0,3)(3,0)(7.5,4.5)(4.5,7.5)(0,3)
\pspolygon[fillstyle=solid,fillcolor=lightlightblue](7.9,4.9)(4.9,7.9)(5.4,8.4)(8.4,5.4)
\pspolygon[fillstyle=solid,fillcolor=pink](0.2,2.8)(2.8,0.2)(2.4,-0.2)(-0.2,2.4)(0.2,2.8)
\pspolygon[fillstyle=solid,linewidth=1.5pt,fillcolor=pink](5.1,7.7)(7.7,5.1)(7.3,4.7)(4.7,7.3)(5.1,7.7)
\pspolygon[fillstyle=solid,fillcolor=pink](6.0,8.6)(8.6,6.0)(8.2,5.6)(5.6,8.2)(6.0,8.6)
\psline{-}(5.4,7.4)(5.9,7.9)
\psline{-}(5.9,6.9)(6.4,7.4)
\psline{-}(6.9,5.9)(7.4,6.4)
\psline{-}(7.4,5.4)(7.9,5.9)
\rput(1.3,1.3){\rput{-90}(0,0){\small$_m$}}
\rput(6.2,6.2){\rput{-90}(0,0){\small$_m$}}
\rput(7.1,7.1){\rput{-90}(0,0){\small$_m$}}
\pspolygon[fillstyle=solid,fillcolor=pink](3.2,0.2)(8.3,5.3)(8.7,4.9)(3.6,-0.2)(3.2,0.2)
\pspolygon[fillstyle=solid,fillcolor=pink](0.2,3.2)(5.3,8.3)(4.9,8.7)(-0.2,3.6)(0.2,3.2)
\rput(5.95,2.55){\rput{-90}(0,0){\small$_n$}}
\rput(2.55,5.95){\rput{-90}(0,0){\small$_n$}}
\psbezier{-}(8.0,5.4)(8.0,5.5)(9.0,5.5)(9.0,5.4)
\psbezier{-}(5.4,8.0)(5.5,8.0)(5.5,9.0)(5.4,9.0)
\rput(9.9,5.2){\rput{-90}(-0.6,0){$u_{n-m}$}}
\rput(5.4,9.25){\rput{-90}(0,.4){$u_{n-1}$}}
\psarc[linewidth=0.025]{-}(7.4,5.4){0.18}{-45}{45}
\psarc[linewidth=0.025]{-}(6.9,5.9){0.18}{-45}{45}
\psarc[linewidth=0.025]{-}(5.4,7.4){0.18}{-45}{45}
\psarc[linewidth=0.025]{-}(4.9,7.9){0.18}{-45}{45}
\rput(7.3,6.0){\tiny$.$}\rput(7.4,5.9){\tiny$.$}\rput(7.5,5.8){\tiny$.$}
\rput(6.55,6.75){\tiny$.$}\rput(6.65,6.65){\tiny$.$}\rput(6.75,6.55){\tiny$.$}
\rput(5.8,7.5){\tiny$.$}\rput(5.9,7.4){\tiny$.$}\rput(6.0,7.3){\tiny$.$}
}
\psarc[linewidth=1.0pt,linecolor=blue](6,0){.4}{45}{135}
\psarc[linewidth=1.0pt,linecolor=blue](6,0){.7}{45}{135}
\psarc[linewidth=1.0pt,linecolor=blue](6,0){1.3}{45}{135}
\psarc[linewidth=1.0pt,linecolor=blue](6,0){1.6}{45}{135}
\psarc[linewidth=1.0pt,linecolor=blue](4.5,7.5){-.4}{45}{135}
\psarc[linewidth=1.0pt,linecolor=blue](4.5,7.5){-.7}{45}{135}
\psarc[linewidth=1.0pt,linecolor=blue](4.5,7.5){-1.3}{45}{135}
\psarc[linewidth=1.0pt,linecolor=blue](4.5,7.5){-1.6}{45}{135}
\psarc[linewidth=1.0pt,linecolor=blue](1.5,4.5){.4}{-45}{45}
\psarc[linewidth=1.0pt,linecolor=blue](1.5,4.5){.7}{-45}{45}
\psarc[linewidth=1.0pt,linecolor=blue](1.5,4.5){2.0}{-45}{45}
\psarc[linewidth=1.0pt,linecolor=blue](1.5,4.5){2.3}{-45}{45}
\psarc[linewidth=1.0pt,linecolor=blue](9,3){-.4}{-45}{45}
\psarc[linewidth=1.0pt,linecolor=blue](9,3){-.7}{-45}{45}
\psarc[linewidth=1.0pt,linecolor=blue](9,3){-2.0}{-45}{45}
\psarc[linewidth=1.0pt,linecolor=blue](9,3){-2.3}{-45}{45}
\rput(7.65,3){\tiny$_{(m-a)}$}
\rput(2.85,4.5){\tiny$_{(m-a)}$}
\rput(6,1.0){\tiny$_{(a)}$}
\rput(4.5,6.5){\tiny$_{(a)}$}
\rput(5.25,3.75){\tiny$_{(\!n\!-\!1\!-\!m\!)}$}
\psbezier[linewidth=1.0pt,linecolor=blue](5.8435, 6.1565)(4.3435, 4.6565)(4.83848, 4.16152)(3.33848, 2.66152)
\psbezier[linewidth=1.0pt,linecolor=blue](6.05563, 5.94437)(4.55563, 4.44437)(5.05061, 3.94939)(3.55061, 2.44939)
\psbezier[linewidth=1.0pt,linecolor=blue](4.6565, 1.3435)(6.1565, 2.8435)(5.66152, 3.33848)(7.16152, 4.83848)
\psbezier[linewidth=1.0pt,linecolor=blue](4.44437, 1.55563)(5.94437, 3.05563)(5.44939, 3.55061)(6.94939, 5.05061)
\rput(3.89,6.69){\rput(0,0){\tiny$.$}\rput(-0.1,-0.1){\tiny$.$}\rput(0.1,0.1){\tiny$.$}}
\rput(2.60,5.40){\rput(0,0){\tiny$.$}\rput(-0.1,-0.1){\tiny$.$}\rput(0.1,0.1){\tiny$.$}}
\rput(6.61,0.81){\rput(0,0){\tiny$.$}\rput(-0.1,-0.1){\tiny$.$}\rput(0.1,0.1){\tiny$.$}}
\rput(7.9,2.1){\rput(0,0){\tiny$.$}\rput(-0.1,-0.1){\tiny$.$}\rput(0.1,0.1){\tiny$.$}}
\rput(2.60,3.60){\rput(0,0){\tiny$.$}\rput(0.1,-0.1){\tiny$.$}\rput(-0.1,0.1){\tiny$.$}}
\rput(4.09,2.16){\rput(0,0){\tiny$.$}\rput(0.1,-0.1){\tiny$.$}\rput(-0.1,0.1){\tiny$.$}}
\rput(5.39,0.81){\rput(0,0){\tiny$.$}\rput(0.1,-0.1){\tiny$.$}\rput(-0.1,0.1){\tiny$.$}}
\rput(5.11,6.69){\rput(0,0){\tiny$.$}\rput(0.1,-0.1){\tiny$.$}\rput(-0.1,0.1){\tiny$.$}}
\rput(6.44, 5.35){\rput(0,0){\tiny$.$}\rput(0.1,-0.1){\tiny$.$}\rput(-0.1,0.1){\tiny$.$}}
\rput(7.9,3.9){\rput(0,0){\tiny$.$}\rput(0.1,-0.1){\tiny$.$}\rput(-0.1,0.1){\tiny$.$}}
\end{pspicture}
\ee 
where we have used (\ref{vw}) to reverse the order of the spectral parameters appearing in the upper-left column 
of elementary face operators. The push-through properties now allow us to remove the projector $P_m$ drawn with a 
thicker boundary. It subsequently follows that only a single configuration in the decomposition of the 
array of elementary face operators appearing in the upper-left part of the previous diagram gives a nonzero contribution, namely
\be
\psset{unit=0.6}
\begin{pspicture}[shift=-2.00](6.05,4.1)(8.0,8.45)
\rput(5.1,4.4){$\underbrace{\quad \hspace{0.2cm} \quad}_{m-a}$}
\rput(7.82,6.83){$_a$}
\rput{90}(7.4,6.84){$\underbrace{\quad \hspace{0.2cm} \quad}$}
\rput{90}(8,0){
\pspolygon[fillstyle=solid,fillcolor=lightlightblue](7.9,4.9)(4.9,7.9)(5.4,8.4)(8.4,5.4)
\psline{-}(5.4,7.4)(5.9,7.9)
\psline{-}(5.9,6.9)(6.4,7.4)
\psline{-}(6.4,6.4)(6.9,6.9)
\psline{-}(6.9,5.9)(7.4,6.4)
\psline{-}(7.4,5.4)(7.9,5.9)
\psarc[linewidth=1.5pt,linecolor=blue]{-}(4.9,7.9){0.354}{-90}{-45}
\psarc[linewidth=1.5pt,linecolor=blue]{-}(4.9,7.9){1.06}{-90}{-45}
\psarc[linewidth=1.5pt,linecolor=blue]{-}(4.9,7.9){1.77}{-90}{-45}
\psarc[linewidth=1.5pt,linecolor=blue]{-}(7.9,4.9){0.354}{135}{180}
\psarc[linewidth=1.5pt,linecolor=blue]{-}(7.9,4.9){1.06}{135}{180}
\psarc[linewidth=1.5pt,linecolor=blue]{-}(7.9,4.9){1.77}{135}{180}
\multiput(4.9,7.9)(0.5,-0.5){3}{\psarc[linewidth=1.5pt,linecolor=blue]{-}(0.5,0.5){0.354}{-135}{-45}
\psarc[linewidth=1.5pt,linecolor=blue]{-}(0.5,-0.5){0.354}{45}{135}
}
\multiput(6.4,6.4)(0.5,-0.5){3}{\psarc[linewidth=1.5pt,linecolor=blue]{-}(0,0){0.354}{-45}{45}
\psarc[linewidth=1.5pt,linecolor=blue]{-}(1,0){-0.354}{-45}{45}
}
}
\end{pspicture} 
\times \prod_{j=n-m}^{n-m+a-1} \hspace{-0.3cm} s_j(u) \hspace{-0.1cm}\prod_{k=n-m+a}^{n-1} \hspace{-0.3cm} s_{1-k}(-u) 
\ee
and subsequently
\be
\psset{unit=0.6364cm}
\begin{pspicture}[shift=-0.9](0,-1)(2,1)
\pspolygon[fillstyle=solid,fillcolor=lightlightblue](0,0)(1,1)(2,0)(1,-1)(0,0)
\psarc[linewidth=0.025]{-}(1,-1){0.21}{45}{135}
\rput(1,0){$u$}\rput(1,0.35){\tiny{$_{(m,n)}$}}
\end{pspicture} 
\,=\,  
\sum_{a = 0}^{m}    \alpha^{m,n-1}_a X^{m,n}_{a} \prod_{j=n-m}^{n-m+a-1}\hspace{-0.3cm} s_{j}(u) \hspace{-0.2cm} 
  \prod_{k=n-m+a}^{n-1}\hspace{-0.4cm} s_{1-k}(-u).
\ee
This completes the proof of \eqref{eq:alpharec1}. 
For $m=n$, we have
\be
\psset{unit=0.6364cm}
\begin{pspicture}(-0.1,-0.1)(2.1,0.7)
\pspolygon[fillstyle=solid,fillcolor=lightlightblue](0,0)(1,1)(2,0)(1,-1)(0,0)
\psarc[linewidth=0.025]{-}(1,-1){0.21}{45}{135}
\rput(1,0){$u$}\rput(1,0.35){\tiny{$_{(n,n)}$}}
\end{pspicture} 
\,=\,  
 \sum_{a = 0}^{n-1} \alpha^{n,n-1}_a 
\begin{pspicture}[shift=-2.7](-0.0,-0.4)(9.5,8.35)
\rput{90}(8,0){
\pspolygon[fillstyle=solid,fillcolor=lightlightblue](0,3.2)(3.2,0)(6.2,3.0)(3.0,6.2)(0,3.2)
\pspolygon[fillstyle=solid,fillcolor=lightlightblue](6.6,3.4)(3.4,6.6)(3.9,7.1)(7.1,3.9)(6.6,3.4)
\pspolygon[fillstyle=solid,fillcolor=pink](0.2,3.0)(3.0,0.2)(2.6,-0.2)(-0.2,2.6)(0.2,3.0)
\pspolygon[fillstyle=solid,linewidth=1.5pt,fillcolor=pink](3.2,6.0)(6.0,3.2)(6.4,3.6)(3.6,6.4)(3.2,6.0)
\pspolygon[fillstyle=solid,fillcolor=pink](4.1,6.9)(6.9,4.1)(7.3,4.5)(4.5,7.3)(4.1,6.9)
\psline{-}(3.9,6.1)(4.4,6.6)
\psline{-}(4.4,5.6)(4.9,6.1)
\psline{-}(5.6,4.4)(6.1,4.9)
\psline{-}(6.1,3.9)(6.6,4.4)
\rput(1.4,1.4){\rput{-90}(0,0){\small$_n$}}
\rput(5.7,5.7){\rput{-90}(0,0){\small$_n$}}
\rput(4.8,4.8){\rput{-90}(0,0){\small$_n$}}
\pspolygon[fillstyle=solid,fillcolor=pink](0.2,3.4)(3.8,7.0)(3.4,7.4)(-0.2,3.8)(0.2,3.4)
\pspolygon[fillstyle=solid,fillcolor=pink](3.4,0.2)(7.0,3.8)(7.4,3.4)(3.8,-0.2)(3.4,0.2)
\rput(1.8,5.4){\rput{-90}(0,0){\small$_n$}}
\rput(5.4,1.8){\rput{-90}(0,0){\small$_n$}}
\psbezier{-}(6.7,3.9)(6.7,4.0)(7.7,4.0)(7.7,3.9)
\psbezier{-}(3.9,6.7)(4.0,6.7)(4.0,7.7)(3.9,7.7)
\rput(8.15,3.85){\rput{-90}(-0.1,0){$u_{0}$}}
\rput(3.95,8.35){\rput{-90}(-0.1,0){$u_{n-1}$}}
\psarc[linewidth=0.025]{-}(6.1,3.9){0.18}{-45}{45}
\psarc[linewidth=0.025]{-}(5.6,4.4){0.18}{-45}{45}
\psarc[linewidth=0.025]{-}(3.9,6.1){0.18}{-45}{45}
\psarc[linewidth=0.025]{-}(3.4,6.6){0.18}{-45}{45}
\rput(5.35,5.15){\tiny$.$}\rput(5.25,5.25){\tiny$.$}\rput(5.15,5.35){\tiny$.$}}
\psarc[linewidth=1.0pt,linecolor=blue](4.8,0){.4}{45}{135}
\psarc[linewidth=1.0pt,linecolor=blue](4.8,0){.7}{45}{135}
\psarc[linewidth=1.0pt,linecolor=blue](4.8,0){1.3}{45}{135}
\psarc[linewidth=1.0pt,linecolor=blue](4.8,0){1.6}{45}{135}
\psarc[linewidth=1.0pt,linecolor=blue](5.0,6.2){-.4}{45}{135}
\psarc[linewidth=1.0pt,linecolor=blue](5.0,6.2){-.7}{45}{135}
\psarc[linewidth=1.0pt,linecolor=blue](5.0,6.2){-1.3}{45}{135}
\psarc[linewidth=1.0pt,linecolor=blue](5.0,6.2){-1.6}{45}{135}
\psarc[linewidth=1.0pt,linecolor=blue](1.8,3.0){.4}{-45}{45}
\psarc[linewidth=1.0pt,linecolor=blue](1.8,3.0){.7}{-45}{45}
\psarc[linewidth=1.0pt,linecolor=blue](1.8,3.0){2.0}{-45}{45}
\psarc[linewidth=1.0pt,linecolor=blue](1.8,3.0){2.3}{-45}{45}
\psarc[linewidth=1.0pt,linecolor=blue](8.0,3.2){-.4}{-45}{45}
\psarc[linewidth=1.0pt,linecolor=blue](8.0,3.2){-.7}{-45}{45}
\psarc[linewidth=1.0pt,linecolor=blue](8.0,3.2){-2.0}{-45}{45}
\psarc[linewidth=1.0pt,linecolor=blue](8.0,3.2){-2.3}{-45}{45}
\psbezier[linewidth=1.0pt,linecolor=blue](6.1435, 1.3435)(4.6435, 2.8435)(5.1565, 3.3565)(3.6565, 4.8565)%to fix
\rput(6.65,3.2){\tiny$_{(n\!-\!a\!-\!1)}$}
\rput(3.15,3.0){\tiny$_{(n\!-\!a\!-\!1)}$}
\rput(4.8,1.0){\tiny$_{(a)}$}
\rput(5.0,5.2){\tiny$_{(a)}$}
\rput(5.4,0.8){\rput(0,0){\tiny$.$}\rput(0.1,0.1){\tiny$.$}\rput(-0.1,-0.1){\tiny$.$}}
\rput(4.2,0.8){\rput(0,0){\tiny$.$}\rput(-0.1,0.1){\tiny$.$}\rput(0.1,-0.1){\tiny$.$}}
\rput(4.4,5.4){\rput(0,0){\tiny$.$}\rput(0.1,0.1){\tiny$.$}\rput(-0.1,-0.1){\tiny$.$}}
\rput(5.6,5.4){\rput(0,0){\tiny$.$}\rput(-0.1,0.1){\tiny$.$}\rput(0.1,-0.1){\tiny$.$}}
\rput(6.95,2.35){\rput(0,0){\tiny$.$}\rput(0.1,0.1){\tiny$.$}\rput(-0.1,-0.1){\tiny$.$}}
\rput(6.95,4.05){\rput(0,0){\tiny$.$}\rput(-0.1,0.1){\tiny$.$}\rput(0.1,-0.1){\tiny$.$}}
\rput(2.85,2.15){\rput(0,0){\tiny$.$}\rput(-0.1,0.1){\tiny$.$}\rput(0.1,-0.1){\tiny$.$}}
\rput(2.85,3.85){\rput(0,0){\tiny$.$}\rput(0.1,0.1){\tiny$.$}\rput(-0.1,-0.1){\tiny$.$}}
\end{pspicture}
\ee
After removing the $P_n$ projector indicated by a thick boundary, we now have two contributing configurations, namely
\vspace{-0.4cm}

\be
\psset{unit=0.6}
\begin{pspicture}[shift=-2.15](4.05,3.2)(8.8,8.95)
\rput(5.85,3.84){$\underbrace{\quad\hspace{0.24cm}\quad}_{n-a-1}$}
\rput{90}(8.6,6.84){$\underbrace{\quad\hspace{0.24cm}\quad}$}
\rput(9.0,6.84){$_a$}
\rput{90}(9,0){
\pspolygon[fillstyle=solid,fillcolor=lightlightblue](7.9,4.9)(4.4,8.4)(4.9,8.9)(8.4,5.4)
\psline{-}(4.9,7.9)(5.4,8.4)
\psline{-}(5.4,7.4)(5.9,7.9)
\psline{-}(5.9,6.9)(6.4,7.4)
\psline[linewidth=1.5pt, linecolor=blue](6.15,6.65)(4.60,5.10)
\psline{-}(6.4,6.4)(6.9,6.9)
\psline{-}(6.9,5.9)(7.4,6.4)
\psline{-}(7.4,5.4)(7.9,5.9)
\psarc[linewidth=1.5pt,linecolor=blue]{-}(4.4,8.4){0.354}{-90}{-45}
\psarc[linewidth=1.5pt,linecolor=blue]{-}(4.4,8.4){1.06}{-90}{-45}
\psarc[linewidth=1.5pt,linecolor=blue]{-}(4.4,8.4){1.77}{-90}{-45}
\psarc[linewidth=1.5pt,linecolor=blue]{-}(7.9,4.9){0.354}{135}{180}
\psarc[linewidth=1.5pt,linecolor=blue]{-}(7.9,4.9){1.06}{135}{180}
\psarc[linewidth=1.5pt,linecolor=blue]{-}(7.9,4.9){1.77}{135}{180}
\multiput(4.4,8.4)(0.5,-0.5){4}{\psarc[linewidth=1.5pt,linecolor=blue]{-}(0.5,0.5){0.354}{-135}{-45}
\psarc[linewidth=1.5pt,linecolor=blue]{-}(0.5,-0.5){0.354}{45}{135}
}
\multiput(6.4,6.4)(0.5,-0.5){3}{\psarc[linewidth=1.5pt,linecolor=blue]{-}(0,0){0.354}{-45}{45}
\psarc[linewidth=1.5pt,linecolor=blue]{-}(1,0){-0.354}{-45}{45}
}
}
\end{pspicture}
\times \prod_{j=0}^{a-1} s_j(u)\prod_{k=a}^{n-1}  s_{1-k}(-u),
\qquad
\begin{pspicture}[shift=-2.15](4.05,3.2)(8.8,8.95)
\rput(5.85,3.84){$\underbrace{\quad\hspace{0.24cm}\quad}_{n-a-1}$}
\rput{90}(8.6,6.84){$\underbrace{\quad\hspace{0.24cm}\quad}$}
\rput(9.0,6.84){$_a$}
\rput{90}(9,0){
\pspolygon[fillstyle=solid,fillcolor=lightlightblue](7.9,4.9)(4.4,8.4)(4.9,8.9)(8.4,5.4)
\psline{-}(4.9,7.9)(5.4,8.4)
\psline{-}(5.4,7.4)(5.9,7.9)
\psline{-}(5.9,6.9)(6.4,7.4)
\psline[linewidth=1.5pt, linecolor=blue](6.15,6.65)(4.60,5.10)
\psline{-}(6.4,6.4)(6.9,6.9)
\psline{-}(6.9,5.9)(7.4,6.4)
\psline{-}(7.4,5.4)(7.9,5.9)
\psarc[linewidth=1.5pt,linecolor=blue]{-}(4.4,8.4){0.354}{-90}{-45}
\psarc[linewidth=1.5pt,linecolor=blue]{-}(4.4,8.4){1.06}{-90}{-45}
\psarc[linewidth=1.5pt,linecolor=blue]{-}(4.4,8.4){1.77}{-90}{-45}
\psarc[linewidth=1.5pt,linecolor=blue]{-}(7.9,4.9){0.354}{135}{180}
\psarc[linewidth=1.5pt,linecolor=blue]{-}(7.9,4.9){1.06}{135}{180}
\psarc[linewidth=1.5pt,linecolor=blue]{-}(7.9,4.9){1.77}{135}{180}
\multiput(4.4,8.4)(0.5,-0.5){3}{\psarc[linewidth=1.5pt,linecolor=blue]{-}(0.5,0.5){0.354}{-135}{-45}
\psarc[linewidth=1.5pt,linecolor=blue]{-}(0.5,-0.5){0.354}{45}{135}
}
\multiput(5.9,6.9)(0.5,-0.5){4}{\psarc[linewidth=1.5pt,linecolor=blue]{-}(0,0){0.354}{-45}{45}
\psarc[linewidth=1.5pt,linecolor=blue]{-}(1,0){-0.354}{-45}{45}
}
}
\end{pspicture} \times \prod_{j=0}^{a} s_j(u)\hspace{-0.1cm}\prod_{k=a+1}^{n-1} \hspace{-0.1cm} s_{1-k}(-u).
\ee
It follows that
\begin{align}
\psset{unit=0.6364cm}
\begin{pspicture}[shift=-0.9](-0,-1)(2,1)
\pspolygon[fillstyle=solid,fillcolor=lightlightblue](0,0)(1,1)(2,0)(1,-1)(0,0)
\psarc[linewidth=0.025]{-}(1,-1){0.21}{45}{135}
\rput(1,0){$u$}\rput(1,0.35){\tiny{$_{(m,n)}$}}
\end{pspicture}
 & \ = \sum_{a = 0}^{n-1} \alpha^{n,n-1}_a \Big( X^{n,n}_a \prod_{j=0}^{a-1} s_j(u)
 \prod_{k=a}^{n-1} s_{1-k}(-u)  + X^{n,n}_{a+1} \prod_{j=0}^{a} s_j(u) \! \prod_{k=a+1}^{n-1}\!  s_{1-k}(-u) 
\Big) \nonumber \\
& = \alpha_0^{n,n} X^{n,n}_0 + \alpha_n^{n,n} X^{n,n}_n + \sum_{a=1}^{n-1} X^{n,n}_a 
 \big(\alpha_a^{n,n-1} + \alpha_{a-1}^{n,n-1}\big) \prod_{j=0}^{a-1}  s_j(u)\prod_{k=a}^{n-1}  s_{1-k}(-u),
\end{align}
where we used (\ref{alpha0r}) to establish the last equality.
This completes the proof of \eqref{eq:alpharec2}.
As already announced, the coefficients $\alpha^{m,n}_a$ for $a=1,\ldots,r-1$ now follow recursively.
\hfill $\square$

%%%%%%%%
\section{Boundary Yang-Baxter equations for fused face operators}
\label{app:YBEs}
%%%%%%%%

First, we note in \eqref{eq:fusedBYBE} that the left BYBE is obtained from the right BYBE by rotating the diagram by $180^\circ$ while 
simultaneously mapping $u \rightarrow u_{n-m}$ and $v \rightarrow v_{n-m}$. 
The objective of this appendix is then to show that 
\bea 
\mathcal L _{m,n} = \mathcal R_{m,n},\qquad\quad 
\mathcal L_{m,n}:= \
\psset{unit=0.6364cm}
\begin{pspicture}[shift=-1.9](0,0)(3,4)
\psarc[linecolor=blue,linewidth=4pt]{-}(2.8,1){0.5}{-90}{90}\psarc[linecolor=white,linewidth=2pt]{-}(2.8,1){0.5}{-90}{90}
\psarc[linecolor=blue,linewidth=4pt]{-}(2.8,3){0.5}{-90}{90}\psarc[linecolor=white,linewidth=2pt]{-}(2.8,3){0.5}{-90}{90}
\psline[linecolor=blue,linewidth=4pt]{-}(0,0.5)(0.6,0.5)\psline[linecolor=white,linewidth=2pt]{-}(0,0.5)(0.6,0.5)
\psline[linecolor=blue,linewidth=4pt]{-}(0,1.5)(0.6,1.5)\psline[linecolor=white,linewidth=2pt]{-}(0,1.5)(0.6,1.5)
\psline[linecolor=blue,linewidth=4pt]{-}(0,2.5)(1.6,2.5)\psline[linecolor=white,linewidth=2pt]{-}(0,2.5)(1.6,2.5)
\psline[linecolor=blue,linewidth=4pt]{-}(0,3.5)(2.8,3.5)\psline[linecolor=white,linewidth=2pt]{-}(0,3.5)(2.8,3.5)
\psline[linecolor=blue,linewidth=4pt]{-}(1.4,0.5)(2.8,0.5)\psline[linecolor=white,linewidth=2pt]{-}(1.4,0.5)(2.8,0.5)
\psline[linecolor=blue,linewidth=4pt]{-}(2.4,1.5)(2.8,1.5)\psline[linecolor=white,linewidth=2pt]{-}(2.4,1.5)(2.8,1.5)
\psline[linecolor=blue,linewidth=4pt]{-}(2.4,2.5)(2.8,2.5)\psline[linecolor=white,linewidth=2pt]{-}(2.4,2.5)(2.8,2.5)
\pspolygon[fillstyle=solid,fillcolor=lightlightblue](0,1)(1,2)(2,1)(1,0)(0,1)
\pspolygon[fillstyle=solid,fillcolor=lightlightblue](1,2)(2,1)(3,2)(2,3)(1,2)
\psline{-}(0,0)(0,4)
\psarc[linewidth=0.025]{-}(0,1){0.21}{-45}{45}\rput(1,1){$u$}\rput(1,1.35){\tiny{$_{(m,n)}$}}
\psarc[linewidth=0.025]{-}(1,2){0.21}{-45}{45}\rput(2,2){$v$}\rput(2,2.35){\tiny{$_{(m,n)}$}}
\end{pspicture} \qquad \qquad 
\mathcal R_{m,n}:= \
\begin{pspicture}[shift=-1.9](0,0)(3,4)
\psset{unit=0.6364cm} 
\psarc[linecolor=blue,linewidth=4pt]{-}(2.8,1){0.5}{-90}{90}\psarc[linecolor=white,linewidth=2pt]{-}(2.8,1){0.5}{-90}{90}
\psarc[linecolor=blue,linewidth=4pt]{-}(2.8,3){0.5}{-90}{90}\psarc[linecolor=white,linewidth=2pt]{-}(2.8,3){0.5}{-90}{90}
\psline[linecolor=blue,linewidth=4pt]{-}(0,2.5)(0.6,2.5)\psline[linecolor=white,linewidth=2pt]{-}(0,2.5)(0.6,2.5)
\psline[linecolor=blue,linewidth=4pt]{-}(0,3.5)(0.6,3.5)\psline[linecolor=white,linewidth=2pt]{-}(0,3.5)(0.6,3.5)
\psline[linecolor=blue,linewidth=4pt]{-}(0,1.5)(1.6,1.5)\psline[linecolor=white,linewidth=2pt]{-}(0,1.5)(1.6,1.5)
\psline[linecolor=blue,linewidth=4pt]{-}(0,0.5)(2.8,0.5)\psline[linecolor=white,linewidth=2pt]{-}(0,0.5)(2.8,0.5)
\psline[linecolor=blue,linewidth=4pt]{-}(1.4,3.5)(2.8,3.5)\psline[linecolor=white,linewidth=2pt]{-}(1.4,3.5)(2.8,3.5)
\psline[linecolor=blue,linewidth=4pt]{-}(2.4,2.5)(2.8,2.5)\psline[linecolor=white,linewidth=2pt]{-}(2.4,2.5)(2.8,2.5)
\psline[linecolor=blue,linewidth=4pt]{-}(2.4,1.5)(2.8,1.5)\psline[linecolor=white,linewidth=2pt]{-}(2.4,1.5)(2.8,1.5)
\pspolygon[fillstyle=solid,fillcolor=lightlightblue](0,3)(1,4)(2,3)(1,2)(0,3)
\pspolygon[fillstyle=solid,fillcolor=lightlightblue](1,2)(2,1)(3,2)(2,3)(1,2)
\psline{-}(0,0)(0,4)
\psarc[linewidth=0.025]{-}(0,3){0.21}{-45}{45}\rput(1,3){\scriptsize$u_{n\!-\!m}$}\rput(1,3.35){\tiny{$_{(n,m)}$}}
\psarc[linewidth=0.025]{-}(1,2){0.21}{-45}{45}\rput(2,2){\scriptsize$v_{n\!-\!m}$}\rput(2,2.35){\tiny{$_{(n,m)}$}}
\end{pspicture} \ \ \ 
\eea 
We refer to this equation as BYBE$_{m,n}$. 
Note that the special case BYBE$_{1,1}$ is identical to the second relation in (\ref{eq:bybes}).
Following the ideas of~\cite{BehrendPearce}, the proof for general $m,n$ is done in two 
steps. First, a recursion in $n$ will show that BYBE$_{1,n}$ results from the YBE,
BYBE$_{1,n-1}$ and BYBE$_{1,1}$. A second recursion, in $m$, will show that the YBE,
BYBE$_{m-1,n}$ and BYBE$_{1,n}$ imply BYBE$_{m,n}$.

%%%%%%%%%%%%%%%%%%%%%%%%%%%%
\paragraph{First recursion: BYBE$_{1,n}$}
%%%%%%%%%%%%%%%%%%%%%%%%%%%%
We start by writing
\be
\psset{unit=0.6364cm}
\begin{pspicture}[shift=-0.9](0,-1)(2,1)
\pspolygon[fillstyle=solid,fillcolor=lightlightblue](0,0)(1,1)(2,0)(1,-1)(0,0)
\psarc[linewidth=0.025]{-}(0,0){0.21}{-45}{45}
\rput(1,0){$u$}\rput(1,0.35){\tiny{$_{(1,n)}$}}
\end{pspicture} \ = \  
\begin{pspicture}[shift=-1.6](0.8,0.8)(4.2,4.2)
\pspolygon[fillstyle=solid,fillcolor=lightlightblue](2,1)(1,2)(2,3)(3,2)(2,1)
\pspolygon[fillstyle=solid,fillcolor=lightlightblue](3,2)(2,3)(3,4)(4,3)(3,2)
\pspolygon[fillstyle=solid,fillcolor=pink](1.1,2.1)(0.85,2.35)(2.65,4.15)(2.9,3.9)(1.1,2.1)\rput(1.875,3.125){\tiny{$_n$}}
\pspolygon[fillstyle=solid,fillcolor=pink](2.1,1.1)(2.35,0.85)(4.15,2.65)(3.9,2.9)(2.1,1.1)\rput(3.125,1.875){\tiny{$_n$}}
\rput(2,2){\small$u_0$}\rput(2,2.35){\tiny{$_{(1,n\!-\!1)}$}}
\rput(3,3){\small$u_{n\!-\!1}$}\rput(3,3.35){\tiny{$_{(1,1)}$}}
\psarc[linewidth=0.025]{-}(1,2){0.21}{-45}{45}
\psarc[linewidth=0.025]{-}(2,3){0.21}{-45}{45}
\end{pspicture} \qquad \qquad
\begin{pspicture}[shift=-0.9](0,-1)(2,1)
\pspolygon[fillstyle=solid,fillcolor=lightlightblue](0,0)(1,1)(2,0)(1,-1)(0,0)
\psarc[linewidth=0.025]{-}(0,0){0.21}{-45}{45}
\rput(1,0){$v$}\rput(1,0.35){\tiny{$_{(1,n)}$}}
\end{pspicture} \ = \  
\begin{pspicture}[shift=-1.6](0.8,0.8)(4.2,4.2)
\pspolygon[fillstyle=solid,fillcolor=lightlightblue](2,1)(1,2)(2,3)(3,2)(2,1)
\pspolygon[fillstyle=solid,fillcolor=lightlightblue](3,2)(2,3)(3,4)(4,3)(3,2)
\pspolygon[fillstyle=solid,fillcolor=pink](1.1,2.1)(0.85,2.35)(2.65,4.15)(2.9,3.9)(1.1,2.1)\rput(1.875,3.125){\tiny{$_n$}}
\pspolygon[fillstyle=solid,fillcolor=pink](2.1,1.1)(2.35,0.85)(4.15,2.65)(3.9,2.9)(2.1,1.1)\rput(3.125,1.875){\tiny{$_n$}}
\rput(2,2){\small$v_1$}\rput(2,2.35){\tiny{$_{(1,n\!-\!1)}$}}
\rput(3,3){\small$v_0$}\rput(3,3.35){\tiny{$_{(1,1)}$}}
\psarc[linewidth=0.025]{-}(1,2){0.21}{-45}{45}
\psarc[linewidth=0.025]{-}(2,3){0.21}{-45}{45}
\end{pspicture}
\ee 
where we have used (\ref{vw}) 
to establish the second equality.
In the diagrammatic representation of $\mathcal L_{1,n}$, we remove the two unnecessary projectors and add an extra tile,
\be 
\mathcal L_{1,n} 
 = \frac1{\eta(w)}\ \ 
 \psset{unit=0.6364cm}
\begin{pspicture}[shift=-2.9](-0.5,0)(5.3,6)
\psline[linecolor=blue,linewidth=1.2pt]{-}(-0.5,0.5)(0.6,0.5)
\psline[linecolor=blue,linewidth=4pt]{-}(-0.5,1.5)(0.6,1.5)\psline[linecolor=white,linewidth=2pt]{-}(-0.5,1.5)(0.6,1.5)
\psline[linecolor=blue,linewidth=1.2pt]{-}(-0.5,2.5)(1.6,2.5)
\psline[linecolor=blue,linewidth=4pt]{-}(-0.5,3.5)(2.6,3.5)\psline[linecolor=white,linewidth=2pt]{-}(-0.5,3.5)(2.6,3.5)
\psline[linecolor=blue,linewidth=1.2pt]{-}(-0.5,4.5)(3.6,4.5)
\psline[linecolor=blue,linewidth=1.2pt]{-}(-0.5,5.5)(4.8,5.5)
\psline[linecolor=blue,linewidth=1.2pt]{-}(4.4,4.5)(4.8,4.5)
\psline[linecolor=blue,linewidth=1.2pt]{-}(4.4,3.5)(4.8,3.5)
\psline[linecolor=blue,linewidth=1.2pt]{-}(4.4,2.5)(4.8,2.5)
\psline[linecolor=blue,linewidth=4pt]{-}(4.4,1.5)(4.8,1.5)\psline[linecolor=white,linewidth=2pt]{-}(4.4,1.5)(4.8,1.5)
\psline[linecolor=blue,linewidth=1.2pt]{-}(2.4,1.5)(3.6,1.5)
\psline[linecolor=blue,linewidth=4pt]{-}(1.4,0.5)(4.8,0.5)\psline[linecolor=white,linewidth=2pt]{-}(1.4,0.5)(4.8,0.5)
\psarc[linecolor=blue,linewidth=4pt]{-}(4.8,1){0.5}{-90}{90}\psarc[linecolor=white,linewidth=2pt]{-}(4.8,1){0.5}{-90}{90}
\psarc[linecolor=blue,linewidth=1.2pt]{-}(4.8,3){0.5}{-90}{90}
\psarc[linecolor=blue,linewidth=1.2pt]{-}(4.8,5){0.5}{-90}{90}
\pspolygon[fillstyle=solid,fillcolor=lightlightblue](0,1)(1,2)(2,1)(1,0)(0,1)
\pspolygon[fillstyle=solid,fillcolor=lightlightblue](1,2)(2,3)(3,2)(2,1)(1,2)
\pspolygon[fillstyle=solid,fillcolor=lightlightblue](2,3)(3,4)(4,3)(3,2)(2,3)
\pspolygon[fillstyle=solid,fillcolor=lightlightblue](3,4)(4,5)(5,4)(4,3)(3,4)
\pspolygon[fillstyle=solid,fillcolor=lightlightblue](3,2)(4,3)(5,2)(4,1)(3,2)
\psline{-}(-0.5,0)(-0.5,6)
\psarc[linewidth=0.025]{-}(0,1){0.21}{-45}{45}
\psarc[linewidth=0.025]{-}(1,2){0.21}{-45}{45}
\psarc[linewidth=0.025]{-}(2,3){0.21}{-45}{45}
\psarc[linewidth=0.025]{-}(3,4){0.21}{-45}{45}
\psarc[linewidth=0.025]{-}(3,2){0.21}{-45}{45}
\rput(1,1){\small$u_0$}\rput(1,1.35){\tiny{$_{(1,n\!-\!1)}$}}
\rput(2,2){\small$u_{n\!-\!1}$}\rput(2,2.35){\tiny{$_{(1,1)}$}}
\rput(4,4){$v_0$}\rput(4,4.35){\tiny{$_{(1,1)}$}}
\rput(3,3){$v_1$}\rput(3,3.35){\tiny{$_{(1,n\!-\!1)}$}}
\rput(4,2){$w$}\rput(4,2.35){\tiny{$_{(1,n\!-\!1)}$}}
\pspolygon[fillstyle=solid,fillcolor=pink](-0.1,1.1)(-0.1,2.9)(-0.5,2.9)(-0.5,1.1)(-0.1,1.1)\rput(-0.3,2){$_n$}
\pspolygon[fillstyle=solid,fillcolor=pink](-0.1,3.1)(-0.1,4.9)(-0.5,4.9)(-0.5,3.1)(-0.1,3.1)\rput(-0.3,4){$_n$}
\end{pspicture}
\qquad\quad
 w = v-u-(n-2)\lambda, \qquad \eta(w) = \prod_{i=0}^{n-2}s_i(w).
\label{eq:w}
\ee
In fact, an extra $(1,n-1)$-fused face with arbitrary spectral parameter $w$ can be added because, after it is decomposed  
as in \eqref{eq:2termexpansion} in terms of two diagrams, one can show using the push-through properties that 
(i) the two projectors can be removed, 
and (ii) the second term is zero. 
The particular choice for $w$ in \eqref{eq:w} allows us to use the YBE and the (assumed) BYBE. 
At every step, the required identity is indicated above the equality sign, and the boxes to be changed are specified by thick 
boundaries. We thus find
\begin{equation*} 
\mathcal L_{1,n}  = \frac1{\eta(w)}\ \ 
\psset{unit=0.6364cm}
 
\ \ = \ \mathcal R_{1,n}
\ee

\noindent which is the required result.

%%%%%%%%%%%%%%%%%%%%%%%%%%
\paragraph{Second recursion: BYBE$_{m,n}$}
%%%%%%%%%%%%%%%%%%%%%%%%%%
This time, we write
\be
\psset{unit=0.6364cm}
%
 \ \ 
\ee 
and use the similar expansion for the $v$-dependent face operator. As before, we add a new face operator,
this time with parameter $w = u+v+(n-2)\lambda$ and function 
$\tilde\eta(w)=\prod_{j=0}^{m-2} s_{-j}(w)$, remove unnecessary projectors and find
\begin{align} 
\mathcal L_{m,n} &= \ \ 
\psset{unit=0.6364cm}
%
 
\ \ = \  \mathcal R_{m,n}.
\nonumber
\end{align} 
\\[-.4cm]
\noindent This concludes the proof.

%%%%%%%%%%%%%%%%%%%%%%%%%
\section{Effective projectors}
\label{app:EffectiveProjectors}
%%%%%%%%%%%%%%%%%%%%%%%%%

Here we recall and prove Proposition~\ref{prop:DTQ}.
\\[.2cm]
{\scshape Proposition~\ref{prop:DTQ}} 
{\em For $\lambda$ generic, the transfer tangles can be expressed in terms of effective projectors as}
\be \Db^{m,n}_0 = \ 
 \psset{unit=0.9}
 %
  
\label{eq:DTQ-app}
\ee  

\noindent{\scshape Proof:}
For the proof on the strip, we use induction to establish a more general relation, namely
\be \Gb^{m,n}(u,v) := \  
\psset{unit=0.65}
%
 
\label{Dxy}
\ee
and note that
$\Gb^{m,n}(u,u_1\!-\!\mu)=\Db^{m,n}(u)$.
The choice of face operators ensures that the propagation of half-arcs from push-through properties is towards the right 
within the top $n$ layers and towards the left 
within the bottom $n$ layers (
%
). Since $Q_1=P_1$ and $Q_2=P_2$, 
the relation (\ref{Dxy}) is trivially true for $n=1,2$. 

The first step in building the induction is to modify the first form in \eqref{Dxy} using \eqref{eq:Prec},
\be 
\Gb^{m,n}(u,v) = \
\psset{unit=0.65}
%
\label{eq:Gmnfirststep}
\ee 
One readily recognizes that the first tangle contains $\Gb^{m,n-1}(u_1,v_0)$ as a sub-tangle, and by the induction assumption,
the full tangle satisfies
\be 
\psset{unit=0.65}
%
 
\ee
As for the second term in \eqref{eq:Gmnfirststep}, the push-through properties and property \eqref{eq:(ii)} of the projectors allow one to
remove the leftmost $P_{n-1}$ projector. The ensuing term can then be written as
\begin{equation*}   
- \frac{s_{n-1}(0)}{s_n(0)} \ 
\psset{unit=0.65}
%
\end{equation*}
\be  
 \hspace{2.8cm}= -\big[q^m(u_1)q^m(-v_{n-3})\big]^N \Gb^{m,n-2}(u_2,v_0)
\ee
where property \eqref{eq:(v)} was used at the last step. 
Again, the induction assumption implies that this can be written in terms of effective projectors. 
One now applies the same evaluation technique to the second form in \eqref{Dxy}, this time using the recursive 
definition \eqref{Qn} for effective projectors. This yields the exact same result as for the first form of 
$\Gb^{m,n}(u,v)$, thereby completing the proof of \eqref{Dxy}. The proposition for $\Tb^{m,n}_0$ is verified similarly. 
\hfill $\square$

%%%%%%%%%%%%%%%%%%%%%
\section{Cabled link states and lattice paths}
\label{app:LatticePaths}
%%%%%%%%%%%%%%%%%%%%%

The dimensions of the linear spans $V^d_{N,m}$ and $\tilde{V}^d_{N,m}$ can be determined by establishing bijections to certain 
families of (generalised Dyck) lattice paths. 
We first consider the situation on the cylinder.
\begin{Proposition}
For $N,m\in\mathbb{N}$ and $d\in\mathbb{N}_0$ subject to (\ref{dNm}), the number of linearly independent $m$-cabled
link states on $N$ nodes with $d$ defects on the cylinder is given by
\be
  \dim \tilde{V}_{N,m}^d=\left(\!\!\!\begin{array}{c} N\\[.1cm] \frac{Nm-d}{2}\end{array}\!\!\!\right)_{\!\!m}.
\label{eq:LinkCountingCylinder}
\ee
\label{prop:LinkCountingCylinder}
\end{Proposition}
\noindent{\scshape Proof:}
We first establish a bijection between the canonical basis of cabled link states and a class of $N$-tuples of integers.
The number of link states is subsequently found by counting these $N$-tuples.

To every node of a cabled link state in the canonical basis of $\tilde{V}_{N,m}^d$, we assign an integer 
\be 
 x_i \in \{+1,-1\},\qquad i = 1, \dots, Nm,
\ee 
setting $x_i=-1$ if the loop segment tied to the $i$-th node connects to another node 
by moving toward the left, and $+1$ otherwise, that is, if the emanating loop 
segment moves to the right or happens to be a defect. 
Because for every $\ell\in\{0,\ldots,N-1\}$, two nodes $i,j$ with $\ell m+1 \le i<j\le (\ell+1)m$
only can be linked via the back of the cylinder, the sequence of integers $x_i$ must be of the form
\be
 \underbrace{\underbrace{(-1)\ldots(-1)}_{\frac{m-y_1}{2}}\underbrace{(+1)\ldots(+1)}_{\frac{m+y_1}{2}}}_m\
 \underbrace{\underbrace{(-1)\ldots(-1)}_{\frac{m-y_2}{2}}\underbrace{(+1)\ldots(+1)}_{\frac{m+y_2}{2}}}_m\ \ldots\
 \underbrace{\underbrace{(-1)\ldots(-1)}_{\frac{m-y_N}{2}}\underbrace{(+1)\ldots(+1)}_{\frac{m+y_N}{2}}}_m,
\ee
where
\be
 y_k\in\{-m,-m+2,\ldots,m\},\qquad k=1,\ldots,N.
\label{yk}
\ee
By construction, we have
\be
 y_k = x_{(k-1)m+1} + x_{(k-1)m+2} + \dots + x_{km}
\ee
and 
\be
 \sum_{k=1}^N y_k = \sum_{i=1}^{Nm} x_i = d.
\label{sumyd}
\ee
Since the structure of link states disallows half-arcs arching over defects, 
the association of the $N$-tuple $(y_1,\ldots,y_N)$ to the link state we started out with is clearly an invertible map.
Indeed, the sequence $x_1,\ldots,x_{Nm}$ associated to the $N$-tuple decomposes into alternating 
subsequences consisting exclusively of $(-1)$'s or $(+1)$'s. There are two possible scenarios.
(i) If there is only one such subsequence, it must consist of $(+1)$'s 
and all associated nodes give rise to defects. (ii) If instead there is more than one subsequence, 
there must be an even number of them, and the leftmost $(-1)$ in every subsequence of $(-1)$'s must be linked to the 
$(+1)$ to its immediate left. Ignoring these pairs of connected nodes, the remaining shorter 
sequence of $x$'s also decomposes
into alternating subsequences of $(-1)$'s or $(+1)$'s, respectively, and we repeat iteratively the steps following (i) or (ii). 
The half-arcs produced in these subsequent iterations will thus arch over already produced half-arcs.
The last step in this
procedure is the first (and only) time option (i) is applied and is reached when the sequence of $x$'s has been reduced to 
$d$ $(+1)$'s. Note that this last step is only executed if $d>0$. By construction, the ensuing link state is the one we started out with.
We have thus established a bijection between the canonical basis of cabled link states in $\tilde{V}_{N,m}^d$ and the set
$Y_{N,m}^d$ of $N$-tuples of integers, $(y_1,\ldots,y_N)\in Y_{N,m}^d$, subject to (\ref{yk}) and (\ref{sumyd}).

These $N$-tuples can be used to characterise a family of lattice paths in 
$\mathbb Z^2$, where the path associated to $(y_1, y_2, \dots, y_N)\in Y_{N,m}^d$ starts at the origin and ends at $(N,d)$, 
with steps $(1,y_1)$, $(1,y_2)$, \dots, $(1,y_N)$ along the way. We thus have a bijection between cabled link states on the cylinder
and the corresponding family of lattice paths labeled by the elements of $Y_{N,m}^d$. Illustrating this bijection,
the path $(3,1,-3,-3,1,-1,3)\in Y_{7,3}^1$ is depicted as
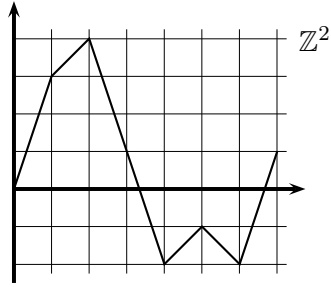
\begin{equation*}
\psset{unit=0.5}
\begin{pspicture}(0,-2)(8,5)
\psline[linewidth=1.5pt]{->}(0,0)(7.75,0)
\psline[linewidth=1.5pt]{->}(0,-2.5)(0,5)
\psline{-}(0,0)(1,3)(2,4)(3,1)(4,-2)(5,-1)(6,-2)(7,1)
\multiput(0,-2)(0,1){7}{\psline[linewidth=0.25pt]{-}(0,0)(7.25,0)}
\multiput(1,0)(1,0){7}{\psline[linewidth=0.25pt]{-}(0,-2.25)(0,4.25)}
\rput(8,4){$\mathbb Z^2$}
\end{pspicture}
\end{equation*}
and corresponds to the link state
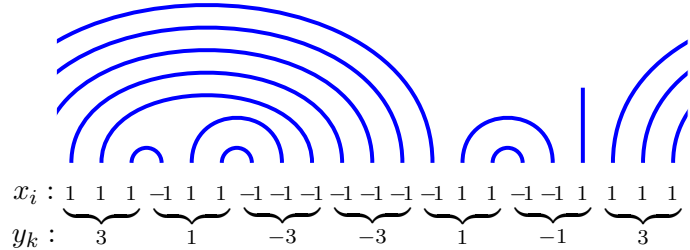
\begin{equation*}
\begin{pspicture}[shift=-1.05](0.0,-1.4)(8,2.3)
\psarc[linewidth=1.5pt,linecolor=blue]{-}(1.0,0){0.2}{0}{180}
\psarc[linewidth=1.5pt,linecolor=blue]{-}(5.8,0){0.2}{0}{180}
\psarc[linewidth=1.5pt,linecolor=blue]{-}(2.2,0){0.2}{0}{180}
\psbezier[linewidth=1.5pt,linecolor=blue]{-}(1.6,0)(1.6,0.8)(2.8,0.8)(2.8,0)
\psbezier[linewidth=1.5pt,linecolor=blue]{-}(0.4,0)(0.4,1.2)(3.2,1.2)(3.2,0)
\psbezier[linewidth=1.5pt,linecolor=blue]{-}(0.0,0)(0.0,1.6)(3.6,1.6)(3.6,0)
\psbezier[linewidth=1.5pt,linecolor=blue]{-}(5.2,0)(5.2,0.8)(6.4,0.8)(6.4,0)
\psline[linewidth=1.5pt,linecolor=blue](6.8,0)(6.8,1.0)
\multiput(0,0)(-8.4,0){2}{
\psbezier[linewidth=1.5pt,linecolor=blue]{-}(8.0,0)(8.0,2.0)(12.4,2.0)(12.4,0)
\psbezier[linewidth=1.5pt,linecolor=blue]{-}(7.6,0)(7.6,2.4)(12.8,2.4)(12.8,0)
\psbezier[linewidth=1.5pt,linecolor=blue]{-}(7.2,0)(7.2,2.8)(13.2,2.8)(13.2,0)
}
\rput(-0.5,-0.43){$x_i:$}
\rput(-0.5,-1.01){$y_k:$}
\psframe[fillstyle=solid,linecolor=white,linewidth=0pt](-2.2,-0.1)(-0.2,2.5)
\psframe[fillstyle=solid,linecolor=white,linewidth=0pt](8.2,-0.1)(13.3,2.5)
\rput(0,0){\rput(0.0,-0.4){$_1$}\rput(0.4,-0.4){$_1$}\rput(0.8,-0.4){$_1$}
\rput(0.4,-0.85){$\underbrace{\hspace{1cm}}_3$}}
\rput(1.2,0){\rput(0.0,-0.415){$_{-\!1}$}\rput(0.4,-0.4){$_1$}\rput(0.8,-0.4){$_1$}
\rput(0.4,-0.85){$\underbrace{\hspace{1cm}}_1$}}
\rput(2.4,0){\rput(0.0,-0.415){$_{-\!1}$}\rput(0.4,-0.415){$_{-\!1}$}\rput(0.8,-0.415){$_{-\!1}$}
\rput(0.4,-0.85){$\underbrace{\hspace{1cm}}_{-3}$}}
\rput(3.6,0){\rput(0.0,-0.415){$_{-\!1}$}\rput(0.4,-0.415){$_{-\!1}$}\rput(0.8,-0.415){$_{-\!1}$}
\rput(0.4,-0.85){$\underbrace{\hspace{1cm}}_{-3}$}}
\rput(4.8,0){\rput(0.0,-0.415){$_{-\!1}$}\rput(0.4,-0.4){$_1$}\rput(0.8,-0.4){$_1$}
\rput(0.4,-0.85){$\underbrace{\hspace{1cm}}_1$}}
\rput(6,0){\rput(0.0,-0.415){$_{-\!1}$}\rput(0.4,-0.415){$_{-\!1}$}\rput(0.8,-0.4){$_1$}
\rput(0.4,-0.85){$\underbrace{\hspace{1cm}}_{-1}$}}
\rput(7.2,0){\rput(0.0,-0.4){$_1$}\rput(0.4,-0.4){$_1$}\rput(0.8,-0.4){$_1$}
\rput(0.4,-0.85){$\underbrace{\hspace{1cm}}_3$}}
\end{pspicture} 
\end{equation*}

To determine the cardinality of the set $Y_{N,m}^d$ and hence the dimension of $\tilde{V}_{N,m}^d$, we relax the parameters
of the space of tuples and thus let $Y_{n,m}^d$ denote the set of $n$-tuples of integers $(y_1, \dots, y_n)$ subject to
\be
 y_k \in \{-m, -m+2, \dots, m\}, \qquad \sum_{k=1}^n y_k = d,\qquad n,\tfrac{1}{2}(nm-|d|),|d|\in\mathbb{N}_0,\qquad m\in\mathbb{N}. 
\label{eq:Ynmd}
\ee 
From the lattice path interpretation, the cardinalities of the sets $Y_{n,m}^d$ are 
easily seen to satisfy the recursive relations
\be
|Y_{n,m}^d| = \sum_{i = -m, -m+2, \dots, m} \hspace{-0.5cm}|Y_{n-1,m}^{d+i}|, \qquad\ |Y_{0,m}^d| = \delta_{d,0},
\label{eq:pathsrec}
\ee
and it is straightforward to verify from the definition \eqref{eq:mnomialgen} 
that
\be
 |Y_{n,m}^d| = \left(\!\!\!\begin{array}{c} n\\ \frac{nm-d}{2}\end{array}\!\!\!\right)_{\!\!m}
\label{eq:dimY}
\ee
satisfies \eqref{eq:pathsrec}.
\hfill $\square$
\medskip

A refinement of the $y$-paths used in the proof of Proposition~\ref{prop:LinkCountingCylinder} is obtained by using the $x$-parameters 
instead to define the steps. The corresponding lattice path thus ends at $(nm,d)$, with steps $(1,x_1)$, $(1,x_2)$, \dots, $(1,x_{nm})$ 
along the way. For instance, the $x$-path refinement of the $y$-path in the example above is given by
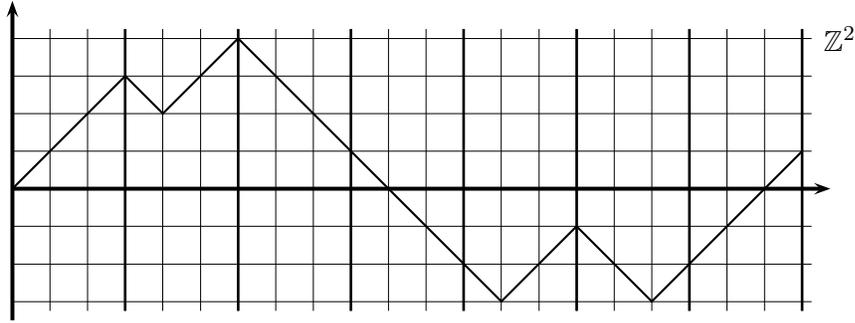
\begin{equation*}
\psset{unit=0.5}
\begin{pspicture}(0,-3)(21,5)
\psline[linewidth=1.5pt]{->}(0,0)(21.75,0)
\psline[linewidth=1.5pt]{->}(0,-3.5)(0,5)
\psline{-}(0,0)(3,3)(4,2)(6,4)(13,-3)(15,-1)(17,-3)(21,1)
\multiput(0,-3)(0,1){8}{\psline[linewidth=0.25pt]{-}(0,0)(21.25,0)}
\multiput(1,0)(1,0){21}{\psline[linewidth=0.25pt]{-}(0,-3.25)(0,4.25)}
\multiput(3,0)(3,0){7}{\psline[linewidth=1pt]{-}(0,-3.25)(0,4.25)}
\rput(22,4){$\mathbb Z^2$}
\end{pspicture}
\end{equation*}

\begin{Proposition}
For $N,m\in\mathbb{N}$ and $d\in\mathbb{N}_0$ subject to (\ref{dNm}), the number of linearly independent $m$-cabled
link states on $N$ nodes with $d$ defects on the strip is given by
\be
 \dim V_{N,m}^d=\left(\!\!\!\begin{array}{c} N\\[.1cm] \frac{Nm-d}{2}\end{array}\!\!\!\right)_{\!\!m}
   -\left(\!\!\!\begin{array}{c} N\\[.1cm] \frac{Nm-d-2}{2}\end{array}\!\!\!\right)_{\!\!m}.
\label{eq:LinkCountingStrip}
\ee
\label{prop:LinkCountingStrip}
\end{Proposition}
\noindent{\scshape Proof:}
The result trivially holds for $d = Nm$. For $d<Nm$, let $\tilde C_{N,m}^d$ denote the linear span of
the set of cabled link states on the cylinder with loop segments crossing the virtual boundary. The number of linearly independent
cabled link states on the strip is then given by
\be
 \dim V_{N,m}^d=\dim \tilde V_{N,m}^d-\dim \tilde C_{N,m}^d.
\ee
There is a simple bijection between the canonical basis of $\tilde C_{N,m}^d$ and that of $\tilde V_{N,m}^{d+2}$.
A link state in the basis of $\tilde C_{N,m}^d$ contains at least one half-arc crossing the virtual boundary. 
By identifying the top one and cutting it into two defects, we produce a unique link state in the basis of $\tilde V_{N,m}^{d+2}$.
Noting that every link state in the basis of $\tilde V_{N,m}^{d+2}$
contains at least two defects, the inverse map follows by replacing
the left- and rightmost defects of a given link state in $\tilde V_{N,m}^{d+2}$ by a half-arc linking the 
corresponding two nodes via the back of the cylinder, thereby producing a unique link state in the basis of $\tilde C_{N,m}^d$.
This bijection implies that $\dim\tilde C_{N,m}^d=\dim\tilde V_{N,m}^{d+2}$ and hence (\ref{eq:LinkCountingStrip}).
\hfill $\square$
\medskip

In terms of $x$-sequences, the cabled link states on the strip are characterised as on the cylinder, but 
with the additional constraint that only nonnegative partial sums can appear,
\be
 \sum_{i=1}^{t}x_i\geq0, \qquad   \forall\, t\in\{1, \dots, Nm\}.
\label{x>0}
\ee
Viewed as a refined lattice path, an $x$-path can thus not extend under the horizontal axis.
In terms of $y$-sequences, this constraint translates into 
\be
 \mathfrak{h}_s\geq0,\qquad \mathfrak{h}_{s-1} + \mathfrak{h}_{s} \ge m,\qquad s=1,\ldots,N
\ee
where the height $\mathfrak{h}_s$ after $s$ steps is defined by
\be
 \mathfrak{h}_0:= 0,\qquad \mathfrak{h}_s:= \sum_{k=1}^s y_k,\qquad s=1,\ldots,N.
\ee
The corresponding $y$-paths are those in $Y_{N,m}^d$ that do not extend below the horizontal axis and for which every pair of 
consecutive heights sum to $m$ or more.
For $m=2$, the $y$-paths on the strip are the generalised Riordan paths discussed in~\cite{PRT2013}.
In this case, \eqref{x>0} implies that these lattice paths cannot remain at height $0$ for two (or more) consecutive steps.

%%%%%%%%%%%%%%%%%%%%%%%
\section{Fusion hierarchies}
\label{app:FusHier}
%%%%%%%%%%%%%%%%%%%%%%%

%%%%%%%%%%%%%%%%
\subsection{On the strip}
\label{app:FusHierStrip}
%%%%%%%%%%%%%%%

Here we recall and prove Proposition~\ref{prop:FusHierD}.

\bigskip

\noindent{\scshape Proposition~\ref{prop:FusHierD}} {\em On the strip, the fusion hierarchy for $m,n\in\mathbb{N}$ is given by}
\bea 
\Db^{m,n}_0 \Db^{m,1}_n &\!\!\!=\!\!\!& \frac{s_{n}(2u-\mu)s_{2n-1}(2u-\mu)}{s_{n-1}(2u-\mu) s_{2n}(2u-\mu)}\,  \Db^{m,n+1}_0   \nn
  &&\hspace{2cm}+\,\frac{s_{n-2}(2u-\mu) s_{2n+1}(2u-\mu)}{s_{n-1}(2u-\mu) s_{2n}(2u-\mu)}
   \big[q^m(u_n)q^m(\mu-u_{n-1})\big]^{\!N}\!\Db^{m,n-1}_{0} 
\label{eq:DmHierApp}
\eea
{\em where $q^m(u)$ is defined in \eqref{eq:qkm}.}
\medskip

\noindent {\scshape Proof:}
Equation \eqref{eq:DmHierApp} will be obtained after applying planar transformations to the tangle $\Db^{m,n+1}_0$. 
For arbitrary $n$, this requires studying diagrams with many layers of $(m,1)$- and $(1,m)$-fused faces. 
We start by noting that the annihilating property \eqref{eq:(ii)} of the WJ projectors allows us to write the left and right boundaries of 
$\Db^{m,n+1}_0$ as 
\be 
\psset{unit=0.55}
 
\label{LRboundaries}
\ee
where the functions
\be
 \alpha_L^{
\psset{unit=0.18cm}
%
}
 = \prod_{i=0}^{n-1}s_0(v_{i})= \prod_{j=n}^{2n-1}s_{-j}(\mu-2u)
 \label{eq:alphaLR}
 \ee
are such that the configuration with all diamond face operators 
given by 
$\psset{unit=0.18cm}
%
$
has weight $1$. The order of the spectral parameters in the diamond boxes in \eqref{LRboundaries} ensures that half-arc propagation in the bulk and 
boundary of the transfer tangle is of the form\, 
$%
$\,. Also, \eqref{LRboundaries} holds for all $v \in \mathbb C$, but setting
\be
 v:= \mu-2u-(2n-1)\lambda
\ee 
will permit the use of the YBE below.
Repeating the argument used in \eqref{eq:transfermatrix2} to write $\Db^{m,n}(u)$ with a single projector, 
we reexpress $\Db^{m,n+1}_0$ as
\be 
\psset{unit=0.85}
 \alpha_L^{
\psset{unit=0.18cm}
%
\ee
Transforming $P_{n+1}$ using its recursive definition \eqref{eq:Prec} yields two terms, the first of which is
\be 
\psset{unit=0.85}
%
 \ = \eta(u,\mu) \, \Db^{m,n}_0 \Db^{m,1}_n,
\ee
where at the first equality, $P_n$ projectors have been reinserted to produce the $(m,n)$-fused faces inversion identity and 
where \eqref{eq:fusedinv} was used at the last equality, with the ensuing factor
\be  
 \eta(u,\mu) = g^{n,1}\big(2u+(2n+1)\lambda-\mu\big) 
  =\Big(\prod_{i=n+1}^{2n} s_i(2u-\mu)\Big)\Big(\prod_{j=n-1}^{2n-2} s_{-j}(\mu-2u)\Big).
\ee
The second term in the decomposition of $
\psset{unit=0.65}
 \alpha_L^{
\psset{unit=0.18cm}
%
\label{eq:sometangle}
\ee
and the next step is to expand the leftmost $P_n$ projector in a linear combination of elementary tangles. Except for the one with $n$ 
horizontal strands, all connectivities have half-arcs propagating as \ 
$%
$  
\ and are annihilated by the second projector by property \eqref{eq:(ii)}. As a consequence, this leftmost projector can be removed. 

Then, using the identity 
\be
\psset{unit=.6}
%
\ee
and property \eqref{eq:(ii)} of $P_n$, it follows that
\be 
\psset{unit=0.55}
%
\qquad \qquad \gamma_L = s_{2-n}(\mu\!-\!2u)\prod_{j=n}^{2n-2}s_{-j}(\mu\!-\!2u)
\label{eq:boundarytrick}
\ee
and that the tangle \eqref{eq:sometangle} can be reexpressed as 
\be 
\psset{unit=0.85}
-\frac{s_{n}(0)}{s_{n+1}(0)} \ 
%
\nonumber
\ee
\be        
 = -\gamma_L\gamma_R \big[q^m(u_{n})q^m(\mu-u_{n-1})\big]^N \Db^{m,n-1}_0
\ee
One should note that in \eqref{eq:E15}, the idea underlying \eqref{eq:boundarytrick} was used for the right boundary, with
\be 
 \gamma_R = s_{2n+1}(2u\!-\!\mu) \prod_{j=n+1}^{2n-1}s_j(2u\!-\!\mu),
\ee
and that property \eqref{eq:(iv)} of the WJ projectors was used at the last equality. 
Finally, 
\be
 \alpha_L^{
 \psset{unit=0.18cm}
%
}
\Db^{m,n+1}_0  
= \eta(u,\mu) \Db^{m,n}_0\Db^{m,1}_n - \gamma_L\gamma_R \big[q^m(u_{n})q^m(\mu-u_{n-1})\big]^{\!N}\Db^{m,n-1}_0
\label{eq:almostfinal}
\ee
which leads to \eqref{eq:DmHierApp} after simplifications of the trigonometric functions. 
As a final remark, we note that all the 
arguments go through for $n=1$ as well, with the final result given by \eqref{eq:almostfinal} with $\Db^{m,0}_0$ replaced by $\Ib^m$.
\hfill $\square$

%%%%%%%%%%%%%%%%
\subsection{On the cylinder}
\label{app:FusHierCylinder}
%%%%%%%%%%%%%%%

Here we recall and prove Proposition~\ref{prop:TmHier}.
\bigskip

\noindent{\scshape Proposition~\ref{prop:TmHier}}
{\em On the cylinder, the fusion hierarchy for $m,n\in\mathbb{N}$ is given by
\be
\Tb_0^{m,n} \Tb^{m,1}_n = \Tb_0^{m,n+1}  + h^m_nh^m_{n-2}\, \Tb_0^{m,n-1},
\label{eq:Tmclosure}
\ee
where $h^m_k$ is defined in (\ref{fkm}).
}
\medskip

\noindent {\scshape Proof:}
We use the form \eqref{eq:transfermatrix2} for $\Tb_0^{m,n+1}$ and rewrite the projector using \eqref{eq:Prec} to obtain
\be
\Tb^{m,n+1}_0 = \   
\psset{unit=0.75}
\begin{pspicture}[shift=-2.40](-0.5,0)(3.2,5.0)
\facegrid{(0,0)}{(3,5)}
\psarc[linewidth=0.025]{-}(0,0){0.16}{0}{90}
\psarc[linewidth=0.025]{-}(0,1){0.16}{0}{90}
\psarc[linewidth=0.025]{-}(0,3){0.16}{0}{90}
\psarc[linewidth=0.025]{-}(0,4){0.16}{0}{90}
\psarc[linewidth=0.025]{-}(2,0){0.16}{0}{90}
\psarc[linewidth=0.025]{-}(2,1){0.16}{0}{90}
\psarc[linewidth=0.025]{-}(2,3){0.16}{0}{90}
\psarc[linewidth=0.025]{-}(2,4){0.16}{0}{90}
\psline[linecolor=blue,linewidth=1.5pt]{-}(-0.3,0.5)(-0.5,0.5)
\psline[linecolor=blue,linewidth=1.5pt]{-}(-0.3,1.5)(-0.5,1.5)
\psline[linecolor=blue,linewidth=1.5pt]{-}(-0.3,2.5)(-0.5,2.5)
\psline[linecolor=blue,linewidth=1.5pt]{-}(-0.3,3.5)(-0.5,3.5)
\psline[linecolor=blue,linewidth=1.5pt]{-}(-0.0,4.5)(-0.5,4.5)
\psline[linecolor=blue,linewidth=1.5pt]{-}(3.0,0.5)(3.2,0.5)
\psline[linecolor=blue,linewidth=1.5pt]{-}(3.0,1.5)(3.2,1.5)
\psline[linecolor=blue,linewidth=1.5pt]{-}(3.0,2.5)(3.2,2.5)
\psline[linecolor=blue,linewidth=1.5pt]{-}(3.0,3.5)(3.2,3.5)
\psline[linecolor=blue,linewidth=1.5pt]{-}(3.0,4.5)(3.2,4.5)
\pspolygon[fillstyle=solid,fillcolor=pink](0,0.1)(0,3.9)(-0.3,3.9)(-0.3,0.1)(0,0.1)
\rput(-0.15,2.0){\scriptsize$_n$}
\rput(0.5,0.75){\tiny{$_{(m,1)}$}}
\rput(0.5,1.75){\tiny{$_{(m,1)}$}}
\rput(0.5,3.75){\tiny{$_{(m,1)}$}}
\rput(0.5,4.75){\tiny{$_{(m,1)}$}}
\rput(2.5,0.75){\tiny{$_{(m,1)}$}}
\rput(2.5,1.75){\tiny{$_{(m,1)}$}}
\rput(2.5,3.75){\tiny{$_{(m,1)}$}}
\rput(2.5,4.75){\tiny{$_{(m,1)}$}}
\rput(0.5,.4){\small$u_0$}
\rput(2.5,.4){\small$u_0$}
\rput(0.5,1.4){\small$u_1$}
\rput(2.5,1.4){\small$u_1$}
\rput(0.5,2.65){$\vdots$}
\rput(2.5,2.65){$\vdots$}
\rput(0.5,3.4){\scriptsize$u_{n\!-\!1}$}
\rput(2.5,3.4){\scriptsize$u_{n\!-\!1}$}
\rput(0.5,4.4){\small$u_{n}$}
\rput(2.5,4.4){\small$u_{n}$}
\rput(1.5,0.5){$\ldots$}
\rput(1.5,1.5){$\ldots$}
\rput(1.5,3.5){$\ldots$}
\rput(1.5,4.5){$\ldots$}
\end{pspicture}  \ - \frac{s_{n}(0)}{s_{n+1}(0)} \
\begin{pspicture}[shift=-2.4](-0.8,0)(3.8,5.0)
\facegrid{(0,0)}{(3,5)}
\psarc[linewidth=0.025]{-}(0,0){0.16}{0}{90}
\psarc[linewidth=0.025]{-}(0,1){0.16}{0}{90}
\psarc[linewidth=0.025]{-}(0,3){0.16}{0}{90}
\psarc[linewidth=0.025]{-}(0,4){0.16}{0}{90}
\psarc[linewidth=0.025]{-}(2,0){0.16}{0}{90}
\psarc[linewidth=0.025]{-}(2,1){0.16}{0}{90}
\psarc[linewidth=0.025]{-}(2,3){0.16}{0}{90}
\psarc[linewidth=0.025]{-}(2,4){0.16}{0}{90}
\psarc[linewidth=1.5pt,linecolor=blue]{-}(-0.3,4){0.5}{90}{-90}
\psarc[linewidth=1.5pt,linecolor=blue]{-}(3.3,4){-0.5}{90}{-90}
\psline[linecolor=blue,linewidth=1.5pt]{-}(-0.3,0.5)(-0.7,0.5)
\psline[linecolor=blue,linewidth=1.5pt]{-}(-0.3,1.5)(-0.7,1.5)
\psline[linecolor=blue,linewidth=1.5pt]{-}(-0.3,2.5)(-0.7,2.5)
\psline[linecolor=blue,linewidth=1.5pt]{-}(-0.0,4.5)(-0.3,4.5)
\psline[linecolor=blue,linewidth=1.5pt]{-}(3.3,0.5)(3.7,0.5)
\psline[linecolor=blue,linewidth=1.5pt]{-}(3.3,1.5)(3.7,1.5)
\psline[linecolor=blue,linewidth=1.5pt]{-}(3.3,2.5)(3.7,2.5)
\psline[linecolor=blue,linewidth=1.5pt]{-}(3.0,4.5)(3.3,4.5)
\pspolygon[fillstyle=solid,fillcolor=pink](0,0.1)(0,3.9)(-0.3,3.9)(-0.3,0.1)(0,0.1)
\rput(-0.15,2.0){\scriptsize$_n$}
\pspolygon[fillstyle=solid,fillcolor=pink](3,0.1)(3,3.9)(3.3,3.9)(3.3,0.1)(3,0.1)
\rput(3.15,2.0){\scriptsize$_n$}
\rput(0.5,0.75){\tiny{$_{(m,1)}$}}
\rput(0.5,1.75){\tiny{$_{(m,1)}$}}
\rput(0.5,3.75){\tiny{$_{(m,1)}$}}
\rput(0.5,4.75){\tiny{$_{(m,1)}$}}
\rput(2.5,0.75){\tiny{$_{(m,1)}$}}
\rput(2.5,1.75){\tiny{$_{(m,1)}$}}
\rput(2.5,3.75){\tiny{$_{(m,1)}$}}
\rput(2.5,4.75){\tiny{$_{(m,1)}$}}
\rput(0.5,.4){\small$u_0$}
\rput(2.5,.4){\small$u_0$}
\rput(0.5,1.4){\small$u_1$}
\rput(2.5,1.4){\small$u_1$}
\rput(0.5,2.65){$\vdots$}
\rput(2.5,2.65){$\vdots$}
\rput(0.5,3.4){\scriptsize$u_{n\!-\!1}$}
\rput(2.5,3.4){\scriptsize$u_{n\!-\!1}$}
\rput(0.5,4.4){\small$u_{n}$}
\rput(2.5,4.4){\small$u_{n}$}
\rput(1.5,0.5){$\ldots$}
\rput(1.5,1.5){$\ldots$}
\rput(1.5,3.5){$\ldots$}
\rput(1.5,4.5){$\ldots$}
\end{pspicture}
\ee
The first term is readily recognised as $\Tb_0^{m,n}\Tb_n^{m,1}$\!. For the second term, we expand the rightmost projector in 
connectivities and find that only the identity connectivity contributes nontrivially. All other connectivities have half-arcs that push
through towards the left and are annihilated when reaching the leftmost projector. The second term can then be written as 
\be
- \frac{s_{n}(0)}{s_{n+1}(0)}    
\psset{unit=0.75}
\begin{pspicture}[shift=-2.4](-1.0,0)(3.55,5.0)
\facegrid{(0,0)}{(3,5)}
\psarc[linewidth=0.025]{-}(0,0){0.16}{0}{90}
\psarc[linewidth=0.025]{-}(0,1){0.16}{0}{90}
\psarc[linewidth=0.025]{-}(0,3){0.16}{0}{90}
\psarc[linewidth=0.025]{-}(0,4){0.16}{0}{90}
\psarc[linewidth=0.025]{-}(2,0){0.16}{0}{90}
\psarc[linewidth=0.025]{-}(2,1){0.16}{0}{90}
\psarc[linewidth=0.025]{-}(2,3){0.16}{0}{90}
\psarc[linewidth=0.025]{-}(2,4){0.16}{0}{90}
\psarc[linewidth=1.5pt,linecolor=blue]{-}(-0.3,4){0.5}{90}{-90}
\psarc[linewidth=1.5pt,linecolor=blue]{-}(3.0,4){-0.5}{90}{-90}
\psline[linecolor=blue,linewidth=1.5pt]{-}(-0.3,0.5)(-0.7,0.5)
\psline[linecolor=blue,linewidth=1.5pt]{-}(-0.3,1.5)(-0.7,1.5)
\psline[linecolor=blue,linewidth=1.5pt]{-}(-0.3,2.5)(-0.7,2.5)
\psline[linecolor=blue,linewidth=1.5pt]{-}(-0.0,4.5)(-0.3,4.5)
\psline[linecolor=blue,linewidth=1.5pt]{-}(3.0,0.5)(3.4,0.5)
\psline[linecolor=blue,linewidth=1.5pt]{-}(3.0,1.5)(3.4,1.5)
\psline[linecolor=blue,linewidth=1.5pt]{-}(3.0,2.5)(3.4,2.5)
\pspolygon[fillstyle=solid,fillcolor=pink](0,0.1)(0,3.9)(-0.3,3.9)(-0.3,0.1)(0,0.1)
\rput(-0.15,2.0){\scriptsize$_n$}
\rput(0.5,0.75){\tiny{$_{(m,1)}$}}
\rput(0.5,1.75){\tiny{$_{(m,1)}$}}
\rput(0.5,3.75){\tiny{$_{(m,1)}$}}
\rput(0.5,4.75){\tiny{$_{(m,1)}$}}
\rput(2.5,0.75){\tiny{$_{(m,1)}$}}
\rput(2.5,1.75){\tiny{$_{(m,1)}$}}
\rput(2.5,3.75){\tiny{$_{(m,1)}$}}
\rput(2.5,4.75){\tiny{$_{(m,1)}$}}
\rput(0.5,.4){\small$u_0$}
\rput(2.5,.4){\small$u_0$}
\rput(0.5,1.4){\small$u_1$}
\rput(2.5,1.4){\small$u_1$}
\rput(0.5,2.65){$\vdots$}
\rput(2.5,2.65){$\vdots$}
\rput(0.5,3.4){\scriptsize$u_{n\!-\!1}$}
\rput(2.5,3.4){\scriptsize$u_{n\!-\!1}$}
\rput(0.5,4.4){\small$u_{n}$}
\rput(2.5,4.4){\small$u_{n}$}
\rput(1.5,0.5){$\ldots$}
\rput(1.5,1.5){$\ldots$}
\rput(1.5,3.5){$\ldots$}
\rput(1.5,4.5){$\ldots$}
\end{pspicture} 
= - \frac{s_{n}(0)}{s_{n+1}(0)}  \big[q^m(u_n)\big]^N\
\begin{pspicture}[shift=-2.4](-0.8,0)(3.45,5.0)
\psarc[linewidth=4pt,linecolor=blue]{-}(3.0,4){-0.5}{90}{-90}\psarc[linewidth=2pt,linecolor=white]{-}(3.0,4){-0.5}{90}{-90}
\facegrid{(0,0)}{(3,5)}
\psarc[linewidth=0.025]{-}(0,0){0.16}{0}{90}
\psarc[linewidth=0.025]{-}(0,2){0.16}{0}{90}
\psarc[linewidth=0.025]{-}(2,0){0.16}{0}{90}
\psarc[linewidth=0.025]{-}(2,2){0.16}{0}{90}
\psarc[linewidth=1.5pt,linecolor=blue]{-}(-0.3,4){0.5}{90}{-90}
\psarc[linewidth=1.5pt,linecolor=blue]{-}(0,4){-0.5}{90}{-90}
\psline[linecolor=blue,linewidth=1.5pt]{-}(-0.3,0.5)(-0.7,0.5)
\psline[linecolor=blue,linewidth=1.5pt]{-}(-0.3,1.5)(-0.7,1.5)
\psline[linecolor=blue,linewidth=1.5pt]{-}(-0.3,2.5)(-0.7,2.5)
\psline[linecolor=blue,linewidth=1.5pt]{-}(-0.0,4.5)(-0.3,4.5)
\psline[linecolor=blue,linewidth=1.5pt]{-}(3.0,0.5)(3.4,0.5)
\psline[linecolor=blue,linewidth=1.5pt]{-}(3.0,1.5)(3.4,1.5)
\psline[linecolor=blue,linewidth=1.5pt]{-}(3.0,2.5)(3.4,2.5)
\pspolygon[fillstyle=solid,fillcolor=pink](0,0.1)(0,3.9)(-0.3,3.9)(-0.3,0.1)(0,0.1)
\rput(-0.15,2.0){\scriptsize$_n$}
\rput(0.5,0.75){\tiny{$_{(m,1)}$}}
\rput(0.5,2.75){\tiny{$_{(m,1)}$}}
\rput(2.5,0.75){\tiny{$_{(m,1)}$}}
\rput(2.5,2.75){\tiny{$_{(m,1)}$}}
\rput(0.5,.4){\small$u_0$}
\rput(2.5,.4){\small$u_0$}
\rput(0.5,1.65){$\vdots$}
\rput(2.5,1.65){$\vdots$}
\rput(0.5,2.4){\scriptsize$u_{n\!-\!2}$}
\rput(2.5,2.4){\scriptsize$u_{n\!-\!2}$}
\psarc[linewidth=4pt,linecolor=blue]{-}(1,3){0.5}{90}{180}\psarc[linewidth=2pt,linecolor=lightlightblue]{-}(1,3){0.5}{90}{180}
\psarc[linewidth=4pt,linecolor=blue]{-}(1,5){0.5}{180}{-90}\psarc[linewidth=2pt,linecolor=lightlightblue]{-}(1,5){0.5}{180}{-90}
\psframe[linewidth=.40pt](0,3)(1,5)
\rput(1.0,4.0){\floopb}\rput(1.0,3.0){\floopa}
\rput(2.0,4.0){\floopb}\rput(2.0,3.0){\floopa}
\rput(1.5,0.5){$\ldots$}
\rput(1.5,2.5){$\ldots$}
\end{pspicture}
\nonumber
\ee
\be            
\psset{unit=0.75}
=-\big[q^m(u_n)\big]^N
\begin{pspicture}[shift=-1.4](-1.1,0)(3.6,3.2)
\facegrid{(0,0)}{(3,3)}
\psarc[linewidth=0.025]{-}(0,0){0.16}{0}{90}
\psarc[linewidth=0.025]{-}(0,2){0.16}{0}{90}
\psarc[linewidth=0.025]{-}(2,0){0.16}{0}{90}
\psarc[linewidth=0.025]{-}(2,2){0.16}{0}{90}
\psline[linecolor=blue,linewidth=1.5pt]{-}(-0.6,0.5)(-0.8,0.5)
\psline[linecolor=blue,linewidth=1.5pt]{-}(-0.6,1.5)(-0.8,1.5)
\psline[linecolor=blue,linewidth=1.5pt]{-}(-0.6,2.5)(-0.8,2.5)
\psline[linecolor=blue,linewidth=1.5pt]{-}(3.0,0.5)(3.2,0.5)
\psline[linecolor=blue,linewidth=1.5pt]{-}(3.0,1.5)(3.2,1.5)
\psline[linecolor=blue,linewidth=1.5pt]{-}(3.0,2.5)(3.2,2.5)
\pspolygon[fillstyle=solid,fillcolor=pink](0,0.1)(0,2.9)(-0.6,2.9)(-0.6,0.1)(0,0.1)
\rput(-0.3,1.5){\tiny$_{n\!-\!1}$}
\rput(0.5,0.75){\tiny{$_{(m,1)}$}}
\rput(0.5,2.75){\tiny{$_{(m,1)}$}}
\rput(2.5,0.75){\tiny{$_{(m,1)}$}}
\rput(2.5,2.75){\tiny{$_{(m,1)}$}}
\rput(0.5,.4){\small$u_0$}
\rput(2.5,.4){\small$u_0$}
\rput(0.5,1.65){$\vdots$}
\rput(2.5,1.65){$\vdots$}
\rput(0.5,2.4){\scriptsize$u_{n\!-\!2}$}
\rput(2.5,2.4){\scriptsize$u_{n\!-\!2}$}
\end{pspicture} 
=-h^m_n h^m_{n-2}\, \Tb_0^{1,n-1}
\ee
as required. Note that property \eqref{eq:(v)} of the WJ projectors was used at the second equality and 
\be         
 \big[q^m(u_n)\big]^N = h^m_n h^m_{n-2}
\ee 
at the last one. 
\hfill $\square$

%%%%%%%%%%%%%%%%%%
\section{Closure of the fusion hierarchies}
\label{app:Closure}
%%%%%%%%%%%%%%%%%%

%%%%%%%%%%%%%%%%%%%%%%%%%%%%%%%
\subsection{Preliminaries}
\label{sec:prelims}
%%%%%%%%%%%%%%%%%%%%%%%%%%%%%%%

For every integer $n\geq2$ and all $\lambda\in\pi\big(\mathbb{R}\!\setminus\!\mathbb{Z}\big)$, 
the fusion hierarchy equations of Section~\ref{Sec:FTY} (proved above in Appendix \ref{app:FusHier}) 
express $(m,n)$-fused transfer tangles in terms of products and sums of $(m,n')$-fused transfer tangles with $n'<n$. In this appendix, 
we show that, for fractional $\lambda=\lambda_{p,p'}$, the fusion hierarchy closes both on the strip and cylinder, 
and find that the $(m,p')$-fused transfer tangle can be expressed as a linear combination of $(m,n')$-fused transfer tangles with $n'<p'$. 
For convenience, in the remainder of this appendix, we parameterise $p'$ as
\be 
 p' =  \ell + 1.
\ee

As will soon become evident, the fusion closure is made possible by properties specific to fractional values of $\lambda$. 
One such property is that the local Boltzmann weights satisfy
\be 
 s_{i + \ell + 1}(u) = (-1)^{\ell+1-p}s_{i}(u), \qquad s_{\ell + 1}(0) = 0
\label{eq:critconds}
\ee 
from which it follows that the following four properties hold:
\begin{itemize}
\item[(a)] $\psset{unit=.8cm}
\begin{pspicture}[shift=-.42](1,1)
\facegrid{(0,0)}{(1,1)}
\psarc[linewidth=0.025]{-}(0,0){0.16}{0}{90}
\rput(.5,.5){$u_\ell$}
\end{pspicture}
\ =\ (-1)^{\ell+1-p}\ 
\begin{pspicture}[shift=-.42](1,1)
\facegrid{(0,0)}{(1,1)}
\psarc[linewidth=0.025]{-}(0,0){0.16}{0}{90}
\rput(.5,.5){$u_{-\!1}$}
\end{pspicture}
$
\item[(b)] $P_{\ell+1}$ does not exist.
\item[(c)] $\begin{pspicture}[shift=-0.2](0,-0.3)(1.40,0.3)
\pspolygon[fillstyle=solid,fillcolor=pink](0,-0.15)(1.2,-0.15)(1.2,0.15)(0,0.15)(0,-0.15)
\rput(0.6,0){$_{\ell}$}
\psarc[linecolor=blue,linewidth=1.5pt]{-}(1.25,0.15){0.15}{0}{180}
\psline[linecolor=blue,linewidth=1.5pt]{-}(1.40,0.15)(1.40,-0.15)
\psarc[linecolor=blue,linewidth=1.5pt]{-}(1.25,-0.15){-0.15}{0}{180}
\end{pspicture}
\  = 0.
$
\item[(d)] The recursive definition of $P_\ell$ in terms of $P_{\ell-1}$ can be simplified to
\be
\begin{pspicture}[shift=-0.05](-0.03,-0.15)(1.2,0.15)
\pspolygon[fillstyle=solid,fillcolor=pink](0,-0.15)(1.2,-0.15)(1.2,0.15)(0,0.15)(0,-0.15)
\rput(0.6,0){$_{\ell}$}
\end{pspicture}
\ = \ 
\begin{pspicture}[shift=-0.05](-0.03,-0.15)(1.2,0.15)
\pspolygon[fillstyle=solid,fillcolor=pink](0,-0.15)(1.0,-0.15)(1.0,0.15)(0,0.15)(0,0.15)
\rput(0.5,0){$_{\ell-1}$}
\psline[linecolor=blue,linewidth=1.5pt]{-}(1.15,-0.17)(1.15,0.17)
\end{pspicture} 
\ -\,\beta \ \,
\begin{pspicture}[shift=-0.45](0,-0.55)(1.24,0.55)
\pspolygon[fillstyle=solid,fillcolor=pink](0,-0.25)(1.0,-0.25)(1.0,-0.55)(0,-0.55)(0,-0.25)
\pspolygon[fillstyle=solid,fillcolor=pink](0,0.25)(1.0,0.25)(1.0,0.55)(0,0.55)(0,0.25)
\rput(0.5,0.4){$_{\ell-1}$}
\rput(0.5,-0.4){$_{\ell-1}$}
\psline[linecolor=blue,linewidth=1.5pt]{-}(0.1,-0.25)(0.1,0.25)
\psline[linecolor=blue,linewidth=1.5pt]{-}(0.25,-0.25)(0.25,0.25)
\rput(0.5,0){$_{\dots}$}
\psline[linecolor=blue,linewidth=1.5pt]{-}(0.75,-0.25)(0.75,0.25)
\psarc[linecolor=blue,linewidth=1.5pt]{-}(1.05,-0.25){0.15}{0}{180}
\psarc[linecolor=blue,linewidth=1.5pt]{-}(1.05,0.25){0.15}{180}{0}
\psline[linecolor=blue,linewidth=1.5pt]{-}(1.2,0.25)(1.2,0.55)
\psline[linecolor=blue,linewidth=1.5pt]{-}(1.2,-0.25)(1.2,-0.55)
\end{pspicture}
\label{eq:rationalrec}
\ee
\end{itemize}

Even though $P_{\ell+1}$ is singular, the limits \eqref{eq:existinglimits1} of the 
transfer tangles $\Db^{m,\ell+1}_k$ and $\Tb^{m,\ell+1}_k$ still exist for $m<p$
and can for example be defined in terms of effective projectors, as explained in Section~\ref{sec:Dcrit}.
Here the closure relations will be shown using only diagrammatic objects which are well defined for $\lambda=\lambda_{p,\ell+1}$.
Indeed, the starting points of the proofs will be $\Db^{m,\ell}_0\Db^{m,1}_{\ell}$ and $\Tb^{m,\ell}_0\Tb^{m,1}_{\ell}$ 
instead of $\Db^{m,\ell+1}_0$ and $\Tb^{m,\ell+1}_0$. Equations \eqref{eq:closure} and \eqref{eq:Tclosurem} are just concise rewritings 
of the results we obtain, using the fusion hierarchy equations \eqref{eq:DmHier} and \eqref{eq:TmHier}. 
For simplicity, we will first prove the closure for $m=1$ and then generalise to $m>1$. 
For $m=1$, the tangles $\Db^{1,\ell}_0\Db^{1,1}_{\ell}$ and $\Tb^{1,\ell}_0\Tb^{1,1}_{\ell}$ will be written in terms of columns of 
elementary face operators, to which projectors are glued, of the form
\be
\psset{unit=.8}
\begin{pspicture}[shift=-2.4](-0.3,0)(1.3,5)
\pspolygon[fillstyle=solid,fillcolor=pink](0,0.1)(0,3.9)(-0.3,3.9)(-0.3,0.1)(0,0.1)
\rput(-0.15,2.0){$_\ell$}
\facegrid{(0,0)}{(1,5)}
\psarc[linewidth=0.025]{-}(0,0){0.16}{0}{90}
\psarc[linewidth=0.025]{-}(0,1){0.16}{0}{90}
\psarc[linewidth=0.025]{-}(0,3){0.16}{0}{90}
\psarc[linewidth=0.025]{-}(0,4){0.16}{0}{90}
\rput(.5,.5){\small $u_0$}
\rput(.5,1.5){\small $u_1$}
\rput(.5,2.6){\small $\vdots$}
\rput(.5,3.5){\small $u_{\ell\!-\!1}$}
\rput(.5,4.5){\small $u_{\ell}$}
\end{pspicture}
\ = \ 
\begin{pspicture}[shift=-2.4](-0.3,0)(1.3,5)
\rput(1.3,0){\pspolygon[fillstyle=solid,fillcolor=pink](0,0.1)(0,3.9)(-0.3,3.9)(-0.3,0.1)(0,0.1)
\rput(-0.15,2.0){$_\ell$}}
\facegrid{(0,0)}{(1,5)}
\psarc[linewidth=0.025]{-}(0,0){0.16}{0}{90}
\psarc[linewidth=0.025]{-}(0,1){0.16}{0}{90}
\psarc[linewidth=0.025]{-}(0,3){0.16}{0}{90}
\psarc[linewidth=0.025]{-}(0,4){0.16}{0}{90}
\rput(.5,3.5){\small $u_0$}
\rput(.5,2.5){\small $u_1$}
\rput(.5,1.6){\small $\vdots$}
\rput(.5,0.5){\small $u_{\ell\!-\!1}$}
\rput(.5,4.5){\small $u_{\ell}$}
\end{pspicture}
\label{eq:Pprop}
\ee 
Because
\be
\psset{unit=.8cm}
\begin{pspicture}[shift=-0.9](-0.5,0.0)(1,2)
\facegrid{(0,0)}{(1,2)}
\psarc[linewidth=0.025]{-}(0,0){0.16}{0}{90}
\psarc[linewidth=0.025]{-}(0,1){0.16}{0}{90}
\psarc[linewidth=1.5pt,linecolor=blue]{-}(0,1){0.5}{90}{-90}
\rput(0.5,.5){$u_0$}
\rput(0.5,1.5){$u_\ell$}
\end{pspicture} \ = (-1)^{\ell+1-p} q^1(u)\ 
\begin{pspicture}[shift=-0.9](-0.5,0.0)(1,2)
\facegrid{(0,0)}{(1,2)}
\psarc[linewidth=1.5pt,linecolor=blue]{-}(0,1){0.5}{90}{-90}
\psarc[linewidth=1.5pt,linecolor=blue]{-}(0,0){0.5}{0}{90}
\psarc[linewidth=1.5pt,linecolor=blue]{-}(1,1){0.5}{90}{-90}
\psarc[linewidth=1.5pt,linecolor=blue]{-}(0,2){0.5}{-90}{0}
\end{pspicture} \ \ 
\label{eq:pushthru0n}
\ee
going from one form to the other in \eqref{eq:Pprop} reverses the order of half-arc propagation for the full column of faces, from 
\begin{pspicture}(0.0,.25)(0.5,0.4)
\psline[linewidth=0.7pt]{->}(0.5,0.35)(0,0.35)
\end{pspicture}
to
\begin{pspicture}(0.0,.25)(0.5,0.4)
\psline[linewidth=0.7pt]{->}(0,0.35)(0.5,0.35)
\end{pspicture}. 
This property will play a key role in the following.

We conclude these preliminaries with the following lemma.

\begin{Lemma} 
For $\lambda = \lambda_{p,\ell+1}$, the $(\ell+2)$-tangle defined by 
\be
\psset{unit=0.65cm}
\begin{pspicture}[shift=-2.5](0,-0.1)(1,4.9)
\pspolygon[fillstyle=solid,fillcolor=lightlightblue](0,0)(1,0)(1,5)(0,5)(0,0)
\psline[linewidth=1.25pt]{-}(0.1,0.1)(0.9,0.1)(0.9,4.9)(0.1,4.9)(0.1,0.1)
\end{pspicture} \ \,  := \ \ 
\begin{pspicture}[shift=-2.5](0,-0.1)(1,4.9)
\facegrid{(0,0)}{(1,5)}
\rput(0,0){\loopb}
\rput(0,1){\loopb}
\rput(0,2){\loopb}
\rput(0,3){\loopb}
\rput(0,4){\loopb}
\end{pspicture} \ \ + (-1)^p \ \left.
\begin{pspicture}[shift=-2.5](0,-0.1)(1,4.9)
\facegrid{(0,0)}{(1,5)}
\rput(0,0){\loopa}
\rput(0,1){\loopa}
\rput(0,2){\loopa}
\rput(0,3){\loopa}
\rput(0,4){\loopa}
\end{pspicture} 
\ \ \right\} \textrm{\scriptsize$\ell\!+\!1$}
\label{eq:bigtile}
\ee
satisfies
\be
\psset{unit=0.65cm}
\begin{pspicture}[shift=-2.5](-0.3,0)(1.5,5.15)
\pspolygon[fillstyle=solid,fillcolor=lightlightblue](0,0)(1,0)(1,5)(0,5)(0,0)
\psline[linewidth=1.25pt]{-}(0.1,0.1)(0.9,0.1)(0.9,4.9)(0.1,4.9)(0.1,0.1)
\pspolygon[fillstyle=solid,fillcolor=pink](0,0.1)(0,3.9)(-0.3,3.9)(-0.3,0.1)(0,0.1)
\pspolygon[fillstyle=solid,fillcolor=pink](1,0.1)(1,3.9)(1.3,3.9)(1.3,0.1)(1,0.1)
\psarc[linecolor=blue,linewidth=1.5pt]{-}(1.3,4){0.5}{-90}{90}
\psline[linecolor=blue,linewidth=1.5pt]{-}(1,4.5)(1.3,4.5)
\rput(-0.15,2.0){\scriptsize $_\ell$}
\rput(1.15,2.0){\scriptsize $_\ell$}
\end{pspicture} 
\ \  = - \beta  \ \ 
\begin{pspicture}[shift=-2.5](-0.3,0)(1.5,5.15)
\facegrid{(0,0)}{(1,5)}
\pspolygon[fillstyle=solid,fillcolor=pink](0,0.1)(0,3.9)(-0.3,3.9)(-0.3,0.1)(0,0.1)
\pspolygon[fillstyle=solid,fillcolor=pink](1,0.1)(1,2.9)(1.6,2.9)(1.6,0.1)(1,0.1)
\psarc[linecolor=blue,linewidth=1.5pt]{-}(1.0,4){0.5}{-90}{90}
\rput(-0.15,2.0){\scriptsize $_\ell$}
\rput(1.3,1.5){\scriptsize $_{\ell\!-\!1}$}
\rput(0,0){\loopb}
\rput(0,1){\loopb}
\rput(0,2){\loopb}
\rput(0,3){\loopa}
\rput(0,4){\loopb}
\end{pspicture} \qquad \qquad \qquad \qquad
\begin{pspicture}[shift=-2.5](-0.3,0)(1.3,5.15)
\pspolygon[fillstyle=solid,fillcolor=lightlightblue](0,0)(1,0)(1,5)(0,5)(0,0)
\psline[linewidth=1.25pt]{-}(0.1,0.1)(0.9,0.1)(0.9,4.9)(0.1,4.9)(0.1,0.1)
\pspolygon[fillstyle=solid,fillcolor=pink](0,0.1)(0,3.9)(-0.3,3.9)(-0.3,0.1)(0,0.1)
\pspolygon[fillstyle=solid,fillcolor=pink](1,0.1)(1,3.9)(1.3,3.9)(1.3,0.1)(1,0.1)
\psarc[linecolor=blue,linewidth=1.5pt]{-}(-0.3,4){-0.5}{-90}{90}
\psline[linecolor=blue,linewidth=1.5pt]{-}(0,4.5)(-0.3,4.5)
\rput(-0.15,2.0){\scriptsize $_\ell$}
\rput(1.15,2.0){\scriptsize $_\ell$}
\end{pspicture} 
\ \  = (-1)^{p+1} \beta  \ \ \
\begin{pspicture}[shift=-2.5](-0.5,0)(1.3,5.15)
\facegrid{(0,0)}{(1,5)}
\pspolygon[fillstyle=solid,fillcolor=pink](1,0.1)(1,3.9)(1.3,3.9)(1.3,0.1)(1,0.1)
\pspolygon[fillstyle=solid,fillcolor=pink](0,0.1)(0,2.9)(-0.6,2.9)(-0.6,0.1)(0,0.1)
\psarc[linecolor=blue,linewidth=1.5pt]{-}(0.0,4){-0.5}{-90}{90}
\rput(1.15,2.0){\scriptsize $_\ell$}
\rput(-0.3,1.5){\scriptsize $_{\ell\!-\!1}$}
\rput(0,0){\loopa}
\rput(0,1){\loopa}
\rput(0,2){\loopa}
\rput(0,3){\loopb}
\rput(0,4){\loopa}
\end{pspicture} 
\label{eq:otherpushthroughs}
\ee
\label{lemma:simple}
\end{Lemma}
{\scshape Proof:} 
To prove the first relation in \eqref{eq:otherpushthroughs}, we examine the two contributions coming from the expansion 
\eqref{eq:bigtile} and write the lefthand side as $X_1 + (-1)^p X_2$. For $X_1$, using 
\eqref{eq:rationalrec},
we expand the rightmost projector and find
\be
\psset{unit=0.65cm}
X_1 =  \ \ 
\begin{pspicture}[shift=-2.5](-0.3,0)(1.5,5.15)
\pspolygon[fillstyle=solid,fillcolor=lightlightblue](0,0)(1,0)(1,5)(0,5)(0,0)
\facegrid{(0,0)}{(1,5)}
\rput(0,0){\loopb}
\rput(0,1){\loopb}
\rput(0,2){\loopb}
\rput(0,3){\loopb}
\rput(0,4){\loopb}
\pspolygon[fillstyle=solid,fillcolor=pink](0,0.1)(0,3.9)(-0.3,3.9)(-0.3,0.1)(0,0.1)
\pspolygon[fillstyle=solid,fillcolor=pink](1,0.1)(1,3.9)(1.3,3.9)(1.3,0.1)(1,0.1)
\psarc[linecolor=blue,linewidth=1.5pt]{-}(1.3,4){0.5}{-90}{90}
\psline[linecolor=blue,linewidth=1.5pt]{-}(1,4.5)(1.3,4.5)
\rput(-0.15,2.0){\scriptsize $_\ell$}
\rput(1.15,2.0){\scriptsize $_\ell$}
\end{pspicture} \ \ = \ \ 
\begin{pspicture}[shift=-2.5](-0.5,0)(1.6,5.15)
\pspolygon[fillstyle=solid,fillcolor=lightlightblue](0,0)(1,0)(1,5)(0,5)(0,0)
\facegrid{(0,0)}{(1,5)}
\rput(0,0){\loopb}
\rput(0,1){\loopb}
\rput(0,2){\loopb}
\rput(0,3){\loopb}
\rput(0,4){\loopb}
\pspolygon[fillstyle=solid,fillcolor=pink](0,0.1)(0,3.9)(-0.3,3.9)(-0.3,0.1)(0,0.1)
\pspolygon[fillstyle=solid,fillcolor=pink](1,0.1)(1,2.9)(1.6,2.9)(1.6,0.1)(1,0.1)
\psarc[linecolor=blue,linewidth=1.5pt]{-}(1.0,4){0.5}{-90}{90}
\rput(-0.15,2.0){\scriptsize $_\ell$}
\rput(1.3,1.5){\scriptsize $_{\ell\!-\!1}$}
\end{pspicture} \ \ - \beta \ \ 
\begin{pspicture}[shift=-2.5](-0.5,0)(3.6,5.15)
\pspolygon[fillstyle=solid,fillcolor=lightlightblue](0,0)(1,0)(1,5)(0,5)(0,0)
\facegrid{(0,0)}{(1,5)}
\rput(0,0){\loopb}
\rput(0,1){\loopb}
\rput(0,2){\loopb}
\rput(0,3){\loopb}
\rput(0,4){\loopb}
\pspolygon[fillstyle=solid,fillcolor=pink](0,0.1)(0,3.9)(-0.3,3.9)(-0.3,0.1)(0,0.1)
\pspolygon[fillstyle=solid,fillcolor=pink](1,0.1)(1,2.9)(1.6,2.9)(1.6,0.1)(1,0.1)
\pspolygon[fillstyle=solid,fillcolor=pink](3,0.1)(3,2.9)(3.6,2.9)(3.6,0.1)(3,0.1)
\psarc[linecolor=blue,linewidth=1.5pt]{-}(1.6,3){0.5}{-90}{90}
\psline[linecolor=blue,linewidth=1.5pt]{-}(1,3.5)(1.6,3.5)
\psline[linecolor=blue,linewidth=1.5pt]{-}(1.6,0.5)(3,0.5)
\psline[linecolor=blue,linewidth=1.5pt]{-}(1.6,1.5)(3,1.5)
\psbezier[linecolor=blue,linewidth=1.5pt]{-}(1,4.5)(2.7,4.5)(2.2,2.5)(3,2.5)
\rput(-0.15,2.0){\scriptsize $_\ell$}
\rput(1.3,1.5){\scriptsize $_{\ell\!-\!1}$}
\rput(3.3,1.5){\scriptsize $_{\ell\!-\!1}$}
\end{pspicture}
\ \ =  \ \ \,
\begin{pspicture}[shift=-2.5](-0.3,0)(1.2,5.15)
\pspolygon[fillstyle=solid,fillcolor=lightlightblue](0,0)(1,0)(1,5)(0,5)(0,0)
\facegrid{(0,0)}{(1,5)}
\rput(0,0){\loopb}
\rput(0,1){\loopb}
\rput(0,2){\loopb}
\psarc[linewidth=1.5pt,linecolor=blue]{-}(0,5){0.5}{-90}{0}
\psarc[linewidth=1.5pt,linecolor=blue]{-}(0,3){0.5}{0}{90}
\pspolygon[fillstyle=solid,fillcolor=pink](0,0.1)(0,3.9)(-0.3,3.9)(-0.3,0.1)(0,0.1)
\rput(-0.15,2.0){\scriptsize $_\ell$}
\end{pspicture}  \ \ - \beta \ \ 
\begin{pspicture}[shift=-2.5](-0.5,0)(1.6,5.15)
\pspolygon[fillstyle=solid,fillcolor=lightlightblue](0,0)(1,0)(1,5)(0,5)(0,0)
\facegrid{(0,0)}{(1,5)}
\rput(0,0){\loopb}
\rput(0,1){\loopb}
\rput(0,2){\loopb}
\rput(0,3){\loopa}
\rput(0,4){\loopb}
\pspolygon[fillstyle=solid,fillcolor=pink](0,0.1)(0,3.9)(-0.3,3.9)(-0.3,0.1)(0,0.1)
\pspolygon[fillstyle=solid,fillcolor=pink](1,0.1)(1,2.9)(1.6,2.9)(1.6,0.1)(1,0.1)
\psarc[linecolor=blue,linewidth=1.5pt]{-}(1.0,4){0.5}{-90}{90}
\rput(-0.15,2.0){\scriptsize $_\ell$}
\rput(1.3,1.5){\scriptsize $_{\ell\!-\!1}$}
\end{pspicture}
\ee
where the two horizontal loop segments in blue represent a total of $\ell-2$ links. 
For $X_2$, in the expansion of the rightmost projector, only a single connectivity survives, 
\be   
\underbrace{(-1)^{\ell-1} \frac{s_1(0)}{s_{\ell}(0)}}_{=(-1)^{p+1}} \
\psset{unit=0.45}
\begin{pspicture}[shift=-2.5](-0.25,0)(2.25,5.0)
\psarc[linewidth=1.5pt, linecolor=blue]{-}(0,1){0.5}{-90}{90}
\psarc[linewidth=1.5pt, linecolor=blue]{-}(2,4){-0.5}{-90}{90}
\psbezier[linewidth=1.5pt, linecolor=blue]{-}(0,2.5)(1,2.5)(1,0.5)(2,0.5)
\rput(0.2,3.675){$\vdots$}
\rput(1.8,1.825){$\vdots$}
\psbezier[linewidth=1.5pt, linecolor=blue]{-}(0,4.5)(1,4.5)(1,2.5)(2,2.5)
\end{pspicture}
\label{eq:survivor}
\ee
and this yields    
\be
\psset{unit=0.65cm}
X_2 =  \ \
\begin{pspicture}[shift=-2.5](-0.3,0)(1.5,5.15)
\pspolygon[fillstyle=solid,fillcolor=lightlightblue](0,0)(1,0)(1,5)(0,5)(0,0)
\facegrid{(0,0)}{(1,5)}
\rput(0,0){\loopa}
\rput(0,1){\loopa}
\rput(0,2){\loopa}
\rput(0,3){\loopa}
\rput(0,4){\loopa}
\pspolygon[fillstyle=solid,fillcolor=pink](0,0.1)(0,3.9)(-0.3,3.9)(-0.3,0.1)(0,0.1)
\pspolygon[fillstyle=solid,fillcolor=pink](1,0.1)(1,3.9)(1.3,3.9)(1.3,0.1)(1,0.1)
\psarc[linecolor=blue,linewidth=1.5pt]{-}(1.3,4){0.5}{-90}{90}
\psline[linecolor=blue,linewidth=1.5pt]{-}(1,4.5)(1.3,4.5)
\rput(-0.15,2.0){\scriptsize $_\ell$}
\rput(1.15,2.0){\scriptsize $_\ell$}
\end{pspicture} \ \ = (-1)^{p+1} \ \ \begin{pspicture}[shift=-2.5](-0.3,0)(1.5,5.15)
\pspolygon[fillstyle=solid,fillcolor=lightlightblue](0,0)(1,0)(1,5)(0,5)(0,0)
\facegrid{(0,0)}{(1,5)}
\rput(0,0){\loopb}
\rput(0,1){\loopb}
\rput(0,2){\loopb}
\psarc[linewidth=1.5pt,linecolor=blue]{-}(0,5){0.5}{-90}{0}
\psarc[linewidth=1.5pt,linecolor=blue]{-}(0,3){0.5}{0}{90}
\pspolygon[fillstyle=solid,fillcolor=pink](0,0.1)(0,3.9)(-0.3,3.9)(-0.3,0.1)(0,0.1)
\rput(-0.15,2.0){\scriptsize $_\ell$}
\end{pspicture}
\ee
Two terms cancel out in $X_1 + (-1)^p X_2$ and this concludes the proof of the first identity.
The second identity in \eqref{eq:otherpushthroughs} is obtained by a left-right reflection of the first, with the factor of $(-1)^p$ 
arising due to the relative sign in \eqref{eq:bigtile}. 
\hfill $\Box$
\medskip

One should note that the proofs is this appendix are for general fractional $\lambda = \lambda_{p,\ell+1}$ and require planar 
calculations involving arrays of face operators whose heights can go up to $2(\ell+1)$. Vertical dots are usually included to indicate high 
columns, but in some cases this would make the diagrams cumbersome, and planar computations are then reported for some large 
column height instead. The previous proof is an example where this reporting
technique has been employed.

%%%%%%%%%%%%%%%%%%%%%%%%%%%%%%%
\subsection{On the strip} 
\label{app:ClosureStrip}
%%%%%%%%%%%%%%%%%%%%%%%%%%%%%%%

\begin{Proposition} 
For $m=1$, the fusion hierarchy on the strip closes as
\be         
 \Db^{1,\ell+1}_0 =  \big[q^1(u_0)q^1(\mu-u_{-1})\big]^N\Db^{1,\ell-1}_1+ 2\, (-1)^{\ell + 1-p} 
  \Big(\prod_{j=0}^{\ell}s_j(u)s_{j+1}(u-\mu)\Big)^N\Ib .
\label{eq:StripClosureApp}
\ee
\end{Proposition}
{\scshape Proof:}
The proof of \eqref{eq:StripClosureApp} is based on a set of recursion relations to be constructed in the following. 
First, we write down the relations and describe the objects they
link together, but postpone their technically involved proofs till the end.
The fusion closure we set out to prove is thus given by
\begin{align}        
 \frac{s_{2\ell}(2u\!-\!\mu)s_{\ell-1}(2u\!-\!\mu)}{s_{2\ell-1}(2u\!-\!\mu)s_{\ell}(2u\!-\!\mu)}\Db^{1,\ell}_0\Db^{1,1}_\ell 
 =\big[q^1(u_{-1})q^1(\mu\!-\!u_{-2})\big]^N\Db^{1,\ell-1}_0+\, &\big[q^1(u_0)q^1(\mu\!-\!u_{-1}) \big]^N\Db^{1,\ell-1}_1
\nonumber
\\&\hspace{-3.5cm}+ 2\,(-1)^{\ell+1-p}  
\Big(\prod_{j=0}^{\ell}s_j(u)s_{j+1}(u-\mu)\Big)^N\Ib.
\label{eq:DDA1}
\end{align}
Because no $P_{\ell+1}$ appears in this expression, all objects are well defined. 
It is also straightforward to show that \eqref{eq:DDA1} and \eqref{eq:DmHier} imply \eqref{eq:StripClosureApp}. 

The first step is to prove that the following identity holds, 
\be 
 s_{2\ell}(2u\!-\!\mu)s_{\ell-1}(2u\!-\!\mu)\Db_0^{1,\ell}\Db_\ell^{1,1} = s_{2\ell-1}(2u\!-\!\mu)s_{\ell}(2u\!-\!\mu)\, \Ab,
\label{eq:shortfuse}
\ee
where the tangle $\Ab$ is defined as
\be 
\psset{unit=0.9}
\Ab := \ 
\begin{pspicture}[shift=-4.9](-2.2,0)(7.7,10)
\multiput(0,0)(1.3,0){4}{\facegrid{(0,0)}{(1,10)}}
\multiput(0,0)(1.3,0){5}{
\pspolygon[fillstyle=solid,fillcolor=pink](0,0.1)(0,3.9)(-0.3,3.9)(-0.3,0.1)(0,0.1)
\rput(-0.15,2){\small$\ell$}
\rput(-0.3,5){\pspolygon[fillstyle=solid,fillcolor=pink](0,0.1)(0,3.9)(0.3,3.9)(0.3,0.1)(0,0.1)
\rput(0.15,2){\small${\ell}$}}
}
\psarc[linewidth=1.5pt,linecolor=blue]{-}(-0.3,5){0.5}{90}{-90}
\psbezier[linewidth=1.5pt,linecolor=blue]{-}(-0.3,3.5)(-1.3,3.5)(-1.3,6.5)(-0.3,6.5)
\psbezier[linewidth=1.5pt,linecolor=blue]{-}(-0.3,2.5)(-1.7,2.5)(-1.7,7.5)(-0.3,7.5)
\psbezier[linewidth=1.5pt,linecolor=blue]{-}(-0.3,1.5)(-2.1,1.5)(-2.1,8.5)(-0.3,8.5)
\psbezier[linewidth=1.5pt,linecolor=blue]{-}(-0.3,0.5)(-2.5,0.5)(-2.5,9.5)(-0.3,9.5)
\rput(2.2,0){
\psarc[linewidth=1.5pt,linecolor=blue]{-}(3,5){0.5}{-90}{90}
\psbezier[linewidth=1.5pt,linecolor=blue]{-}(3,3.5)(4,3.5)(4,6.5)(3,6.5)
\psbezier[linewidth=1.5pt,linecolor=blue]{-}(3,2.5)(4.4,2.5)(4.4,7.5)(3,7.5)
\psbezier[linewidth=1.5pt,linecolor=blue]{-}(3,1.5)(4.8,1.5)(4.8,8.5)(3,8.5)
\psbezier[linewidth=1.5pt,linecolor=blue]{-}(3,0.5)(5.2,0.5)(5.2,9.5)(3,9.5)}
\psline[linewidth=1.5pt,linecolor=blue]{-}(-0.3,4.5)(0,4.5)
\psline[linewidth=1.5pt,linecolor=blue]{-}(-0.3,9.5)(0,9.5)
\pspolygon[fillstyle=solid,fillcolor=pink](0,0.1)(0,3.9)(-0.3,3.9)(-0.3,0.1)(0,0.1)
\rput(-0.15,2){\small$\ell$}
\multiput(4.9,4)(0,5){2}{\psline[linewidth=1.5pt,linecolor=blue]{-}(0,0.5)(0.3,0.5)}
\multiput(0,0)(3.9,0){2}{\rput(0.5,.5){\small$u_0$}
\psarc[linewidth=0.025]{-}(0,0){0.16}{0}{90}
}
\multiput(0,1)(3.9,0){2}{\rput(0.5,.5){\small$u_1$}
\psarc[linewidth=0.025]{-}(0,0){0.16}{0}{90}
}
\multiput(0,3)(3.9,0){2}{\rput(0.5,.5){\scriptsize$u_{\ell\!-\!1}$}
\psarc[linewidth=0.025]{-}(0,0){0.16}{0}{90}
}
\multiput(0,4)(3.9,0){2}{\rput(0.5,.5){\small$u_\ell$}
\psarc[linewidth=0.025]{-}(0,0){0.16}{0}{90}
}
\multiput(0,5)(3.9,0){2}{\rput(0.5,.5){\footnotesize$u_1\!-\!\mu$}
\psarc[linewidth=0.025]{-}(1,0){0.16}{90}{180}
}
\multiput(0,6)(3.9,0){2}{\rput(0.5,.5){\footnotesize$u_2\!-\!\mu$}
\psarc[linewidth=0.025]{-}(1,0){0.16}{90}{180}
}
\multiput(0,8)(3.9,0){2}{\rput(0.5,.5){\footnotesize$u_\ell\!-\!\mu$}
\psarc[linewidth=0.025]{-}(1,0){0.16}{90}{180}
}
\multiput(0,9)(3.9,0){2}{\rput(0.5,.5){\tiny$u_{\ell\!+\!1}\hs-\hs\mu$}
\psarc[linewidth=0.025]{-}(1,0){0.16}{90}{180}
}
\rput(0.5,2.67){$\vdots$}\rput(0.5,7.67){$\vdots$}\rput(4.4,2.67){$\vdots$}\rput(4.4,7.67){$\vdots$}
\rput(1.8,0.5){$\ldots$}\rput(1.8,1.5){$\ldots$}\rput(1.8,3.5){$\ldots$}\rput(1.8,4.5){$\ldots$}\rput(1.8,5.5){$\ldots$}
\rput(1.8,6.5){$\ldots$}\rput(1.8,8.5){$\ldots$}\rput(1.8,9.5){$\ldots$}
\rput(3.1,0.5){$\ldots$}\rput(3.1,1.5){$\ldots$}\rput(3.1,3.5){$\ldots$}\rput(3.1,4.5){$\ldots$}\rput(3.1,5.5){$\ldots$}
\rput(3.1,6.5){$\ldots$}\rput(3.1,8.5){$\ldots$}\rput(3.1,9.5){$\ldots$}
\multiput(1,4)(1.3,0){3}{\psline[linewidth=1.5pt,linecolor=blue]{-}(0,0.5)(0.3,0.5)}
\multiput(1,9)(1.3,0){3}{\psline[linewidth=1.5pt,linecolor=blue]{-}(0,0.5)(0.3,0.5)}
\end{pspicture}
\label{eq:A10}
\ee 
The proof of \eqref{eq:shortfuse} is rather technical and given below.

The next step is to establish recursion relations between two (new) families of objects, $\Ab_{r}$ and 
$\Db^{1,\ell-1}_{k,r}$, $r = 0, \dots, N$, for which $\Ab= \Ab_{0}$ and $\Db^{1,\ell}_{k}= \Db^{1,\ell}_{k,0}$ are special cases. The new 
lower index $r$ indicates that the 
$r$ leftmost columns of boxes are removed from the diagrammatic definition of $\Ab_{0}$ and 
$\Db^{1,\ell}_{k,0}$ and replaced by $r$ vertical strands. 
The left boundary is then displaced toward the right to act as the left boundary for
the reduced diagram of face operators which is composed of $N-r$ columns, that is,
\be 
\psset{unit=.9cm}
\Db^{1,\ell}_{0,r} := \ \
\begin{pspicture}[shift=-1.58](-0.55,-0.7)(6.5,2)
\psarc[linewidth=4pt,linecolor=blue]{-}(2,1){0.5}{90}{-90}\psarc[linewidth=2pt,linecolor=white]{-}(2,1){0.5}{90}{-90}
\psarc[linewidth=4pt,linecolor=blue]{-}(6,1){0.5}{-90}{90}\psarc[linewidth=2pt,linecolor=white]{-}(6,1){0.5}{-90}{90}
\facegrid{(2,0)}{(6,2)}
\psarc[linewidth=0.025]{-}(2,0){0.16}{0}{90}
\psarc[linewidth=0.025]{-}(3,1){0.16}{90}{180}
\psarc[linewidth=0.025]{-}(5,0){0.16}{0}{90}
\psarc[linewidth=0.025]{-}(6,1){0.16}{90}{180}
\rput(2.5,0.75){\tiny{$_{(1,\ell)}$}}\rput(2.5,1.75){\tiny{$_{(\ell,1)}$}}
\rput(5.5,0.75){\tiny{$_{(1,\ell)}$}}\rput(5.5,1.75){\tiny{$_{(\ell,1)}$}}
\rput(3.5,0.5){$\ldots$}
\rput(3.5,1.5){$\ldots$}
\rput(4.5,0.5){$\ldots$}
\rput(4.5,1.5){$\ldots$}
\psline[linewidth=1.5pt,linecolor=blue]{-}(1.05,0)(1.05,2)
\psline[linewidth=1.5pt,linecolor=blue]{-}(0.25,0)(0.25,2)
\psline[linewidth=1.5pt,linecolor=blue]{-}(-0.55,0)(-0.55,2)
\rput(2.5,.4){$u_0$}
\rput(2.5,1.4){\footnotesize$u_{\ell}\!-\!\mu$}
\rput(5.5,.4){$u_0$}
\rput(5.5,1.4){\footnotesize$u_{\ell}\!-\!\mu$}
\rput(0.25,-0.5){$\underbrace{\quad \qquad \quad}_r$}
\rput(4,-0.5){$\underbrace{\quad \qquad \quad \qquad \qquad \quad}_{N-r}$}
\end{pspicture}
\ee 
By summing over configurations of face operators in the first column of $\Ab_r$, we will find, in \eqref{eq:firstcolumn} -- \eqref{eq:C10}, 
\begin{align}     
 \Ab_{r} =&\Ab_{r+1} \Big(\prod_{j=0}^{\ell}s_j(u)s_{j+1}(u-\mu)\Big) \nonumber \\
  &+ \big[q^1(u_0)q^1(\mu\!-\!u_{-1})\big]^{N-r} \Big(\Db_{1,r}^{1,\ell-1} 
 - \Db_{1,r+1}^{1,\ell-1}\,s_\ell(u)s_{\ell-1}(u\!-\!\mu) \prod_{j=1}^{\ell-2}s_j(u)s_{j+1}(u\!-\!\mu) \Big)
\nonumber \\
 & +\big[q^1(u_{-1})q^1(\mu\!-\!u_{-2})\big]^{N-r} \Big(\Db_{0,r}^{1,\ell-1} 
  - \Db_{0,r+1}^{1,\ell-1}\,s_{\ell-1}(u)s_{\ell}(u\!-\!\mu) \prod_{j=0}^{\ell-3}s_j(u)s_{j+1}(u\!-\!\mu) \Big).
\label{eq:striphorrec}
\end{align}
For $r=N$, it is straightforward to show that
\be
\Ab_{N} = \xi_\ell \Ib = \frac{s^2_{\ell+1}(0)}{s_{\ell}(0)}\Ib = 0, \qquad  \Db_{k,N}^{1,\ell-1} = \zeta_\ell \Ib = s_{\ell}(0)\Ib = (-1)^{\ell-p}\Ib, 
\label{eq:AnDkN}
\ee
where, in general, for $n \in \mathbb N$ and $\lambda$ generic,
\be
\xi_n = 
\begin{pspicture}[shift=-2.65](-1.8,-0.1)(2.1,5.2)
\pspolygon[fillstyle=solid,fillcolor=pink](0,-0.1)(0,2.1)(0.3,2.1)(0.3,-0.1)(0,-0.1)\rput(0.15,1){\small$_n$}
\pspolygon[fillstyle=solid,fillcolor=pink](0,2.7)(0,4.9)(0.3,4.9)(0.3,2.7)(0,2.7)\rput(0.15,3.8){\small$_n$}
\psline[linewidth=1.5pt,linecolor=blue]{-}(0.3,2.4)(0,2.4)
\psline[linewidth=1.5pt,linecolor=blue]{-}(0.3,5.2)(0,5.2)
\psarc[linewidth=1.5pt,linecolor=blue]{-}(0,2.6){0.2}{90}{-90}
\psbezier[linewidth=1.5pt,linecolor=blue]{-}(0,2)(-0.5,2)(-0.5,3.2)(0,3.2)
\psbezier[linewidth=1.5pt,linecolor=blue]{-}(0,1.6)(-0.8,1.6)(-0.8,3.6)(0,3.6)
\psbezier[linewidth=1.5pt,linecolor=blue]{-}(0,0.8)(-1.4,0.8)(-1.4,4.4)(0,4.4)
\psbezier[linewidth=1.5pt,linecolor=blue]{-}(0,0.4)(-1.7,0.4)(-1.7,4.8)(0,4.8)
\psbezier[linewidth=1.5pt,linecolor=blue]{-}(0,0.0)(-2.0,0.0)(-2.0,5.2)(0,5.2)
\psarc[linewidth=1.5pt,linecolor=blue]{-}(0.3,2.6){-0.2}{90}{-90}
\psbezier[linewidth=1.5pt,linecolor=blue]{-}(0.3,2)(0.8,2)(0.8,3.2)(0.3,3.2)
\psbezier[linewidth=1.5pt,linecolor=blue]{-}(0.3,1.6)(1.1,1.6)(1.1,3.6)(0.3,3.6)
\psbezier[linewidth=1.5pt,linecolor=blue]{-}(0.3,0.8)(1.7,0.8)(1.7,4.4)(0.3,4.4)
\psbezier[linewidth=1.5pt,linecolor=blue]{-}(0.3,0.4)(2.0,0.4)(2.0,4.8)(0.3,4.8)
\psbezier[linewidth=1.5pt,linecolor=blue]{-}(0.3,0.0)(2.3,0.0)(2.3,5.2)(0.3,5.2)
\rput(-0.15,1.3){$\vdots$}\rput(-0.15,4.05){$\vdots$}
\rput(0.45,1.3){$\vdots$}\rput(0.45,4.05){$\vdots$}
\rput(0.93,2.08){$.$}\rput(1.05,2.00){$.$}\rput(1.17,1.92){$.$}
\rput(0.93,3.12){$.$}\rput(1.05,3.2){$.$}\rput(1.17,3.28){$.$}
\rput(-0.63,2.08){$.$}\rput(-0.75,2.00){$.$}\rput(-0.87,1.92){$.$}
\rput(-0.63,3.12){$.$}\rput(-0.75,3.2){$.$}\rput(-0.87,3.28){$.$}
\end{pspicture} = \frac{s^2_{n+1}(0)}{s_n(0)},
\qquad
\zeta_n = 
\begin{pspicture}[shift=-2.25](-1.8,0.3)(1.8,4.8)
\pspolygon[fillstyle=solid,fillcolor=pink](-0.3,0.3)(-0.3,2.5)(0.3,2.5)(0.3,0.3)(-0.3,0.3)\rput(0.0,1.4){\small$_{n\!-\!1}$}
\psline[linewidth=1.5pt,linecolor=blue]{-}(0.3,2.8)(-0.3,2.8)
\psline[linewidth=1.5pt,linecolor=blue]{-}(0.3,3.2)(-0.3,3.2)
\psline[linewidth=1.5pt,linecolor=blue]{-}(0.3,3.6)(-0.3,3.6)
\psline[linewidth=1.5pt,linecolor=blue]{-}(0.3,4.4)(-0.3,4.4)
\psline[linewidth=1.5pt,linecolor=blue]{-}(0.3,4.8)(-0.3,4.8)
\psarc[linewidth=1.5pt,linecolor=blue]{-}(-0.3,2.6){0.2}{90}{-90}
\psbezier[linewidth=1.5pt,linecolor=blue]{-}(-0.3,2)(-0.8,2)(-0.8,3.2)(-0.3,3.2)
\psbezier[linewidth=1.5pt,linecolor=blue]{-}(-0.3,1.6)(-1.1,1.6)(-1.1,3.6)(-0.3,3.6)
\psbezier[linewidth=1.5pt,linecolor=blue]{-}(-0.3,0.8)(-1.7,0.8)(-1.7,4.4)(-0.3,4.4)
\psbezier[linewidth=1.5pt,linecolor=blue]{-}(-0.3,0.4)(-2,0.4)(-2,4.8)(-0.3,4.8)
\psarc[linewidth=1.5pt,linecolor=blue]{-}(0.3,2.6){-0.2}{90}{-90}
\psbezier[linewidth=1.5pt,linecolor=blue]{-}(0.3,2)(0.8,2)(0.8,3.2)(0.3,3.2)
\psbezier[linewidth=1.5pt,linecolor=blue]{-}(0.3,1.6)(1.1,1.6)(1.1,3.6)(0.3,3.6)
\psbezier[linewidth=1.5pt,linecolor=blue]{-}(0.3,0.8)(1.7,0.8)(1.7,4.4)(0.3,4.4)
\psbezier[linewidth=1.5pt,linecolor=blue]{-}(0.3,0.4)(2.0,0.4)(2.0,4.8)(0.3,4.8)
\rput(-0.45,1.3){$\vdots$}\rput(-0.35,4.07){$\vdots$}
\rput(0.45,1.3){$\vdots$}\rput(0.35,4.07){$\vdots$}
\rput(0.93,2.08){$.$}\rput(1.05,2.00){$.$}\rput(1.17,1.92){$.$}
\rput(0.93,3.12){$.$}\rput(1.05,3.2){$.$}\rput(1.17,3.28){$.$}
\rput(-0.93,2.08){$.$}\rput(-1.05,2.00){$.$}\rput(-1.17,1.92){$.$}
\rput(-0.93,3.12){$.$}\rput(-1.05,3.2){$.$}\rput(-1.17,3.28){$.$}
\end{pspicture}  = s_n(0).
\label{eq:a340}
\ee
The recursion relation \eqref{eq:striphorrec} allows us to write an equality tying $\Ab$, $\Db^{1,\ell-1}_{0}$ and $\Db^{1,\ell-1}_{1}$ 
with $\Ab_N$, $\Db_{0,N}^{1,\ell-1}$ and $\Db_{1,N}^{1,\ell-1}$ which reads, after simplification,
\be       
 \Ab - \big[q^1(u_{-1})q^1(\mu\!-\!u_{-2})\big]^N  \Db^{1,\ell-1}_0 -  \big[q^1(u_0)q^1(\mu\!-\!u_{-1})\big]^N \Db^{1,\ell-1}_1 
  = 2\,(-1)^{\ell+1-p}\Big(\prod_{j=0}^\ell s_j(u)s_{j+1}(u\!-\!\mu)\Big)^N\Ib.
\ee
With \eqref{eq:shortfuse}, this concludes the proof of \eqref{eq:StripClosureApp}. The remaining
technical computations are the proofs of \eqref{eq:shortfuse} and \eqref{eq:striphorrec}, which are done in the next two 
paragraphs.

%%%%%%%%
\paragraph{Proof of equation (\ref{eq:shortfuse})}
%%%%%%%%

We start with the form \eqref{eq:E5} for $\eta(u,\mu) \, \Db^{1,\ell}_0 \Db^{1,1}_\ell$,
\be 
\psset{unit=0.9}

\label{eq:beginningofproofF13}
\ee 
where three new $P_\ell$ projectors have been added. This is permitted by push-through 
properties and the annihilation property \eqref{eq:(ii)} of 
the projectors. Diamond face operators are then expanded as
\be
\psset{unit=0.6364cm}
%
}$ 
are obtained by setting $n=\ell$ in \eqref{eq:alphaLR}. This gives four terms, three of which are zero. 
For example, the term corresponding to 
$\alpha_L^{
 \psset{unit=0.18cm}
%
\ee 
A single connectivity now contributes to the lower-left $P_\ell$ projector,
\be   
\underbrace{\frac{(-1)^{\ell-1} }{s_{\ell}(0)}}_{(-1)^{p+1}} \ 
\psset{unit=0.45}
%
\ee
while all the others have half-arcs that propagate and are annihilated by the remaining $P_\ell$ projector. 
Up to the factor $(-1)^{p+1}[q^1(u_\ell)]^N$, this gives
\be       
\psset{unit=0.9}
%
\label{eq:endofproofF13}
\ee
which is zero due to property (c) given in Section~\ref{sec:prelims}. Similarly, the tangles corresponding to 
$\alpha_L^{
 \psset{unit=0.18cm}
%
}$
are also zero, because half-arcs propagate all the way around the tangle, 
 \ 
$%
$\,, 
and are annihilated by the last $P_\ell$ projector. The only contribution to
$\eta(u,\mu)\Db_0^{1,\ell}\Db_\ell^{1,1}$ is $\alpha_L^{
 \psset{unit=0.18cm}
%
} \Ab$, with $\Ab$ defined in \eqref{eq:A10}, and this simplifies to \eqref{eq:shortfuse}.

%%%%%%%%%%%%%%%%%%%%%%%%%%%%%%
\paragraph{Proof of equation (\ref{eq:striphorrec})}
%%%%%%%%%%%%%%%%%%%%%%%%%%%%%%

In the computations to follow, only the $L:=N-r$ columns of boxes will be considered, while the $r$ identity strands 
located on the left simply act as spectators. These will therefore be omitted from the diagrams in our planar computations. 
We start by expanding, in $\Ab_{r}$, the bottom $\ell$ elementary faces in the leftmost column of face operators,
\be
\psset{unit=0.55cm}

\label{eq:firstcolumn}
\ee
where the last tangle is defined in \eqref{eq:bigtile}. We thus write
\be
\Ab_{r} = \Bb_1 + \Bb_2 + \Bb_3 
\label{eq:AC123}
\ee
and set out to study each term individually. For $\Bb_1$,
\be
\psset{unit=0.9}
\Bb_1 = x_1\ 
%
 
\ee
where the $P_\ell$ projector at the bottom left of the tangle has been replaced by its only contributing connectivity \eqref{eq:survivor} 
and all the other projectors (except two) have been removed using push-through properties and property \eqref{eq:(ii)} of the projectors. 
Then,
\be        
\psset{unit=0.9}
\Bb_1 
= x_1(-1)^{p+1} \big[q^1(\mu\!-\!u_{-2})]^L
%
 
\ee
where this time, the $P_\ell$ projector has been replaced by a left-right reflection of the connectivity \eqref{eq:survivor}.
The next step is to use \eqref{eq:rationalrec} on the remaining projector, and this yields two tangles,
\be
\Bb_1 = x_1\big[q^1(\mu\!-\!u_{-1})\big]^{L}\big[q^1(u_{0})\big]^{L-1} (\Mb_1+\Mb_2).
\ee
The first one is
\be \Mb_1= \!\!
\psset{unit=0.9}
%
  
\label{eq:tangle1}
\ee
while the second one is
\be \Mb_2 = - \beta
\psset{unit=0.9}
%
  
\label{eq:tangle2}
\ee
where at the last step we replaced the leftmost $P_{\ell-1}$ projector by its unique contributing connectivity,
\be  
\underbrace{(-1)^{\ell-2} \frac{s_1(0)}{s_{\ell-1}(0)}}_{=(-1)^{p}\beta^{-1}} \
\psset{unit=0.45}
%
\label{eq:survivor2}
\ee
A final form for $\Bb_1$ is obtained after making two remarks. The first one
is that an extra projector can be added at no cost in the resulting tangles of both \eqref{eq:tangle1} and \eqref{eq:tangle2}, 
\begin{align} 
\hspace{-0.4cm}\Mb_1=\!\!\!  
\psset{unit=0.9}
%
 
\nonumber \\[.15cm]
\end{align}
A particular linear combination of these gives $\Db^{1,\ell-1}_{0,r}$, that is
\be
\Db^{1,\ell-1}_{0,r} = \Mb_1 \prod_{j=0}^{\ell-2} s_{1-j}(-u) +\Mb_2 \prod_{j=0}^{\ell-2} s_{j}(u).
\ee
 The other remark is that 
\be
\psset{unit=0.9}
%
\ee
Putting these results together, we find
\begin{align}
 \Bb_1=\,& \big[q^1(u_{-1})q^1({\mu\!-\!u_{-2}}) \big]^{L}
 \nonumber\\[.15cm]
 & \Big(\Db^{1,\ell-1}_{0,r} - \beta s_{\ell-1}(u) 
   \Big(\prod_{j=0}^{\ell-3}s_j(u)s_{j+1}(u-\mu)\Big)
   \big[s_{\ell-1}(u\!-\!\mu)\Db^{1,\ell-1}_{0,r+1} - s_{\ell-2}(u\!-\!\mu) \Cb_{0,r} \big]\Big).
\label{eq:C1}
\end{align}
The same ideas are used for $\Bb_2$, which in the end reads
\be   
 \Bb_2 = \big[q^1(u_{0})q^1({\mu\!-\!u_{-1}}) \big]^L\Big(\Db^{1,\ell-1}_{1,r} - \beta\,\Cb_{1,r} s_{\ell}(u)s_{\ell+1}(u-\mu) 
  \prod_{j=1}^{\ell-2}s_j(u)s_{j+1}(u-\mu) \Big)
\label{eq:C2}
\ee
where $\Cb_{1,r}:= \Cb_{0,r}(u+\lambda)$. 

For $\Bb_3$, we proceed to sum over the remaining $\ell$ face operators of the leftmost column,
\be
\psset{unit=0.55cm}
\begin{pspicture}[shift=-2.4](-0.3,0)(1.3,5.0)
\facegrid{(0,0)}{(1,5)}
\multiput(-0.3,0)(1.3,0){2}{\pspolygon[fillstyle=solid,fillcolor=pink](0,0.1)(0,3.9)(0.3,3.9)(0.3,0.1)(0,0.1)\rput(0.15,2.0){\scriptsize $_\ell$}}
\psarc[linewidth=0.025]{-}(1,0){0.16}{90}{180}
\psarc[linewidth=0.025]{-}(1,1){0.16}{90}{180}
\psarc[linewidth=0.025]{-}(1,3){0.16}{90}{180}
\psarc[linewidth=0.025]{-}(1,4){0.16}{90}{180}
\rput(0.55,0.5){$_{u_0}$}
\rput(0.55,1.5){$_{u_1}$}
\rput(0.5,2.675){$\vdots$}
\rput(0.55,3.5){$_{u_{\ell\!-\!1}}$}
\rput(0.55,4.5){$_{u_\ell}$}
\end{pspicture} 
\  =   s_{\ell+1}(u\!-\!\mu) \prod_{j=0}^{\ell-1}s_{-j}(\mu\!-\!u)
\begin{pspicture}[shift=-2.4](-0.5,0)(1.5,5)
\facegrid{(0,0)}{(1,5)}
\multiput(-0.3,0)(1.3,0){2}{\pspolygon[fillstyle=solid,fillcolor=pink](0,0.1)(0,3.9)(0.3,3.9)(0.3,0.1)(0,0.1)\rput(0.15,2.0){\scriptsize $_\ell$}}
\rput(0,0){\loopb}
\rput(0,1){\loopb}
\rput(0,2){\loopb}
\rput(0,3){\loopb}
\rput(0,4){\loopa}
\end{pspicture} + 
s_{\ell}(\mu\!-\!u) \prod_{j=1}^{\ell}s_{j}(u\!-\!\mu)
\begin{pspicture}[shift=-2.4](-0.5,0)(1.5,5)
\facegrid{(0,0)}{(1,5)}
\multiput(-0.3,0)(1.3,0){2}{\pspolygon[fillstyle=solid,fillcolor=pink](0,0.1)(0,3.9)(0.3,3.9)(0.3,0.1)(0,0.1)\rput(0.15,2.0){\scriptsize $_\ell$}}
\rput(0,0){\loopa}
\rput(0,1){\loopa}
\rput(0,2){\loopa}
\rput(0,3){\loopa}
\rput(0,4){\loopb}
\end{pspicture} 
+ 
(-1)^p\prod_{j=1}^{\ell+1} s_j(u\!-\!\mu)\ \ 
\begin{pspicture}[shift=-2.4](-0.3,0)(1.5,5.00)
\multiput(-0.3,0)(1.3,0){2}{\pspolygon[fillstyle=solid,fillcolor=pink](0,0.1)(0,3.9)(0.3,3.9)(0.3,0.1)(0,0.1)\rput(0.15,2.0){\scriptsize $_\ell$}}
\pspolygon[fillstyle=solid,fillcolor=lightlightblue](0,0)(1,0)(1,5)(0,5)(0,0)
\psline[linewidth=1.25pt]{-}(0.1,0.1)(0.9,0.1)(0.9,4.9)(0.1,4.9)(0.1,0.1)
\end{pspicture}
\ee
This leads to the separate contributions
\be 
\Bb_3 = \Bb_4 + \Bb_5 + \Bb_6
\label{eq:CC456}
\ee
which are tackled independently. Using the same arguments as before, one finds 
\be   
 \Bb_4 = -\beta s_{\ell-1}(u) s_{\ell+1}(u\!-\!\mu) 
  \Big(\prod_{j=0}^{\ell-3}s_j(u)s_{j+1}(u\!-\!\mu)\Big)\big[q^1(u_{-1})q^1(\mu\!-\!u_{-2})\big]^{L} 
   \big( \Db^{1,\ell-1}_{0,r+1} -\beta\, \Cb_{0,r} \big)
\label{eq:C4}
\ee
and also
\be
\Bb_5 = 0.
\label{eq:C5}
\ee
This last equality results from
\be
\psset{unit=0.55}
\begin{pspicture}[shift=-4.9](-2.2,0)(1.5,10)
\facegrid{(0,5)}{(1,10)}
\multiput(-0.3,0)(1.3,0){2}{\pspolygon[fillstyle=solid,fillcolor=pink](0,0.1)(0,3.9)(0.3,3.9)(0.3,0.1)(0,0.1)\rput(0.15,2.0){\scriptsize $_\ell$}}
\multiput(-0.3,5)(1.3,0){2}{\pspolygon[fillstyle=solid,fillcolor=pink](0,0.1)(0,3.9)(0.3,3.9)(0.3,0.1)(0,0.1)\rput(0.15,2.0){\scriptsize $_\ell$}}
\rput(0,5){\loopa}\rput(0,6){\loopa}\rput(0,7){\loopa}\rput(0,8){\loopa}\rput(0,9){\loopb}
\pspolygon[fillstyle=solid,fillcolor=lightlightblue](0,0)(1,0)(1,5)(0,5)(0,0)
\psline[linewidth=1.25pt]{-}(0.1,0.1)(0.9,0.1)(0.9,4.9)(0.1,4.9)(0.1,0.1)
\psarc[linewidth=1.5pt,linecolor=blue]{-}(-0.3,5){0.5}{90}{-90}
\psbezier[linewidth=1.5pt,linecolor=blue]{-}(-0.3,3.5)(-1.3,3.5)(-1.3,6.5)(-0.3,6.5)
\psbezier[linewidth=1.5pt,linecolor=blue]{-}(-0.3,2.5)(-1.7,2.5)(-1.7,7.5)(-0.3,7.5)
\psbezier[linewidth=1.5pt,linecolor=blue]{-}(-0.3,1.5)(-2.1,1.5)(-2.1,8.5)(-0.3,8.5)
\psbezier[linewidth=1.5pt,linecolor=blue]{-}(-0.3,0.5)(-2.5,0.5)(-2.5,9.5)(-0.3,9.5)
\psline[linewidth=1.5pt,linecolor=blue]{-}(0,4.5)(-0.3,4.5)
\psline[linewidth=1.5pt,linecolor=blue]{-}(0,9.5)(-0.3,9.5)
\end{pspicture} \ = (-1)^{p+1}\
\begin{pspicture}[shift=-4.9](-2.2,0)(1.5,10)
\facegrid{(0,5)}{(1,10)}
\multiput(-0.3,0)(1.3,0){2}{\pspolygon[fillstyle=solid,fillcolor=pink](0,0.1)(0,3.9)(0.3,3.9)(0.3,0.1)(0,0.1)\rput(0.15,2.0){\scriptsize $_\ell$}}
\multiput(1,5)(1.3,0){1}{\pspolygon[fillstyle=solid,fillcolor=pink](0,0.1)(0,3.9)(0.3,3.9)(0.3,0.1)(0,0.1)\rput(0.15,2.0){\scriptsize $_\ell$}}
\rput(0,5){\loopa}\rput(0,6){\loopa}\rput(0,7){\loopa}\rput(0,8){\loopa}\rput(0,9){\loopb}
\pspolygon[fillstyle=solid,fillcolor=lightlightblue](0,0)(1,0)(1,5)(0,5)(0,0)
\psline[linewidth=1.25pt]{-}(0.1,0.1)(0.9,0.1)(0.9,4.9)(0.1,4.9)(0.1,0.1)
\psarc[linewidth=1.5pt,linecolor=blue]{-}(-0.3,8){0.5}{90}{-90}
\rput(0,-1){\psarc[linewidth=1.5pt,linecolor=blue]{-}(-0.3,5){0.5}{90}{-90}
\psbezier[linewidth=1.5pt,linecolor=blue]{-}(-0.3,3.5)(-1.3,3.5)(-1.3,6.5)(-0.3,6.5)
\psbezier[linewidth=1.5pt,linecolor=blue]{-}(-0.3,2.5)(-1.7,2.5)(-1.7,7.5)(-0.3,7.5)}
\psbezier[linewidth=1.5pt,linecolor=blue]{-}(-0.3,0.5)(-2.5,0.5)(-2.5,9.5)(-0.3,9.5)
\multiput(0,0)(0,1){6}{\psline[linewidth=1.5pt,linecolor=blue]{-}(0,4.5)(-0.3,4.5)}
\end{pspicture} \ = \beta \ 
\begin{pspicture}[shift=-4.9](-2.7,0)(1.5,10)
\facegrid{(0,0)}{(1,10)}
\multiput(1,0)(1.3,0){1}{\pspolygon[fillstyle=solid,fillcolor=pink](0,0.1)(0,3.9)(0.3,3.9)(0.3,0.1)(0,0.1)\rput(0.15,2.0){\scriptsize $_\ell$}}
\multiput(-0.8,0)(1.3,0){1}{\pspolygon[fillstyle=solid,fillcolor=pink](0,0.1)(0,2.9)(0.8,2.9)(0.8,0.1)(0,0.1)
\rput(0.4,1.5){\scriptsize$_{\ell\!-\!1}$}}
\multiput(1,5)(1.3,0){1}{\pspolygon[fillstyle=solid,fillcolor=pink](0,0.1)(0,3.9)(0.3,3.9)(0.3,0.1)(0,0.1)\rput(0.15,2.0){\scriptsize $_\ell$}}
\rput(0,0){\loopa}\rput(0,1){\loopa}\rput(0,2){\loopa}\rput(0,3){\loopb}\rput(0,4){\loopa}
\rput(0,5){\loopa}\rput(0,6){\loopa}\rput(0,7){\loopa}\rput(0,8){\loopa}\rput(0,9){\loopb}
\rput(-0.5,0){
\psarc[linewidth=1.5pt,linecolor=blue]{-}(-0.3,8){0.5}{90}{-90}
\rput(0,-1){\psarc[linewidth=1.5pt,linecolor=blue]{-}(-0.3,5){0.5}{90}{-90}
\psbezier[linewidth=1.5pt,linecolor=blue]{-}(-0.3,3.5)(-1.3,3.5)(-1.3,6.5)(-0.3,6.5)
\psbezier[linewidth=1.5pt,linecolor=blue]{-}(-0.3,2.5)(-1.7,2.5)(-1.7,7.5)(-0.3,7.5)}
\psbezier[linewidth=1.5pt,linecolor=blue]{-}(-0.3,0.5)(-2.5,0.5)(-2.5,9.5)(-0.3,9.5)
}
\multiput(0,0)(0,1){7}{\psline[linewidth=1.5pt,linecolor=blue]{-}(0,3.5)(-0.8,3.5)}
\end{pspicture}\ = \beta \ 
\begin{pspicture}[shift=-4.9](-1.9,0)(1.5,10)
\facegrid{(0,0)}{(1,10)}
\multiput(1,0)(1.3,0){1}{\pspolygon[fillstyle=solid,fillcolor=pink](0,0.1)(0,3.9)(0.3,3.9)(0.3,0.1)(0,0.1)\rput(0.15,2.0){\scriptsize $_\ell$}}
\multiput(1,5)(1.3,0){1}{\pspolygon[fillstyle=solid,fillcolor=pink](0,0.1)(0,3.9)(0.3,3.9)(0.3,0.1)(0,0.1)\rput(0.15,2.0){\scriptsize $_\ell$}}
\rput(0,0){\loopa}\rput(0,1){\loopa}\rput(0,2){\loopa}\rput(0,3){\loopb}\rput(0,4){\loopa}
\rput(0,5){\loopa}\rput(0,6){\loopa}\rput(0,7){\loopa}\rput(0,8){\loopa}\rput(0,9){\loopb}
\rput(0.3,0){
\psarc[linewidth=1.5pt,linecolor=blue]{-}(-0.3,8){0.5}{90}{-90}
\rput(0,-1){\psarc[linewidth=1.5pt,linecolor=blue]{-}(-0.3,5){0.5}{90}{-90}
\psbezier[linewidth=1.5pt,linecolor=blue]{-}(-0.3,3.5)(-1.3,3.5)(-1.3,6.5)(-0.3,6.5)
\psbezier[linewidth=1.5pt,linecolor=blue]{-}(-0.3,2.5)(-1.7,2.5)(-1.7,7.5)(-0.3,7.5)}
\psbezier[linewidth=1.5pt,linecolor=blue]{-}(-0.3,0.5)(-2.5,0.5)(-2.5,9.5)(-0.3,9.5)
}
\end{pspicture} \ =  0
\ee
where, in order, we replaced the top $P_\ell$ by the connectivity \eqref{eq:survivor}, used \eqref{eq:otherpushthroughs} and 
property \eqref{eq:(iv)} of the projectors.
Finally, to compute $\Bb_6$, we use the identity
\be
\psset{unit=0.45}
\begin{pspicture}[shift=-5.9](-2.2,0)(1.5,12)
\multiput(-0.3,0)(1.3,0){2}{\pspolygon[fillstyle=solid,fillcolor=pink](0,0.1)(0,4.9)(0.3,4.9)(0.3,0.1)(0,0.1)\rput(0.15,2.5){\scriptsize $_\ell$}}
\multiput(-0.3,6)(1.3,0){2}{\pspolygon[fillstyle=solid,fillcolor=pink](0,0.1)(0,4.9)(0.3,4.9)(0.3,0.1)(0,0.1)\rput(0.15,2.5){\scriptsize $_\ell$}}
\multiput(0,0)(0,6){2}{\pspolygon[fillstyle=solid,fillcolor=lightlightblue](0,0)(1,0)(1,6)(0,6)(0,0)
\psline[linewidth=1.05pt]{-}(0.1,0.1)(0.9,0.1)(0.9,5.9)(0.1,5.9)(0.1,0.1)}
\psarc[linewidth=1.5pt,linecolor=blue]{-}(-0.3,6){0.5}{90}{-90}
\psbezier[linewidth=1.5pt,linecolor=blue]{-}(-0.3,4.5)(-1.3,4.5)(-1.3,7.5)(-0.3,7.5)
\psbezier[linewidth=1.5pt,linecolor=blue]{-}(-0.3,3.5)(-1.7,3.5)(-1.7,8.5)(-0.3,8.5)
\psbezier[linewidth=1.5pt,linecolor=blue]{-}(-0.3,2.5)(-2.1,2.5)(-2.1,9.5)(-0.3,9.5)
\psbezier[linewidth=1.5pt,linecolor=blue]{-}(-0.3,1.5)(-2.5,1.5)(-2.5,10.5)(-0.3,10.5)
\psbezier[linewidth=1.5pt,linecolor=blue]{-}(-0.3,0.5)(-2.9,0.5)(-2.9,11.5)(-0.3,11.5)
\psline[linewidth=1.5pt,linecolor=blue]{-}(0,5.5)(-0.3,5.5)
\psline[linewidth=1.5pt,linecolor=blue]{-}(0,11.5)(-0.3,11.5)
\end{pspicture}  \ = (-1)^p\ 
\begin{pspicture}[shift=-5.9](-2.0,0)(1.5,12)
\multiput(1,0)(1.3,0){1}{\pspolygon[fillstyle=solid,fillcolor=pink](0,0.1)(0,4.9)(0.3,4.9)(0.3,0.1)(0,0.1)\rput(0.15,2.5){\scriptsize $_\ell$}}
\multiput(1,6)(1.3,0){1}{\pspolygon[fillstyle=solid,fillcolor=pink](0,0.1)(0,4.9)(0.3,4.9)(0.3,0.1)(0,0.1)\rput(0.15,2.5)
{\scriptsize $_\ell$}}
\facegrid{(0,0)}{(1,12)}
\rput(0,0){\loopb}
\rput(0,1){\loopb}
\rput(0,2){\loopb}
\rput(0,3){\loopb}
\rput(0,4){\loopb}
\rput(0,5){\loopb}
\rput(0,6){\loopa}
\rput(0,7){\loopa}
\rput(0,8){\loopa}
\rput(0,9){\loopa}
\rput(0,10){\loopa}
\rput(0,11){\loopa}
\rput(0.3,0){\psarc[linewidth=1.5pt,linecolor=blue]{-}(-0.3,6){0.5}{90}{-90}
\psbezier[linewidth=1.5pt,linecolor=blue]{-}(-0.3,4.5)(-1.3,4.5)(-1.3,7.5)(-0.3,7.5)
\psbezier[linewidth=1.5pt,linecolor=blue]{-}(-0.3,3.5)(-1.7,3.5)(-1.7,8.5)(-0.3,8.5)
\psbezier[linewidth=1.5pt,linecolor=blue]{-}(-0.3,2.5)(-2.1,2.5)(-2.1,9.5)(-0.3,9.5)
\psbezier[linewidth=1.5pt,linecolor=blue]{-}(-0.3,1.5)(-2.5,1.5)(-2.5,10.5)(-0.3,10.5)
\psbezier[linewidth=1.5pt,linecolor=blue]{-}(-0.3,0.5)(-2.9,0.5)(-2.9,11.5)(-0.3,11.5)}
\end{pspicture}
\ \ + 
\begin{pspicture}[shift=-5.9](-1.8,0)(1.5,12)
\multiput(1,0)(1.3,0){1}{\pspolygon[fillstyle=solid,fillcolor=pink](0,0.1)(0,4.9)(0.3,4.9)(0.3,0.1)(0,0.1)\rput(0.15,2.5){\scriptsize $_\ell$}}
\multiput(1,6)(1.3,0){1}{\pspolygon[fillstyle=solid,fillcolor=pink](0,0.1)(0,4.9)(0.3,4.9)(0.3,0.1)(0,0.1)\rput(0.15,2.5)
{\scriptsize $_\ell$}}
\facegrid{(0,0)}{(1,12)}
\rput(0,2){\loopb}
\rput(0,3){\loopb}
\rput(0,4){\loopb}
\rput(0,5){\loopa}
\rput(0,6){\loopa}
\rput(0,7){\loopa}
\rput(0,10){\loopb}
\rput(0,11){\loopa}
\psarc[linewidth=1.5pt,linecolor=blue]{-}(1,0){0.5}{90}{180}
\psarc[linewidth=1.5pt,linecolor=blue]{-}(1,2){0.5}{180}{-90}
\psarc[linewidth=1.5pt,linecolor=blue]{-}(1,8){0.5}{90}{180}
\psarc[linewidth=1.5pt,linecolor=blue]{-}(1,10){0.5}{180}{-90}
\rput(0.3,0){\psarc[linewidth=1.5pt,linecolor=blue]{-}(-0.3,5){0.5}{90}{-90}
\psarc[linewidth=1.5pt,linecolor=blue]{-}(-0.3,11){0.5}{90}{-90}
\psbezier[linewidth=1.5pt,linecolor=blue]{-}(-0.3,3.5)(-1.3,3.5)(-1.3,6.5)(-0.3,6.5)
\psbezier[linewidth=1.5pt,linecolor=blue]{-}(-0.3,2.5)(-1.7,2.5)(-1.7,7.5)(-0.3,7.5)}
\end{pspicture}
 \ + (-1)^ps_3(0)\ \ \
\begin{pspicture}[shift=-5.9](-1.6,0)(1.5,12)
\multiput(1,0)(1.3,0){1}{\pspolygon[fillstyle=solid,fillcolor=pink](0,0.1)(0,4.9)(0.3,4.9)(0.3,0.1)(0,0.1)\rput(0.15,2.5){\scriptsize $_\ell$}}
\multiput(1,6)(1.3,0){1}{\pspolygon[fillstyle=solid,fillcolor=pink](0,0.1)(0,4.9)(0.3,4.9)(0.3,0.1)(0,0.1)\rput(0.15,2.5)
{\scriptsize $_\ell$}}
\facegrid{(0,0)}{(1,12)}
\rput(0,2){\loopb}
\rput(0,3){\loopb}
\rput(0,4){\loopb}
\rput(0,5){\loopb}
\rput(0,6){\loopa}
\rput(0,7){\loopa}
\rput(0,8){\loopa}
\psarc[linewidth=1.5pt,linecolor=blue]{-}(1,0){0.5}{90}{180}
\psarc[linewidth=1.5pt,linecolor=blue]{-}(1,2){0.5}{180}{-90}
\psarc[linewidth=1.5pt,linecolor=blue]{-}(1,11){0.5}{180}{-90}
\psarc[linewidth=1.5pt,linecolor=blue]{-}(1,9){0.5}{90}{180}
\rput(0,11){\loopa}
\rput(0.3,0){\psarc[linewidth=1.5pt,linecolor=blue]{-}(-0.3,6){0.5}{90}{-90}
\psbezier[linewidth=1.5pt,linecolor=blue]{-}(-0.3,4.5)(-1.3,4.5)(-1.3,7.5)(-0.3,7.5)
\psbezier[linewidth=1.5pt,linecolor=blue]{-}(-0.3,3.5)(-1.7,3.5)(-1.7,8.5)(-0.3,8.5)
\psbezier[linewidth=1.5pt,linecolor=blue]{-}(-0.3,2.5)(-2.4,2.5)(-2.4,11.5)(-0.3,11.5)
}
\end{pspicture}
\ + \  
\begin{pspicture}[shift=-5.9](-1.8,0)(1.5,12)
\multiput(1,0)(1.3,0){1}{\pspolygon[fillstyle=solid,fillcolor=pink](0,0.1)(0,4.9)(0.3,4.9)(0.3,0.1)(0,0.1)\rput(0.15,2.5){\scriptsize $_\ell$}}
\multiput(1,6)(1.3,0){1}{\pspolygon[fillstyle=solid,fillcolor=pink](0,0.1)(0,4.9)(0.3,4.9)(0.3,0.1)(0,0.1)\rput(0.15,2.5)
{\scriptsize $_\ell$}}
\facegrid{(0,0)}{(1,12)}
\rput(0,0){\loopb}
\rput(0,1){\loopb}
\rput(0,2){\loopb}
\rput(0,3){\loopb}
\rput(0,4){\loopb}
\rput(0,5){\loopa}
\rput(0,6){\loopa}
\rput(0,7){\loopa}
\rput(0,8){\loopa}
\rput(0,9){\loopa}
\psarc[linewidth=1.5pt,linecolor=blue]{-}(1,10){0.5}{90}{180}
\psarc[linewidth=1.5pt,linecolor=blue]{-}(1,12){0.5}{180}{-90}
\rput(0.3,0){\psarc[linewidth=1.5pt,linecolor=blue]{-}(-0.3,5){0.5}{90}{-90}
\psbezier[linewidth=1.5pt,linecolor=blue]{-}(-0.3,3.5)(-1.3,3.5)(-1.3,6.5)(-0.3,6.5)
\psbezier[linewidth=1.5pt,linecolor=blue]{-}(-0.3,2.5)(-1.7,2.5)(-1.7,7.5)(-0.3,7.5)
\psbezier[linewidth=1.5pt,linecolor=blue]{-}(-0.3,1.5)(-2.1,1.5)(-2.1,8.5)(-0.3,8.5)
\psbezier[linewidth=1.5pt,linecolor=blue]{-}(-0.3,0.5)(-2.5,0.5)(-2.5,9.5)(-0.3,9.5)}
\end{pspicture}
\label{eq:uglyidentity}
\ee
whose proof will be the topic of the paragraph below. This separates the computation of $\Bb_6$ into the four parts
\be
\Bb_6 = \Bb_7 + \Bb_8 + \Bb_9 + \Bb_{10},
\label{eq:CC78910}
\ee
each of which can be computed using the techniques presented earlier. This yields
\be
\Bb_7 =\Big(\prod_{j=0}^\ell s_j(u)s_{j+1}(u\!-\!\mu)\Big)\Ab_{r+1},  \qquad \Bb_8 = 0, \label{eq:C78}
\ee \vspace{-0.4cm}
\begin{align}
 \Bb_9 &= s_3(0) \,s_{\ell-1}(u)s_{\ell}(u\!-\!\mu) \Big(\prod_{j=0}^{\ell-3} s_j(u)s_{j+1}(u\!-\!\mu)\Big) 
   \big[q^1(u_{-1})q^1(\mu\!-\!u_{-2})]^{L} \big(\Db^{1,\ell-1}_{0,r+1} -\beta\, \Cb_{0,r} \big),
\label{eq:C9} 
\\ 
  \Bb_{10} &= (-1)\,s_{\ell}(u)s_{\ell+1}(u\!-\!\mu) \Big(\prod_{j=1}^{\ell-2} s_j(u)s_{j+1}(u\!-\!\mu)\Big)
    \big[ q^1(u_{0})q^1(\mu\!-\!u_{-1})  \big]^{2L} \big(\Db^{1,\ell-1}_{1,r+1} -\beta\, \Cb_{1,r} \big).\label{eq:C10}
\end{align}

Combining \eqref{eq:AC123}, \eqref{eq:C1}, \eqref{eq:C2}, \eqref{eq:CC456}-\eqref{eq:C5} and 
\eqref{eq:CC78910}-\eqref{eq:C10}, one finds that the coefficients of $\Cb_{0,r}$ and $\Cb_{1,r}$ cancel out and 
that the final result for $\Ab_r$ indeed is 
given by \eqref{eq:striphorrec}. The only missing piece of the puzzle is the proof of equation \eqref{eq:uglyidentity}.

%%%%%%%%%%%%%%%%%%%%%%%%%%%
\paragraph{Proof of equation (\ref{eq:uglyidentity})}
%%%%%%%%%%%%%%%%%%%%%%%%%%%
We write the lefthand side as 
\be
\psset{unit=0.45}

\label{eq:willgetugly}
\ee
By virtue of property $(c)$ given in Section \ref{sec:prelims}, the last term is readily seen to be zero. The first term is not and expands as 
\be
\psset{unit=0.55}
%
 \ee
For the second term,
\be
\psset{unit=0.54}
%
\ee
Finally, the last term in \eqref{eq:willgetugly} is the longest to compute,
\be
\psset{unit=0.55}
%
\ee
Both diagrams on the right can be simplified,
\be - \beta \
\psset{unit=0.55}
%
 
\label{eq:lastP}
\ee
Two connectivities contribute to the remaining $P_{\ell-1}$ projector in the second tangle of \eqref{eq:lastP},
\be 
\psset{unit=0.45}
%
\ee
Adding up all these contributions gives \eqref{eq:uglyidentity}, as already announced.
\hfill $\square$
\begin{Proposition}
On the strip, the fusion hierarchy for $m\in \mathbb N$ closes as
\be       
 \Db^{m,\ell+1}_0-\big[q^m(u_0)q^m(-u_{-2})\big]^N\Db^{m,\ell-1}_1= 2\,(-1)^{\ell+1-p} \Big(\prod_{j=0}^{\ell}s_j(u) \Big)^{2Nm}\Ib^m,
\label{eq:closureDm-app}
\ee
where $q^m(u)$ is defined in \eqref{eq:qkm}.
\label{prop:closureTm}
\end{Proposition}
\noindent{\scshape Outline of the proof:} 
The generalisation from $m=1$ to $m>1$ is almost straightforward, 
so we will only give an outline of the steps of the proof without repeating the diagrammatic arguments.  Because, as discussed in 
Section~\ref{Sec:CabledLinkStates}, the tangles $\Db^{m,n}_k$ are in the subalgebra $FTL_{N,m}(\beta)$, it suffices to prove 
that \eqref{eq:closureDm-app} is satisfied with $\Db$ replaced by $\Dbb$, see \eqref{eq:Dbar}. Now, the two transfer tangles 
$\Dbb^{m,\ell}_0$ and $\Dbb^{1,\ell}_0$, seen as elements of $TL_{Nm}(\beta)$, are quite similar. The only distinction is that, 
in $\Dbb^{m,\ell}_0$, columns of face operators have their spectral parameters shifted by an integer multiple of $\lambda$, and thus 
appear as
\be
\psset{unit=0.65cm}
\begin{pspicture}[shift=-2.4](-0.3,0)(1.3,8.0)
\facegrid{(0,0)}{(1,8)}
\multiput(0,0)(0,4){2}{
\multiput(-0.3,0)(1.3,0){2}{\pspolygon[fillstyle=solid,fillcolor=pink](0,0.1)(0,2.9)(0.3,2.9)(0.3,0.1)(0,0.1)\rput(0.15,1.5){\scriptsize $_\ell$}}}
\psarc[linewidth=0.025]{-}(0,0){0.16}{0}{90}
\psarc[linewidth=0.025]{-}(0,1){0.16}{0}{90}
\psarc[linewidth=0.025]{-}(0,3){0.16}{0}{90}
\psarc[linewidth=0.025]{-}(1,4){0.16}{90}{180}
\psarc[linewidth=0.025]{-}(1,5){0.16}{90}{180}
\psarc[linewidth=0.025]{-}(1,7){0.16}{90}{180}
\rput(0.5,0.5){\scriptsize$u_{-i}$}
\rput(0.5,1.5){\scriptsize$u_{1\!-\!i}$}
\rput(0.5,2.67){$\vdots$}
\rput(0.5,3.5){\scriptsize$u_{\ell\!-\!i}$}
\rput(0.5,4.5){\scriptsize$u_{i}$}
\rput(0.5,5.5){\scriptsize$u_{1\!+\!i}$}
\rput(0.5,6.67){$\vdots$}
\rput(0.5,7.5){\scriptsize$u_{\ell\!+\!i}$}
\end{pspicture} 
\ee
for some $i \in \{0, \dots, m-1\}$. The proof of Proposition~\ref{prop:closureT1} can be adjusted and now involves modified versions of 
$\Ab_{r}$, $\Db_{k,r}^{m,\ell-1}$, $\Cb_{0,r}$ and $\Cb_{1,r}$ with varying spectral parameters. 
In the end, by summing over configurations 
of the leftmost column of face operators in $\Ab_{r}$, one finds a recursion relation relating these objects from 
which \eqref{eq:closureDm-app} stems after simplifications.

%%%%%%%%%%%%%%%%%%
\subsection{On the cylinder} 
\label{app:ClosureCylinder}
%%%%%%%%%%%%%%%%%%

In Proposition~\ref{prop:closureT1} below, we prove Proposition~\ref{prop:ClosureTm} for $m=1$, recalling that 
$\lambda$ is assumed fractional, $\lambda = \lambda_{p,p'}$, and that $p'$ is parametrised as
$p' = \ell + 1$.

\begin{Proposition}
On the cylinder, the fusion hierarchy for $m=1$ closes as 
\be
 \Tb_0^{m,\ell+1} =h^1_0h^1_{\ell-1}\, \Tb_1^{m,\ell-1}+2\,\ir^{N(\ell+1-p)}\Big(\prod_{j=0}^{\ell} h^1_{j}\Big)\Jb^1
\label{prop:ClosureTmApp}
\ee
where $\Jb^1$\! is a $u$-independent tangle.
\label{prop:closureT1}
\end{Proposition}
{\scshape Proof:}
The fusion closure we set out to prove is
\be
 \Tb_0^{1,\ell}\Tb_\ell^{1,1} = h^1_\ell h^1_{\ell-2} \Tb_0^{1,\ell-1} + h^1_0 h^1_{\ell-1} \Tb_1^{1,\ell-1} 
 +2\,\ir^{N(\ell+1-p)}\Big(\prod_{j=0}^{\ell} h^1_{j}\Big)\Jb^1, \qquad h^1_k = \big[\!-\!\ir \,s_k(u)\big]^N
\label{eq:closureT-app}
\ee
which, from \eqref{eq:TmHier}, is equivalent to \eqref{prop:ClosureTmApp}.
For every $r \in \{ 0, \dots, N\}$, we introduce the three new objects 
$\Ub_r$, $\Vb_r$ and $\Wb_r$ defined as
\be 
\psset{unit=0.85cm}
\Ub_r:= \
\begin{pspicture}[shift=-3.35](-0.3,-0.9)(9.1,5.0)
\rput(1.8,-0.6){$\underbrace{\quad \hspace{2.3cm}\quad}_{r}$}
\rput(6.35,-0.6){$\underbrace{\quad \hspace{3.4cm}\quad}_{N-r}$}
\psline[linewidth=1.5pt,linecolor=blue]{-}(0,4.5)(-0.3,4.5)
\multiput(-0.3,0)(1.3,0){8}{\pspolygon[fillstyle=solid,fillcolor=pink](0,0.1)(0,3.9)(0.3,3.9)(0.3,0.1)(0,0.1)
\rput(0.15,2.0){\scriptsize $_\ell$}}
\multiput(-0,0)(1.3,0){3}
{
\psline[linewidth=1.5pt,linecolor=blue]{-}(1,4.5)(1.3,4.5)
\pspolygon[fillstyle=solid,fillcolor=lightlightblue](0,0)(1,0)(1,5)(0,5)(0,0)
\psline[linewidth=1.25pt]{-}(0.1,0.1)(0.9,0.1)(0.9,4.9)(0.1,4.9)(0.1,0.1)
}
\multiput(3.9,0)(1.3,0){4}
{\facegrid{(0,0)}{(1,5)}
\psline[linewidth=1.5pt,linecolor=blue]{-}(1,4.5)(1.3,4.5)
\rput(0.5,.5){\small $u_0$}
\rput(0.5,1.5){\small $u_1$}
\rput(0.5,2.6){\small $\vdots$}
\rput(0.5,3.5){\small $u_{\ell-1}$}
\rput(0.5,4.5){\small $u_\ell$}
\psarc[linewidth=0.025]{-}(0,0){0.16}{0}{90}
\psarc[linewidth=0.025]{-}(0,1){0.16}{0}{90}
\psarc[linewidth=0.025]{-}(0,3){0.16}{0}{90}
\psarc[linewidth=0.025]{-}(0,4){0.16}{0}{90}
}
\end{pspicture}
\label{eq:Mb}
\ee 
\be 
\psset{unit=0.85cm}
\Vb_r:= \
\begin{pspicture}[shift=-2.3](-0.6,-0.9)(11.2,3.0)
\rput(2.1,-0.6){$\underbrace{\quad \hspace{2.8cm}\quad}_{r}$}
\rput(7.7,-0.6){$\underbrace{\quad \hspace{4.2cm}\quad}_{N-r}$}
\multiput(-0.6,0)(1.6,0){8}
{
\pspolygon[fillstyle=solid,fillcolor=pink](0,0.1)(0,2.9)(0.6,2.9)(0.6,0.1)(0,0.1)
\rput(0.3,1.5){\scriptsize $_{\ell\!-\!1}$}
}
\multiput(-0,0)(1.6,0){3}
{
\facegrid{(0,0)}{(1,3)}
\multiput(0,0)(0,1){3}{\rput(0,0){\loopb}}
}
\multiput(4.8,0)(1.6,0){4}
{\facegrid{(0,0)}{(1,3)}
\rput(0.5,.5){\small $u_0$}
\rput(0.5,1.6){\small $\vdots$}
\rput(0.5,2.5){\small $u_{\ell-2}$}
\psarc[linewidth=0.025]{-}(0,0){0.16}{0}{90}
\psarc[linewidth=0.025]{-}(0,2){0.16}{0}{90}
}
\end{pspicture}
\label{eq:Pb}
\ee 
\be 
\psset{unit=0.85cm}
\Wb_r:= \
\begin{pspicture}[shift=-2.3](-0.6,-0.9)(11.2,3.0)
\rput(2.1,-0.6){$\underbrace{\quad \hspace{2.8cm}\quad}_{r}$}
\rput(7.7,-0.6){$\underbrace{\quad \hspace{4.2cm}\quad}_{N-r}$}
\multiput(-0.6,0)(1.6,0){8}
{
\pspolygon[fillstyle=solid,fillcolor=pink](0,0.1)(0,2.9)(0.6,2.9)(0.6,0.1)(0,0.1)
\rput(0.3,1.5){\scriptsize $_{\ell\!-\!1}$}
}
\multiput(-0,0)(1.6,0){3}
{
\facegrid{(0,0)}{(1,3)}
\multiput(0,0)(0,1){3}{\rput(0,0){\loopa}}
}
\multiput(4.8,0)(1.6,0){4}
{\facegrid{(0,0)}{(1,3)}
\rput(0.5,.5){\small $u_1$}
\rput(0.5,1.6){\small $\vdots$}
\rput(0.5,2.5){\small $u_{\ell\!-\!1}$}
\psarc[linewidth=0.025]{-}(0,0){0.16}{0}{90}
\psarc[linewidth=0.025]{-}(0,2){0.16}{0}{90}
}
\end{pspicture}
\label{eq:Qb}
\ee 
Below, we prove that they satisfy the recursion relation
\begin{align}
 \Ub_r = \Big(\prod_{j=0}^\ell& s_j(u) \Big)\Ub_{r+1} + (-\beta)^r\big[s_\ell(u)s_{2-\ell}(-u)\big]^{N-r}
  \Big(\Vb_r - \Vb_{r+1}\times \beta s_{\ell-1}(u) \prod_{j=0}^{\ell-3}s_j(u)\Big) 
\nonumber \\
 &+\big((-1)^{p+1} \beta\big)^r\big[s_0(u)s_{1-\ell}(-u)\big]^{N-r}\Big(\Wb_r - \Wb_{r+1}\times (-1)^p\beta s_{\ell}(u) 
  \prod_{j=1}^{\ell-2}s_j(u)\Big).
\label{eq:recMPQ}
\end{align}
This allows us to write an equality relating 
\be 
 \Ub_0 = \Tb_0^{1,\ell}\Tb_\ell^{1,1}, \quad \Vb_0 = \Tb_0^{1,\ell-1} \quad \textrm{and} \quad \Wb_0 = \Tb_1^{1,\ell-1}
\ee 
with $\Ub_N$, $\Vb_N$ and $\Wb_N$, namely
\begin{align}
\Ub_0 - \big(s_\ell(u)s_{2-\ell}(-u)\big)^N \Vb_0& - \big(s_0(u)s_{1-\ell}(-u)\big)^N \Wb_0 = \nonumber \\& \Big( \prod_{j=0}^\ell s^N_j(u)\Big) \times   \Big( \Ub_N - (-\beta)^N (\Vb_N + (-1)^{pN} \Wb_N)\Big).
\end{align}
Defining 
\be 
 \Jb^1:= \frac{\ir^{Np}}{2}\Big(\Ub_N - (-\beta)^N (\Vb_N + (-1)^{pN} \Wb_N)\Big),
\label{Jdef}
\ee 
and noting that $\Ub_N$, $\Vb_N$ and $\Wb_N$ are all independent of $u$, we 
obtain equation \eqref{eq:closureT-app}.

%%%%%%%%%%%%%%%%%%%%%%%%%%%%
\paragraph{Proof of equation (\ref{eq:recMPQ})}
%%%%%%%%%%%%%%%%%%%%%%%%%%%% 

From the definition \eqref{eq:Mb} for fractional $\lambda=\lambda_{p,\ell+1}$, 
we expand the leftmost column of elementary face operators using \eqref{eq:firstcolumn} and find
\be
 \Ub_r = s_\ell(u) \Big(\prod_{i=0}^{\ell-1}s_{1-i}(-u)  \Big)\Xb_1 + s_{1-\ell}(-u) \Big( \prod_{i=0}^{\ell-1}s_{i}(u)   \Big)\Xb_2 
 +\Big(\prod_{i=0}^\ell s_i(u) \Big)\Ub_{r+1}.
\label{eq:Mi}
\ee 
To evaluate $\Xb_1$, we first use \eqref{eq:otherpushthroughs} $r$ times, then remove all $P_\ell$ projectors except the leftmost, 
and finally use the push-through property $N-r-1$ times:
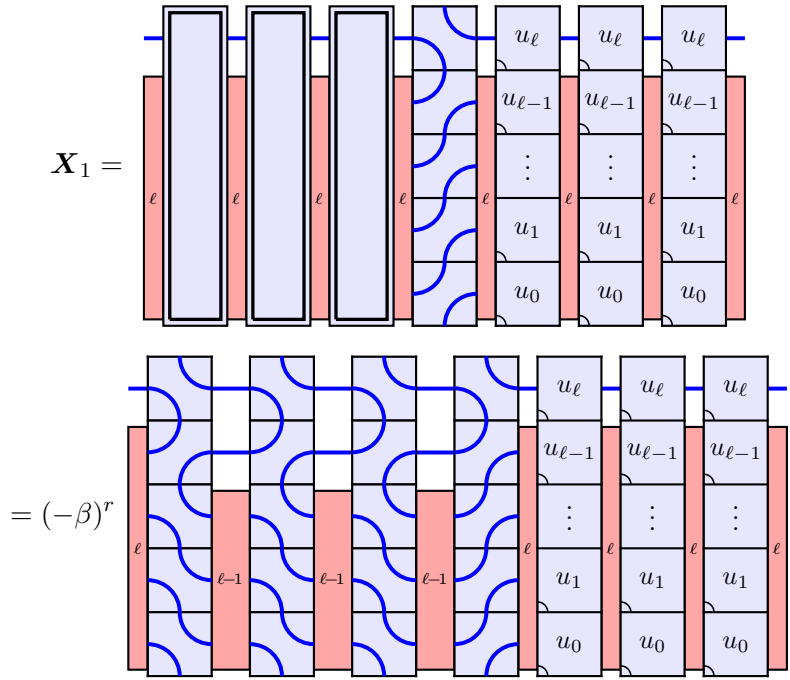
\be 
\psset{unit=0.85cm}
\Xb_1 = 
\begin{pspicture}[shift=-2.4](-0.5,0)(9.4,5)
\multiput(-0.3,0)(1.3,0){8}{\pspolygon[fillstyle=solid,fillcolor=pink](0,0.1)(0,3.9)(0.3,3.9)(0.3,0.1)(0,0.1)
\psline[linewidth=1.5pt,linecolor=blue]{-}(0,4.5)(0.3,4.5)
\rput(0.15,2.0){\scriptsize $_\ell$}}
\multiput(-0,0)(1.3,0){3}
{
\pspolygon[fillstyle=solid,fillcolor=lightlightblue](0,0)(1,0)(1,5)(0,5)(0,0)
\psline[linewidth=1.25pt]{-}(0.1,0.1)(0.9,0.1)(0.9,4.9)(0.1,4.9)(0.1,0.1)
}
\rput(3.9,0){\facegrid{(0,0)}{(1,5)}}
\rput(3.9,0){\loopa}
\rput(3.9,1){\loopa}
\rput(3.9,2){\loopa}
\rput(3.9,3){\loopa}
\rput(3.9,4){\loopb}
\multiput(5.2,0)(1.3,0){3}
{\facegrid{(0,0)}{(1,5)}
\rput(0.5,.5){\small $u_0$}
\rput(0.5,1.5){\small $u_1$}
\rput(0.5,2.6){\small $\vdots$}
\rput(0.5,3.5){\small $u_{\ell-1}$}
\rput(0.5,4.5){\small $u_\ell$}
\psarc[linewidth=0.025]{-}(0,0){0.16}{0}{90}
\psarc[linewidth=0.025]{-}(0,1){0.16}{0}{90}
\psarc[linewidth=0.025]{-}(0,3){0.16}{0}{90}
\psarc[linewidth=0.025]{-}(0,4){0.16}{0}{90}
}
\end{pspicture}
\nonumber
\ee
\be 
\psset{unit=0.85cm}
= (-\beta)^r
\begin{pspicture}[shift=-2.4](-0.5,-0.0)(10.4,5.0)
\multiput(1,0)(1.6,0){3}{\pspolygon[fillstyle=solid,fillcolor=pink](0,0.1)(0,2.9)(0.6,2.9)(0.6,0.1)(0,0.1)
\psline[linewidth=1.5pt,linecolor=blue]{-}(0,4.5)(0.6,4.5)
\psline[linewidth=1.5pt,linecolor=blue]{-}(0,3.5)(0.6,3.5)
\rput(0.3,1.5){\scriptsize $_{\ell\!-\!1}$}}
\multiput(5.8,0)(1.3,0){4}{\pspolygon[fillstyle=solid,fillcolor=pink](0,0.1)(0,3.9)(0.3,3.9)(0.3,0.1)(0,0.1)
\psline[linewidth=1.5pt,linecolor=blue]{-}(0,4.5)(0.3,4.5)
\rput(0.15,2.0){\scriptsize $_\ell$}}
\multiput(-0.3,0)(1.3,0){1}{\pspolygon[fillstyle=solid,fillcolor=pink](0,0.1)(0,3.9)(0.3,3.9)(0.3,0.1)(0,0.1)
\psline[linewidth=1.5pt,linecolor=blue]{-}(0,4.5)(0.3,4.5)
\rput(0.15,2.0){\scriptsize $_\ell$}}
\multiput(0,0)(1.6,0){3}
{
\rput(0,0){\facegrid{(0,0)}{(1,5)}}
\rput(0,0){\loopb}
\rput(0,1){\loopb}
\rput(0,2){\loopb}
\rput(0,3){\loopa}
\rput(0,4){\loopb}
}
\rput(4.8,0){\facegrid{(0,0)}{(1,5)}}
\rput(4.8,0){\loopa}
\rput(4.8,1){\loopa}
\rput(4.8,2){\loopa}
\rput(4.8,3){\loopa}
\rput(4.8,4){\loopb}
\multiput(6.1,0)(1.3,0){3}
{\facegrid{(0,0)}{(1,5)}
\rput(0.5,.5){\small $u_0$}
\rput(0.5,1.5){\small $u_1$}
\rput(0.5,2.6){\small $\vdots$}
\rput(0.5,3.5){\small $u_{\ell-1}$}
\rput(0.5,4.5){\small $u_\ell$}
\psarc[linewidth=0.025]{-}(0,0){0.16}{0}{90}
\psarc[linewidth=0.025]{-}(0,1){0.16}{0}{90}
\psarc[linewidth=0.025]{-}(0,3){0.16}{0}{90}
\psarc[linewidth=0.025]{-}(0,4){0.16}{0}{90}
}
\end{pspicture}
\nonumber
\ee
\be   
\psset{unit=0.85cm}
= (-\beta)^r \big[s_\ell(u)s_{2-\ell}(-u)\big]^{N-r-1}
\begin{pspicture}[shift=-2.4](-0.4,-0.0)(9.4,5.0)
\multiput(1,0)(1.6,0){3}{\pspolygon[fillstyle=solid,fillcolor=pink](0,0.1)(0,2.9)(0.6,2.9)(0.6,0.1)(0,0.1)
\psline[linewidth=1.5pt,linecolor=blue]{-}(0,4.5)(0.6,4.5)
\psline[linewidth=1.5pt,linecolor=blue]{-}(0,3.5)(0.6,3.5)
\rput(0.3,1.5){\scriptsize $_{\ell\!-\!1}$}}
\multiput(5.8,0)(1.3,0){1}{\pspolygon[fillstyle=solid,fillcolor=pink](0,0.1)(0,3.9)(0.3,3.9)(0.3,0.1)(0,0.1)
\psline[linewidth=1.5pt,linecolor=blue]{-}(0,4.5)(0.3,4.5)
\rput(0.15,2.0){\scriptsize $_\ell$}}
\multiput(0,0)(1.6,0){3}
{
\rput(0,0){\facegrid{(0,0)}{(1,5)}}
\rput(0,0){\loopb}
\rput(0,1){\loopb}
\rput(0,2){\loopb}
\rput(0,3){\loopa}
\rput(0,4){\loopb}
}
\rput(4.8,0){\facegrid{(0,0)}{(1,5)}}
\rput(4.8,0){\loopa}
\rput(4.8,1){\loopa}
\rput(4.8,2){\loopa}
\rput(4.8,3){\loopa}
\rput(4.8,4){\loopb}
\multiput(6.1,0)(1.0,0){3}
{\facegrid{(0,0)}{(1,5)}
\rput(0.5,.5){\small $u_0$}
\rput(0.5,1.6){\small $\vdots$}
\rput(0.5,2.5){\small $u_{\ell-2}$}
\rput(0,3){\loopa}
\rput(0,4){\loopb}
\psarc[linewidth=0.025]{-}(0,0){0.16}{0}{90}
\psarc[linewidth=0.025]{-}(0,2){0.16}{0}{90}
}
\end{pspicture}
\ee
The final step is to rewrite the remaining $P_\ell$ projector using \eqref{eq:rationalrec}. 
This gives
\be   
 \Xb_1 = (-\beta)^r\big[s_\ell(u)s_{2-\ell}(-u)\big]^{N-r-1} \big( \Zb_1 - \beta \,\Zb_2 \big)
\ee 
where
\be 
\psset{unit=0.85cm}
\Zb_1 = \ \
\begin{pspicture}[shift=-2.4](0,-0)(9.4,5)
\multiput(1,0)(1.6,0){4}{\pspolygon[fillstyle=solid,fillcolor=pink](0,0.1)(0,2.9)(0.6,2.9)(0.6,0.1)(0,0.1)
\psline[linewidth=1.5pt,linecolor=blue]{-}(0,4.5)(0.6,4.5)
\psline[linewidth=1.5pt,linecolor=blue]{-}(0,3.5)(0.6,3.5)
\rput(0.3,1.5){\scriptsize $_{\ell\!-\!1}$}}
\multiput(5.8,0)(1.3,0){0}{\pspolygon[fillstyle=solid,fillcolor=pink](0,0.1)(0,3.9)(0.3,3.9)(0.3,0.1)(0,0.1)
\psline[linewidth=1.5pt,linecolor=blue]{-}(0,4.5)(0.3,4.5)
\rput(0.15,2.0){\scriptsize $_\ell$}}
\multiput(0,0)(1.6,0){3}
{
\rput(0,0){\facegrid{(0,0)}{(1,5)}}
\rput(0,0){\loopb}
\rput(0,1){\loopb}
\rput(0,2){\loopb}
\rput(0,3){\loopa}
\rput(0,4){\loopb}
}
\rput(4.8,0){\facegrid{(0,0)}{(1,5)}}
\rput(4.8,0){\loopa}
\rput(4.8,1){\loopa}
\rput(4.8,2){\loopa}
\rput(4.8,3){\loopa}
\rput(4.8,4){\loopb}
\multiput(6.4,0)(1.0,0){3}
{\facegrid{(0,0)}{(1,5)}
\rput(0.5,.5){\small $u_0$}
\rput(0.5,1.6){\small $\vdots$}
\rput(0.5,2.5){\small $u_{\ell-2}$}
\rput(0,3){\loopa}
\rput(0,4){\loopb}
\psarc[linewidth=0.025]{-}(0,0){0.16}{0}{90}
\psarc[linewidth=0.025]{-}(0,2){0.16}{0}{90}
}
\end{pspicture}
\nonumber
\ee
\be
\psset{unit=0.85cm}
= \ \ 
\begin{pspicture}[shift=-1.5](-0.6,-0.0)(10.2,3.2)
\multiput(-0.6,0)(1.6,0){8}{\pspolygon[fillstyle=solid,fillcolor=pink](0,0.1)(0,2.9)(0.6,2.9)(0.6,0.1)(0,0.1)
\rput(0.3,1.5){\scriptsize $_{\ell\!-\!1}$}}
\multiput(0,0)(1.6,0){3}
{
\rput(0,0){\facegrid{(0,0)}{(1,3)}}
\rput(0,0){\loopb}
\rput(0,1){\loopb}
\rput(0,2){\loopb}
}
\rput(4.8,0){\facegrid{(0,0)}{(1,3)}}
\rput(4.8,0){\loopa}
\rput(4.8,1){\loopa}
\rput(4.8,2){\loopa}
\multiput(6.4,0)(1.6,0){3}
{\facegrid{(0,0)}{(1,3)}
\rput(0.5,.5){\small $u_0$}
\rput(0.5,1.6){\small $\vdots$}
\rput(0.5,2.5){\small $u_{\ell-2}$}
\psarc[linewidth=0.025]{-}(0,0){0.16}{0}{90}
\psarc[linewidth=0.025]{-}(0,2){0.16}{0}{90}
}
\end{pspicture}
\qquad \ \;=\frac{ \Vb_r - \Vb_{r+1} \prod_{i=0}^{\ell-2}s_i(u)}{\prod_{k=0}^{\ell-2} s_{1-k}(-u)}
\ee
and
\be 
\psset{unit=0.85cm}
\Zb_2 = \ \
\begin{pspicture}[shift=-2.4](0,0)(11.3,5)
\multiput(1,0)(1.6,0){4}{\pspolygon[fillstyle=solid,fillcolor=pink](0,0.1)(0,2.9)(0.6,2.9)(0.6,0.1)(0,0.1)
\psline[linewidth=1.5pt,linecolor=blue]{-}(0,4.5)(0.6,4.5)
\psline[linewidth=1.5pt,linecolor=blue]{-}(0,3.5)(0.6,3.5)
\rput(0.3,1.5){\scriptsize $_{\ell\!-\!1}$}}
\rput(7.7,0){\pspolygon[fillstyle=solid,fillcolor=pink](0,0.1)(0,2.9)(0.6,2.9)(0.6,0.1)(0,0.1)
\rput(0.3,1.5){\scriptsize $_{\ell\!-\!1}$}}
\psarc[linewidth=1.5pt,linecolor=blue]{-}(6.4,3){0.5}{-90}{90}
\psarc[linewidth=1.5pt,linecolor=blue]{-}(7.7,3){-0.5}{-90}{90}
\psline[linewidth=1.5pt,linecolor=blue]{-}(6.4,0.5)(7.7,0.5)
\psline[linewidth=1.5pt,linecolor=blue]{-}(6.4,1.5)(7.7,1.5)
\psline[linewidth=1.5pt,linecolor=blue]{-}(7.7,3.5)(8.3,3.5)
\psline[linewidth=1.5pt,linecolor=blue]{-}(6.4,4.5)(8.3,4.5) 
\multiput(0,0)(1.6,0){3}
{
\rput(0,0){\facegrid{(0,0)}{(1,5)}}
\rput(0,0){\loopb}
\rput(0,1){\loopb}
\rput(0,2){\loopb}
\rput(0,3){\loopa}
\rput(0,4){\loopb}
}
\rput(4.8,0){\facegrid{(0,0)}{(1,5)}}
\rput(4.8,0){\loopa}
\rput(4.8,1){\loopa}
\rput(4.8,2){\loopa}
\rput(4.8,3){\loopa}
\rput(4.8,4){\loopb}
\multiput(8.3,0)(1.0,0){3}
{\facegrid{(0,0)}{(1,5)}
\rput(0.5,.5){\small $u_0$}
\rput(0.5,1.6){\small $\vdots$}
\rput(0.5,2.5){\small $u_{\ell-2}$}
\rput(0,3){\loopa}
\rput(0,4){\loopb}
\psarc[linewidth=0.025]{-}(0,0){0.16}{0}{90}
\psarc[linewidth=0.025]{-}(0,2){0.16}{0}{90}
}
\psline[linewidth=1.5pt](5.8,0.1)(5.8,2.9)(6.4,2.9)(6.4,0.1)(5.8,0.1)
\end{pspicture}
\label{Z2}
\ee
To proceed further, we note that the $P_{\ell-1}$ projector drawn with thicker boundaries 
in (\ref{Z2}) has only one contributing connectivity, \eqref{eq:survivor2}, and this allows us to write
\be 
\psset{unit=0.85cm}
\Zb_2= (-1)^p \beta^{-1} 
\begin{pspicture}[shift=-1.4](-0.8,0)(11.3,3)
\multiput(-0.6,0)(1.6,0){8}{\pspolygon[fillstyle=solid,fillcolor=pink](0,0.1)(0,2.9)(0.6,2.9)(0.6,0.1)(0,0.1)
\rput(0.3,1.5){\scriptsize $_{\ell\!-\!1}$}}
\multiput(0,0)(1.6,0){3}
{
\rput(0,0){\facegrid{(0,0)}{(1,3)}}
\rput(0,0){\loopb}
\rput(0,1){\loopb}
\rput(0,2){\loopb}
}
\rput(4.8,0){\facegrid{(0,0)}{(1,3)}}
\rput(4.8,0){\loopb}
\rput(4.8,1){\loopb}
\rput(4.8,2){\loopb}
\multiput(6.4,0)(1.6,0){3}
{\facegrid{(0,0)}{(1,3)}
\rput(0.5,.5){\small $u_0$}
\rput(0.5,1.6){\small $\vdots$}
\rput(0.5,2.5){\small $u_{\ell-2}$}
\psarc[linewidth=0.025]{-}(0,0){0.16}{0}{90}
\psarc[linewidth=0.025]{-}(0,2){0.16}{0}{90}
}
\end{pspicture}
= (-1)^p \beta^{-1} \Vb_{i+1}.
\ee
Together, these rewritings of $\Zb_1$ and $\Zb_2$ yield
\be   
 s_\ell(u) \Big(\prod_{i=0}^{\ell-1}s_{1-i}(-u)  \Big)\Xb_1 = (-\beta)^r \big[s_\ell(u) s_{2-\ell}(-u)\big]^{N-r} \Big(\Vb_r - \Vb_{r+1}
 \,\beta s_{\ell-1}(u) \prod_{j=0}^{\ell-3}s_j(u) \Big).\label{eq:B1}
\ee
For $\Xb_2$, the same calculation is done by first reversing the order of 
the spectral parameters using \eqref{eq:Pprop}, such that
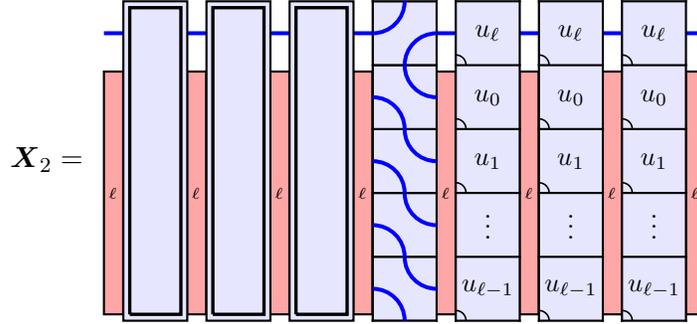
\be 
\psset{unit=0.85cm}
\Xb_2 = 
\begin{pspicture}[shift=-2.4](-0.5,0)(9.4,5)
\multiput(-0.3,0)(1.3,0){8}{\pspolygon[fillstyle=solid,fillcolor=pink](0,0.1)(0,3.9)(0.3,3.9)(0.3,0.1)(0,0.1)
\psline[linewidth=1.5pt,linecolor=blue]{-}(0,4.5)(0.3,4.5)
\rput(0.15,2.0){\scriptsize $_\ell$}}
\multiput(-0,0)(1.3,0){3}
{
\pspolygon[fillstyle=solid,fillcolor=lightlightblue](0,0)(1,0)(1,5)(0,5)(0,0)
\psline[linewidth=1.25pt]{-}(0.1,0.1)(0.9,0.1)(0.9,4.9)(0.1,4.9)(0.1,0.1)
}
\rput(3.9,0){\facegrid{(0,0)}{(1,5)}}
\rput(3.9,0){\loopb}
\rput(3.9,1){\loopb}
\rput(3.9,2){\loopb}
\rput(3.9,3){\loopb}
\rput(3.9,4){\loopa}
\multiput(5.2,0)(1.3,0){3}
{\facegrid{(0,0)}{(1,5)}
\rput(0.5,3.5){\small $u_0$}
\rput(0.5,2.5){\small $u_1$}
\rput(0.5,1.6){\small $\vdots$}
\rput(0.5,0.5){\small $u_{\ell-1}$}
\rput(0.5,4.5){\small $u_\ell$}
\psarc[linewidth=0.025]{-}(0,0){0.16}{0}{90}
\psarc[linewidth=0.025]{-}(0,2){0.16}{0}{90}
\psarc[linewidth=0.025]{-}(0,3){0.16}{0}{90}
\psarc[linewidth=0.025]{-}(0,4){0.16}{0}{90}
}
\end{pspicture}
\nonumber\ee
Then, with arguments similar to the ones above, we find
\be   
 s_{1-\ell}(-u) \Big( \prod_{i=0}^{\ell-1}s_{i}(u)   \Big)\Xb_2 = \big((-1)^{p+1}\beta\big)^r \big[s_0(u) s_{1-\ell}(-u)\big]^{N-r} 
  \Big(\Wb_r - \Wb_{r+1} 
 (-1)^p\beta s_{\ell}(u) \prod_{j=1}^{\ell-2}s_j(u) \Big).\label{eq:B2}
\ee
Combining equations \eqref{eq:Mi}, \eqref{eq:B1} and \eqref{eq:B2} completes the proof of equation \eqref{eq:recMPQ}.
\hfill $\square$
\begin{Proposition}
On the cylinder, the fusion hierarchy for $m\in \mathbb N$ closes as
\be 
 \Tb_0^{m,\ell}\Tb_\ell^{m,1} - h^m_\ell h^m_{\ell-2} \Tb_0^{m,\ell-1} - h^m_0 h^m_{\ell-1} \Tb_1^{m,\ell-1} 
 = 2\,\ir^{Nm(\ell+1-p)}\Big(\prod_{j=0}^{\ell} h^m_{j}\Big)\Jb^m,
\label{eq:closureTm-app}
\ee
where $\Jb^m$\! is a $u$-independent tangle and $h^m_j$ is defined in \eqref{fkm}. 
\label{prop:closureTmapp}
\end{Proposition}
\noindent{\scshape Outline of the proof:} 
For $m>1$, the generalisation turns out to be straightforward albeit tedious.  Following the discussion in 
Section~\ref{Sec:CabledLinkStates}, it suffices to prove \eqref{eq:closureTm-app} with $\Tb$ replaced by $\Tbb$, see \eqref{eq:Tbar}. 
As tangles, $\Tbb^{m,n}_0, \Tbb^{1,n}_0 \in \mathcal EPTL_{Nm}(\alpha,\beta)$ are very similar, with the difference that rows of face 
operators in $\Tbb^{m,n}_0$ have spectral parameters shifted by an integer multiple of $\lambda$ and appear as
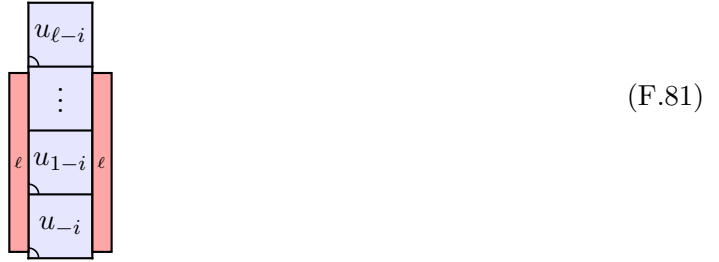
\be
\psset{unit=0.85cm}
\begin{pspicture}[shift=-2.4](-0.3,0)(1.3,4.0)
\facegrid{(0,0)}{(1,4)}
\multiput(-0.3,0)(1.3,0){2}{\pspolygon[fillstyle=solid,fillcolor=pink](0,0.1)(0,2.9)(0.3,2.9)(0.3,0.1)(0,0.1)\rput(0.15,1.5){\scriptsize $_\ell$}}
\psarc[linewidth=0.025]{-}(0,0){0.16}{0}{90}
\psarc[linewidth=0.025]{-}(0,1){0.16}{0}{90}
\psarc[linewidth=0.025]{-}(0,3){0.16}{0}{90}
\rput(0.5,0.5){$u_{-i}$}
\rput(0.5,1.5){$u_{1-i}$}
\rput(0.5,2.6){$\vdots$}
\rput(0.5,3.5){$u_{\ell-i}$}
\end{pspicture} 
\ee
for some $i \in \{0, \dots, m-1\}$. The steps in the proof of Proposition~\ref{prop:closureT1} are then repeated: 
Summing over configurations of a
single column of face operators produces a recursion relation relating adapted versions of $\Ub_{r}$, $\Vb_{r}$ and $\Wb_{r}$, 
eventually leading to \eqref{eq:closureTm-app}.

%%%%%%%%%%%%%%%%%%%%%
%

\end{document}